\documentclass{aa}  

\usepackage{subfigure}

\usepackage{graphicx,amsmath}

\usepackage{txfonts}
 \DeclareMathOperator{\erf}{erf}
  
\begin{document}

   \title{Probing elliptical orbital configuration of the close binary of supermassive black holes with differential interferometry}

   \subtitle{}

\titlerunning{Differential interferometry of close binary of supermassive black holes in an elliptical configuration}

\author{Andjelka B. Kova{\v c}evi{\'c}  \inst{1},
	Yu-Yang Songsheng \inst{2,3},
	Jian-Min Wang \inst{2,3,4}, 
	Luka  {\v C}. Popovi{\'c}  \inst{3,1}  
}

\institute{Department of astronomy, Faculty of mathematics, University of Belgrade 
	Studentski trg 16, Belgrade, 11000, Serbia\\
	\email{andjelka@matf.bg.ac.rs}
	\and
	Key Laboratory for Particle Astrophysics, Institute of High Energy Physics,
	Chinese Academy of Sciences, 19B Yuquan Road, Beijing 100049, China\\
	\email{songshengyuyang@ihep.ac.cn, wangjm@ihep.ac.cn}
	\and
		School of Astronomy and Space Science, University of Chinese Academy of Sciences,
	19A Yuquan Road, Beijing 100049, China
	\and
	National Astronomical Observatories of China, Chinese Academy of Sciences, 20A
	Datun Road, Beijing 100020, China
		\and
	Astronomical observatory Belgrade 
	Volgina 7, P.O.Box 74 11060, Belgrade,  11060, Serbia \\
	\email{lpopovic@aob.rs}
}
\authorrunning{Kova{\v c}evi{\'c} A. B. et al.}


   \date{Received ; accepted}

 
  \abstract
   {Detections of electromagnetic signatures from the close binaries of supermassive  black holes (CB-SMBH) is still a grand observational  challenge. The Very Large Telescope Interferometer (VLTI), and the Extremely Large Telescope (ELT) will  be a robust astrophysics suite offering the opportunity of probing the structure and dynamics of CB-SMBH  at high spectral and  angular resolution. }
{Here, we explore and illustrate the application of differential interferometry on unresolved the CB-SMBH systems in elliptical orbital configurations. We also investigate  peculiarities  of interferometry signals for a single SMBH with clouds in elliptical orbital motion.}
{ Photocenter displacements between each SMBH and regions in their disc-like broad line regions (BLR) appear as small interferometric differential phase variability. To investigate the application of interferometric phases for the detection of CB-SMBH systems, we simulate a series of differential interferometry signatures, based on our model comprising ensembles of clouds surrounding each of supermassive black hole in a CB-SMBH. Setting model to the parameters of a single SMBH with elliptical cloud motion, we also calculated a series of  differential interferometry observables for this case.}
 {We found various deviations from the canonical S-shaped of CB-SMBH phases profile for elliptically configured CB-SMBH systems. The amplitude and specific shape of the interferometry observables depend on orbital configurations of the CB-SMBH system. We get distinctive results when considering antialigned angular momenta of cloud orbits regarding total CB-SMBH angular momentum. We also show that their velocity distributions differ from the aligned cloud orbital motion. Some simulated spectral lines from our model closely resemble observations of Pa$\alpha$ line got from near-infrared AGN surveys. We found differences between differential phases ‘zoo’ of a single SMBH and CB-SMBH systems. The differential phases ‘zoo’  for a single SMBH comprises deformed S shape. We also showed how their differential phase shape, amplitude, and slope evolve with various sets of cloud orbital parameters and observer position.}
   {We calculated an extensive atlas of the interferometric observables, revealing distinctive signatures for the elliptical configuration of close supermassive binary black holes. We also provide interferometry atlas for the case of a single SMBH with clouds in an elliptical motion, differing from those of CB-SMBH. These maps can be useful  for extracting exceptional features of the BLR structure from future high-resolution observations of CB-SMBH systems, but also of a single SMBH with clouds in elliptical orbital setup.}
   \keywords{ Techniques: interferometric--quasars: supermassive black holes --quasars: emission lines, Pa$\alpha$ 
               }

   \maketitle
%

\section{Introduction}

Observations of  active galactic nuclei (AGN) probe the  extreme limits of  physical conditions in the universe set by the event horizon of  a supermassive black hole (SMBH) in their centers. 
Our view of AGNs was  limited until technologically advanced observatories, and monitoring campaigns uncover their electromagnetic spectrum.
Particularly, information about broad line regions (BLRs) surrounding SMBH comes mainly from reverberation mapping (RM) campaigns \citep{1998ApJ...501...82P,2002ApJ...581..197P,10.1086/423269, 2000ApJ...533..631K,10.1086/512094, 2001A&A...376..775S,2004A&A...422..925S,
2008A&A...486...99S,2010A&A...509A.106S,
2010A&A...517A..42S,2012ApJS..202...10S,
2013A&A...559A..10S,2016ApJS..222...25S,
2017MNRAS.466.4759S,2019MNRAS.485.4790S,2008ApJ...689L..21B,2009ApJ...705..199B,
2009ApJ...704L..80D,2011ApJ...743L...4B,2013ApJ...769..128B,2015ApJS..217...26B,
2011A&A...528A.130P,2014A&A...572A..66P,
2012ApJ...755...60G,2014ApJ...793..108W,2014ApJ...782...45D,10.1088/0004-637X/806/1/22,2016ApJ...825..126D,2018ApJ...856....6D,2016ApJ...818...30S,2017ApJ...851...21G, 2019ApJ...870..123E,10.3847/1538-4357/aaf806, 10.1038/s41550-019-0979-5}. 
Most of these data suffer from 'Static Illusion', because what we can observe with this method perhaps changes on timescales larger than centuries  \citep[$10^{4}-10^{7}$yr,][]{10.1007/978-94-010-0320-9}. The problem is most apparent in the periodicity detection in the light curves of AGNs.
AGNs mostly appear stationary with notable exceptions, such as  changing look AGNs \citep[see, e.g.][]{10.3847/1538-4357/aab88b, 2019MNRAS.485.4790S, Ilic20}.

\noindent RM campaigns indicate that the largest BLR  radii can be a few hundreds of light days, this  is equivalent to angular sizes of  $\sim 100 \mu \mathrm{as}$ for nearby AGNs, which is 
 well below the canonical resolution of optical
interferometry instruments  \citep[e.g., $\sim 3 \mathrm{mas}$ for VLTI/GRAVITY,][]{ 10.1051/0004-6361/201730838}.
Still, interferometry could  offer a qualitatively new view of these objects.  It is possible to obtain optical interferometry information from non-resolved sources  measuring the displacement of object photocenter with wavelength. It was used first by  speckle interferometry, then adopted by long baseline interferometry and finally has been one of the design parameters of the VLTI focal instrument AMBER \citep[see][and references therein]{doi.org/10.1117/12.926595}.
AMBER was  the first generation  near-infrared instrument of VLTI, with one of the  goal to study AGNs \citep{10.1051/0004-6361:20066496}.
Several authors \citep{10.1038/nature02531,10.1117/12.551881,10.1051/0004-6361/200913512,10.1051/0004-6361/201219213} have done pioneering optical-interferometric  observations of some Seyfert I AGNs. \cite{doi.org/10.1117/12.926595} reported the first optical interferometry observations of the BLR of a quasar  just a decade ago.
The GRAVITY instrument at the VLTI had made a significant breakthrough  when its differential phase
precision $ <1^{\circ}\sim 10\, \mu \mathrm{as}$  allowed \cite{ 10.1038/s41586-018-0731-9}  to probe and  detect successfully disc-like BLR in 3C 273\footnote{the differential phase precision  differs from  resolving the object.}. 
A recent high signal-to-noise, decadal reverberation mapping campaign by  \cite{doi.org/10.3847/1538-4357/ab1099}  yielded  H$\beta$ time lag  (to the continuum  5100\, \AA)  of $\sim 146.8$ light days in the rest frame, which agrees well with the Pa$\alpha$ region (145 light days) measured by the GRAVITY.

However, the Event Horizon Telescope (EHT), the first planetary very long baseline interferometric array, obtained the first image of the  shadow of a spinning Kerr SMBH in the M87 galaxy \citep{10.3847/2041-8213/ab0ec7 }. 
\cite{2019MNRAS.488L..90S} pointed out that M87 at the center of  the cluster of galaxies has the possibility of its  SMBH  mass growth through mergers of neighboring galaxies \citep{2006ApJS..163....1H}. Their analysis implies the existence of a binary companion such as an intermediate-mass black hole \citep{10.1086/307952}. \cite{2019MNRAS.488L..90S}  argued that the EHT long-term monitoring campaign of M87 at $\sim 1\, \mu \mathrm{as}$ positional accuracy would detect small mass SMBH companion of M87.
  The differential interferometry is among the very promising techniques for detection of close binaries of supermassive  black holes   \citep[CB-SMBH][]{10.3847/1538-4357/ab2e00}.  
   Also, the dispersion, in AGN mass-luminosity relation (based on RM measurements of SMBH masses)  amount to about a factor of three \citep{10.1086/423269}; this systematic uncertainty decreases when using RM and differential interferometry measurements.  \cite{10.3847/1538-4357/aacdfa}  have recently proposed   kinematic signatures based on 2D transfer functions, that can be derived from RM campaigns,  as a promising avenue to address the problem of CB-SMBH detection \citep{   
   		2020ApJS..247....3S}.
   
   The process of synergy between RM and interferometry investigations of AGN has already begun. The long-term RM project `The Monitoring AGNs with H$\beta$ Asymmetry' (MAHA), using the Wyoming Infrared Observatory 2.3 m telescope, explores the geometry and kinematics of the gas responsible for complex H$\beta$ emission-line profiles that also provides the opportunity to search for evidence of CB-SMBH \citep[see][]{10.3847/1538-4357/aaed2c}.
  Also,  \cite{10.3847/1538-4357/ab2e00} have already presented the first atlas of expected differential phases of circular CB-SMBH systems. They found that the current GRAVITY setup can detect differential phase curves  because of the circular orbital motion of  CB-SMBH.
   In this work, we consider the central question of the modeling interferometry observables of eccentric orbital settings of clouds in a single SMBH and  eccentric orbital configurations of CB-SMBH systems.
   As  for excitation of binary orbital elongation, advanced  N-body or hydrodynamical simulations have  showed  major mergers of gas-rich disc galaxies with central SMBHs \citep{10.1126/science.1141858, 10.1111/j.1365-2966.2008.14147.x, 10.1088/0004-637X/719/1/851,  10.1111/j.1365-2966.2011.18927.x}. For binary systems with a circumbinary accretion disc, the binary  can reach a limiting eccentricity of  $\sim$ 0.6 \citep{10.1111/j.1365-2966.2011.18927.x}  through the inspiral. \citet{10.1088/0004-637X/719/1/851} proposed that scattering of bound and unbound stars in the galactic bulge  excites high eccentric  binary systems.  Then circularization can occur when the gravitational wave shrinking time-scale is smaller than the gas or star  induced binary migration. But, once again, a significant amount of eccentricity can be gained when the binary occupies the frequency band of gravitational wave observations. 
   The excitation of binary eccentricity also has physical consequences \citep{10.1111/j.1365-2966.2011.18927.x}. Binaries with larger eccentricity will  coalesce on a shorter time-scale due to faster GW energy loss \citep{10.1103/PhysRev.131.435}.  Accretion flows leaking through the cavity, and fueling  SMBHs can induce  a more prominent periodicity signal \citep{10.1086/310200}, which can increase the binary identification via time-variability.
  In this paper, we extend and refine the search for specific features of differential interferometry   $\boldsymbol{\xi}(\lambda)$ and spectroscopic $\Xi(\lambda)$ functions to the case of CB-SMBH systems with elliptical orbits and the BLR configurations. The results are in the form of an atlas showing the evolution of $\xi(\lambda)$ and $\Xi(\lambda)$ functions with varying parameters of orbital configurations.

The outline of the paper is: in Section \ref{SLD}, we briefly overview  our CB-SMBH phenomenological model  and introduce interferometric observables.
 General inferences about differential phase shape for single SMBH and CB-SMBH, based  on a first order approximation, are given in Appendix \ref{appendix:generall}.
The chosen  results of the simulations,  also resembling real-AGN spectral lines,  are described in two separate Subsections   \ref{RD}  and  \ref{RD1},  related to single SMBH and CB-SMBH, respectively.   Detailed results are included  in the Appendices \ref{appendix:atlas}, 	\ref{appendix:aligned} and \ref{appendix:nonaligned} for  single SMBH, aligned and anti-aligned CB-SMBH, respectively.
In Section \ref{discuss},  we discuss anticipated results from observations, limitations of the present model, and some implications of random clouds' motion on interferometric observables. Finally, Section \ref{conc} concludes the paper by summarizing our results.

\section{Spectroastrometry: Spectral lines and Differential phase}\label{SLD}
Here we present the elliptical CB-SMBH interferometry-oriented model.
We first briefly summarize our dynamical CB-SMBH model and introduce equations of the observable quantities of elliptical CB-SMBH interferometry. 
\subsection{Recapitulation of CB-SMBH model}\label{recapitulation}

  \begin{figure*}
	\begin{minipage}[b]{0.49\textwidth}
		\includegraphics[trim=80 30 80 50,clip,width=\textwidth]{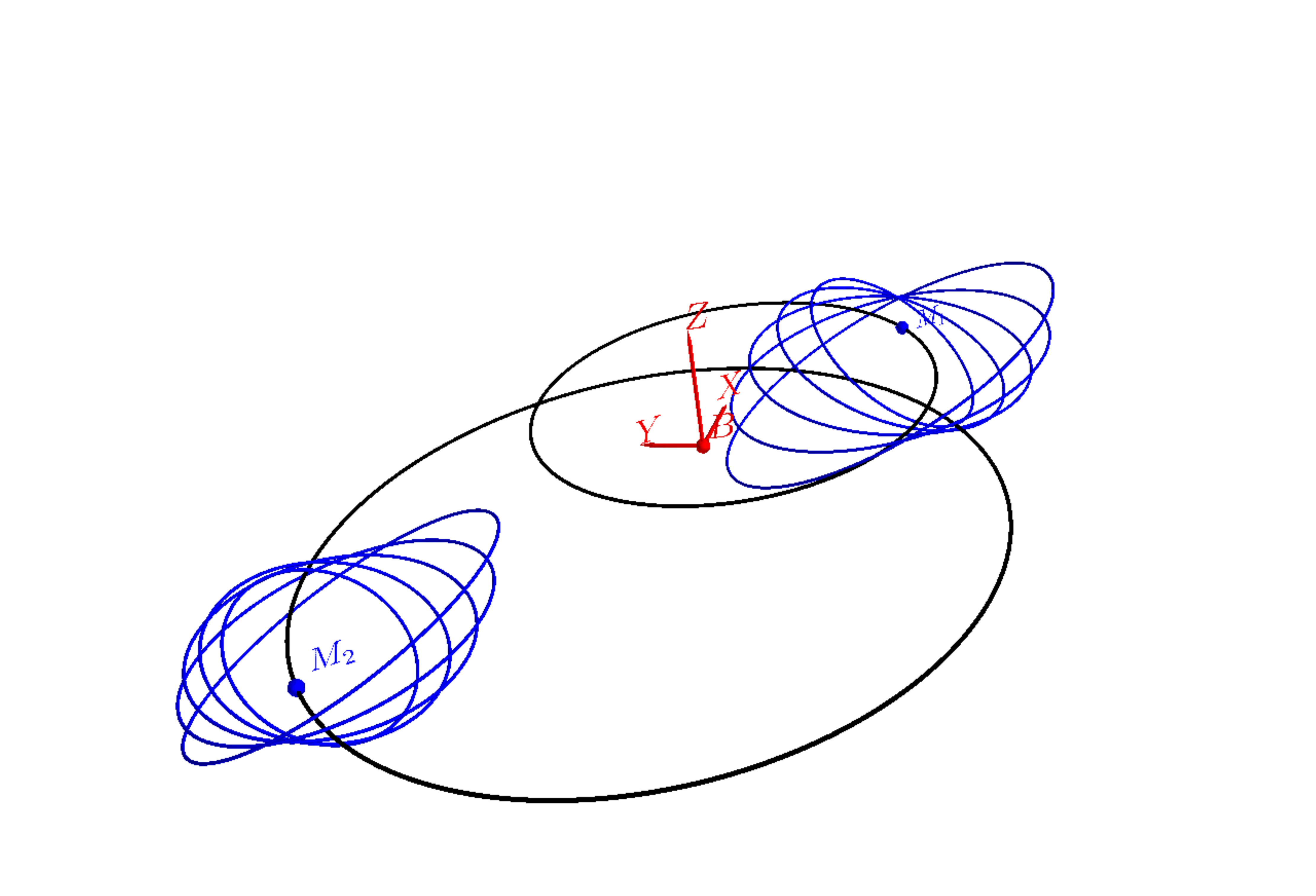}
		\hspace*{125pt} (a)
	\end{minipage}
	\hfill
	\begin{minipage}[b]{0.5\textwidth}
		\includegraphics[trim=75 30 80 50,clip,width=\textwidth]{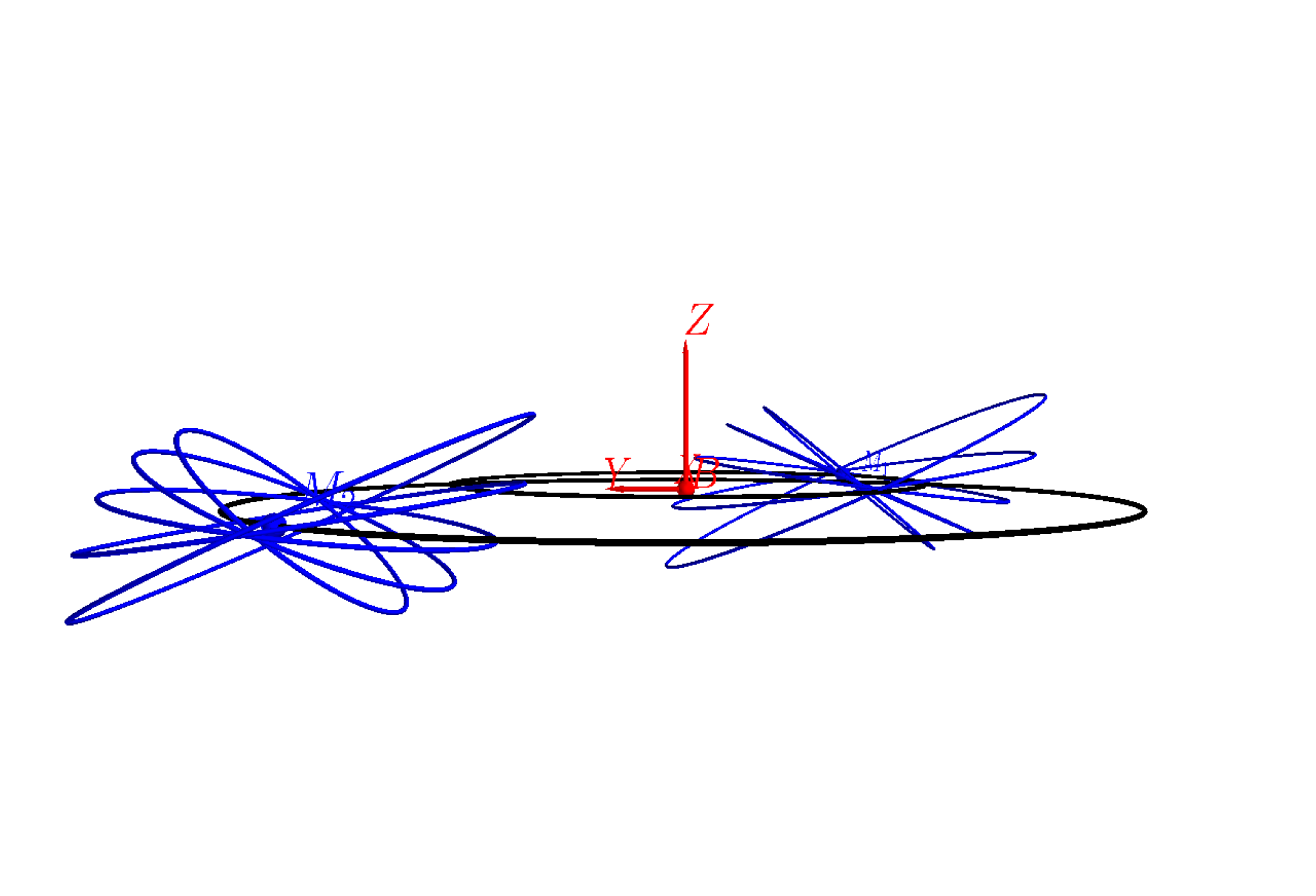} 
		\hspace*{125pt} (b)
	\end{minipage}
	\caption{Realization of a coplanar elliptical CB-SMBH system with clouds in non-coplanar  orbits in both BLRs calculated from the  model. $M_{1} \text{ and }M_{2}$ are the locations of the SMBHs in a binary system. Blue ellipses are every 20th of the 100 orbits of clouds in each  BLR. SMBH parameters are $M_{1}=6\cdot 10^{7} M{\odot},\, M_{2}=4\cdot 10^{7} M{\odot},\, A_{0}=100,\, \Omega_{1}=\Omega_{2}=3^{\circ},\, \omega_{1}=1^{\circ}, \,\omega_{2}=181^{\circ},\, e_{1}=e_{2}=0.5$. Clouds parameters are $\Omega_{c1}=\Omega_{c2}=180^{\circ},\, \omega_{c1}=2^{\circ},\, \omega_{c2}=182^{\circ},\, i=(-40^{\circ},40^{\circ}),\, e=0.5$, but their orbital planes have different inclinations. The reference plane ($BXY$) is the plane of the relative orbit of $M_2$  to $M_{1}$, and the origin of the coordinate system is the barycenter  ($B$) of the CB-SMBH system.   (a) An elevation view of the CB-SMBH system. (b) A side view to the YBZ coordinate plane.}
	\label{fig:cosis}
	\hfill
\end{figure*}

 \begin{table*}
 	\begin{footnotesize}
	\caption{ Model parameters and their range of values used in  simulations.   }             
	\label{table:model}  
	    
	\centering          
	\begin{tabular}{c c c c c}     
		\hline\hline       
		Parameters & Description &Range or monotony & Values in simulations\\ 
		&       &     & &             \\
		\hline 
			($M_{1}, M_{2}$)&SMBHs masses in $10^{7}M{\odot}$&& (6, 4)\\
		q                            &   $\frac{M_{2}}{M_{1}}$&(0,1) &$\frac{2}{3}$\\
		$\mathcal{A}_{0}$& Semimajor axis of SMBHs relative orbit [ld]& &(100, 175)\\
		($a_{1}, a_{2}$)&SMBHs semimajor axes
		&&($\frac{q\mathcal{A}_{0}}{1+q},\frac{\mathcal{A}_{0}}{1+q}$) \\
		($e_{1}= e_{2}$)&SMBHs orbital barycentric eccentricities&(0,1)& (0.05, 0.5)\\
		($i_{1},i_{2}$)&SMBHs orbital barycentric inclinations \tablefootmark{a}  &&($0^{\circ}, \,45^{\circ}$  )\\
		($\Omega_{k}$,$\omega_{k}, k=1,2$)& longitude of ascending node and &  ($0^{\circ}, \,360^{\circ}$)  &  ($0^{\circ}, 360^{\circ}$  )\\   
			& argument of pericenter of SMBHs barycentric orbits&   &  \\      
			(f$_{k}, k=1,2$)& true anomaly of  SMBHs  orbits&  ($0^{\circ},\, 360^{\circ}$)  &  ($0^{\circ},90^{\circ}, 180^{\circ}$  )\\    
		($a^{c}_{1}, a^{c}_{2}$)&semimajor axes of clouds orbits&& ((20, 40), (16, 32))\\
		&  in BLRs around SMBHs [ld]&   &  \\     
			($e^{c}_{1},e^{c}_{2}$)&clouds orbital  eccentricities&(0,1) & (0.05, 0.5),  $\Gamma_{s}  (0.3,1),\, \mathcal{R}(1)$\\
			($i^{c}_{1},i^{c}_{2}$)&clouds orbital  inclinations&&($0^{\circ}, 45^{\circ}$  )$\vee$ ($90^{\circ}, 175^{\circ}$)\tablefootmark{b}\\
		
		($\Omega^{c}_{1}$,$\omega^{c}_{k}, k=1,2$)&longitude of ascending node and &  ($0^{\circ}, 360^{\circ}$  )  &  ($0^{\circ}, 360^{\circ}$  )\\                 
	& argument of pericenter of cloud orbits&   &  \\     

		\hline
		Auxiliary parameters	 & &  & \\  
		\hline       
		$\Im$&varied range of clouds orbital inclinations&   &  $\Im=i^{c}_{min}+i^{c}_{max}$\\
			& within single simulation&   &  \\     
		$i_0$& angular position of the observer &     & ($0^{\circ},45^{\circ}$)\\
		
		$\boldsymbol{J}_{bin}$ & CB-SMBH angular momentum&  $ \boldsymbol{J}_{bin}   \lessgtr 0$& $\boldsymbol{J}_{bin} > 0$ \\
		$\boldsymbol{J}^{c}_{k}, k=1,2$ & BLRs angular momenta&  $ \boldsymbol{J}_{bin}\cdot \boldsymbol{J}^{c}_{k}   \lessgtr 0$ & $ \boldsymbol{J}_{bin}\cdot \boldsymbol{J}^{c}_{k}   \lessgtr 0$  \\
		\hline
	\end{tabular}
\tablefoot{
\tablefoottext{a}{Degeneration of barycentric orbits can arise  for  different reasons such as third body gravitational interaction known to be significant in some cases, quadrupole or relativistic precession. We take into account non-coplanar CB-SMBHs as one of the degenerate cases. However, the latter two effects are  irrelevant for qualitative orbital characteristics at short observational time scales. In any situation, there is a  freedom to select the coordinate system so that the parameterization is convenient and reasonable \citep{10.1086/367593}}\\
\tablefoottext{b}{The first  and the second range of clouds orbital inclinations are equivalent to the angular momenta conditions  $\boldsymbol{J}_{bin}\cdot \boldsymbol{J}^{c}_{k}  >0$
	and  $\boldsymbol{J}_{bin}\cdot \boldsymbol{J}^{c}_{k}  <0$ , respectively.}
}
\end{footnotesize}
\end{table*}

To study, characterize, and illustrate the evolution of differential phases of a single SMBH with elliptical clouds motion in a disc-like BLR surrounding it as well as CB-SMBH system and clouds in their BLRs on elliptical orbits, we used  the  model which closely resembles that in \cite{10.1051/0004-6361/201936398}.
Here, we only briefly recall the key aspects and some modifications and refer the reader to the above paper for further details.
In this paper, we adopt the following notation: a bold font variable refers to a 3$\times$1 vector, and unless otherwise specified, the indices $k=1,2$ are used to discern parameters of the primary and secondary component. We identify the respective quantities for the cloud in the BLR with the subscript c, which can be followed by index $k=1,2$ if a cloud is in the BLR of primary or secondary.  Solely used subscript c  indicate that all clouds in the BLR have the same value of a particular parameter.
We define $M_1$ and $M_2$ to be the primary  and secondary SMBH masses ($M_{1}>M_{2}$), respectively and 
$$q=\frac{M_{2}}{M_{1}}<1$$ to be the CB-SMBH mass ratio. 

 All  model parameters are   defined in   \cite{10.1051/0004-6361/201936398}, but we briefly describe them here for completeness.
The input parameters are five orbital elements defining the size and shape ($a, e, i, \Omega, \omega$)  of orbit and time $t$. The  output parameters are position ($\boldsymbol{r}(t)$) and velocity  ($\boldsymbol{\dot{r}}(t)$) of  SMBHs and each cloud in the BLRs, obtained by solving Kepler's equation:
\begin{equation}
\left\{a,e,i,\Omega,\omega,\mathcal{M},t\right\}_{k}\Rightarrow \mathrm{Kepler's \, Eqn.}\Rightarrow\left\{\boldsymbol{r}(t),\boldsymbol{\dot{r}}(t)\right\}_{q}, k=1,2
\end{equation}

\noindent where $\mathcal{M}$ is mean anomaly. The   barycentric  vector $\boldsymbol{n}$ defines the line of the sight.   Then the binary inclination angle to the observer is  $\cos{i_0}=\boldsymbol{n}\cdot\boldsymbol{J}_{bin}$ where $\boldsymbol{J}_{bin}$ is  the normalized orbital angular momentum vector of the CB-SMBH system. 
There is no reason to expect AGN not  to be oriented randomly,  meaning all orientations are equally likely. We will estimate the probable inclination angle of such an object. The relative likelihood that  AGN inclination is within a differential range  between $d i$ and $i+ d i$ of a specific inclination angle $i $ is proportional to the area on the unit sphere covered by that range of angles. The solid angle defined by the inclination range $i $ to $i+di$ equals $w=2\pi \sin i$, then we can write the probability density function as  a geometric probability  in the form of a ratio of  $w$ and the solid angle of complete sphere 
 \begin{equation*}
 d(Prob)=\frac{w}{4\pi}=\frac{2\pi \sin i d i }{4\pi}=\frac{d\cos i}{2}.
 \end{equation*}
   From this,  we got that  it is more likely to observe highly inclined  AGN whereas the  probability of observing an object at  inclination  $i_0$  is $\sim \sin i_0$.

For the case of single SMBH, we adopted physical parameters given in \cite{10.1051/0004-6361/201936398}, while for the CB-SMBH system, we used physical parameters given in \citet[][see Table 1]{10.3847/1538-4357/ab2e00}.
 For the BLRs, we adopted a disc-like model. The BLR kinematic structures for about two dozens  of AGNs, inferred from velocity-resolved RM, indicate a virialized disc, inflow, and outflow BLR geometries \citep[see][and references therein]{10.3847/1538-4357/ab5790}. 
The latest finding of \cite{10.3847/1538-4357/ab5790} indicates that the BLR of Mrk 79, which is sub-Eddington accreting AGN, probably originates from a disc wind launched from the accretion disc.   We will focus  on the most straightforward  BLR elliptical disc-like geometry here to simulate differential observables.

\begin{figure*}[ht!]
 	\begin{center}

 		\subfigure[$A_{0}=175 \mathrm{ld},i_{0}=40^{\circ},i=80^{\circ}, \mathrm{ld},\Omega_{1}=300^{\circ}, \Omega_{2}=  50^{\circ},  e_{k}=\newline\hspace*{1.5em} 0.5, k=1,2,\omega_{1}=70^{\circ}, \omega_{2}=250^{\circ}, i_{c1}=i_{c2}=40^{\circ}, \Omega_{c1}=\newline\hspace*{1.5em}\Omega_{c2}=300^{\circ},  \omega_{c1}=70^{\circ},  \omega_{c2}=250^{\circ},e_{c}=0.5 $ 
 		]	{%
 			\label{fig:doubleel71}
 			\includegraphics[trim = 1.0mm 0mm 0mm 0mm, clip, width=0.45\textwidth]{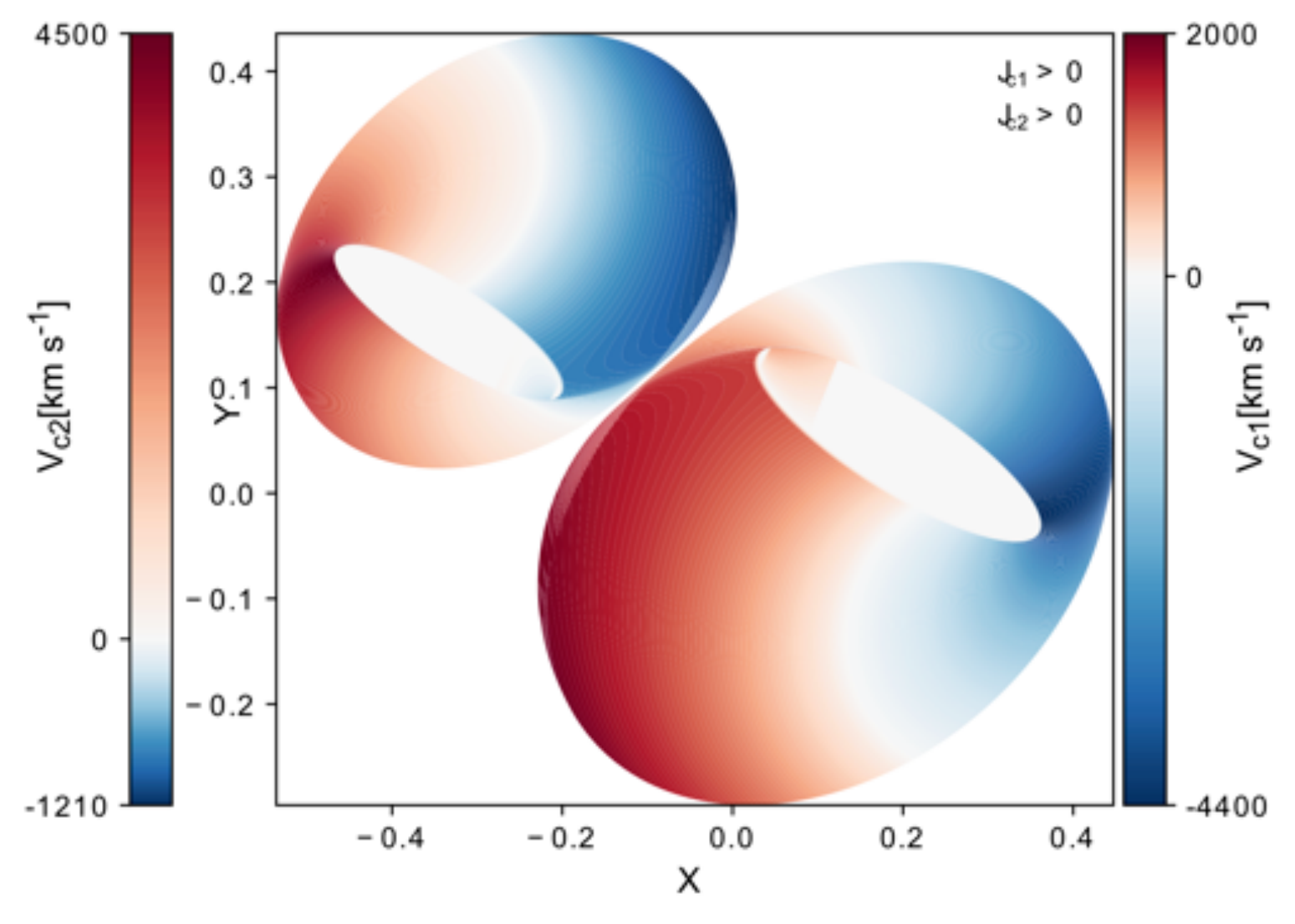}
 		}%
 		\hspace{-0.8em}
 		\subfigure[$A_{0}=175 \mathrm{ld} , i_{0}=40^{\circ},i=80^{\circ}, ,\Omega_{1}=300^{\circ}, \Omega_{2}=  50^{\circ},  e_{k}=0.5, k=1,2 , \omega_{1}=70^{\circ}, \omega_{2}=250^{\circ}, i_{c1}=(90^{\circ},175^{\circ}),i_{c2}=40^{\circ}, \Omega_{c1}=\Omega_{c2}=300^{\circ},  \omega_{c1}=70^{\circ},  \omega_{c2}=250^{\circ},e_{c}=0.5$ 
 		]{%
 			\label{fig:doubleel72}
 			\includegraphics[trim = 3mm 0mm 0mm 0mm, clip,width=0.45\textwidth]{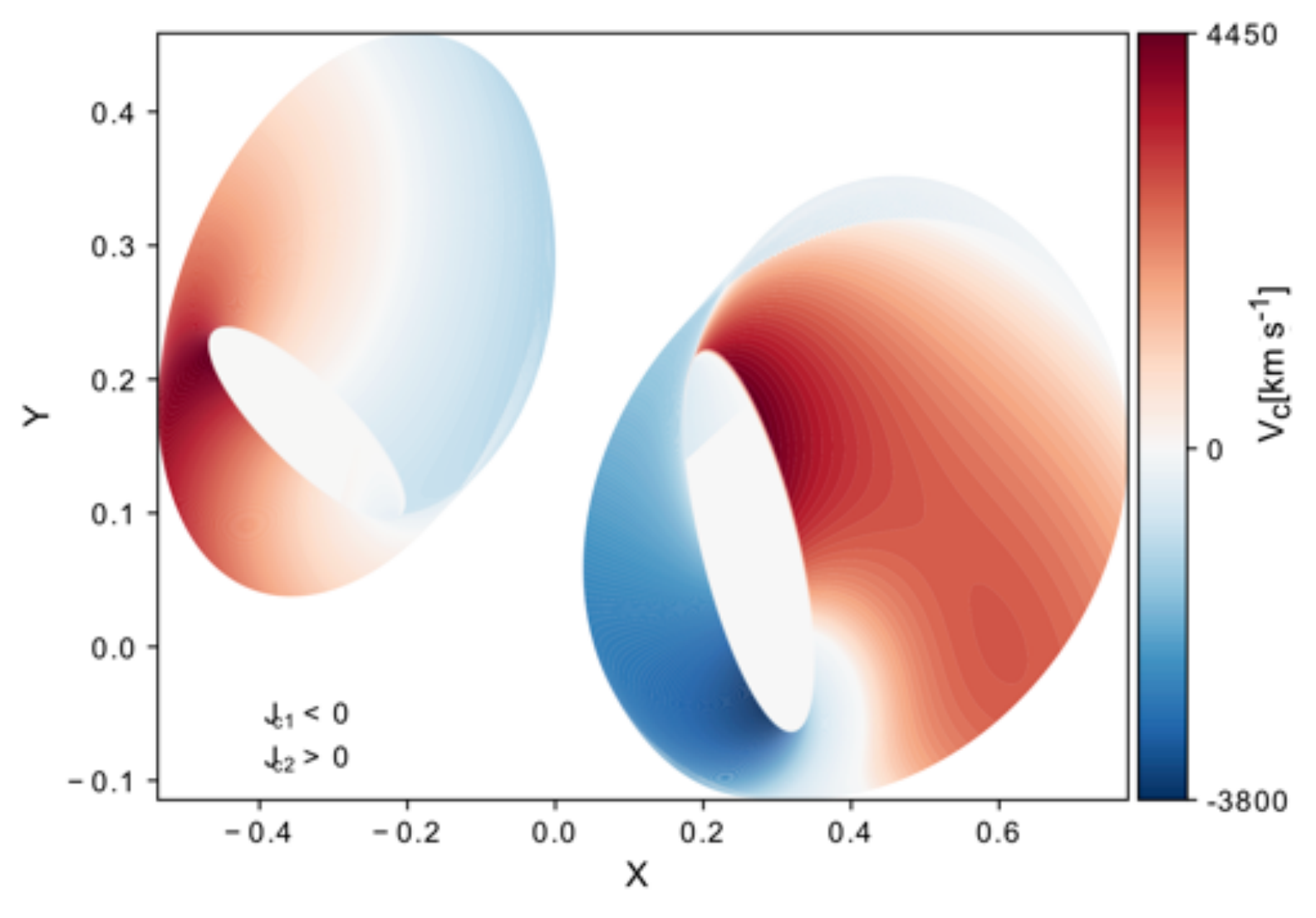}

 		}	\\ 
 		\vspace{-1.em}
 		\subfigure[$A_{0}=175, i_{0}=40^{\circ},i=80^{\circ}, \mathrm{ld},\Omega_{1}=300^{\circ}, \Omega_{2}=  50^{\circ},  e_{k}=\newline\hspace*{1.5em}0.5, k=1,2 , \omega_{1}=70^{\circ} \omega_{2}=250^{\circ}, i_{c1}=40^{\circ},i_{c2}=(90^{\circ},175^{\circ}),\newline\hspace*{1.5em} \Omega_{c1}=\Omega_{c2}=300^{\circ},  \omega_{c1}=60^{\circ},  \omega_{c2}=240^{\circ},e_{c}=0.5$  ]
 		{%
 			\label{fig:doubleel73}
 			\includegraphics[trim = 3.0mm 0mm 0mm 0mm, clip, width=0.45\textwidth]{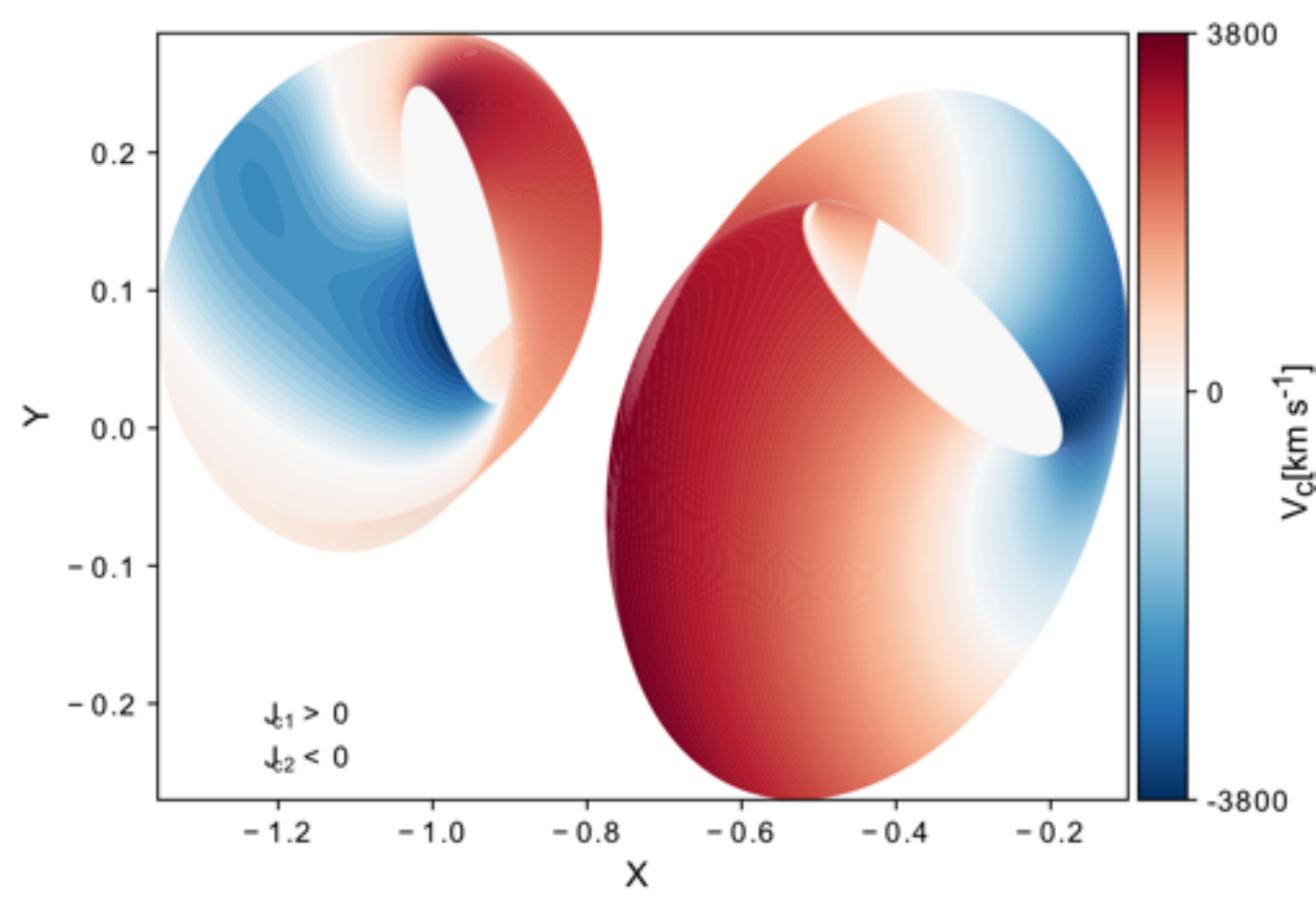}
 		}%
 		\hspace{-0.8em}
 		\subfigure[$A_{0}=175, i_{0}=40^{\circ},i=80^{\circ}, \mathrm{ld},\Omega_{1}=300^{\circ}, \Omega_{2}=  50^{\circ},  e_{k}=\newline\hspace*{1.5em}0.5, k=1,2 , \omega_{1}=70^{\circ} \omega_{2}=250^{\circ}, i_{c1}=i_{c2}=(90^{\circ},175^{\circ}), \Omega_{c1}=\newline\hspace*{1.5em}\Omega_{c2}=300^{\circ},  \omega_{c1}=60^{\circ},  \omega_{c2}=240^{\circ},e_{c}=0.5$
 		]{%
 			\label{fig:doubleel74}
 			\includegraphics[trim = 0.0mm 1mm 0mm 0mm, clip,width=0.45\textwidth]{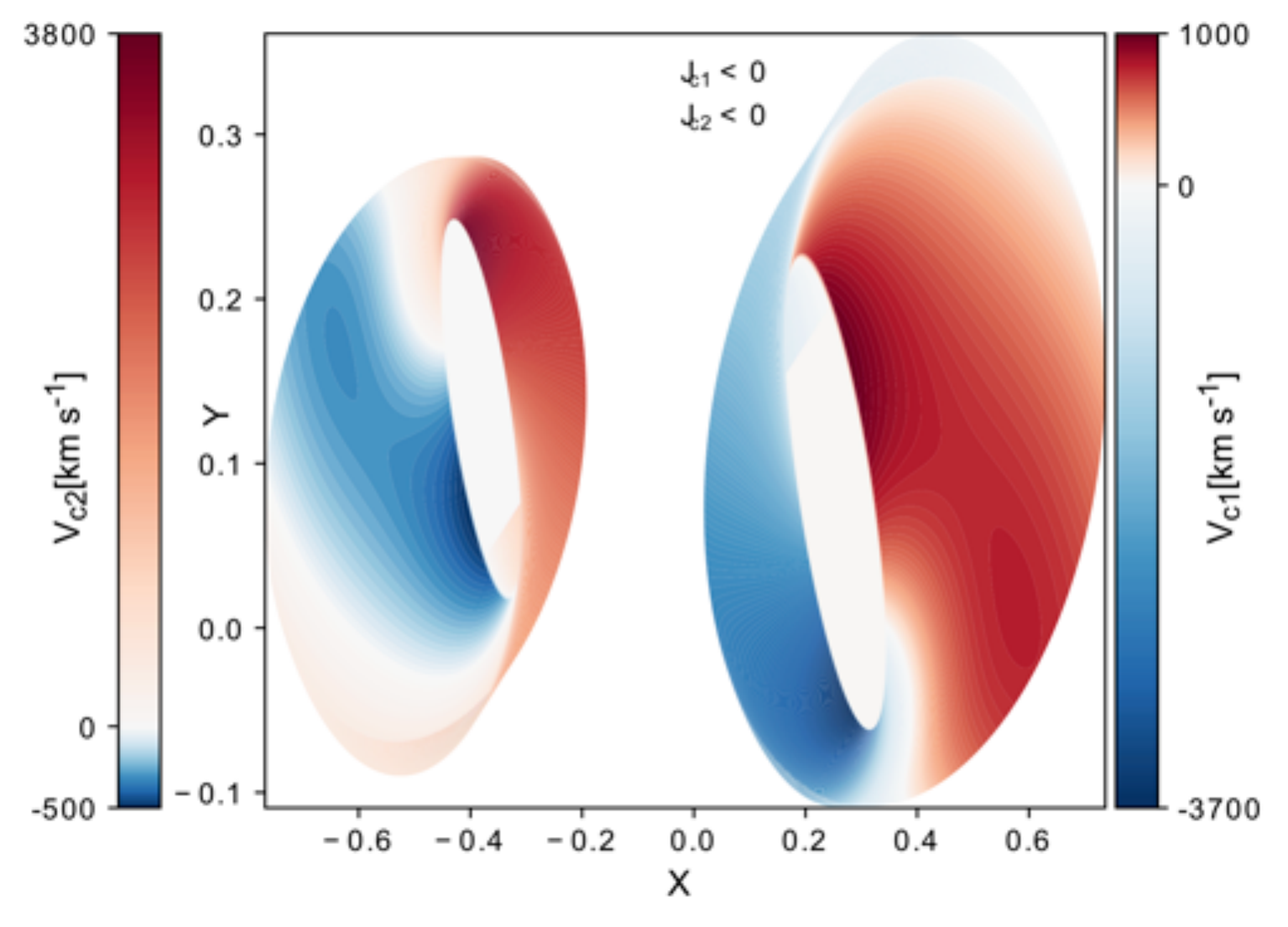}
 		}
 		
 		\vspace{-1.9em}
 	\end{center}
 	\caption{%
 		Radial velocity maps of simulated images of the BLRs  around the primary and secondary components in CB-SMBH. Abscissas (X) and ordinates (Y) are normalized by the $\mathcal{A}_{0}=100 \mathrm{ld}$ .
 		Clouds orbital angular momenta alignment with CB-SMBH orbital angular momentum is indicated in legends.
 	}%
 	
 	\label{fig:doubleel7}
 \end{figure*}

Returning briefly to the subject of circular binaries (with orbital eccentricities  $e_{i}=0,i=1,2$, and semi-major axes $a_{i},i=1,2$),  we point out the time independence of the relative distance of two SMBH
 \begin{equation}
  |\boldsymbol{r}(e_{1}=0,e_{2}=0)|=a,
 \end{equation}
 \noindent where  $a=a_{1}+a_{2}$.
 In circular case, the velocities of  SMBHs and their relative velocity are also time-independent.
 However, in the case of elliptical configurations of clouds in the BLR of SMBH and elliptical  CB-SMBH system, the positions  and velocities depend on time and osculating orbital elements. Our model accounts for all these parameters, as explained in \cite{10.1051/0004-6361/201936398}.
 We used  the parameter ranges in Table  \ref{table:model} to generate  different  combinations of parameters, and run  simulations  with 100 clouds in each BLR. The clouds orbital positions  are uniformly sampled at every time instance ${t_{n}}, n=1,..,1000$. 
 All simulations  were performed at the initial CB-SMBH orbital phase, if not stated otherwise. As an illustrative example, one realization of elliptical  geometrical configuration  of CB-SMBH, depicted from the elevated and the side, is shown in  Fig.  \ref{fig:cosis}.
 Moreover, Fig. \ref{fig:doubleel7} illustrates  how our model captures dynamics and consequently  the   distribution of radial velocities of clouds in two BLRs occurring for different clouds' orbital  angular momenta alignments  to the CB-SMBH orbital angular momentum using randomly chosen orbital parameters. 
 We set the radial velocity  distributions at   clouds positions in each BLR  on the same plot for comparison.
 For each cloud  in the disc-like BLR, we determine its combined   velocities of orbital motion sampled at 1000 points of its orbit and its angular momenta, taking into account its central SMBH motion.
 We then project the clouds  velocity field onto the plane of the observer for a value of  $i_0$  to determine the line-of-sight projection of the combined velocity at each position in the inclined disc-like BLRs. 
 The maximum of  $V_{1}$ and $V_{2}$ radial velocities of the primary and secondary components in CB-SMBH occur  at  passaging the ascending node in their true orbits.
 The epoch of passage through the ascending node corresponds to  the  instant when  radial velocity of the secondary component  reaches maximum (or when the radial velocity of the primary component  reaches minimum). The asymmetry of distributions is a consequence of the asymmetry of elliptical orbital motions of SMBHs and clouds. There is  a distinction between radial velocity  distributions  (see Fig. \ref{fig:doubleel7}).  Specifically,  we see the topological difference between  velocity fields  for aligned (Fig. \ref{fig:doubleel71})  and anti-aligned orbital configurations of clouds (Figs. \ref{fig:doubleel72}-\ref{fig:doubleel74}).
Clouds in anti-aligned orbits have inclinations between $90^{\circ}$ and $175^{\circ}$ consequently. Velocity field for anti-aligned clouds orbital momenta shows elongated features that are strained.   In all these cases, clouds orbital inclinations are linearly spaced between given boundaries. Velocity fields  are the closed surfaces. \footnote{ Compact and without boundary surface is  closed.  Notable examples are the sphere and torus. An example of non-closed surfaces is a cylinder.}, preserving topological volume and surface coherency\footnote{We use coherence  in the sense that levels of velocity intensity do not fluctuate widely.}. We also tested the cases where inclinations are randomly distributed within given ranges, and such velocity fields of the  BLRs are not volume-preserving in the topological sense.

For aligned BLRs, the absolute value of velocity increases toward the outer left and right side lobes of disc-like BLRs (regarding the smaller BLR diameter). However, with anti-aligned BLRs, the absolute value of velocity increases toward frontal and backward lobes (oriented in the directions of larger BLR diameters).
 
 As discussed above, the general  form of equations for calculation of brightness distribution of singular SMBH and CB-SMBH system is the same as in \cite{10.3847/1538-4357/ab2e00}, but 
 the difference is  that the input and output variables  depend on the time.
  Replacing the double integral by a summation on the  coordinates of clouds, and time  with single SMBH gives a discrete form of equations. For the CB-SMBH system,  the summation is on the coordinates of clouds in the  BLRs, both SMBHs and time.
 
  \subsection{Differential phase}\label{difp}
 
 The  primary advantage of using differential phase measurements is that their uncertainties are reduced in the case of narrow   spectral lines  and they are not contaminated by wavelength-independent errors \citep[see][]{10.3847/1538-4357/aa79f1}.
 We modeled the differential phases  from the computed brightness distribution according to the prescription given by \citep{10.3847/1538-4357/ab2e00} based on  the Zernike-van Cittert’s theorem.
 In particular, from this theorem, we obtain:
 \begin{equation}
 {Q(\boldsymbol{u},\lambda)}=\iint I(\boldsymbol{\sigma},\lambda)e^{-2\pi i\, \boldsymbol{\sigma\cdot u}} d\alpha\, d\beta
 \label{eq:flux}
 \end{equation}
 \noindent where $ {Q}(\boldsymbol{u},\lambda)$ is a coherent flux of the object, $I(\boldsymbol{\sigma},\lambda)$ is the intensity of the object emission in    the observer direction,  transferred  the spectral channel of band $\delta \lambda$ and centered on the wavelength $\lambda$; $\boldsymbol{\sigma}=(\alpha,\beta)$ is the celestial angular coordinates of an emitting point in the object, and $\boldsymbol{u}=\frac{\boldsymbol{B}}{\lambda}=(u,v)$ is the interferometry baseline vector.
 By defining the moments of intensity distribution \citep[see details in][]{10.3847/1538-4357/aa79f1}.
 
 \begin{equation}
 \mu_{lm}=\iint I{\delta\lambda}(\boldsymbol{\sigma},\lambda) {\alpha}^{l} {\beta}^{m}\,d\alpha\, d\beta
 \end{equation}
 \noindent one can see that  $\mu_{00}$ is total intensity of the object, while the normalized first order moments give the centroid positions:
 \begin{equation}
 \begin{aligned}
 p_{\alpha}(\lambda)&=\frac{\mu_{10}}{\mu_{00}}=\frac{\iint \alpha I(\boldsymbol{\sigma},\lambda) d\alpha d\beta}{\iint  I(\boldsymbol{\sigma},\lambda) d\alpha d\beta}, \\
 p_{\beta}(\lambda)&=\frac{\mu_{01}}{\mu_{00}}=\frac{\iint \beta I_(\boldsymbol{\sigma},\lambda) d\alpha d\beta}{\iint  I(\boldsymbol{\sigma},\lambda) d\alpha d\beta}
 \end{aligned}
 \label{eq:photocent}
 \end{equation}

 Consequently, we have that $Q(0,\lambda)=\mu_{00}=\Xi(\lambda)$. The emission line ($\Xi(\lambda)$)  is the zero order moment of brightness distribution of the object.
 
 Also, from the Zernike-van Cittert’s theorem, the fringe phase is:
 \begin{equation}
 \begin{split}
 -Arg\bigg( \frac{Q(\boldsymbol{u},\lambda)}{\mu_{00}}\bigg)\sim 
 \frac{\int \sin(2\pi \boldsymbol{\sigma}\cdot \boldsymbol{u}) I(\boldsymbol{\sigma},\lambda) d\alpha d\beta }{\int \cos(2\pi \boldsymbol{\sigma}\cdot \boldsymbol{u}) I(\boldsymbol{\sigma},\lambda) d\alpha d\beta}\\ \sim 2\pi  \frac{\int I(\boldsymbol{\sigma},\lambda)\,  \boldsymbol{\sigma} d\alpha d\beta }{\int  I(\boldsymbol{\sigma},\lambda) d\alpha d\beta} \boldsymbol{u}\\
 \sim 2\pi\boldsymbol{u}\cdot
 \boldsymbol{\xi}(\lambda)\\=2\pi u \xi_{\alpha}(\lambda)+2\pi v \xi_{\beta}(\lambda)
  \end{split}
 \label{eq:complex}
 \end{equation}

 Let us now consider an unresolved source ($\boldsymbol {u\cdot \sigma}<1$), for which is possible to relate  vectorial components given in  Eq. \ref{eq:complex} to  the photocenter $p_{\alpha}, p_{\beta}$ (Eq. \ref{eq:photocent}) of the intensity distribution \citep[see details in][]{10.1051/0004-6361:20011047}:
 \begin{equation}
 \xi_{\alpha}(\lambda)=\frac{\mu_{10}}{\mu_{00}}=p_{\alpha}(\lambda) \\
 \xi_{\beta}(\lambda)=\frac{\mu_{01}}{\mu_{00}}=p_{\beta}(\lambda)
 \label{eq:complex2}
 \end{equation}
 
 The terms in Eq. \ref{eq:complex2}  are the fringe phase components. They  are  proportional to the photocenter of intensity distribution and  equivalent to the first moment of the intensity distribution.
 However, the fringe phase can be disrupted by  the atmospheric turbulence  if  only one baseline (two telescopes) is used.  This problem  can be reduced with the differential phase $\Delta \phi$  \citep[see, e.g.][and references therein]{10.1051/0004-6361/201220689}.  The differential phase is defined as the difference between the fringe phases obtained  in two spectral channels centered on the wavelengths $\lambda$ and $\lambda_r$, respectively 
 \begin{equation}
 \Delta \phi=-2 \pi \boldsymbol{u}\cdot( \boldsymbol{\xi}(\lambda)-\boldsymbol{\xi}(\lambda_r))
 \label{eq:complex2x}
 \end{equation}
 This quantity is  relevant for line profiles when accounting  for the kinematics of the source, and the continuum region is the natural choice as a reference so that  $\boldsymbol{\xi}(\lambda_r)=0$ \citep[see][]{10.1051/0004-6361:20040051}.
 Thus, the first vectorial component of differential  phase (Eq.  \ref{eq:complex2}) is usually  observed and modeled in the literature. We have also followed this approach.
 It has been shown both empirically  \citep{10.1038/s41586-018-0731-9} and theoretically \citep{10.3847/1538-4357/ab2e00} that  differential interferometry produces a variation of BLR photocenter $\xi(\lambda)$.   It is an unexploited  astrophysical parameter  in AGN investigation, with an interpretation  similar  to the BLR spectral lines $\Xi(\lambda)$. Line profiles reveal a large scale spatial and physical effects of the BLRs.  But with $\xi(\lambda)$ we can check and  add new details to the spectroscopic information  to get  a complete picture of the object. The information  in $\xi(\lambda)$ is always new, because the weights of contribution of different parts of  the object in $\Xi(\lambda)$ and $\xi(\lambda)$ are different  \citep{1989ASIC..274..249P}.
 
  All the interferometric data are usually quantitatively interpreted within the frame of a single, self-consistent physical model, which should produce observables. 
  Modern spectro-interferometers  can measure the variation of the interferometric phase across an emission line of AGN \citep{10.1038/s41586-018-0731-9}. 
   This first measurement of  phase  displays a simple S-shaped profile  comprising a single reversal, arising from the line emission from a  circular rotating disc-like BLR. 
  The S form  is  one of the most common observed  differential phase  shapes, and here we  briefly explained it.

  Assuming that the source has Gaussian spatial brightness distribution, $\boldsymbol{G}(\boldsymbol{\sigma},\lambda)$, and integrating $\xi_{\alpha}(\lambda)$ (see Eq. \ref{eq:complex2}) in both directions  $\beta$ and $\alpha$,  we  get that differential phase shapes is $$\xi_{\alpha}(\lambda)\sim  Gaussian(\alpha) \erf(\beta)$$  where $\erf$ is  a 'error function' of sigmoid shape which causes  differential phase (Eq.  \ref{eq:complex2x}) to be  of S shape.  In general,  \cite{10.1051/0004-6361:20030072} showed that the differential phase can be related to the skewness and diameter of the object flux  distribution.

  \cite{10.3847/1538-4357/ab2e00} showed  the same reasoning for interferometric observables is  valid for the CB-SMBH case.  The interferometric model of the CB-SMBH system is a composition of two sources (their BLRs) considered either as  point-like or disc-like models with  assumed morphologies. Let we denoted the brightness of disc-like BLR components as  $I_{k}(\boldsymbol{\sigma_k},\lambda)$ and their barycentric positions $(\boldsymbol{\sigma}_{k}), k=1,2$.

 The total brightness distribution of such system can be written as
 \begin{equation} 	
 I(\boldsymbol{\sigma},\lambda)= I_{1}(\boldsymbol{\sigma_1}, \lambda)+ I_{2}(\boldsymbol{\sigma_2
 },\lambda)  .
 \label{eq:complex3}
 \end{equation}
 
 Taking into account  Eq.\ref{eq:flux}, Eq. \ref{eq:complex3}, additional and translational  property of Fourier transform  one can get that coherent flux is simply
 \begin{equation}
 {Q(\boldsymbol{u},\lambda)}= Q_{1}(\boldsymbol{u},\lambda) +Q_{2}( \boldsymbol{u},\lambda)=
 F_{1}e^{-2\pi i\, \boldsymbol{\sigma_{1}\cdot u}} +F_{2}e^{-2\pi i\, \boldsymbol{\sigma_{2}\cdot u}}
 \label{eq:complex4}
\end{equation}
where  $F_1$ and $F_2$ are the fluxes of components.
Normalization gives

\begin{equation}
Q_{n}(\boldsymbol{u},\lambda)=
\frac{F_{1}e^{-2\pi i\, \boldsymbol{\sigma_{1}\cdot u}}+F_{2}
	e^{-2\pi i\, \boldsymbol{\sigma_{2}\cdot u}}}{F_{1}+F_{2}}
\label{eq:complex5}
\end{equation}

\noindent We can expand the complex exponential terms in Eq. \ref{eq:complex5} in a Taylor series:
\begin{equation}
e^{-2\pi i\, \boldsymbol{\sigma_{k}\cdot u}}\sim 1-2\pi i\, (\boldsymbol{\sigma_{k}\cdot u})-\mathcal{O}((\boldsymbol{\sigma_{k}\cdot u})^2), \, k=1,2
\label{eq:complex51}
\end{equation}

\noindent Taking into account that $\arctan(x)\sim x$ when $ |x|<1$ and   $|\boldsymbol{\sigma_{k}\cdot u}|<1, \, k=1,2$, we can estimate 

\begin{equation}
Arg(Q_{n}(\boldsymbol{u},\lambda))\sim \arctan (\frac{Im (Q_{n}(\boldsymbol{u},\lambda))}{Re (Q_{n}(\boldsymbol{u},\lambda))} ) \sim 2\pi \boldsymbol{u}\cdot
\frac{F_{1} \boldsymbol{\sigma_{1}}-F_{2} \boldsymbol{\sigma_{2}}}{F_{1} +F_{2}}
\label{eq:complex6}
\end{equation}

If the binary components are on circular orbits, then $\sigma_{1}=\sigma$ and $\sigma_{2}=\sigma$, where $\sigma$ is a component barycentric distance. Under this condition, the Eq. \ref{eq:complex6} shows that 
$\xi(\lambda)$ signal is proportional to the binary separation $\sigma$.
However,  with CB-SMBH on elliptical orbits, we can expect $\xi(\lambda)$ signal to be  more complex.   Especially, since spectral lines are  compound functions of the form
$F_{k}={\Xi}_{k}(\lambda, \boldsymbol{r}^{c}_{km}, \boldsymbol{V}^{c}_{km}), k=1,2,m=1,100$ where 
$\boldsymbol{r}^{c}_{km},\boldsymbol{V}^{c}_{km}$ are composite  vector field, representing at each  position ($\boldsymbol{r^{c}}_{km}$) of the $m$-th cloud  in the  $k$-th BLR   its  velocity $\boldsymbol{V}^{c}_{km}$ given in CB-SMBH barycentric coordinate system \citep[see][]{10.1051/0004-6361/201936398}.

Moreover,  Eq. \ref{eq:complex3} -\ref{eq:complex6} show us that the primary difference between the binary and single  model is that the brightness distribution is no longer symmetric, 
 what makes the differential phase  to be a complex   function.

\subsection{Observational constrains}
The differential phase of the circular  CB-SMBH system depends on  extrinsic parameters (the inclination of a CB-SMBH system 
to the observer) and intrinsic parameters (an opening angle of the BLR, SMBH masses and their mutual distance, and orientation of total angular momentum) as it was  shown in \citep{10.3847/1538-4357/ab2e00}.
Here we extend analysis on non-coplanar  CB-SMBH system with varying orbital elements,  and  randomly distributed clouds  orbital inclinations, eccentricities, and semi-major axes.
 We choose clouds in randomly inclined orbits as an  interesting case based on  \cite{10.1086/309750}    model of  C IV emitting region of NGC 5548 with an ensemble of clouds on randomly inclined and spherically distributed orbits around the continuum source. However, this  spectral line is observed as blue-shifted with certain blue-asymmetric features, perhaps indicating material outflow.  Because of these characteristics, the C IV emitting region is distinctive from those related to H$\beta$ and Pa$\alpha$.
Moreover, emission line profiles are indicators of  the BLR gas kinematics. No one of the functional forms (i.e., Gaussian, Lorentzian, even Power law) is connected directly to the ordered motion of clouds in the BLR. Except for double-peaked profiles, all spectroscopic line shapes are consistent with randomly oriented orbits in a gravitationally bound cloud system \citep[see]
[]{2013peag.book.....N}. Randomly distributed  clouds' orbital semi-major axes and eccentricities lead to their random position, velocities and  orbital angular momenta.
We calculated differential phases for a 100-meter baseline with a position angle  of $90^{\circ}$ to the direction of the  semi-major axis of the CB-SMBH   system. The differential phases correspond to the  emission Pa${\alpha}$ line, which rest-frame  wavelength is $\lambda_{0}=1.875\mu \mathrm{m}$ for objects at redshift of $z\sim 0.1$.
These parameters correspond to  sources (such as 3C 273), which are red-shifted enough so that the Pa$\alpha$ line is  in the K band.  This band is convenient because the emission is at least two times stronger in Pa$\alpha$  than in e.g.  Br$\gamma$ \citep{doi.org/10.1117/12.926595}.
Moreover,  as  shown by  \cite{10.1086/522373}, the strongest Paschen hydrogen broad emission lines Pa$\alpha$ and Pa$\beta$ are seen to be without  strong blends  implying  good measurement of their widths. The line   attenuation by dust is much more reduced in the near-IR lines than in  the optical and UV.

\section{Results}
 Here we present our investigation of the effects of variation of SMBH and clouds orbital elements   on the amplitude and shape of Pa${\alpha} $ and corresponding differential phase.  We present the results  divided in the domains of observables of the single SMBH and CB-SMBH system.  We select  representative cases that resemble examples of “real-world” AGN discussed in SubSect 4.1.  For each referred AGN,  we show the models whose spectrum resembles the observations, and the predicted differential phases for that model. The plots from all the combinations of parameters (i.e. the atlas of spectra and differential phases) are moved  to appendices.	As expected, the phase signatures occur under specific conditions, and their diagnostic potential is clearer for some parameters than for others. The set of models being compared is composed of  those in which one (or all) of the parameters could  have different values.
	For reference, interferometric observables marked with a blue colour in plots were simulated for  values at the lower end of the parameters ranges, if  not stated otherwise.

\subsection{Interferometric signatures of  a single SMBH} \label{RD}
  Here, we explore potential effects,  induced by   changes in the elliptic orbital parameters of clouds in single BLR, on  the interferometric observables.  Before conducting our numerical simulations, we analyzed generalized forms of the first order approximation of differential phase (Appendix \ref{appendix:generalsingle}).  The predicted deformation of differential phase  S shape is asymmetric    indicating clouds elliptical orbital motion (see Fig. \ref{fig:ilustris}).
  
\begin{figure*}[ht!]
	\begin{center}
  \subfigure[$\Omega_{c} \in \mathcal{U}(10^{\circ}, 180^{\circ})$ , $\omega_{c}=110^{\circ}$]{%
  	\label{fig:singleel12}
  	\includegraphics[trim = 0.0mm 0mm 5mm 0mm, clip,width=0.3\textwidth]{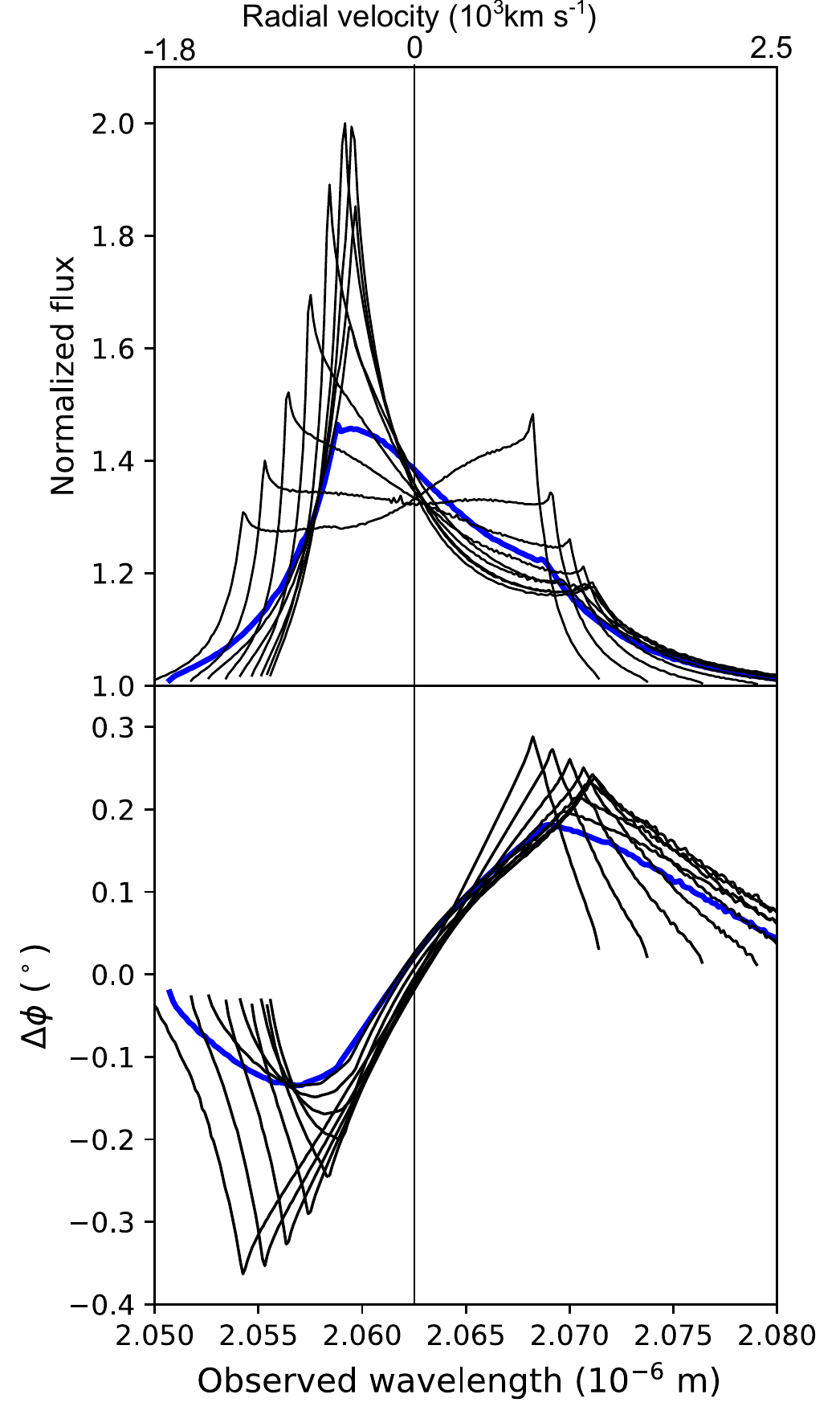}
  }
  \hspace{-1.2em}
  \subfigure[$\Omega_{c}=100^{\circ}$, $\omega_{c}=\mathcal{U}(10^{\circ}, 180^{\circ})$]{%
  	\label{fig:singleel13}
  	\includegraphics[trim = 1.0mm 0mm 0mm 0mm, clip, width=0.3\textwidth]{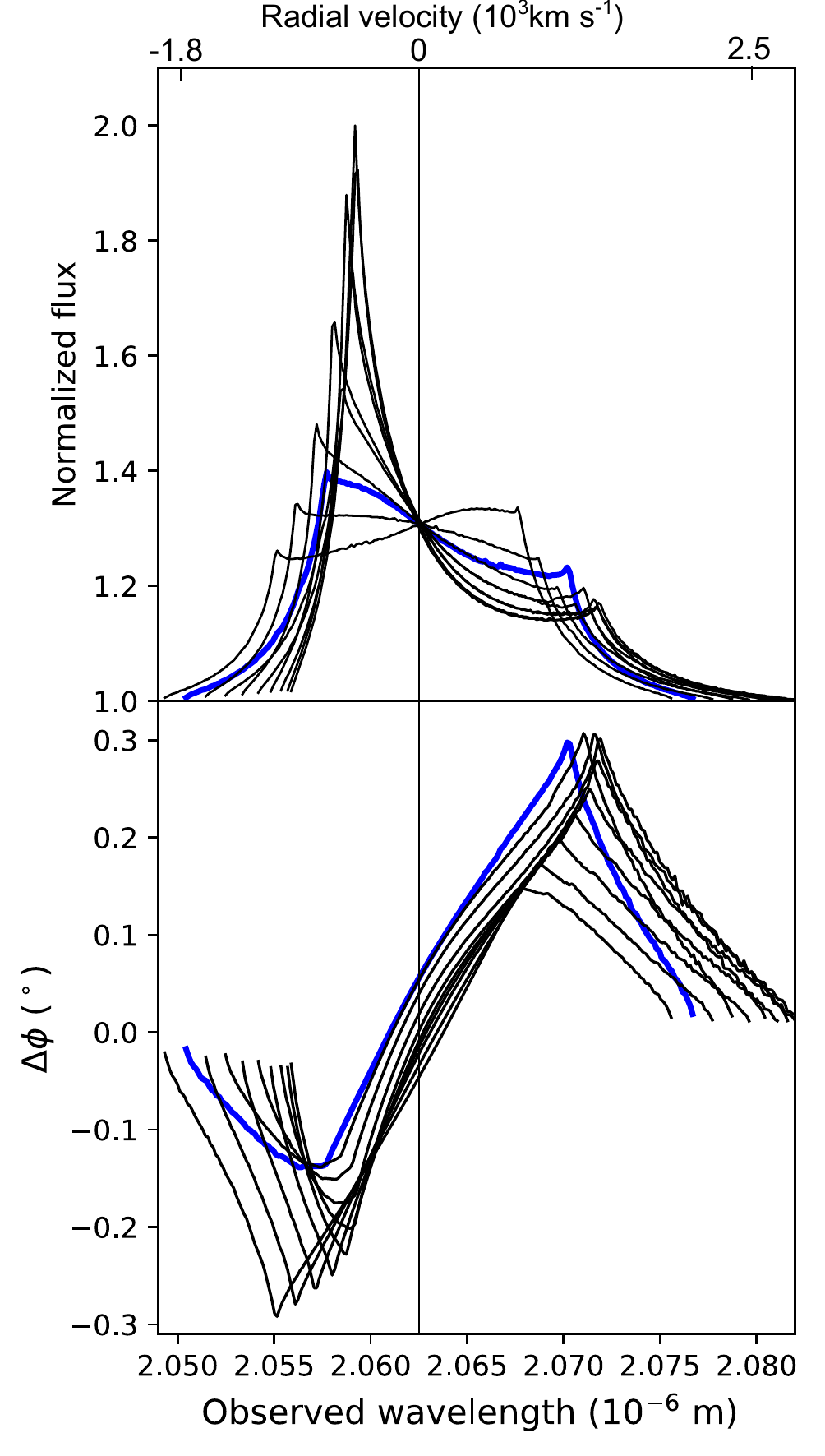}
  	
  }
	\subfigure[$\omega_{c}=200^{\circ}$ ]
{%
	\label{fig:singleel24}
	\includegraphics[trim = 1.0mm 0mm 0mm 0mm, clip,width=0.3\textwidth]{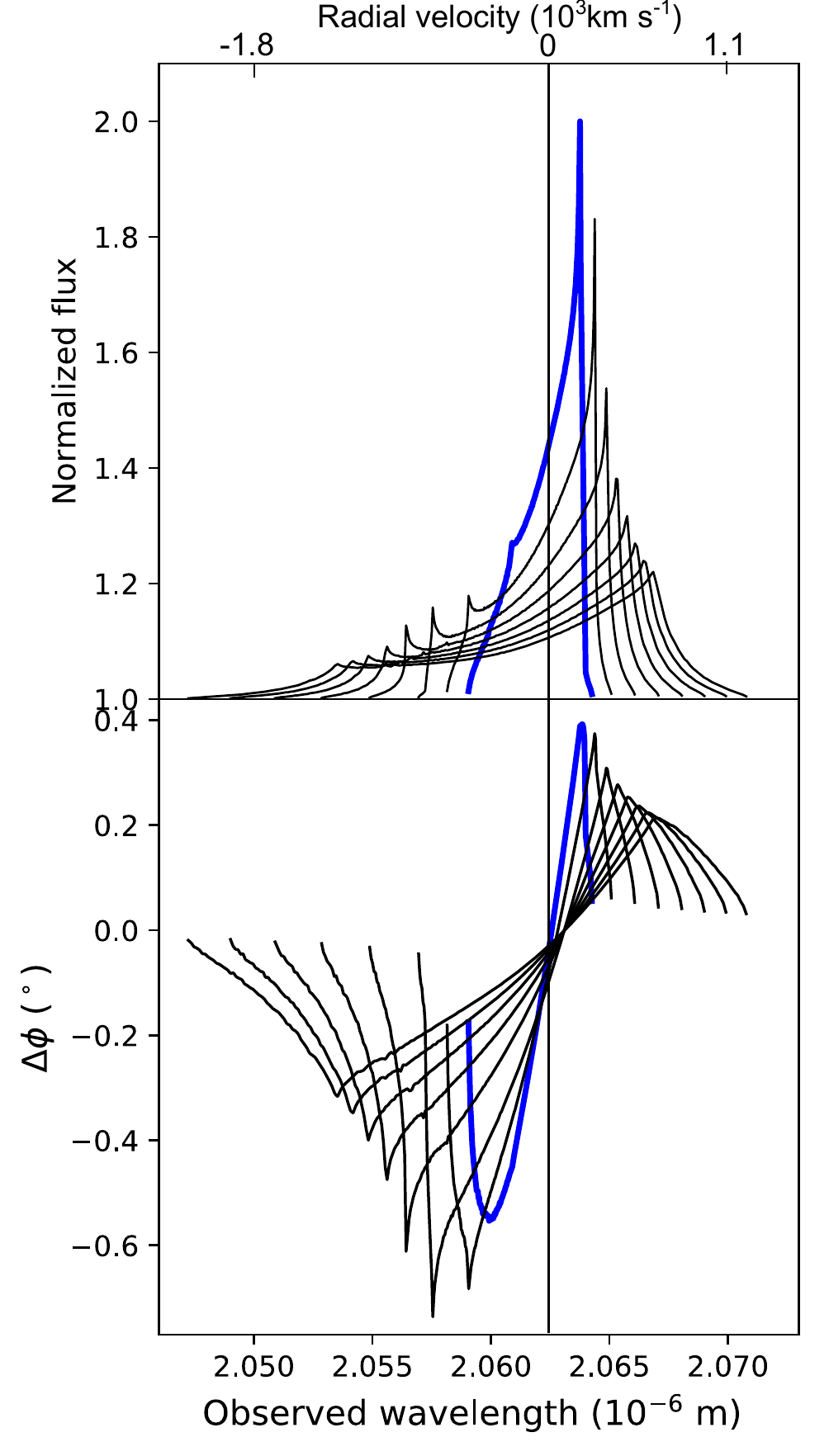}
}%
\end{center}
  	\caption{%
  	Evolution of the Pa$\alpha$ emission line (upper subplots) and corresponding differential  phase ($\Delta \phi$, lower subplots) as a function of the wavelength  for different values of clouds' orbital parameters in the  different models of single SMBH. Common parameters for all models: (a) and (b) clouds' orbital inclination $i_{c} \in [-5^{\circ},5^{\circ}]$, clouds' orbital eccentricity $e_{c}=0.5$ and $i_{0}=45^{\circ}$.  (c) clouds' orbital inclination $i_{c}=\mathcal{U}(-7.5^{\circ},7.5^{\circ})$ and 
  	$i_{0}=\mathcal{U}(10^{\circ},45^{\circ})$.  Varying parameters are given in subcaptions.   $\Im=\mathcal {U} (l,r)$ stands for collection of  inclination ranges from $\mathcal{U}(-l,l)$ up to $\mathcal{U}(-r,r)$.   For reference, interferometric observables marked with blue colour were modeled for values at the lower end of the parameters ranges.
  }%
  
  \label{fig:singleelg1}
\end{figure*}

  Then we  simulated the Pa${\alpha}$ lines  and corresponding differential phases by considering the orbital dynamics controlled by orbital elements of clouds and different  values of the $i_{0}$ parameter.  The curves are given as a function of the wavelength.
  For a better comparison, the scale of twin x-axis  at the top of each plot  is presented in radial velocity.
  First, we adopted that the clouds move on orbits with    uniformly distributed inclinations $i_{c}=\mathcal{U}(-5^{\circ},5^{\circ})$ and varying other parameters. The most dramatic evolutions of observables occur when varying either $\Omega_c$ or $\omega_c$ between $10^{\circ}$ and$180^{\circ}$ (see Fig. \ref{fig:singleel12}).
  For fixed  clouds orbital eccentricity $e=0.5$ larger values of   $\Omega$  and $\omega$ cause an increase of the amplitude of the differential phase (see  Figs. \ref{fig:singleel12} and \ref{fig:singleel13}). Additionally, these two orbital angle parameters   have more significant influence on the amplitude of differential phase when eccentricity is smaller  as it is  shown in Fig. \ref{fig:ilustris}.

  Further, these simulations capture another characteristic of the spectra: the flat-topped spectra can appear not only  for circular motion but also  because of clouds' elliptical  motion (see Figs. \ref{fig:singleel12} and \ref{fig:singleel13}). The shapes of the flat top depend on the angles $\Omega$ and $\omega$ of the elliptical orbits of clouds.
  Compared to simulations for small clouds' orbital eccentricity,  we recorded  the most notable variations  of differential phase shape  associated with variation of angle $i_{0}$, fixed high orbital eccentricity  ($e=0.5$) and orbital inclination  $i_{c}=U(-7.5^{\circ},7.5^{\circ})$ (see Fig.
   \ref{fig:singleel24}).
   The inclination of the observer $i_0$  and clouds orbital inclination affects  similarly the slope of the differential phase  (see Fig. 
  \ref{fig:singleel24}). Here  we also  identify a characteristic pattern of  alternation of the amplitudes in the left and  the right wing of differential phase  because of variation of  the clouds' orbital pericenter $\omega$ (Fig.
  \ref{fig:singleel24}).

\begin{figure*}[ht!]
	\begin{center}

		\subfigure[$ i_{0}=\mathcal{U}(-10^{\circ},45^{\circ}),\delta i_{0}=5^{\circ},  \mathcal{C}, \Omega_{c}=  100^{\circ}, \newline \hspace*{1.5em}  \omega_{c}=270^{\circ},  e_{c}\in\Gamma_{s}  (0.3,1)$]
		{%
			\label{fig:singledistr41}
			\includegraphics[trim = 2.5mm 0mm 0mm 0mm, clip, width=0.35\textwidth]{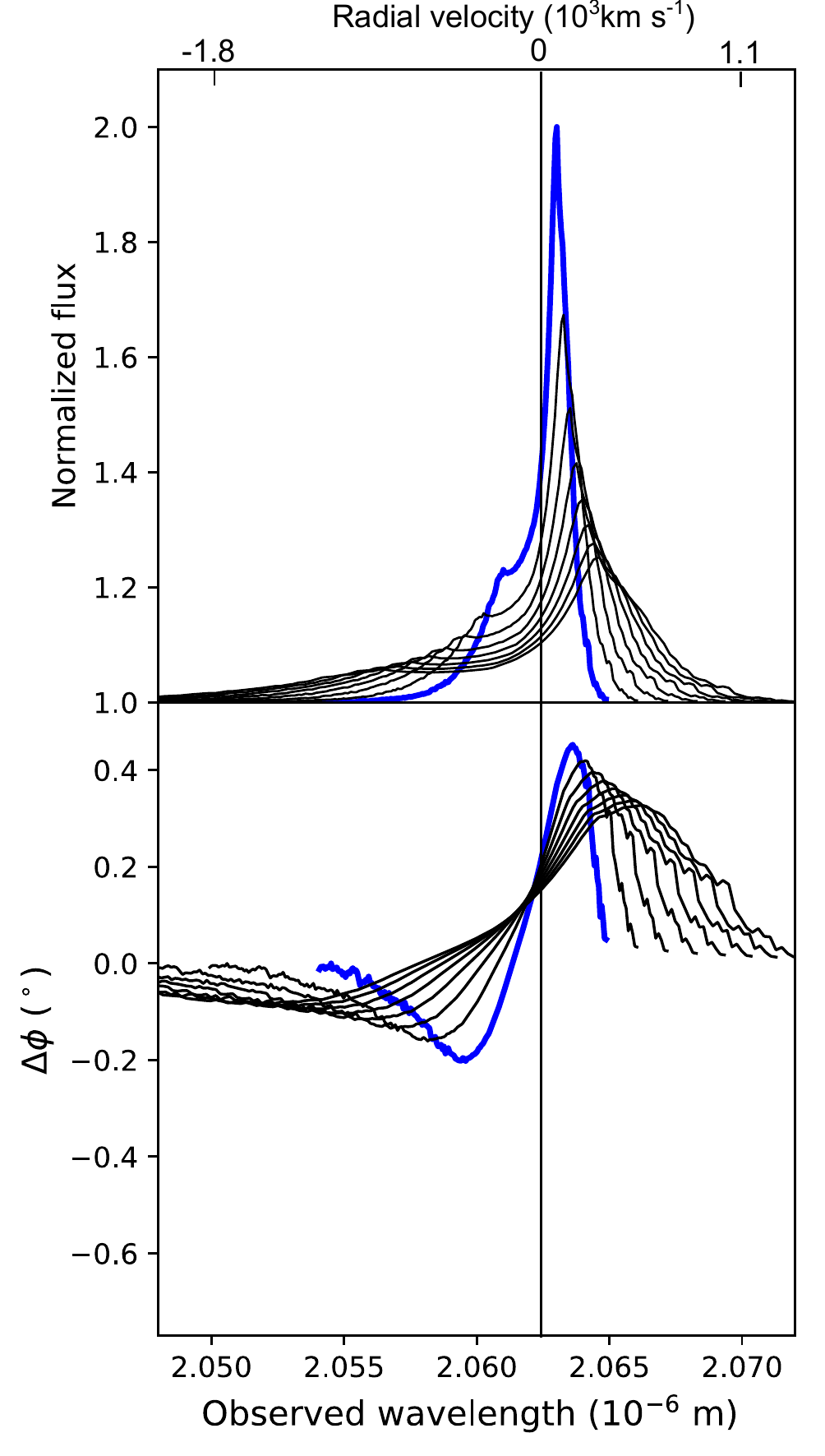}
		}%
 \hspace{-1.2em}
\subfigure[$  i_{0}=\mathcal{U}(-10^{\circ},45^{\circ}),\delta i_{0}=5^{\circ}, i_{c}=\mathcal{U}(-20^{\circ}, 20^{\circ}),\newline \hspace*{1.5em} \Omega_{c}=  100^{\circ},   \omega_{c}=270^{\circ},    e_{c}\in \mathcal{R}(1)$
]{%
	\label{fig:singledistr44}
	\includegraphics[trim = 2.5mm 0mm 0mm 0mm, clip,width=0.35\textwidth]{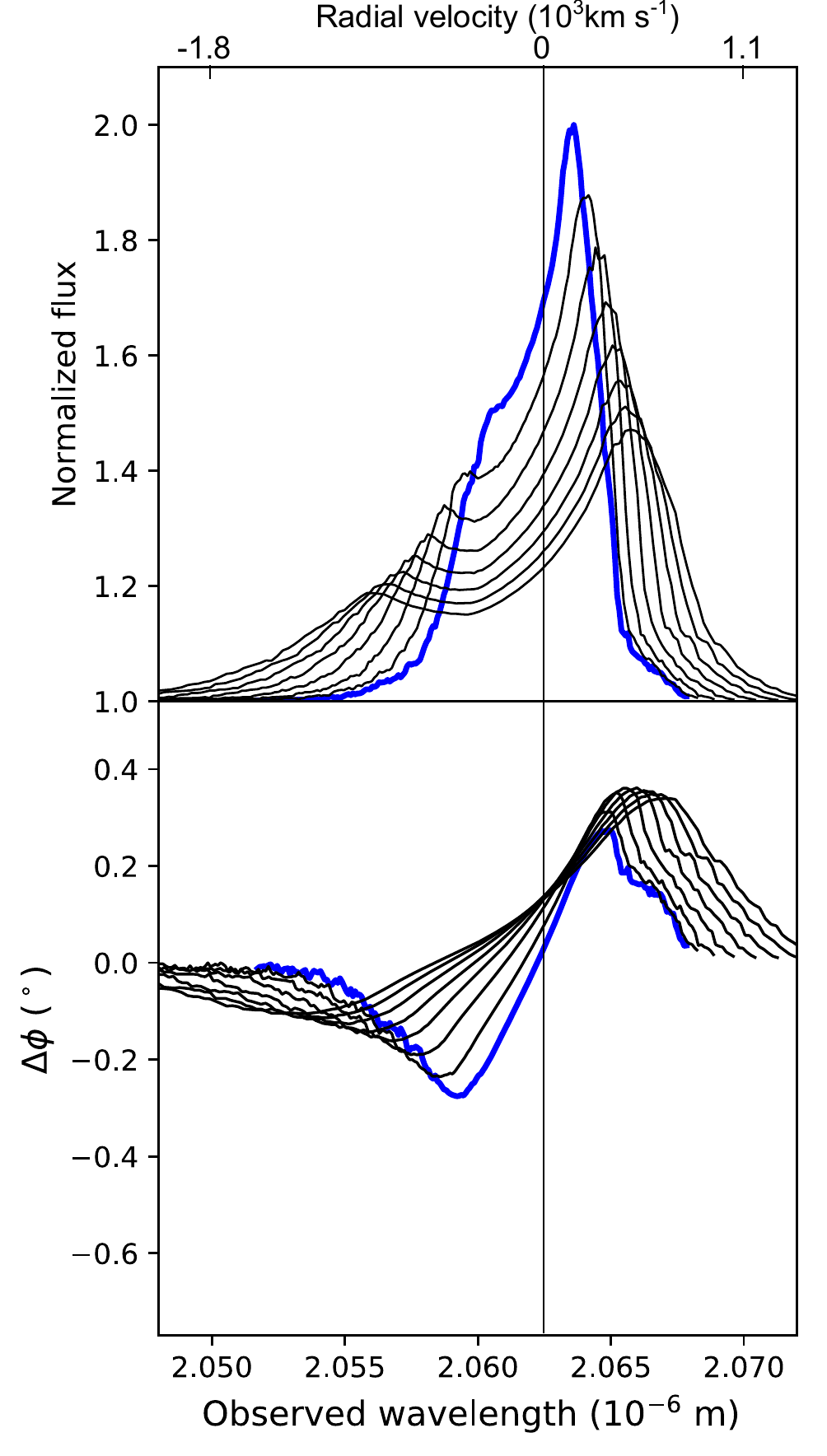}			
}

	\end{center}
\caption{%
Same as 	Fig. \ref {fig:singleelg1} but for random samples of clouds' orbital eccentricities generated from  $\Gamma_{s}  (0.3,1)$ and $\mathcal{R}(1)$ distributions. 
Varying parameters are listed in subcaptions,  $\mathcal{C}$ stands for coplanar clouds orbits.   
}%

\label{fig:singledistrg4}
\end{figure*}

All previous simulations assumed constant or homogenous distribution of clouds' orbital eccentricities.
 As a valuable study case, the novel  simulations  for non-homogenous  clouds' orbital elongation have been considered where some interesting effects affecting  differential phase  can occur.
The detailed distribution  of orbits is unknown   because clouds  have a wide distribution of masses. The elongation of these orbits changes with variation of clouds column densities:  high column densities clouds   allow for mildly elongated  even circular orbits (the right-skewed eccentricity distribution), but low column density clouds require highly eccentric ones  \citep[the left-skewed eccentricity distribution][]{10.1111/j.1365-2966.2010.17698.x}. 
 We can now construct these distributions, reflecting  assumptions about skewness based on clouds' column density, which will be used in our simulations.
Similar to the  variable defined for construction of a skewed random spatial distribution of clouds in BLR models \citep[see][and references therein]{10.1038/s41550-019-0979-5},  we sampled clouds orbital eccentricities  from the left-skewed distribution given as: 
\begin{equation}
 \mathcal{ E}=S \langle e\rangle+\frac{\langle e\rangle (1-S)}{\alpha}\Gamma(\alpha,\beta)
  \label{eq:skg}
  \end{equation}
    where 
  S is a random number from an uniform distribution between 0 and 1, $\langle e\rangle \sim 0.8$  is the mean value of the random variable,   and parameters of $\Gamma$ distributions are $\alpha=0.3, \beta=1$.
 The random variable in Eq.  \ref{eq:skg} is a shifted and scaled $\Gamma$ distribution, thus we will refer to it as $\Gamma_{s}$.
Randomly chosen  samples of eccentricities  increase steadily   with a peak around  $e\sim 0.65-0.8$. 

After  introducing the left-skewed eccentricity distribution, we  outline the right-skewed distribution   using Rayleigh distribution $\mathcal{R}(1)$
which probability density function is
\begin{equation}
\mathcal{R}(p)=\frac{x}{p^2}e^{\frac{-x^2}{2p^2}}
\label{eq:skg2}
\end{equation}

 In contrast to  $\Gamma_{s}$,  random samples of clouds' orbital eccentricities  drawn from $\mathcal{R}(1)$ have a peak at $e\sim 0.2-0.4$. 
  For each of these distributions of  eccentricities  and $\omega>180^{\circ}$ of non-coplanar clouds, we observed various broad spectral lines and differential phases (compare Fig. \ref{fig:singledistr41}  to Fig. \ref{fig:singledistr44}). Conversely,  larger differential phases amplitudes  are seen  for coplanar  orbital configurations of clouds.  Clouds' orbital  setups drawn from $\mathcal{R}$ yield  differential phases with larger amplitudes  than more elongated  orbital layouts (following shifted  $\Gamma_{s}$ distribution).
 But the effects of observer position $i_0$ and clouds' orbital inclinations are the same as   in the models with a homogenous distribution of clouds' eccentricities. 
 
Distinctly,  for models with clouds' orbital parameter $\omega\lesssim50^{\circ}$, the spectral lines  and differential phases are narrower for observer angles smaller than $30^{\circ}$ (see Fig. \ref{fig:singledistr44}).
We found  global geometrical pattern that  differential phases are deformed regarding an invariant point in its right wing for coplanar  orbital configurations  and different observer positional angle (compare Figs. \ref{fig:singledistr41} and  \ref{fig:singledistr44}).  For any value of wavelength  that is close enough to the wavelength of  the invariant point, the differential phase values are close to the fixed point. 

Based on our simulations, the shape of the differential phase  obtained from the emission of clouds on elongated orbits is distinctive from circular ones. The typical S-shape  signature of a rotating disk is clearly distorted. The peaks are deformed  due to the superposition of trigonometric functions of angles controlling the orbital shape  arising from the first term in the Eq. \ref{eq:complex8} (see also  Fig. \ref{fig:ilustris} upper plot). Also, an increasing cloud orbital inclination produces  differential phases with smaller amplitudes. These distortions are an illustrative proof of the presence of an asymmetry in the disk. \cite{10.1051/0004-6361/201116798} showed that losing simple S shape,  when stellar  environments  are dominated by rotation,   occurs for  larger baselines. 
 We summarize the simulation results in Table \ref{table:sumsingle}, which shows some general qualitative morphological characteristics  of the differential phase  for  various  simulation parameters.  Appendix \ref{appendix:atlas} provides detailed atlas of interferometric observables.

\begin{table*}
	\caption{Qualitative summary of simulation parameters effects on the morphology of differential phase (DP) curves   for a single SMBH. The columns are: orbital and auxiliary parameters, range or monotony of parameters used in simulations, the effect on DP amplitude, and DP slope between two peaks.  }             
	\label{table:sumsingle}      
	\centering          
	\begin{tabular}{c c c c  }     
		\hline\hline       
		Orbital & Range or monotony & Amplitude &  Slope\\ 
		element&       &                     \\
		\hline                    
	   (e,i,$\Omega$,$\omega$) & $(0.1-0.5),(5^{\circ}-40^{\circ}), >\frac{\pi}{2},  >\frac{\pi}{2} $ & decrease   &decrease   \\
	   	(e,coplanar orbits,$i_0$) &$(e\sim \Gamma_{s}  (0.3,1)\lor \mathcal{R}(1) ,\mathcal{C},(10^{\circ}-45^{\circ} ))$  & increase,fixed point in the right wing & increase \\
	   (e,non-coplanar orbits, $i_0$) &$(e\sim \Gamma(0.3,1)\lor \mathcal{R}(1),\lnot\mathcal{C}, (10^{\circ}-45^{\circ} ))$  & decrease, no fixed point  & decrease \\
	   $i$&increasing &decrease&decrease\\
	   	$\Omega$ & increasing  & increase &   \\
		$\omega$ & $ <\frac{\pi}{2}$ & prominent right wing &   \\
			$\omega$ & $ >\frac{\pi}{2}$ & prominent left wing &   \\
	
		\hline
	Auxiliary parameters	 & &  & \\  
				\hline               
				$i_0$& increasing & decrease    & decrease \\
					\hline
	\end{tabular}
\end{table*}

\subsection{ Interferometric signatures of CB-SMBH} \label{RD1}

\cite{10.1038/s41586-018-0731-9} assumed that all clouds have ordered motion when they measured the BLR size and black hole mass of 3C 273.  But  as pointed out by \cite{10.1088/0004-637X/804/1/57}, and shown for circular CB-SMBH by \cite{10.3847/1538-4357/ab2e00}, the phase curves could encode  the distribution of angular momentum of clouds in the BLR. If the GRAVITY provides the angular momentum distribution of clouds, then the clue for BLR formation will be found as well \citep{10.1038/s41550-017-0264-4}.
In our simulations,  all clouds of individual BLRs have their own either clockwise or anticlockwise rotation. Then the total angular momentum of the BLR  is the vector sum of the all clouds' orbital angular momenta.
The direction of the total BLRs angular momenta (either aligned or counter-aligned with the CB-SMBH system orbital angular momentum) has a strong effect on both interferometric observables  as shown in further paragraphs.
The orbital angular momentum of CB-SMBH is simply
$$\boldsymbol{J}_{bin}=\sum_{i=1.2} M_{i} \boldsymbol{V}^{i}_{\bullet} \times \boldsymbol{r}^{i}_{\bullet}$$
We define the orientation of the cloud orbits (with inclination $i_c$)  to the invariable plane of the  CB-SMBH system, which is perpendicular to the total orbital angular momentum  $\boldsymbol{J}_{bin}$.
 In the next we assume  the net angular momenta  of BLRs and the orbital angular momentum of CB-SMBH are independent of one another, so there are  cases of CB-SMBH with all angular momenta aligned or when one or both of BLR net angular momenta are misaligned. In this section we will consider two sets of results: for aligned and anti-aligned CB-SMBH. Before simulation, we first predict the pattern of differential phases from  first order approximation given in Appendix \ref{appendix:general2}.
Prediction surfaces  comprise a complex system of two ridges and valleys (see Fig. \ref{fig:ilustris2}).  Because of such topology,  deformed double S shapes are expected.
We have presented more detailed atlases in Appendices \ref{appendix:aligned} for aligned  and \ref{appendix:nonaligned} for anti-aligned CB-SMBH. 
\subsubsection{CB-SMBH system with aligned angular momenta}

\begin{figure*}[ht!]
	\begin{center}
	\subfigure[$i_{0}=45^{\circ}$;$i_{k}=\mathcal{U}(90^{\circ},10^{\circ}), \delta i_{k}=-10^{\circ},\newline \hspace*{1.5em} \Omega_{1}= 200^{\circ},   \Omega_{2}=  100^{\circ}, \omega_{k}=210^{\circ},  \newline \hspace*{1.5em} e_{k}=0.5, k=1,2$; $\Im= \mathcal{U} (5^{\circ},45^{\circ}),\newline \hspace*{1.5em} \delta \Im=5^{\circ},\Omega_{c}=200^{\circ}$, 
$\omega_{c1}= 150^{\circ}, \newline \hspace*{1.5em}\omega_{c2}=210^{\circ},  e_{c}=0.5$]
{%
	\label{fig:doubleel24}
	\includegraphics[trim=3mm 3mm 1mm 0mm, clip,width=0.3\textwidth]{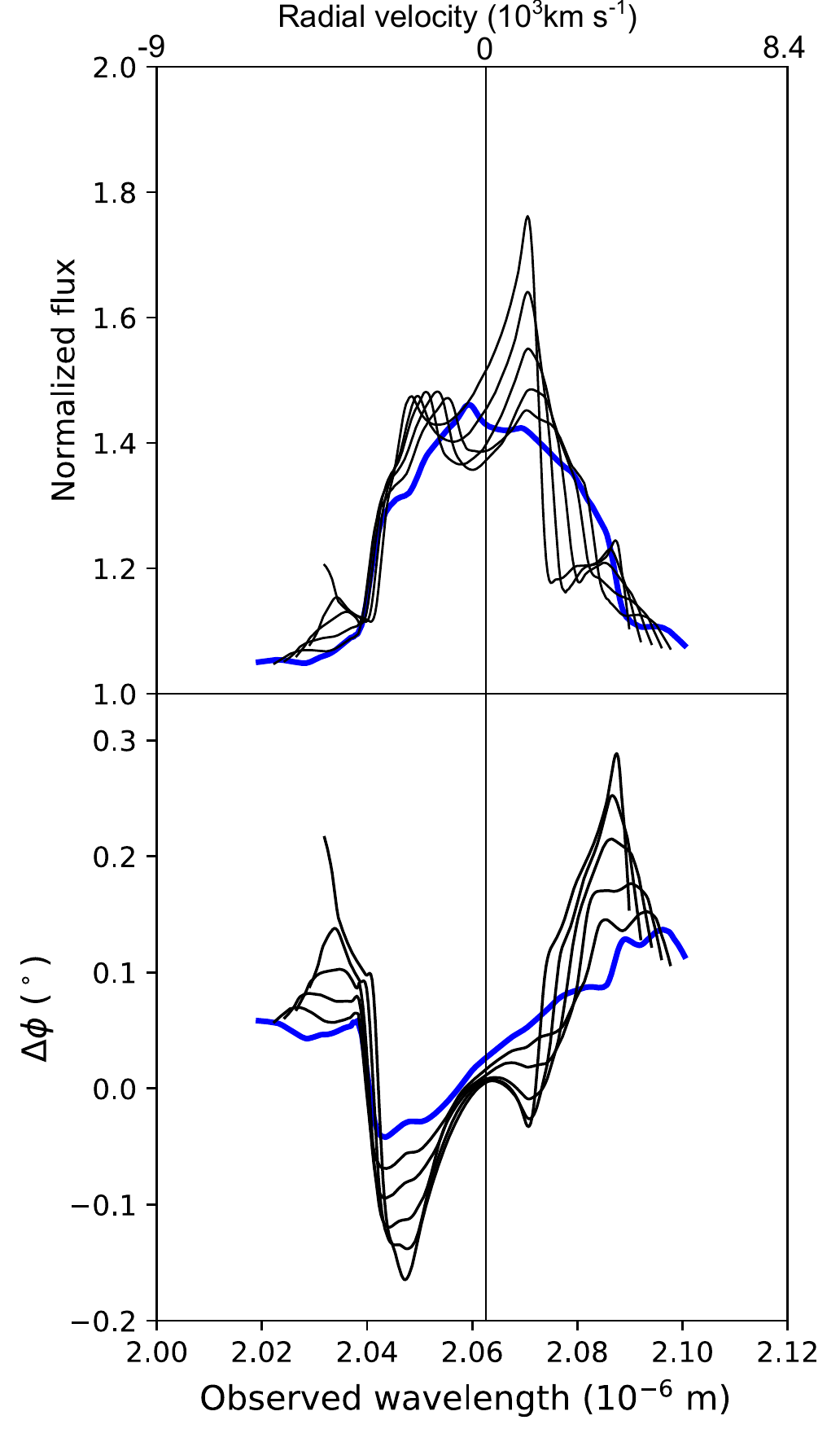}
}%
	\hspace{-0.8em}
\subfigure[$i_{0}=\mathcal{U}(15^{\circ},45^{\circ}), \delta i_{0}=5, \mathcal{C}, \Omega_{k}=\mathcal{U} (170^{\circ}, 230^{\circ}), \newline\hspace*{1.5em}\delta \Omega_{k}=20^{\circ} \omega_{k}=\mathcal{U}(270^{\circ}, 170^{\circ}),\delta \omega_{k}=-20^{\circ}, \newline\hspace*{1.5em}   e_{k}=(0.1,0.5),\delta e_{k}=0.1, k=1,2, i_{c}=\mathcal{U}(-5^{\circ},5^{\circ}),\newline\hspace*{1.5em} \Omega_{c}=\mathcal{U}(20^{\circ},160^{\circ}), \delta \Omega_{c}=20$, $\omega_{c}=\mathcal{U}(120^{\circ}, 20^{\circ}), \newline\hspace*{1.5em}  \delta \omega_{c}=20^{\circ}, e_{c}=e_{k}$
]{%
	\label{fig:doubleel36}
	\includegraphics[trim=3mm 0mm 1mm 0mm, clip,width=0.4\textwidth]{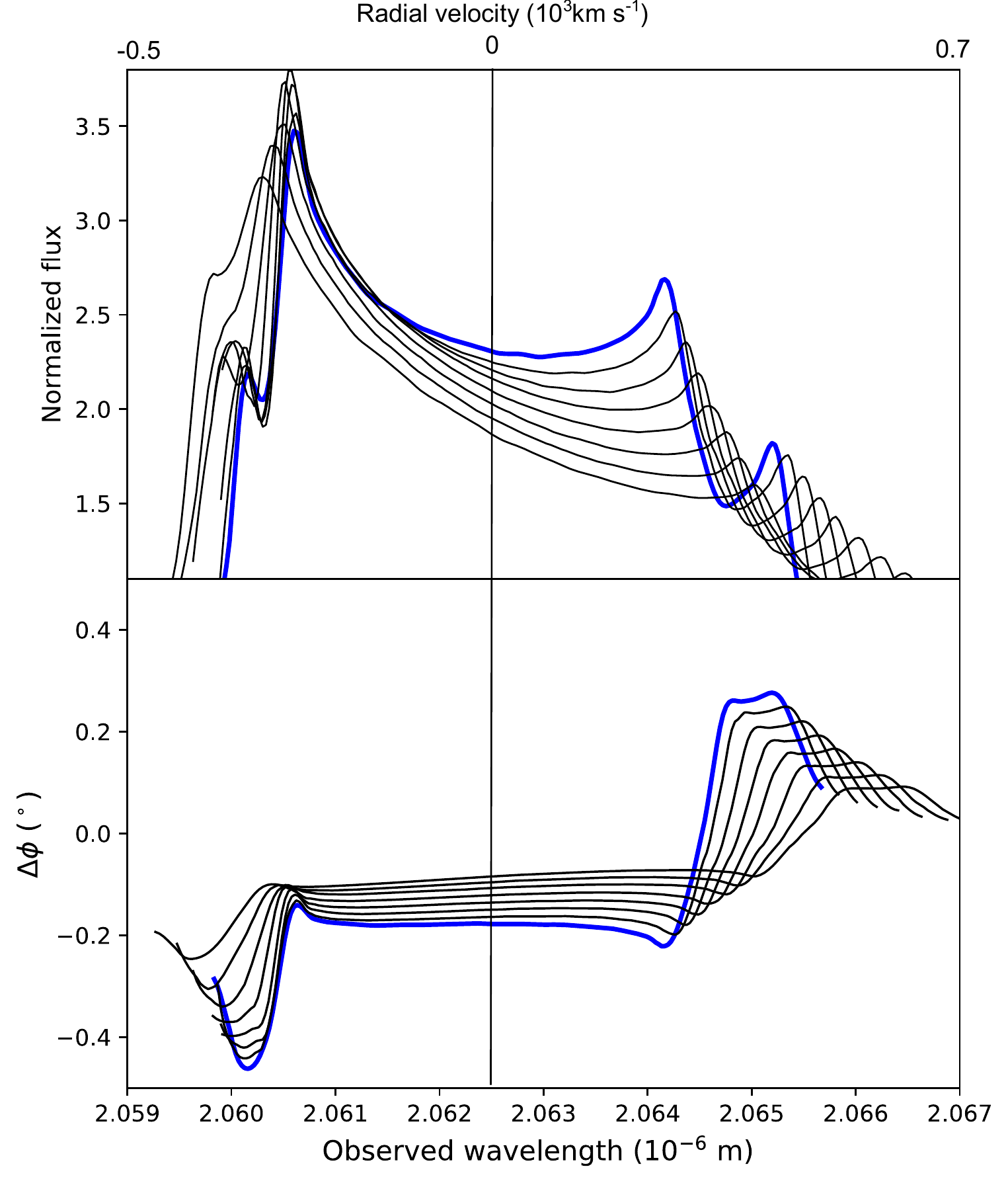}
}%
\vspace{-1.9em}
\end{center}
\caption{%
Evolution of the Pa$\alpha$ emission line (upper subplots) and corresponding differential  phase ($\Delta \phi$, lower subplots) as a function of the wavelength  for different values of model parameters for  aligned CB-SMBH system.  Orbital elements that are the same for all clouds in both BLR are designed by subscript c.
Varying parameters are listed in sub-captions, $\mathcal{C}$ stands for the coplanar  CB-SMBH system.
}%

\label{fig:doubleeln}
\end{figure*}

 In these simulations, we focused  on CB-SMBH systems which all clouds'  orbital angular momenta in both BLRs are aligned with the CB-SMBH system angular momentum $\boldsymbol{J}_{bin}\cdot\boldsymbol{J}_{ci}>0, i=1,2$. 
 This condition is equivalent to the condition that   clouds in both BLRs have  prograde  motion, i.e.  clouds' orbital inclinations are $i_{c}<90^{\circ}$.
 Fig. \ref{fig:doubleel24} shows some  typical  interferometry observables for  aligned CB-SMBH systems.  Even when the  the spectral lines  of both SMBH are well blended  in blue-coloured models in Fig. \ref{fig:doubleel24},    corresponding differential phases still show two peaks. Hence small velocity differences can be measured with differential interferometry \citep[see, e.g.][]{1992ESOC...39..425T}. 
The phases  are most sensitive to the asymmetric brightness distribution of the object.
Even if the spectra might have differed a very little  in peculiar aspects, the corresponding differential phases differ  clearly.
We found that there is a  distinction between differential phases `zoo' of a single SMBH and CB-SMBH systems.

\noindent Namely, the single SMBH phases `zoo' consists of deformed but still  recognizable S shape, which indicates the rotational motion of clouds in the BLR. The differential phases of CB-SMBH show a complex structured signal looking as two blended and deformed S-shaped signals. Differential phase variations are related to the orbital motions of clouds and SMBHs. These patterns are  asymmetrical about the line center.

 For fixed values of SMBHs and clouds' orbital eccentricities,  variations of the ranges of clouds orbital inclinations lead to an appearance of  ridges in the wings of differential phases 
(see Fig. \ref{fig:doubleel24}).
When the orbits of larger SMBH and clouds in its BLR have a smaller angle of pericenter, the plateaus between peaks of differential phases are less deformed (compare Fig. \ref{fig:doubleel24}.
We turn now to the effects in the differential phase  for  various positions of SMBHs orbits in the orbital plane. 
Simulations for different combinations of variations of parameters  revealed a wide range of distinct  patterns (Fig. \ref{fig:doubleel36}). For  simultaneous increase of observer angular position,  SMBHs and clouds'  orbital inclination,  we observed a  broadening  of  spectral lines and differential phases. Specifically, a plateau between peaks of differential phases flattens and broadens. The more dramatic change of position of differential phase wings is associated with the change of longitude of ascending nodes of SMBHs between $230^{\circ}-330^{^\circ}$ (Fig. \ref{fig:doubleel36}).

  \subsubsection{CB-SMBH system with antialigned angular momenta}

  Next  we look  for the effects of retrograde orbital motion of clouds in the  BLRs (i.e., $i_{c}>90^{\circ}$) or equivalently with a general class of systems that satisfy the condition $\boldsymbol{J}_{bin}\cdot\boldsymbol{J}_{ci}<0, i=1,2$.
Predicted deformations of the differential phase shows two whirls from the opposite sides of the ridge  specific for this kind of simulation (see the bottom left plot in Fig. \ref{fig:ilustris2}). The shape of  whirls  vary,  depending on the values of  the  SMBHs and clouds longitudes of node and pericenter.  More or less prominent additional deformation in the form of depression in the right wing of the surface appears (the bottom left plot in Fig. \ref{fig:ilustris2}). When this  deformation is prominent,  we can find specific double deformed S shape of differential phase as  in Fig. 	\ref{fig:doubleel5}.
  
\begin{figure*}[ht!]
    	\begin{center}
    	\subfigure[$\mathcal{C}, \Omega_{1}=  100^{\circ}, \Omega_{2}=  300^{\circ}, \omega_{1}=250^{\circ},\newline \hspace*{1.5em} \omega_{2}=70^{\circ},  e_{k}=0.5, k=1,2$; $i_{c1}= rnd(10^{\circ},\newline \hspace*{1.5em}45^{\circ}), i_{c2}=rnd(90^{\circ},175^{\circ}), \Omega_{c}=\omega_{c}=\newline \hspace*{1.5em}rnd(0.1^{\circ},359^{\circ}, e_{c}=0.5$]
    {%
    	\label{fig:doubleel44}
    	\includegraphics[trim = 2.0mm 4mm 2mm 0mm, clip, width=0.315\textwidth]{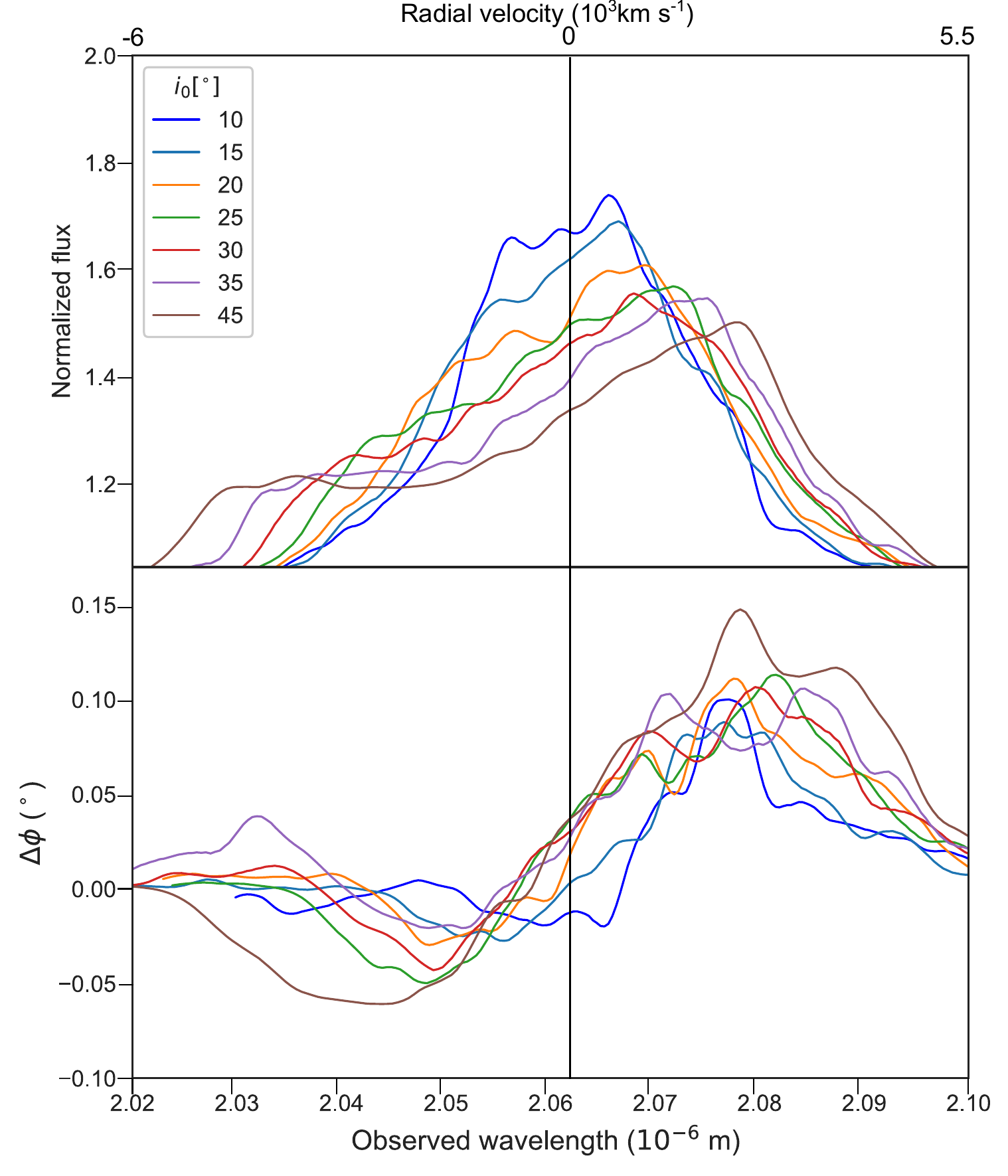}
    }%
    \hspace{-0.8em}
    \subfigure[$\mathcal{C}, \Omega_{1}=  100^{\circ}, \Omega_{2}=  300^{\circ}, \omega_{1}=250^{\circ},\newline \hspace*{1.5em}  \omega_{2}=70^{\circ},  e_{k}=0.5, k=1,2$; $i_{c1}=rnd(10^{\circ},\newline \hspace*{1.5em} 45^{\circ}),  i_{c2}=rnd(90^{\circ},175^{\circ}), \Omega_{c1}=300^{\circ},  \Omega_{c2}\newline \hspace*{1.5em}=100^{\circ},  \omega_{c1}= 170^{\circ},  \omega_{c2}=350^{\circ} \newline \hspace*{1.5em}  e_{c}=0.5$
    ]{%
    	\label{fig:doubleel45}
    	\includegraphics[trim = 0.0mm 0mm 0mm 0mm, clip,width=0.315\textwidth]{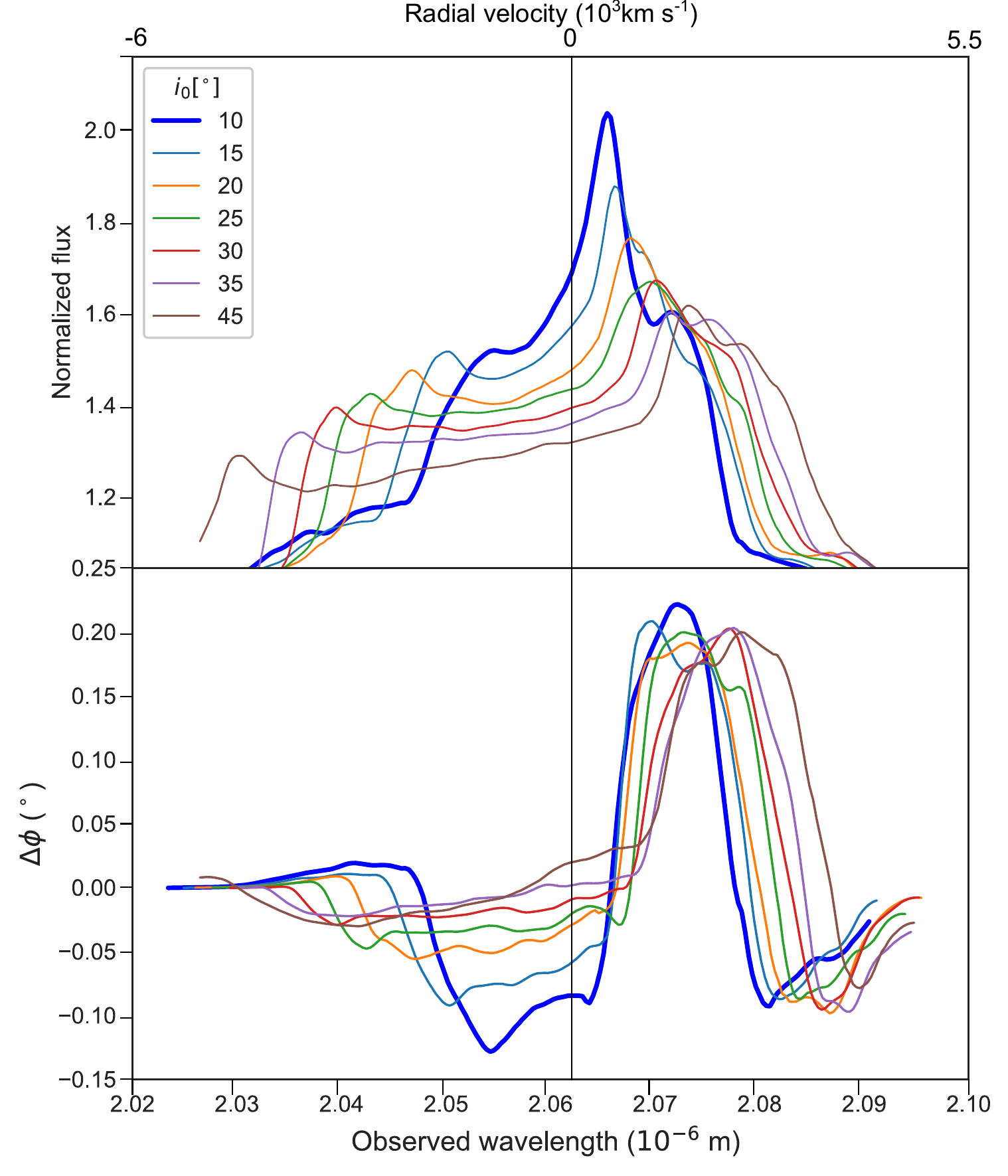}
    }
    \hspace{-0.8em}
    \subfigure[$\mathcal{C}, \Omega_{1}=  100^{\circ}, \Omega_{2}=  300^{\circ}, \omega_{1}=250^{\circ}, \newline \hspace*{1.5em}\omega_{2}=	
    70^{\circ}$; $i_{c1}=rnd(10^{\circ},45^{\circ}),  i_{c2}=,\newline \hspace*{1.5em} rnd(90^{\circ},175^{\circ}), \Omega_{c1}=300^{\circ},\newline \hspace*{1.5em}  \Omega_{c2}=100^{\circ},  \omega_{c1}=170^{\circ},  \omega_{c2}=350^{\circ}$
    ]{%
    	\label{fig:doubleel46}
    	\includegraphics[trim = 0.0mm 2mm 0mm 0mm, clip,width=0.315\textwidth]{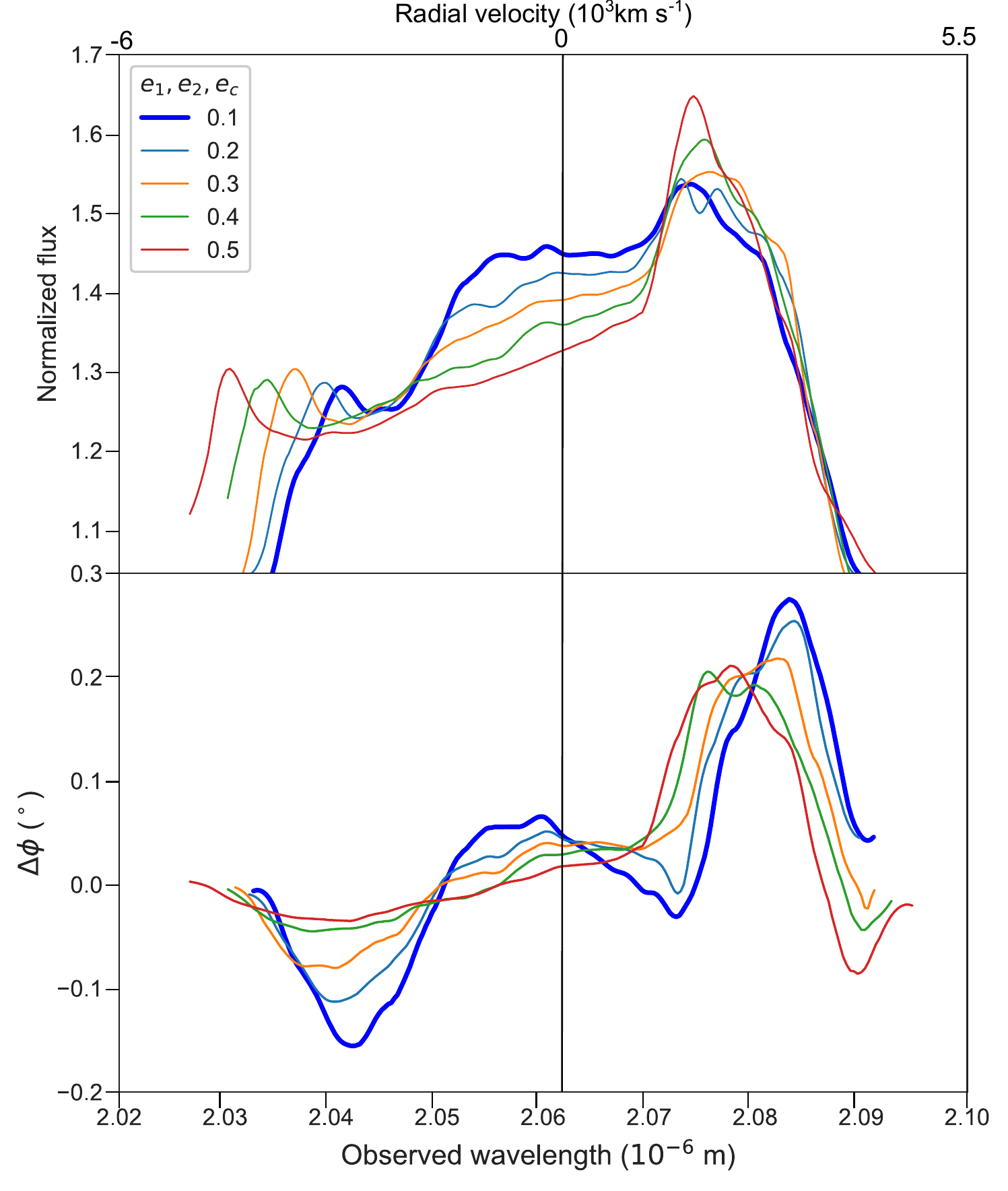}
    }%
    \vspace{-1.9em}
\end{center}
\caption{%
Same as Fig. \ref{fig:doubleeln} but for anti-aligned angular momenta of clouds in the BLR of less massive SMBH. 
}%

\label{fig:doubleelg4}
\end{figure*}

We  consider the case  when   the CB-SMBH are coplanar while clouds in the BLR of less massive component have anti-aligned angular momenta. The inclinations of orbits of clouds
  are randomly distributed between $90^{\circ}$ and $175^{\circ}$ (see Figs. \ref{fig:doubleel44}-\ref{fig:doubleel46}).   Similarly to consideration of angle distributions of AGN in Subsect. \ref{recapitulation}, one can expect more frequently highly inclined  clouds' orbits. This is a consequence  of the form of  distribution function for $i_c$  which is given by $P (i_c) di_{c} = \sin i_{c} di$. From this one can see that  the average $i_c$ is about $45^{\circ}$. Moreover,  {\it a priori} probability that $\sin i_{c} >\sin{45}^{\circ}$  is about 87 percent.

 The grid of models presented in Fig. \ref{fig:doubleel44}-\ref{fig:doubleel46} shows that if a more substantial  number parameters vary simultaneously, the shapes of differential phases will be more complicated.  We can see that the central whirl  has deformed wings
  because of larger  ascending node and pericenter  ($>\pi/2$) of SMBHs and clouds orbits  as well  for highly inclined clouds' orbital motion (see Figs. \ref{fig:doubleel44}-\ref{fig:doubleel46}).
  Based on the theory of orbital motion,  the variation of orbital nodes is influenced only by perturbations that are normal to the orbital plane. For example, in the case of a spinning SMBH, there may be rocket effect orthogonal to the orbital plane \citep[see][]{1989ComAp..14..165R}. 

 \begin{figure*}[ht!]
	\begin{center}
		
\subfigure[$\mathcal{C},\Omega_{k}=  100^{\circ}, k=1,2, \omega_{1}= 110^{\circ},\omega_{2}=290^{\circ}, i_{c1}=\newline\hspace*{1.5em}\mathcal{U}(-5^{\circ},5^{\circ}), i_{c2}=\mathcal{C},  \Omega_{c1}=200^{\circ}, \Omega_{c2}=10^{\circ}, \omega_{c1}=150^{\circ}, \newline\hspace*{1.5em}\omega_{c2}=330^{\circ},  e_{c1}=e_{c2}=rnd \mathcal{R}(1)$
]{%
	\label{fig:doubleel93}
	\includegraphics[trim = 2.mm 0mm 3mm 0mm, clip, width=0.4\textwidth]{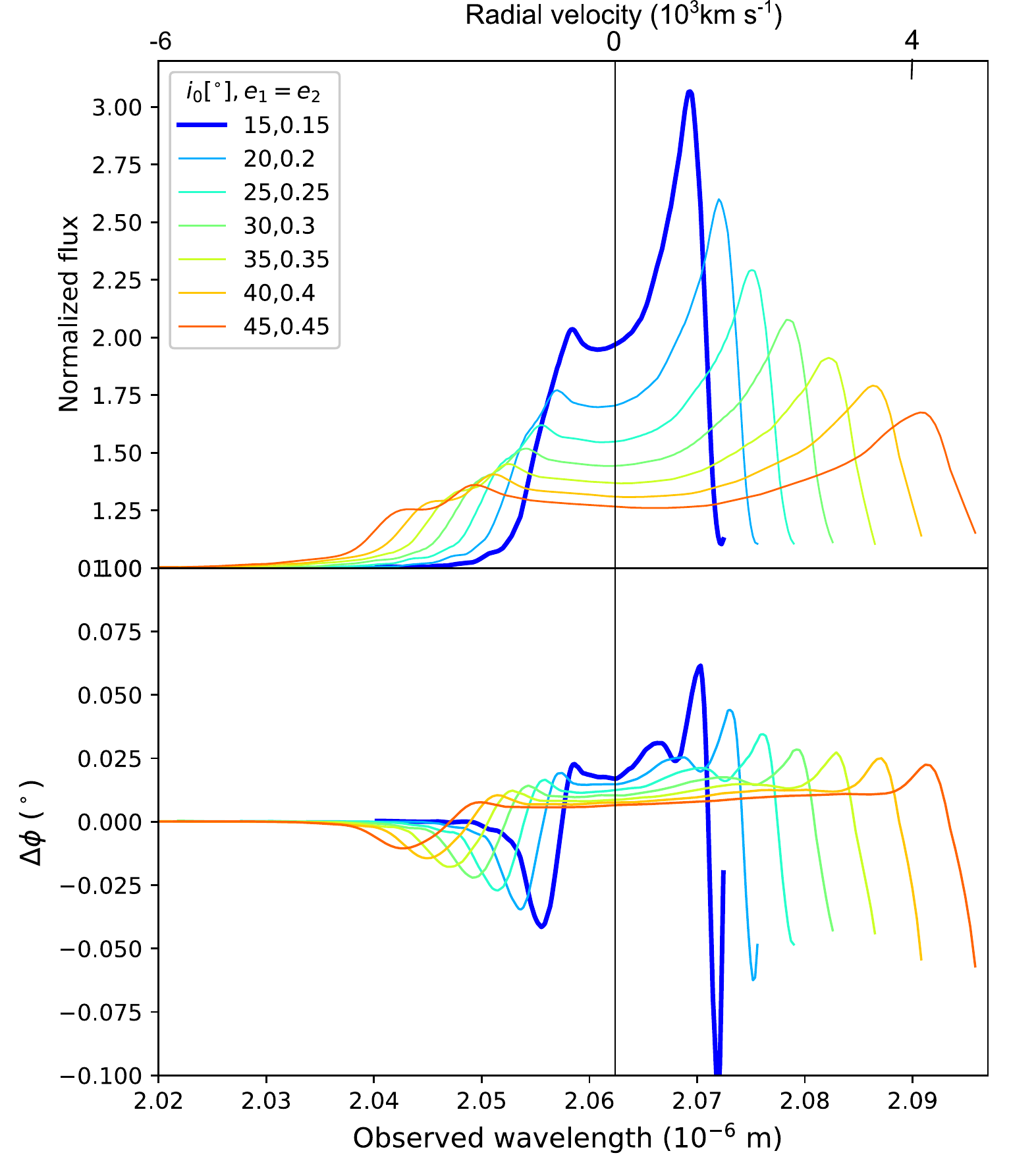}
	
}	
		\hspace{-0.3em}
	\subfigure[$\mathcal{C},\Omega_{k}=  100^{\circ}, k=1,2, \omega_{1}= 110^{\circ}, \omega_{2}=290^{\circ},   \Omega_{c1}=\newline\hspace*{1.5em} 200^{\circ}, \Omega_{c2}=10^{\circ}, \omega_{c1}=150^{\circ}, \omega_{c2}=330^{\circ}, i_{c1}=i_{c2}=\newline\hspace*{1.5em}\mathcal{U}(-5^{\circ},5^{\circ}), e_{c1}=0.5,e_{c2}=rnd \Gamma_{s}  (0.3,1)$
	]{%
		\label{fig:doubleel102}
		\includegraphics[trim = 2.0mm 0mm 0mm 0mm, clip,width=0.4\textwidth]{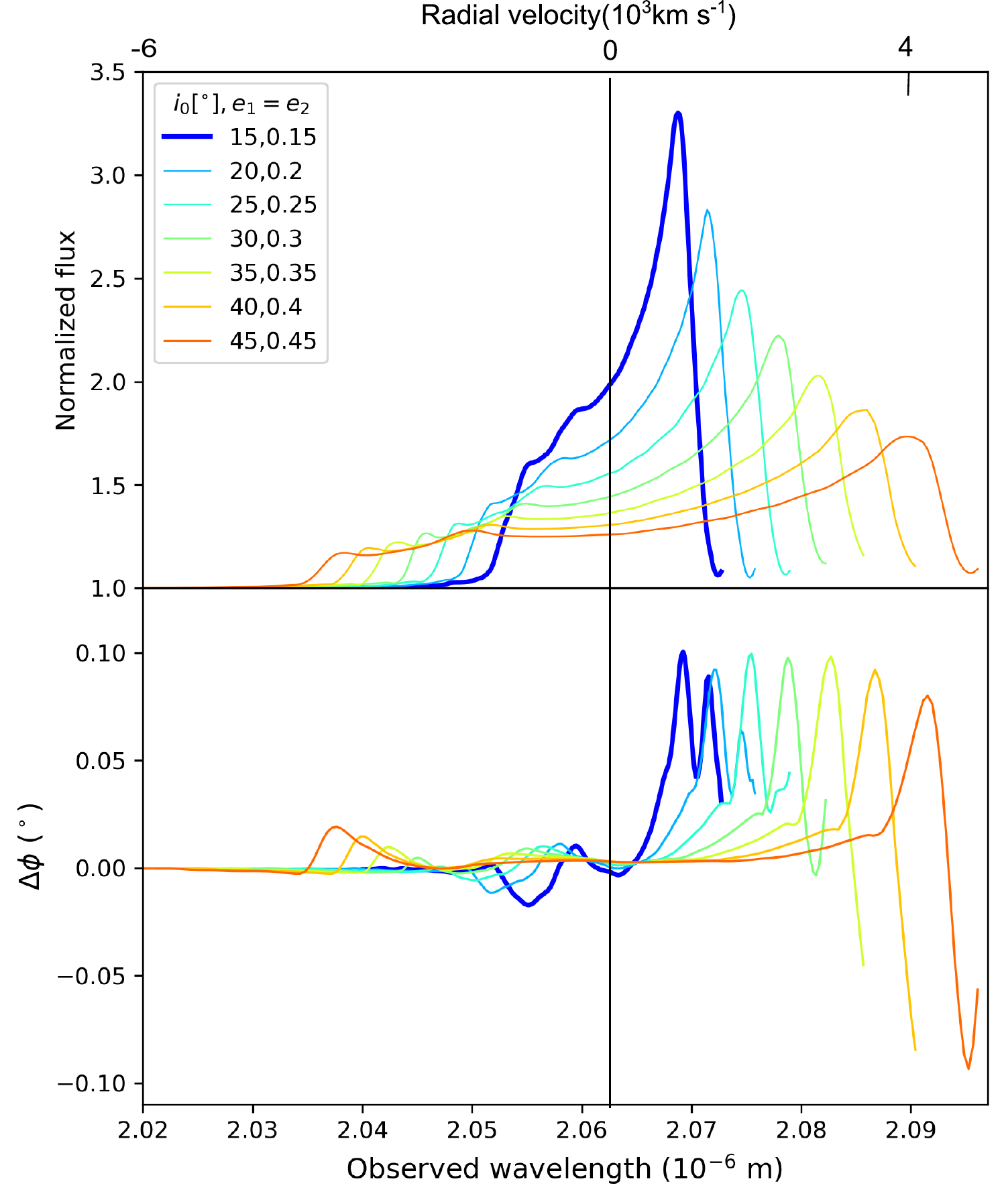}
	}
	
	\vspace{-0.5em}
\end{center}
\caption{%
Evolution of the Pa$\alpha$ emission line (upper subplots) and corresponding differential phase ($\Delta \phi$, lower subplots) as a function of the wavelength for aligned CB-SMBH system when clouds orbital eccentricities are drawn from non-uniform distributions. (a) clouds orbital eccentricites are drawn from Rayleigh distribution (b)  clouds orbital eccentricities are drawn from  scaled and shifted $\Gamma_{s}(0.3,1)$ distribution.
}%

\label{fig:doubleelnn}
\end{figure*}

 Similarly to simulations of non-uniformly elongated clouds' orbits in the case of single SMBH  (see )Subsection \ref{RD}), we   consider  the same skewed distributions ($\Gamma_s$ and Rayleigh $\mathcal{R}$)  of clouds' eccentricities in both BLRs of CB-SMBH system .
 A general prediction of $\mathcal{R}$ distribution effects on the differential phase over the grid of the ascending node and the true anomaly of more massive SMBH is given in the bottom right plot of Fig. \ref{fig:ilustris2}. We observe a rough and unsmooth surface because of the randomly sampled clouds orbital eccentricities from Rayleigh distribution.
 Detailed Fig. \ref{fig:doubleel93} summarizes simulations for different combinations of Rayleigh distribution of clouds' eccentricities and fixed SMBHs and clouds other orbital parameters. 
 We performed simulations separately for $\Gamma_{s}  (0.3,1)$  distribution of clouds orbital eccentricities. The differential phases are smooth  but with subtle differences to the previous case. 
 The central part of differential phase is depressed for coplanar clouds'  motion in both BLRs (see Fig.  \ref{fig:doubleel102}). 
  For fixed orbital eccentricities of clouds ($e=0.5$) around larger SMBH,  irrespective of the coplanarity of their orbital planes,  the right wing of differential phase is prominent (Figs. \ref{fig:doubleel102}).
  It seems that subtle distinctions between orbital characteristics of clouds'  motion could be important  rather than to assume they are broadly similar. 
 
 More extensive atlas of expected interferometric observables probing the elliptical configuration of anti-aligned CB-SMBH systems is given in Appendix \ref{appendix:nonaligned}.
 We summarize the results for the simulation in Table \ref{table:sumcb} that reports the effects of orbital parameters of the aligned and anti-aligned  CB-SMBH system on the differential phase curves. The overall impact on the differential phase is more significant than for the single SMBH case presented in  Table \ref{table:sumsingle}. We point out the following two effects.
Eccentricities of SMBHs orbits,  together with     SMBH orbital orientation angles, complicate differential phase curves so that even double S-like shapes can appear. 
 \begin{table*}
 \begin{footnotesize}
 	\caption{Qualitative summary of simulation parameters effects on the morphology of differential phase (DP) curves for CB-SMBH systems. The columns are: CB-SMBH and clouds' orbital and auxiliary parameters; range or monotony of parameters used in simulations; the effects on DP amplitude, its central part, wings, and slope,  respectively.   }             
 	\label{table:sumcb}      
 	\centering          
 	\begin{tabular}{c c c c c c}     
 		\hline\hline       
 		Orbital & Range or monotony & Amplitude &central part &  wings &slope\\ 
 		elements&       &     & &       &         \\
 		\hline 
 		($e_{1}, e_{2}$)&(increasing,increasing)&decreasing& &deforming&\\
 		($e_{c1}, e_{c2}$)&($\mathcal{R}(1)$, $\mathcal{R}(1)$ )&decreasing and noise& &deforming and noise&\\
 	 	($e_{c1}, e_{c2}, i_{ck}, k=1,2$)&(${\Gamma_{s}}(0.3,1)$, ${\Gamma_{s}  }(0.3,1), \mathcal{C}$)&smooth&depressed &smooth&\\
 		($e_{c1}, e_{c2}, i_{c1}, i_{c2}$)&(${\Gamma_{s}  }(0.3,1)$, ${\Gamma_{s}  }(0.3,1), \mathcal{C}, \lnot\mathcal{C}$)&smooth and decreasing& &smooth&\\
 	($e_{c1}, e_{c2},  k=1,2$)&(0.5, ${\Gamma_{s}  }(0.3,1), \mathcal{C}$)&smooth& &prominent right wing&\\
 		($\Omega_{k}$,$\omega_{k}, k=1,2$)&  $>\pi/2$&   & deforming& &\\                   
 		($i_{k}, k=1,2$) & increasing &      &    && decreasing\\
 		($i_{ck}, k=1,2$) & increasing & $(0^{\circ}, 45^{\circ})$ &    && decreasing\\
 	    $\boldsymbol{J}^{c}_{k}, k=1,2$ & $\boldsymbol{J}_{bin}\cdot \boldsymbol{J}^{c}_{k}<0$\tablefootmark{a}&  & central whirl  &  & \\
 	 		
 		\hline
 		Auxiliary parameters	 & &  & \\  
 		\hline               
 		$i_0$& increasing & decrease    & decrease \\
 		\hline
 	\end{tabular}
 \tablefoot{
 \tablefoottext{a}{ Conditions:  $\boldsymbol{J}_{bin}\cdot \boldsymbol{J}^{c}_{k}  >0$ is equivalent to clouds orbital inclinations less then $90^{\circ}$; $\boldsymbol{J}_{bin}\cdot \boldsymbol{J}^{c}_{k}  <0$ is equivalent to clouds orbital inclinations larger then $90^{\circ}$.}}
\end{footnotesize}
 \end{table*}

   \section{Discussion}\label{discuss}

In this work, we probed the detection of interferometric signatures of single SMBH and CB-SMBH systems with elliptical orbital configurations. The differential phase  was computed at the wavelength of Pa$\alpha$ spectral line.
As expected, these signatures in differential phases occur under  specific conditions, and their diagnostic potential is more evident for some parameters than for others. To quantify this, we calculated the evolution of the spectral line and differential phase as a function of wavelength  and radial velocity for distinct sets of models.  Further discussion focuses on three points:  anticipated results for specific objects-which can be crucial to observers, the limitations of the present model in the light of selections of future observational targets  and   the information loss   because of random cloud’s motion.

\subsection{Anticipated results for future observations}

The results of simulations also show interesting qualitative similarities of some synthetic  Pa$\alpha$ lines with those observed and reported in the literature. The corresponding differential phases   might be used as a starting point to predict what would be the interferometric signatures  of these specific targets.

{\it Type 1 AGN and single SMBH models expected signatures}  With single SMBH models,  more or less asymmetric synthetic flat top lines given in Fig. \ref{fig:singleel12} and \ref{fig:singleel13}
show   similar gradients and shapes with  flat top lines reported in the literature by \citet[][see their Fig. 2]{10.1093/mnras/stu031} for IRAS 1750-508, PDS 456, PG 0026$+$129, Mrk 335. These objects are classified as type 1 AGN. Their expected differential phases  could be  of S-form for single SMBH, but with asymmetric  amplitudes and widths  of peaks implicating elongated cloud' orbits. 
  Furthermore,  the sharpness of the differential phases peaks became more prominent as   
  $\Omega_{c}$  or $\omega_c$ is increased.
  The spectral line  in Fig.  \ref{fig:singleel24} resembles Pa$\alpha$  line of SDSSJ05530.0-0.85704.0 \citep {10.1088/0004-637X/724/1/386}, which is also type 1 AGN. Here anticipated differential phase would be broader as inclination position of the observer is larger, the sharpness of differential phases is more prominent than in the previous case, because of larger $\omega_{c}=200^{\circ}$ and slightly more inclined cloud orbits. 
  Similarly, in the third series of plots  of  Fig. \ref{fig:singleel3}, the evolution of observables is peculiar. Here, models keep orbital eccentricity of clouds  constant but vary the inclination angle from $i=5^{\circ}$  to $i = 40^{\circ}$. We see that amplitude changes of phases are prominent and the position of maxima of phases, which gives an additional  tool for interpreting the observations.
  
The blue lines reported in  Figs. \ref{fig:singleel11}-\ref{fig:singleel13}, \ref{fig:singleel21} and \ref{fig:singleel31} show  the profiles in the waveform  with a more prominent left peak. A comparable pattern was found in PG 0844$+$349 \citep[][see their Fig. A2]{10.1093/mnras/stu031}. It suggests that predicted differential phases are deformed,  when either  $\Omega_c$ or $\omega_c$  reaches minimal value  of $\sim 10^{\circ}$. The width of the phase  increases with increase in clouds' orbital inclinations.
We find that at this wavelength, there is a clear signal in the differential phase for different parameters.
In general, single SMBH produces differential phases that are more or less symmetric relative to the core of spectral lines depending on the excitation of eccentricity and parameters controlling the position of clouds orbit.  

{\it Type 1 AGN (NLSy1, quasars) and aligned CB-SMBH models expected signatures}

Two SMBH and their BLRs induce more rich and complex differential phase patterns.  There are many configurations for which the aligned CB-SMBH  differs among themselves and to single SMBH.
Features of synthetic spectra given in Figs. \ref{fig:doubleel22} and  \ref{fig:doubleel23} show similarity with those found in 3C 390.3 \citep[][see their Fig. A1]{10.1093/mnras/stu031}. This object is well known as a double-peaked line emitter in the optical band.
The double-peaked profiles can be also   associated with accretion disc emission \citep{10.1086/191856, 10.1086/379540, 10.1086/511032}. But, if the binary model is appropriate for this object, then associated differential phases will have a complex double S-shaped structure.   This model reflect the possibility of a non-coplanar CB-SMBH, with highly inclined orbits of clouds in both BLRs.   If  a single SMBH model with a BLR d of 95 light days  is true  \citep[see][]{2010A&A...517A..42S}  a differential signal would be $0.9 ^{\circ}$.
Even that  spectral lines of objects can share some characteristics, our model predicts that corresponding differential phases are distinct because of different SMBH and clouds orbital parameters.
The optically bright quasar PG 1211$+$143 \citep{10.1093/mnras/stu031} has a convex Pa$\alpha$ shape slightly depressed of  the center, as in our synthetic case present in {\tiny }Fig. \ref{fig:doubleel21}. The predicted differential phase would resemble asymmetric double S shape, caused by non-coplanarity of the CB-SMBH system and high values of clouds orbital elements $\Omega_c$ and $\omega_c$.
Moreover, the high rise spectral line in Fig. \ref{fig:doubleel24} is also observed in the spectrum of SDSSJ032213.8+005513.4  \citep[][see their Fig. 1] {10.1088/0004-637X/724/1/386}. If the CB-SMBH model is appropriate for this object, then the expected differential phase would be similar in morphology with the previous case but  distinct in their details  because of different values of $\Omega_2$.
In addition, asymmetric two horn  feature (blue line)  found in Fig. \ref{fig:doubleel35},  is also highly consistent with observed line  in Mrk 79 \citep[][see their Fig. 13]{10.1086/522373}. Anticipated differential phases  differ from those seen in previous cases  because of large values $230^{\circ}<\Omega_{1},\Omega_{2}<330^{\circ}$ and the decreasing   $\omega_{k},k=1,2$.

{\it Type 1 AGN (binary black hole candidates, quasars) and anti-aligned CB-SMBH models expected signatures}

 Significant changes in differential phases shape  for anti-aligned CB-SMBH, are given in Fig. \ref{fig:doubleel44}.
The spectral line marked as  a blue  curve in Fig. \ref{fig:doubleel45} is  like  Pa$\alpha$ line  in SDSSJ01530.0-085704.0 \citep[see Fig. 2 in][]{10.1088/0004-637X/724/1/386},   and in NGC 7469 \citep[see Fig. 13 in ][black line]{10.1086/522373}.  The expected differential phase for these two objects would have complex double S shape with prominent wings, which is a consequence of anti-aligned orbital angular momenta of clouds  in the second BLR and randomized $\Omega_c$ of clouds orbits in both BLRs.  Even double  peaks in  simulated spectral line merge for larger inclinations of the CB-SMBH system, differential phases preserve more or less deformed double S shape.
In the core of  NGC 7469, there is a dust torus at a distance of 65-87 light days from central engine, based on K-band time lags \citep{10.1086/499326}. Thus, expected differential phases could be of the order of $2.17^{\circ}-3.04^{\circ}$. 
However, the peculiar spectral line (blue model) in Fig. \ref{fig:doubleel46} with convex core resembles  Pa$\alpha$ line observed in the binary black hole candidate NGC 4151 \citep[see Fig. 4 in][]{10.1086/522373}. Here, the expected differential phase has a smaller central part and prominent right wing,  caused  by anti-aligned orbital angular momenta of clouds  in the second BLR.  But $\Omega_c$  of cloud orbits in both BLRs are not randomized. This object, located at 19 Mpc,  is among the nearest galaxies that  contains an actively growing black hole \citep{10.1088/2041-8205/719/2/L208}.  Its proximity and claimed evidence for the first spectroscopically resolved sub-parsec  CB-SMBH \citep{2012ApJ...759..118B}  offer one of the best chances also for interferometric studies.  \cite{2012ApJ...759..118B}  found CB-SMBH to have  eccentric orbit $e\sim 0.4$ with period of 15.2 years, argument of longitude of pericenter $\omega\sim 95^{\circ}$, semi-major axes of $(a_{1}+a_{2}) \sin i\sim 0.002+0.008$ pc,  and masses of components $M_{1}\sim 3 \times 10^{7} M{\odot}$ , $M_{2}\sim 8.5 \times10^{6} M{\odot}$. 

These results imply that the expected signal in the differential phase would be of the order $\sim 0.5^{\circ}$ using Eq. \ref{eq:complex6}.
 CB-SMBH with a small orbital period of the order of 1 yr, as discussed in the next subsection \ref{limit}, we will be able to observe them at different orbital phases. Fig. \ref{fig:doubleel8}  shows a set of models depicting  interferometric observables calculated at three different  CB-SMBH orbital phases $\Theta(t)=0,0.25,0.5$ where $\Theta(t)=\frac{t-\tau}{P}$, $t$ is time, $\tau$ is time of pericenter passage and $P$ is a binary period. Interferometric observables appeared to be unfavorable in all the cases. However, careful inspection shows that interferometric signal varies with orbital phase (i.e., time). These  synthetic spectral lines look like those observed in PG 1307+085, PG 0804+761, PG 1211+143, PG 0026+129 \citep{10.1093/mnras/stu031}.
In summary,  we show that using differential phases morphology  can help us distinguish  between models. Though some spectral lines of single SMBH can be morphologically indistinguishable from  those in CB-SMBH, their differential phases differ in various manners, and they are likely to be one of the essential tools for CB-SMBH detection.

\subsection{Limitation of present model}\label{limit}

 In our model  we assumed  that  the  CB-SMBH orbital period  is longer than the dynamical timescale of each BLR, which holds for separated  BLRs. In opposite  the orbital motion of SMBHs will perturb the BLRs, complicating  line profiles  and  phase curves.  For example, periodic flares in the light curves of OJ 287 have been successfully predicted by the model of the secondary impacting the accretion disc of the primary SMBH, twice during one orbit \citep{10.3847/2041-8213/ab79a4}. 

We assumed the physical parameters of the BLRs emissivity and clouds motion as an unaltered if the orbital period is longer than observational time-baseline. But, if the orbital period of the binary system is of the order of  a time baseline of observation, e.g.  $\sim$ 1 year orbital period   in the photometric light curve of  Mrk 231  \citep{10.1093/mnras/staa737},  then these quantities can differ from being invariant.

 If the CB-SMBH merged to a later phase, a  viable way of overcoming the 'final parsec problem' is interaction with a disc surrounding  binary  \citep[see, e.g.][]{10.1086/497108/pdf,doi.org/10.1086/523869}. Dissipative torques  could align the circumbinary disc plane  with the binary orbital plane  while  circumbinary disc is rotating in the same sense as the binary (aligned angular momenta).  Detailed simulations \citep[see, e.g.][]{10.1086/588837}  showed that  such triple-disc CB-SMBH systems have   tidally deformed  BLR discs  by the time-varying binary potential because of  the orbital eccentricity.
Also \citet{10.1111/j.1365-2966.2009.15179.x}  show that  tidal interaction  could transform circumbinary disc  into a decretion  regime, which removes  angular momentum outward, but with small  inward mass transfer.  This could prevent the binary separation from reducing to the  $10^{-2} \mathrm{pc}$.
However, it is highly likely that at some point, there will be a retrograde coplanar disc surrounding the binary,  remaining in accretion mode.   Material from such disc is gravitationally captured by the binary while  reducing its angular momentum.   Then the  eccentricity  threshold between the circular and non-circular binary system is determined by the surface density distribution of the circumbinary disc \citep[see e.g.][and references therein]{doi.org/10.1111/j.1365-2966.2010.17952.x}.  \cite{10.1093/mnras/staa1985} explained recent   ALMA  observation of counter-rotating disc surrounding the core of NGC 1068   as a binary system with a counter-rotating circumnuclear disc. Also, ALMA observed a gas hole in the core of  NGC 1365, NGC 1566, and NGC 1672  \citep{10.1051/0004-6361/201834560}.  Similar  observations  are  expected to resolve the signatures of counter-rotating circumnuclear discs.

Also,  we  supposed  both components in CB-SMBH to have  same a non-negligible Eddington ratio ($\lambda_{Edd}\sim 0.1$). However, if  luminosity variations are much smaller 
( $\lambda_{Edd}\lesssim 0.01$), then  we can expect a small periodically varying flux component with an amplitude  proportional to the black hole mass $\mathbf{M_{\bullet}}$ \citep{10.1088/0004-637X/700/2/1952}
$$\sim \frac{\lambda_{Edd}}{0.01}\left(\frac{M_{\bullet}}{3\cdot 10^{7} M_{\odot}}\right)\times10^{-15}\mathrm{erg\, s^{-1} cm^{-2}},$$ so that  object would be too faint for analysis.

Regarding  near-infrared observables, they  might be detectable if a dusty torus surrounds the binary core. For example, geometrically thick BLR of single SMBH in  3C 273 detected by the GRAVITY is roughly consistent with dusty torus  \citep{doi.org/10.3847/1538-4357/ab1099}.  Thus torus inner edge would be irradiated by, and thus mirror, the variable central UV and X-ray sources.  

Estimated error of the differential phase ($\delta \sigma_{\phi}$),  can be obtained from  Eq. \ref{eq:complex6}
as a function of  errors on the fluxes of both components ($\delta F_{i}, i=1,2$)  and  $\delta \sigma_{i},i=1,2$ (Eq. \ref{eq:posled}):
\begin{equation}
\delta \sigma_{\phi}\sim \frac{\frac{\delta F_1}{F_1}\sigma_{1}+\delta\sigma_{1}-\frac{\delta F_2}{F_1}\sigma_{2}+\frac{F_2}{F1}\delta\sigma_{2}}{1+\frac{F_2}{F_1}}
\label{eq:greska}
\end{equation}

To estimate  contribution of terms in the error budget, we first consider that larger component is dominant in CB-SMBH  $\frac{F_2}{F_1}<<1$ implying that also  $\frac{\delta F_2}{F_1}<<1$. Then Eq.\ref{eq:greska} simplifies to 
\begin{equation}
\delta \sigma_{\phi}\sim  \frac{\delta F_1}{F_1}\sigma_{1}+\delta\sigma_{1}
\end{equation}
Taking into account that  $ \frac{\delta F_1}{F_1}\sim 0.05 (5\%)$,  the angular scale of nearby  CB-SBHBs is $\sim o(10 \mu as)$, measured photocenter offset by GRAVITY,  and that $\delta\sigma_{1}\sim o(1\mu a as)$ is of the order of precision of GRAVITY  photocenter offset  measurement of 3C 273, we get
 $$\delta  \sigma_{\phi}\sim 0.05\sigma_{1}+ o(1\mu as)\sim 0.05 o(10 \mu as) + o(1 \mu as)$$.  We see that differential phase error  is dominated by photocenter offset measurement of the more massive component. However, if  $\frac{F_2}{F_1}\sim1$ then $$\delta  \sigma_{\phi}\sim 0.025(\sigma_{1}-\sigma_{2}) +\frac{\delta(\sigma_{1}-\sigma_{2})}{2}\sim 0.025o(10 \mu as) + 0.5o(1 \mu as) $$, which is smaller  than   previous estimate. But we  did not consider  the sources of noise  such as the photon noise, the background thermal noise and the detector readout noise.
A crude analysis of Eq. \ref{eq:complex6} suggests that the cadence of fluxes of both BLRs and photocenter offset measurements are crucial for a well-sampled differential phase curve.
Also, we assumed that the object wavelength dependence is only  because of the variation of the fluxes of both  BLRs.  If the object intensity distribution depends on the wavelength in a more sophisticated way, the changes of  the phase  arise because  of effects from the spatial frequency variation and the wavelength dependency of the object. 

\subsection{Differential phase loss of information}
Interferometric signals decorrelate (become incoherent) due to either instrumental and/or physical characteristics of observed object such as random motion of clouds in the BLR \citep[see analysis in][]{doi.org/10.3847/1538-4357/ab3c5e}.  The differential phase curve becomes sensitive to the numbers of the clockwise and counter-clockwise rotating clouds in the BLR, which is essential for GRAVITY measurements of the BLR.

 Information loss can be explained by moments  introduced in Subsection \ref{difp}. Namely, if  the velocity distributions of clockwise and counter-clockwise moving clouds resemble roughly  $N(0, \sigma^{2})$, then the differential phase as the first-order moment  (barycenter)  will be zero. In turn, the second-order moment of such distribution will be $\sim \sigma^{2}\neq0$. If the disordered motion of clouds in the BLRs  comes from turbulence  \citep[see][]{10.1093/mnras/stu1809} then  information recovered from the second-order moment would correspond to the parameter of distribution of turbulence velocity field $\sim \sigma^{2}=\sigma^{2}_{turb}$. Thus higher-order moments  can contain interferometric information.

\section{Conclusions}{\label{conc}}

We used the interferometry-oriented model described here, to estimate signatures of the elliptical orbital motion of clouds in the BLR of a single SMBH and elliptical orbital setup of CB-SMBH  on spectro-interferometric observables. Besides,  we investigate how these observables evolve with a variety of different combinations of orbital and geometrical parameters. 
Important findings and conclusions are summarized as follows:
\begin{enumerate}
\item{The differential phases 'zoo'  for a single SMBH  comprises  deformed but still recognizable S shape, indicating rotational and  asymetrical motion of clouds in BLR. We showed the evolution of differential phase shape, amplitude and slope with various sets of cloud orbital parameters and observer position. These maps can be useful  for extracting exceptional features of the BLR structure from future high-resolution observations. }

\item{ We found various deviations from the canonical S-shaped phases profile for elliptically configured CB-SMBH systems.  There are notable differences between differential phases 'zoo' of single SMBH and CB-SMBH systems. The differential phases of CB-SMBH look as  two blended and deformed S-shaped signals, asymmetrical about line center, which variability depend on the orbital motion of clouds and SMBHs.}
 \item{The shape and amplitude of the phases of CB-SMBH systems depend on presumably orbital characteristics of SMBHs and clouds in their BLRs.
 	Among the many sets of model parameters explored, we found that the signal is sensitive to the position of orbital nodes, inclinations, eccentricities, and argument of pericenter along with standardly expected effects of geometrical inclination of an observer. The right-skewed distributions of clouds orbital eccentricities cause noise effects in the form of small random fluctuations in the differential phase curve. 

We found some examples of synthetic spectral lines of a single SMBH,  which are indistinguishable from those obtained from the CB-SMBH system, but with differing differential phases. Thus, the differential phases are markers for identifying signals of CB-SMBH. }
\item{Observationally, the variability of the differential phase is most substantial for lower inclinations of an observer. As much as the central part of the spectral lines is disfigured, the net effect is that the differential phase peaks move away from the line center. The plateau between differential phase peaks is more prominent. The opposite is valid when there are higher contributions of projected lower velocities in spectral lines.
	The reversed situation is occurring when  line peaks are closer together,  then differential phase peaks move closer to the center of the line.   }
\item{ The velocity distributions for anti-aligned clouds' orbital momenta  in CB-SMBH show elongated features that are strained by the surface of positions of clouds.   Velocity fields manifest in the closed surface, preserving topological volume and spatial coherency. We also tested the cases of randomly distributed inclinations, but the velocity fields of such BLRs are not volume-preserving in the topological sense. For synchronous alignment of angular momenta of the  BLR clouds, the absolute value of clouds' velocities increases toward outer side lobes of disc-like BLRs.  For  anti-aligned BLRs, the absolute value of velocity increases toward sections close to apocenter and pericenter.}

\end{enumerate}

\begin{acknowledgements}
 The authors gratefully acknowledge valuable comments and suggestions  of the Reviewer, which helped us to improve and clarify our work.
      AK and L{\v C}P  are supported by Ministry of Education, Science and Technological development of Republic Serbia through the  \emph{ Astrophysical Spectroscopy of Extragalactic Objects} project number 176001. 
      JMW and YYS are supported by  \emph{National Key R\&D Program of China}
      through grant 2016YFA0400701, by NSFC through grants NSFC-11991050, -11873048, and
      by
      Grant QYZD-SSW-SLH007 from the \emph{Key Research Program of Frontier Sciences, CAS}, by
      \emph{the Strategic Priority Research Program of the Chinese Academy of Sciences} grant no.
      XDB23010400.

\end{acknowledgements}

%

\begin{thebibliography}{}



\bibitem[Gravity Collaboration et al. (2017)] {10.1051/0004-6361/201730838}
Gravity Collaboration,  Abuter, R.,  Accardo, M.,  Amorim, A.,  Anugu, N.,  {\' A}vila, G. et al., 2017, A\&A, 602, id. A94


\bibitem[Akiyama et al. (2019)]
{10.3847/2041-8213/ab0ec7 }
Event Horizon Telescope collaboration Akiyama K., et al. 2019, ApJL, 875, L1

\bibitem[Armitage and Natarajan (2005)]{10.1086/497108/pdf}
Armitage P. J. Natarajan P. 2005, ApJ,634, 921




\bibitem[Artymowicz  and Lubow (1996)]
{10.1086/310200}
Artymowicz, P., Lubow, S. H. 1996, ApJ, 467, L77


\bibitem[Barth et al. (2011)]
{2011ApJ...743L...4B}
Barth, A. J., Pancoast, A., Thorman, S. J. et al. 2011, ApJ, 743, 4

\bibitem[Barth et al. (2013)]
{2013ApJ...769..128B}
Barth, A. J., Pancoast, A., Bennert, V. N. et al. 2013, ApJ, 769, 128

\bibitem[Barth et al. (2015)]
{2015ApJS..217...26B}
Barth, A. J., Bennert, V. N., Canalizo, G. et al. 2015, ApJS, 217, 26

\bibitem[Begelman et al. (1990)]{10.1038/287307a0}
Begelman, M. C., Blandford, R. D., Rees, M. J. 1990, Nature 287, 307

\bibitem[Bentz et al. (2008)]
{2008ApJ...689L..21B}
Bentz, M. C., Walsh, J. L., Barth, A. J. et al. 2008, ApJL, 689, L21

\bibitem[Bentz et al. (2009)]
{2009ApJ...705..199B}
Bentz, M. C., Walsh, J. L., Barth, A. J. et al. 2009, ApJ, 705, 199

\bibitem[Bon et al. (2012)]{2012ApJ...759..118B}
Bon, E., Jovanovi{\'c}, P., Marziani, P., Shapovalova, A. I.,  Bon, N., Borka -Jovanovi{\'c}, V., Borka, D.,  Sulentic, J., Popovi{\'c}, L. {\v C}. 2012, ApJ, 759, id. 118, 9


\bibitem[Colpi et al. (2009)]{10.1086/307952}
Colpi, M., Mayer, L., Governato, F. 1999, ApJ, 525, 720

\bibitem[Combes et al. (2019)]{10.1051/0004-6361/201834560}
 Combes, F.,  Garc{\'i}a-Burillo, S.,  Audibert, A., Eckart, A.,  et al. 2019,  A\&A, 623, A79 


\bibitem[Cuadra et al. (2010)]{10.1111/j.1365-2966.2008.14147.x}
Cuadra, J.,   Armitage, P. J.,  Alexander, R. D.,  Begelman,  M.C. 2009, MNRAS, 393, 1423 

\bibitem[Delaa et al. (2013)]{10.1051/0004-6361/201220689}
Delaa, O. Zorec, J.,  Domiciano de Souza, D.,  Mourard, D.,   Perrau, K., et al. 2013, A\&A, 555, A100


\bibitem[Denney et al. (2009)]
{2009ApJ...704L..80D}
Denney, K. D., Peterson, B. M., Pogge, R. W. et al. 2009, ApJL,704, L80


\bibitem[Domiciano de Souza et al. (2004)] {10.1051/0004-6361:20040051}
Domiciano de Souza, A., Zorec, J., Jankov, S., et al. 2004, A\&A, 418, 781

\bibitem[Du et al. (2014)]
{2014ApJ...782...45D}
Du, P., Hu, C., Lu, K.-X., Wang, F., Qui, J., Li, Y.-R., Bai, J.-M., Kaspi, S., Netzer, H., Wang, J.-M. 2014,ApJ, 782, 45

\bibitem[Du et al. (2015)]
{10.1088/0004-637X/806/1/22}
Du, P., Hu, C., Lu, K.-X., Huang, Y.-K., Cheng, C., Qiu, J., Li, Y.-R.,  et al. 2015, ApJ, 806, 22



\bibitem[Du et al. (2016)]
{2016ApJ...825..126D}
Du, P., Lu, K.-X., Zhang, Z.-X., Huang, Y.-K., Wang, K., Hu, C., Qiu, J., Li, Y.-R. et al.  2016, ApJ, 825, 126

\bibitem[Du et al. (2018a)]{10.3847/1538-4357/aaed2c}
Du, P., Brotherton, M. S., Wang, K., Huang, Z.-P., Hu, C.,  Kasper, D. H., Chick, W. T. et al. 2018, 
ApJ,  869,  id. 142

\bibitem[Du et al. (2018b)]
{2018ApJ...856....6D}
Du, P., Zhang, Z.-X., Wang, K.,  Huang, Y.-K., Zhang, Y., Lu, K.-X., Hu, C., Li, Y.-R. et al. 2018, ApJ, 856, 6

\bibitem[Edelson et al. (2019)]
{2019ApJ...870..123E}
Edelson, R., Gelbord, J., Cackett, E., Peterson, B. M., Horne, K., Barth, A. J. et al. 2019, ApJ,  870,  2, id. 123


\bibitem[Elvis (2001)]
{10.1007/978-94-010-0320-9}
Elvis, M., 2001,  in The Century of Space Science  (eds J.A. Bleeker, J. Geiss, and  M. Huber),  Kluwer Academiic Publishers, 529

\bibitem[Eracleous and Halpern (1994)]
{10.1086/191856}
Eracleous, M., Halpern, J. P. 1994, ApJS, 90, 1

\bibitem[Eracleous and Halpern (2003)]
{10.1086/379540}
Eracleous, M.,  Halpern, J. P. 2003, ApJ, 599, 886

\bibitem[Gezari et al (2007)]{10.1086/511032}
Gezari, S., Halpern, J. P., Eracleous, M.  2007, ApJS, 169, 167

\bibitem[Grier et al. (2012)]
{2012ApJ...755...60G}
Grier, C. J., Peterson, B. M., Pogge, R. W. et al. 2012, ApJ, 755, 60

\bibitem[Grier et al. (2017)]
{2017ApJ...851...21G}
Grier, C. J., Trump, J. R., Shen, Y. et al. 2017, ApJ, 851, 21

\bibitem[Haiman et al. (2009)]{10.1088/0004-637X/700/2/1952} 
Haiman, Z., Kocsis, B., Menou, K. 2009, ApJ, 700, 1952

\bibitem[Hayasaki et al. (2008)]{10.1086/588837} 
Hayasaki, K., Mineshige, S.,   Ho, L. C.  2008, \apj,  682,  1134

\bibitem[Hopkins et al. (2006)] {2006ApJS..163....1H}
Hopkins, P. F., Hernquist, L., Cox, T. J., Di Matteo, T., Robertson, B., Springel, V. 2006, ApJS, 163, 1

\bibitem[Ili{\'c} et al. (2020)] {Ilic20}
Ili{\'c}, D.,  Oknyansky, V.,  Popovi{\'c}, L. {\v C}.,  Tsygankov, S. S.,  Belinski, A. A., Tatarnikov, A. M.,  Dodin, A. V.,  Shatsky, N. I.,  Ikonnikova, N. P., Raki{\'c}, N.,  Kova{\v c}evi{\'c}, A.,  Mar{\v c}eta-Mandi{\'c}, S. et al.  2020, accepted in A\&A


\bibitem[Jaffe(2004)]{10.1038/nature02531}
Jaffe, W., Meisenheimer, K. R{\"o}ttgering, H. J. A., Leinert, Ch., Richichi, A.  et al. 2004, Nature, 429,6987, 47

\bibitem[Jankov et al. (2001)] {10.1051/0004-6361:20011047}
Jankov, S., Vakili, F., Domiciano de Souza Jr., A.,  Janot-Pacheco,
E. 2001, A\&A, 377, 721


\bibitem[Kaspi et al. (2000)]
{2000ApJ...533..631K}
Kaspi, S., Smith, P. S., Netzer, H. et al. 2000, ApJ, 533, 631

\bibitem[Kaspi et al. (2007)]{10.1086/512094}
Kaspi, S., Brandt, W. N., Maoz, D. et al. 2007, ApJ, 659, 997

\bibitem[Kim et al. (2010)]{10.1088/0004-637X/724/1/386}
Kim, D., Im, M., Kim, M.  2010, ApJ, 724, 386


\bibitem[Kishimoto(2009)]{10.1051/0004-6361/200913512}
Kishimoto, M., H{\"o}nig, S. F., Antonucci, R., Kotani, T., Barvainis, R., Tristram, K. R. W., Weigelt, G.  2009, A\&A,  507, L57	


\bibitem[Klioner (2003)]{10.1086/367593}
Klioner, S. 2003, ApJ, 125, 1580

\bibitem[Kova{\v c}evi{\'c} et al. (2020)]{10.1051/0004-6361/201936398}
Kova{\v c}evi{\'c}, A.,   Wang, J.-M.,  Popovi{\'c}, {\v C}. L.  2020, A\&A, 635, id.A1, 19

\bibitem[Kova{\v c}evi{\' c} et al.(2020)]{10.1093/mnras/staa737}
Kova{\v c}evi{\' c}, A.~B.,  Yi, T.,  Dai, X.,  Yang, X.,  {\v C}vorovi{\'c}-Hajdinjak, I., Popovi{\' c}, L. {\v C}. 2020, MNRAS,  494, 4069


\bibitem[Krause et al. (2011)]{10.1111/j.1365-2966.2010.17698.x}
Krause, M., Burkert, A., Schartmann, M.  2011, MNRAS, 411, 550


\bibitem[Laine  et al. (2020)]{10.3847/2041-8213/ab79a4}
Laine,  S.,   Dey, L.,   Valtonen, M.,  Gopakumar, A.,  Zola, S.,  Komossa,  S.  et al.  2020,  ApJL 894 L1


\bibitem[Landt et al. (2014)]{10.1093/mnras/stu031}
Landt, H., Ward, M. J.,  Elvis, M.,  Karovska, M. 2014,  MNRAS,  439,  1051 

\bibitem[Landt et al. (2013)]{10.1093/mnras/stt421}
Landt, H., Ward, M. J.,  Peterson, B. M., Bentz, M. 2013,  MNRAS, 432, 113


\bibitem[Landt et al. (2008)]{10.1086/522373}
Landt, H., Bentz, M. C., Ward, M. J., Elvis, M., Peterson, B. M., et al. 2008, ApJS, 174, 282

\bibitem[Lodato et al. (2009)]{10.1111/j.1365-2966.2009.15179.x} 
Lodato, G., Nayakshin, S., King, A. R., Pringle, J. E. 2009, MNRAS, 398, 1392


\bibitem[Lachaume (2003)]{10.1051/0004-6361:20030072} 
Lachaume, R., 2003, A\&A, 400, 795



\bibitem[Lu et al. (2019)]
{10.3847/1538-4357/ab5790}
Lu, K.-X., Bai, J.-M.,  Zhang, Z.-X.,  Du, P.,  Hu, C.,  Kim, M., Wang, J.-M.,  Ho, L. C.,  Li, Y.-R. et al. 2019, ApJ, 887,   id. 13

\bibitem[MacFadyen and Milosavljevi{\'c}(2008)]{doi.org/10.1086/523869}
MacFadyen, A. I. Milosavljevi{\'c}, M. 2008, ApJ, 672, 83




\bibitem[Mayer et al. (2007)]
{10.1126/science.1141858}
Mayer, L., Kazantzidis, S., Madau, P., Colpi, M., Quinn, T., Wadsley, J. 2007, Science, 316, 1874  


\bibitem[Meilland et al. (2011)]{10.1051/0004-6361/201116798}
Meilland, A., Delaa, O.,  Stee, Ph.,  Kanaan, S.,  Millour, F. et al., 2011, A\&A 532, A80  

\bibitem[Netzer (2013)]{2013peag.book.....N}
Netzer, H. 2013, The physics and evolution of active galactic nuclei, Cambridge university press

\bibitem[Nixon et al. (2011)]{doi.org/10.1111/j.1365-2966.2010.17952.x}
Nixon,  C. J.,  Cossins, P. J.,  King, A. R., Pringle, J. E. 2011, \mnras,  412, 1591 

\bibitem[Pancoast et al. (2014)]{10.1093/mnras/stu1809}
Pancoast, A., Brewer, B. J., Treu, T.  2014, MNRAS, 445, 3055

\bibitem[Pancoast et al. (2019)]
{10.3847/1538-4357/aaf806}
Pancoast, A., Skielboe, A., Pei, L., Bennert, V. N.;et al. 2019, ApJ,   871,  id. 108

\bibitem[Peters and Mathews (1963)]
{10.1103/PhysRev.131.435}
Peters, P. C., Mathews, J. 1963, Phys. Rev., 131, 435



\bibitem[Peterson et al. (1998)]
{1998ApJ...501...82P}
Peterson, B. M., Wanders, I., Bertram, R. et al. 1998, ApJ, 501, 82

\bibitem[Peterson et al. (2002)]
{2002ApJ...581..197P}
Peterson, B. M., Berlind, P., Bertram, R. et al. 2002,ApJ, 581, 197



\bibitem[Peterson et al. (2004)]{10.1086/423269}
Peterson, B. M.,  Ferrarese, L.,   Gilbert, K. M.,   Kaspi, S. et al. 2004, ApJ, 613, 682

\bibitem[Petrov(2006)]{10.1051/0004-6361:20066496}
Petrov,  R. G.,   Malbet, F.,   Weigelt, G.,   Antonelli, P.,  Beckmann, U.,   Bresson, Y., et al. 2007, A\&A, 464, 1 


\bibitem[Petrov(2012)]{doi.org/10.1117/12.926595}
Petrov,  R. G.,   Millour, F.,  Lagarde, S.,  Vannier, M.,  Rakshit, S.,   Marconi, A.,    Weigelt, G., 2012,in Proc. SPIE 8445, Optical and Infrared Interferometry III,  ed.
F. Delplancke, J. K. Rajagopal, \& F. Malbet, 84450W-1


\bibitem[Petrov (1989)]{1989ASIC..274..249P}
Petrov, R. G. 1989, in  
Diffraction-Limited Imaging with Very Large Telescopes, Proceedings of the NATO Advanced Study Institute, held in Cargese, September 13-23, 1988, Dordrecht: Kluwer, 1989, edited by D. M. Alloin and J. M. Mariotti. NATO Advanced Science Institutes (ASI) Series C,  274, 249

 

\bibitem[Popovi{\'c} et al. (2011)]{2011A&A...528A.130P}
Popovi{\'c}, L. {\v C}., Shapovalova, A. I., Ili{\'c}, D., Kova{\v c}evi{\'c}, A., Kollatschny, W. et al. 2011, A\&A, ,  528, A130 

\bibitem[Popovi{\'c} et al. (2014)]
{2014A&A...572A..66P}
Popovi{\'c}, L. {\v C}., Shapovalova, A. I.,  Ili{\'c}, D., Burenkov, A. N., Chavushyan, V. H., Kollatschny, W, Kova{\v c}evi{\'c}, A. et al. 2014, A\&A,  572, A66



\bibitem[Redmount and  Rees (1989)]
{1989ComAp..14..165R}
Redmount, I. H.,  Rees, M. J. 1989, Comments on Astrophys., 14,  165 

\bibitem[Roedig et al. (2011)]
{10.1111/j.1365-2966.2011.18927.x}
Roedig, C.,   Dotti, M.,  Sesana, A.,   Cuadra, J.,  Colpi,  M.  2011, MNRAS, 415, 3033



\bibitem[Safarzadeh et al. (2019)]
{2019MNRAS.488L..90S}
Safarzadeh, M.,  Loeb, A.,  Reid, M. 2019, MNRAS, 488, L90





\bibitem[Sesana (2010)]
{10.1088/0004-637X/719/1/851}
Sesana,  A. 2010,  ApJ,   719, 851


\bibitem[Shapovalova et al. (2001)]
{2001A&A...376..775S}
Shapovalova, A. I., Burenkov, A. N., Carrasco, L., Chavushyan, V. H. et al. 2001, A\&A, 376,  775

\bibitem[Shapovalova et al. (2004)]
{2004A&A...422..925S}
Shapovalova, A. I., Doroshenko, V. T., Bochkarev, N. G., Burenkov, A. N. et al. 2004, A\&A, 422, 92

\bibitem[Shapovalova et al. (2008)]
{2008A&A...486...99S}
Shapovalova, A. I., Popovi{\'c} L. {\v C}., Collin, S., Burenkov, A. N. et al. 2008, A\&A,486,  99


\bibitem[Shapovalova et al. (2010a)]
{2010A&A...509A.106S}
Shapovalova, A. I., Popovi{\'c}, L. {\v C}., Burenkov, A. N., Chavushyan, V. H., Ili{\'c}, D., Kova{\v c}evi{\'c}, A. et al. 2010a, A\&A,  509, A106

\bibitem[Shapovalova et al. (2010b)]
{2010A&A...517A..42S}
Shapovalova, A. I., Popovi{\'c}, L. {\v C}., Burenkov, A. N., Chavushyan, V. H., Ili{\'c}, D., Kollatschny, W., Kova{\v c}evi{\'c}, A. et al. 2010b, A\&A, 517, A42

\bibitem[Shapovalova et al. (2012)]
{2012ApJS..202...10S}
Shapovalova, A. I.,Popovi{\'c}, L. {\v C}., Burenkov, A. N., Chavushyan, V. H., Ili{\'c}, D., Kova{\v c}evi{\'c}, A, Kollatschny, W., Kova{\v c}evi{\'c},  J. et al. 2012, ApJS,  202,  10

\bibitem[Shapovalova et al. (2013)]
{2013A&A...559A..10S}
Shapovalova, A. I., Popovi{\'c}, L. {\v C}, Burenkov, A. N., Chavushyan, V. H.,  Ilii{\'c}, D., Kollatschny, W.,  Kova{\v c}evi{\'c}, A. et al. 2013, A\&A, 559, A10

\bibitem[Shapovalova et al. (2016)]
{2016ApJS..222...25S}
Shapovalova, A. I., Popovi{\'c}, L. {\v C}, Chavushyan, V. H. Burenkov, A. N., Ili{\'c}, D., Kollatschny, W., Kova{\v c}evi{\'c}, A. et al. 2016, ApJS,  222,  id. 25

\bibitem[Shapovalova et al. (2017)]
{2017MNRAS.466.4759S} 
Shapovalova, Alla I, Popovi{\'c}, L. {\v C}., Chavushyan, V. H., Afanasiev, V. L., Ili{\'c},  D., Kova{\v c}evi{\'c}, A. et al. 2017, MNRAS,  466, 4759

\bibitem[Shapovalova et al. (2019)]
{2019MNRAS.485.4790S}
Shapovalova, A. I., Popovi{\'c}, L. {\v C}, Afanasiev, V. L., Ili{\'c}, D., Kova{\v c}evi{\'c}, A., Burenkov, A. N., Chavushyan, V. H., Mar{\v c}eta-Mandi{\'c} , S. et al. 2019, MNRAS,  485, 4790 

\bibitem[Shen et al. (2016)]
{2016ApJ...818...30S}
Shen, Y., Horne K., Grier, C. J., Peterson, B. M., Denney, K. D. et al. 2016, ApJ, 818, 30
\bibitem[Songsheng et al. (2019a)]
{10.3847/1538-4357/ab2e00}
Songsheng, Y. Y.,  Wang, J.-M.,  Li, Y.-R.,   Du, P. 2019a, ApJ,  881, 140 

\bibitem[Songsheng et al. (2019b)] {doi.org/10.3847/1538-4357/ab3c5e}
Songsheng, Y. Y.,  Wang, J.-M.,  Li, Y.-R. 2019b, ApJ,  883, 184 


\bibitem[Songsheng et al. (2020)]{2020ApJS..247....3S}
Songsheng, Y.-Y., Xiao, M., Wang, J.-M., Ho, L. C. 2020, ApJS, 247, 3

\bibitem[Sugunama et al. (2006)] {10.1086/499326}
Suganuma, M., Yoshii, Y., Kobayashi, Y., Minezaki, T., Enya, K. et al. 2006, ApJ, 639, 46  10.1086/499326

\bibitem[Shull et al. (2012)]
{10.1088/0004-637X/752/2/162}
 Shull, J. M.,  Stevans, M.,  Danforth,  C. W. 2012, ApJ, 752, 162


\bibitem[Stern et al. (2005)]{10.1088/0004-637X/804/1/57}
Stern, J., Hennawi, J. F.,  Pott, J.-U. 2015, ApJ, 804, 57

\bibitem[Gravity Collaboration (2018)]{10.1038/s41586-018-0731-9}
Gravity Collaboration, Sturm, E.,  Dexter, J.,  Pfuhl, O.,  Stock, M. R.,  Davies, R. I., Lutz, D. et al. 2018, Nature,  563,  7733, 657




\bibitem[Swain(2004)]{10.1117/12.551881}
Swain, M. R. 2004, in  Proceedings of SPIE Volume 5491. Edited by Wesley A. Traub. Bellingham, WA: The International Society for Optical Engineering, 2004., 1		












\bibitem[Tokovinin (1992)]{1992ESOC...39..425T}
Tokovinin, A. 1992,in Proceedings of  ESO Conference on High-Resolution Imaging by Interferometry II. Ground-Based Interferometry at Visible and Infrared Wavelengths,  Garching bei Munchen, Germany, October 15-18, 1991. Editors, J.M. Beckers, F. Merkle,  European Southern Observatory, Garching bei Munchen, Germany, 425










\bibitem[Vannier et al. (2006)]{10.1111/j.1365-2966.2006.10015.x}
Vannier, M., Petrov, R. G., Lopez, B., Millour, F. 2006, MNRAS, 367, 825



\bibitem[Waisberg et al. (2017)]{10.3847/1538-4357/aa79f1}
Waisberg, I.,   Dexter,  J., Pfuhl,  O., Abuter, R.,  Amorim, A.,  Anugu, N. et al 2017, ApJ,  844, 72 

\bibitem[Wanders et al. (1995)]{10.1086/309750}
Wanders, I., Goad, M. R., Korista, K. T., Peterson, B. M., Horne, K., Ferland, G. J.,et al. 1995, ApJL, 453, L87

\bibitem[Wang et al. (2010)]{10.1088/2041-8205/719/2/L208}
Wang, J., Fabbiano, G., Risaliti, G.,   Elvis, M., et  al. 2010, ApJL, 719,  L208

\bibitem[Wang et al. (2014)]
{2014ApJ...793..108W}
Wang, J.-M., Du, P., Hu, C., Netzer, H.  et al. 2014, ApJ,793,108


\bibitem[Wang et al. (2017)] {10.1038/s41550-017-0264-4}
Wang, J.-M., Du, P., Brotherton, M. S., et al. 2017, NatAs, 1, 775






\bibitem[Wang et al. (2020a)]
{10.1038/s41550-019-0979-5}
Wang, J.-M., Songsheng, Y.-Y., Li, Y.-R., Du, P., Zhang, Z.-X.  2020a, Nat. Astron., 4, 517

\bibitem[Wang et al. (2020b)]{10.1093/mnras/staa1985}
Wang, J. -M., Songsheng, Y. -Y., Li, Y. -R., Du, P., Zhe, Y. 2020b, MNRAS, 497, 1020


\bibitem[Wang et al. (2018a)] {10.3847/1538-4357/aacdfa}
Wang, J.-M., Songsheng, Y.-Y, Li, Y.-R, Zhe, Y., 2018, ApJ, 862, 171

\bibitem[Wang et al. (2018b)]
{10.3847/1538-4357/aab88b}
Wang, J., Xu, D. W., Wei, J. Y.  2018, ApJ, 858, 49


\bibitem[Weigelt et al. (2012)]{10.1051/0004-6361/201219213}	
Weigelt, G., Hofmann, K. -H., Kishimoto, M.,  H{\"o}nig, S., Schertl, D. et al., 2012, A\&A,  541, L9 


\bibitem[Zhang et al. (2019)]{doi.org/10.3847/1538-4357/ab1099}
 Zhang, Z. X., Du, P.  Smith, P. S.,  Zhao, Y.,  Hu, C.,   Xiao, M.,   Li, Y. R.,  Huang, Y. H., 
 Wang, K.,  Bai, J., M.,   Ho, L. C.,  Wang, J. M., 2019, ApJ, 876,14

 
\end{thebibliography}
%

\begin{appendix}
	\section{General inferences about differential phase based  on a first order approximation } \label{appendix:generall}
	
	\subsection{Single SMBH}\label{appendix:generalsingle}
	 First, we infer some generic patterns about differential phase shape,  to assess more specific results of numerical simulations. 
	We recall that spectral line of the source can be approximated as  a function of the form
	${\Xi}(\lambda, \boldsymbol{r}^{c}_{k}, \boldsymbol{V}^{c}_{k})$ where 
	$\boldsymbol{r}^{c}_{k},\boldsymbol{V}^{c}_{k}$ are composite  vector fields of  the $k$-th cloud' position and velocity, given in SMBH centered  coordinate system \citep[see][]{10.1051/0004-6361/201936398}.  
	Considering a single-source model, a first-order approximation of $\xi({\lambda})$ based on Eq. \ref{eq:complex5}  is:
	\begin{equation}
	\xi({\lambda})\sim \arctan \frac{\sin (2\pi \boldsymbol{u}\cdot \boldsymbol{\sigma})}{\cos(2\pi \boldsymbol{u}\cdot \boldsymbol{\sigma} )}\sim{\sin (2\pi \boldsymbol{u}\cdot\boldsymbol{\sigma})}\sim 2\pi \boldsymbol{u}\cdot\boldsymbol{\sigma}
	\label{eq:complex7}
	\end{equation}
	When  $\boldsymbol {u}\cdot\boldsymbol{\sigma}$   gets smaller below the resolution limit, the phase tends to a linear expression as it is given by the third term in Eq. \ref{eq:complex7} \citep{10.1111/j.1365-2966.2006.10015.x}. 
	Thus the shape of the differential phase is defined by the projected vectorial field of positions  of clouds.
	For non-resolved sources, the term $2\pi \boldsymbol{u}\cdot \boldsymbol{\sigma}$ never reaches  the value of $\frac{\pi}{2}$ and the maximum  value of  phase  occrus when $\boldsymbol {u}\cdot\boldsymbol{\sigma}$ is largest.
	Thus, the  elements of clouds elliptical  orbit appear in the   scalar product:
	\begin{equation}
	\begin{split}
	\boldsymbol {u}\cdot\boldsymbol{\sigma} =  u \frac{r_{c}}{d}\left[\cos \Omega_{c} \cos(\omega_{c} +f_{c})-\sin \Omega_{c} \sin (\omega_{c}+f_{c})\cos i_{c}\right] \\
	+v \frac{ r_{c}}{d} \left[\sin \Omega_{c} \cos(\omega_{c} +f_{c})+\cos \Omega_{c} \sin (\omega_{c}+f_{c})\cos i_{c}\right]
	\end{split}         
	\label{eq:complex8}
	\end{equation}
	\noindent where $d$ is a distance between the observer and SMBH.
	The intensity of the radius vector of clouds' orbit  is 
	\begin{equation}
	r_{c}=\frac{a_{c} (1-e_{c}^2)}{1+e_{c} \cos f_{c}}
	\end{equation}
	\noindent where the true anomaly $f_c$ is computed as  outlined in \cite{10.1051/0004-6361/201936398}.
	Since we  assume that the interferometric baseline is  perpendicular to the rotation axis  \citep[$PA=90^{\circ}$, see in][and references therein]{10.3847/1538-4357/ab2e00} then in our calculations will not appear second term containing v-coordinate of the vector $\boldsymbol{u}$.
	
\begin{figure*}
		\includegraphics[trim=10 10 10 40,clip,width=0.5\textwidth]{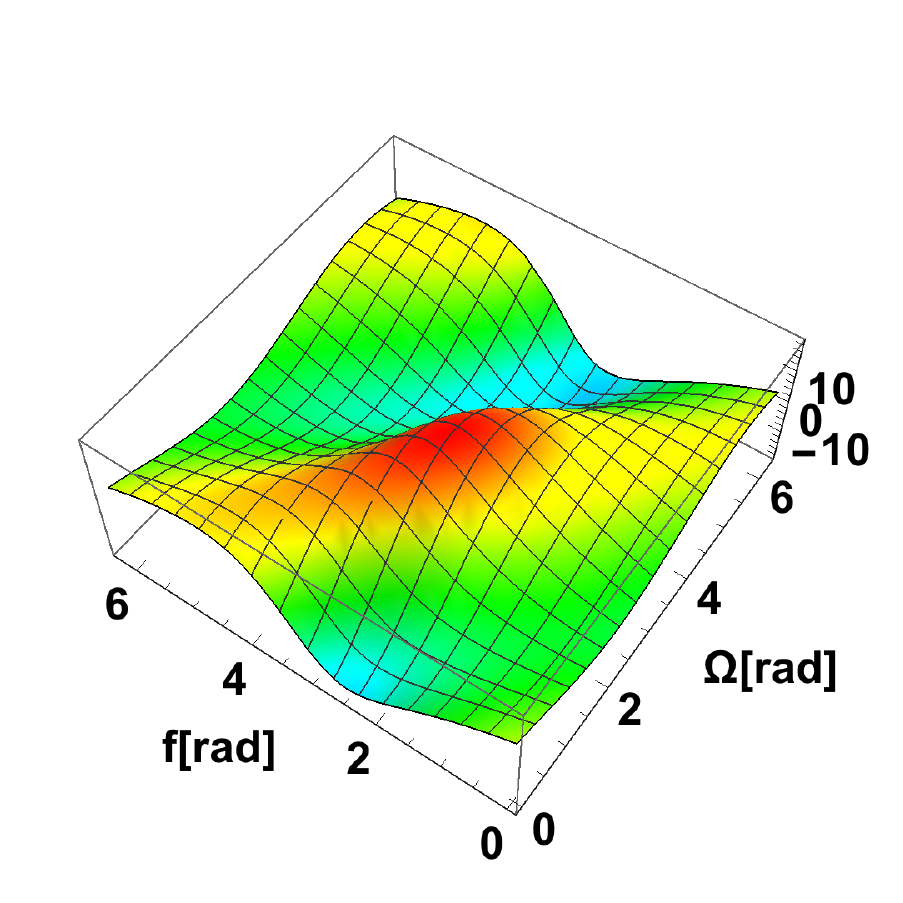} 
		\includegraphics[trim=10 10 10 40,clip,width=0.5\textwidth]{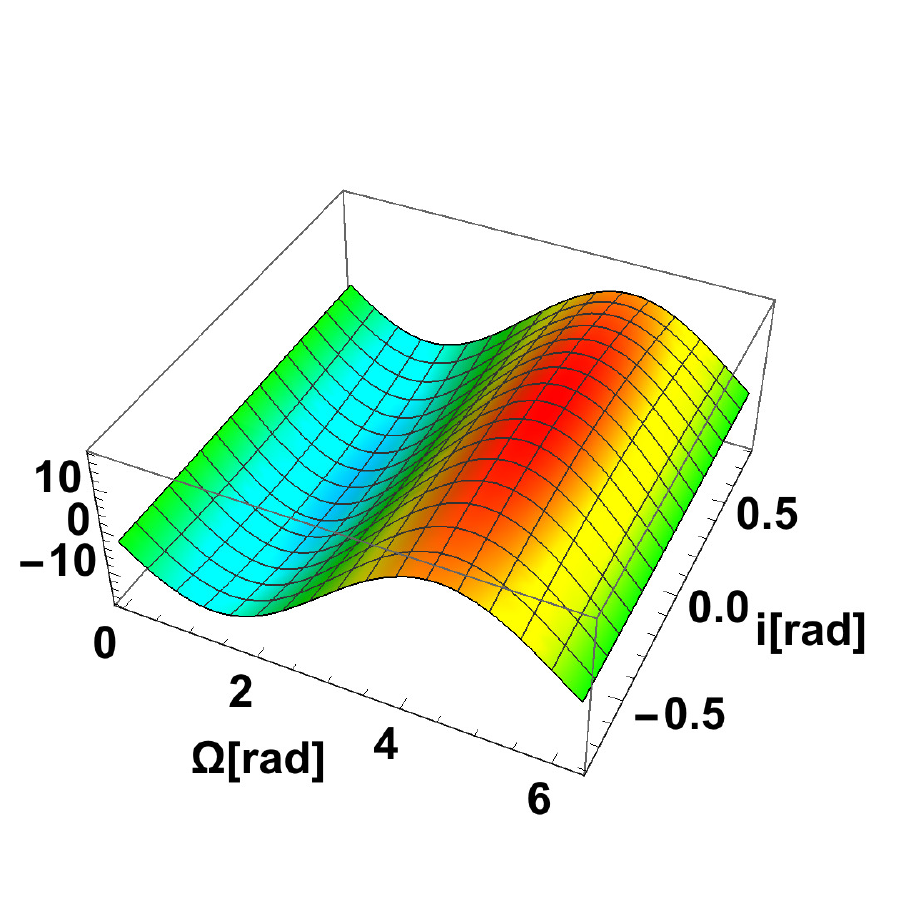}\\
		\includegraphics[trim=5 5 10 40,clip,width=0.5\textwidth]{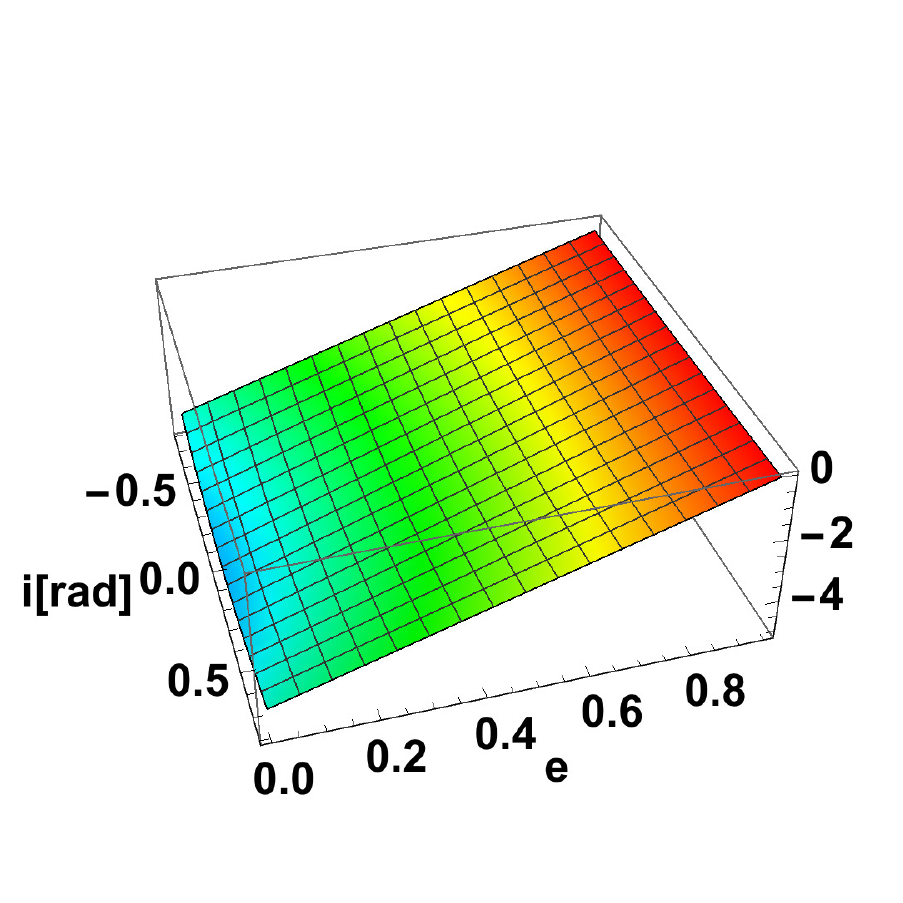}
		\caption{Interactive effects of clouds' orbital elements on the values of the first parameter in Eq.  \ref{eq:complex8} present as the  surface mesh. Upper left:  on the x-axis is cloud's ascending node $\Omega$, on the y-axis is   true anomaly (f),   on the z-axis are values of the calculated parameter  for fixed elements  $\omega=\pi/3, i=\pi/3, e=0.5, a=15\, \mathrm{ld}$. Upper right:  on the x-axis is cloud's orbital inclination, and   fixed elements are $\omega=4.66\,  \mathrm{rad}, f=3.38\, \mathrm{rad}, e=0.19$. Bottom plot: on  the x-axis is  eccentricity,  and fixed elements are $\omega=11\pi/18, \Omega=10\pi/18, f=0\, \mathrm{rad}$. }
		\label{fig:ilustris}
		\hfill
	\end{figure*}

	Having defined quantity $ \boldsymbol {u}\cdot\boldsymbol{\sigma}$ in terms of clouds orbital parameters  and angular position of interferometric baseline, we will now move on to discuss the interactive effects of  clouds orbital  elements on the value of the first term in  Eq. \ref{eq:complex8} using surface plots.
	In Fig. \ref{fig:ilustris} upper left plot, we present the surface mesh  for $\omega=\pi/3,\, i=\pi/3,\, e=0.5,\, a=15\, \mathrm{ld}$ which is plotted over  $\Omega$ - $f$ grid. The surface is smooth with  regions of local  extrema and  a central peak, but  some slopes are skewed. 
	
	An increase in the surface's amplitude results from an  increase of  the eccentricity of clouds' orbits. This is evident   even  for smaller clouds' orbital inclinations (see Fig. \ref{fig:ilustris}, bottom plot).
	 An increasing clouds' orbital inclination causes peak slopes to  decrease in the diagonal direction (see the red region in the upper  left plot  of  Fig. \ref{fig:ilustris}). Deformed 
	side intersections  have an S-like shape. The effects of fixed  parameters $\omega=4.66\,  \mathrm{rad}, f=3.38\, \mathrm{rad}, e=0.19$ resemble a wavelike, or hilly, structure with  S side profile  (see Fig. \ref{fig:ilustris} upper left plot), but with deformed peaks (see red and blue regions). 
	
	It is useful, for presenting results that follow, to  get an analytic estimate of the first term in Eq. \ref{eq:complex8} for  some specific values of model parameters. When clouds' orbital inclinations are $\sim 0^{\circ}$, the term  becomes
	$\sim \frac{r}{d} \cos(\Omega+\omega+f) $.  If  $\Omega\sim 0^{\circ}$, the term  turns into  $\sim \frac{r}{d} \cos(\omega+f) $.
	At the  initial time  (i.e.,  $f=0$), the term becomes $\sim \frac{r}{d} \cos (\Omega) \cos(\omega)-\sin \Omega \sin (\omega)\cos i$. So its amplitude depends on values of $\Omega$ and $\omega$,  and having the largest values for smaller eccentricity (see Fig. \ref{fig:ilustris}, bottom plot).

	Let us explore  the analytic forms under  degenerate cases, when some orbital parameters are not defined. The simplest case is an inclined circular orbit   for which we can not define f and $\omega$. Then the  so-called argument of latitude $\eta$ is used and   the first term  becomes
	$$\sim \frac{a}{d} \cos \Omega \cos \eta-\sin\Omega \sin \eta \cos i $$. 
	For circular orbits with zero inclination, the argument of latitude  $\eta$ is also undefined since  the line of nodes is undetermined. Thus, the  true longitude L can be used instead,  so the first term asymptotically behaves as $\sim \frac{a}{d} \cos L$.

	\subsection{CB-SMBH }\label{appendix:general2} 
	
	\renewcommand{\thesubtable}{A\alph{subtable}}

\begin{figure*}
		\includegraphics[trim=1 0 0 0,clip,width=0.5\textwidth]{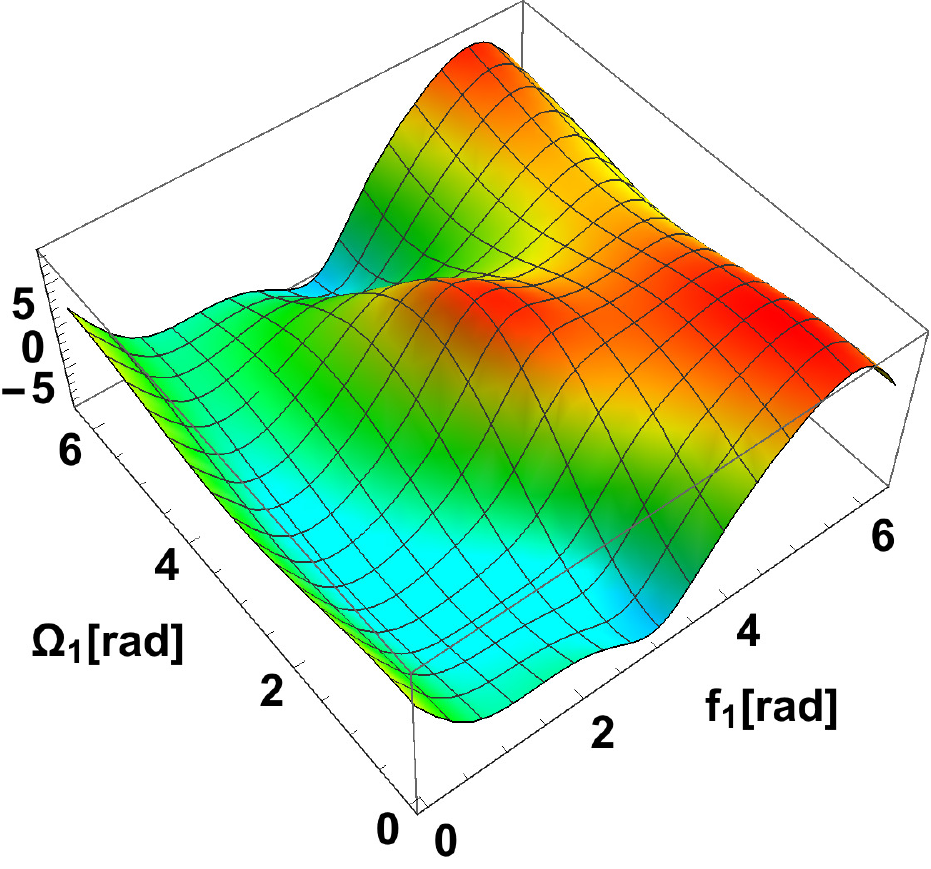} 
		\includegraphics[trim=8 0 9 40,clip,width=0.5\textwidth]{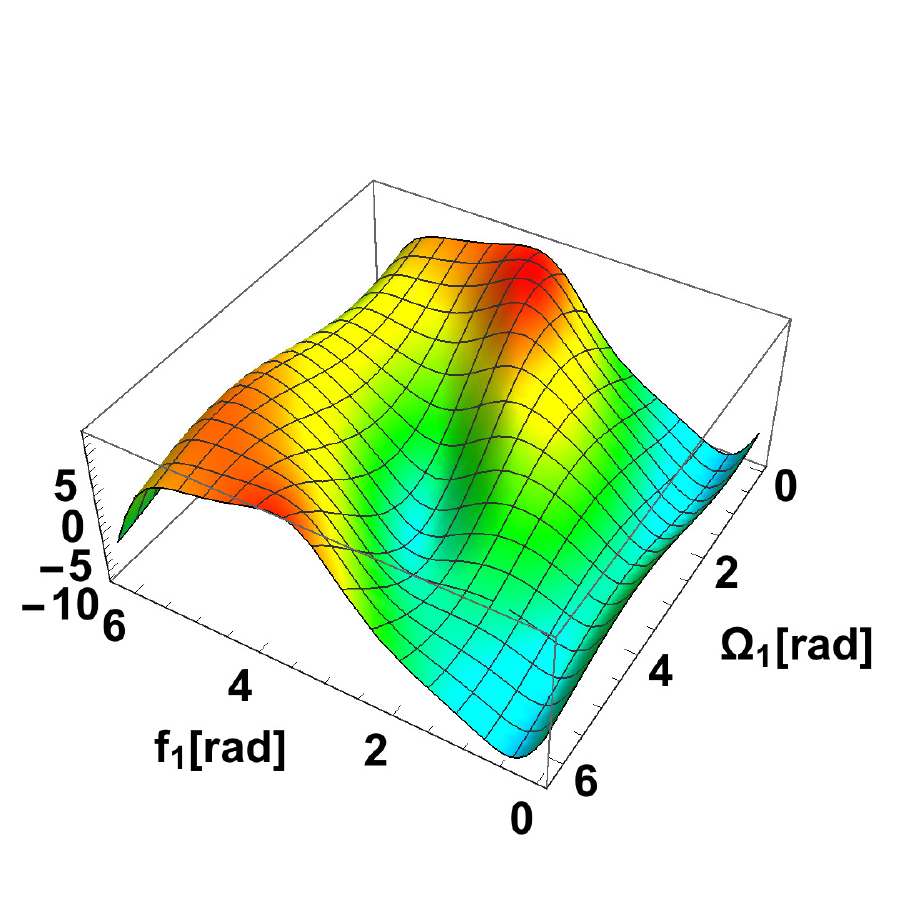} \
		\includegraphics[trim=8 0 9 40,clip,width=0.5\textwidth]{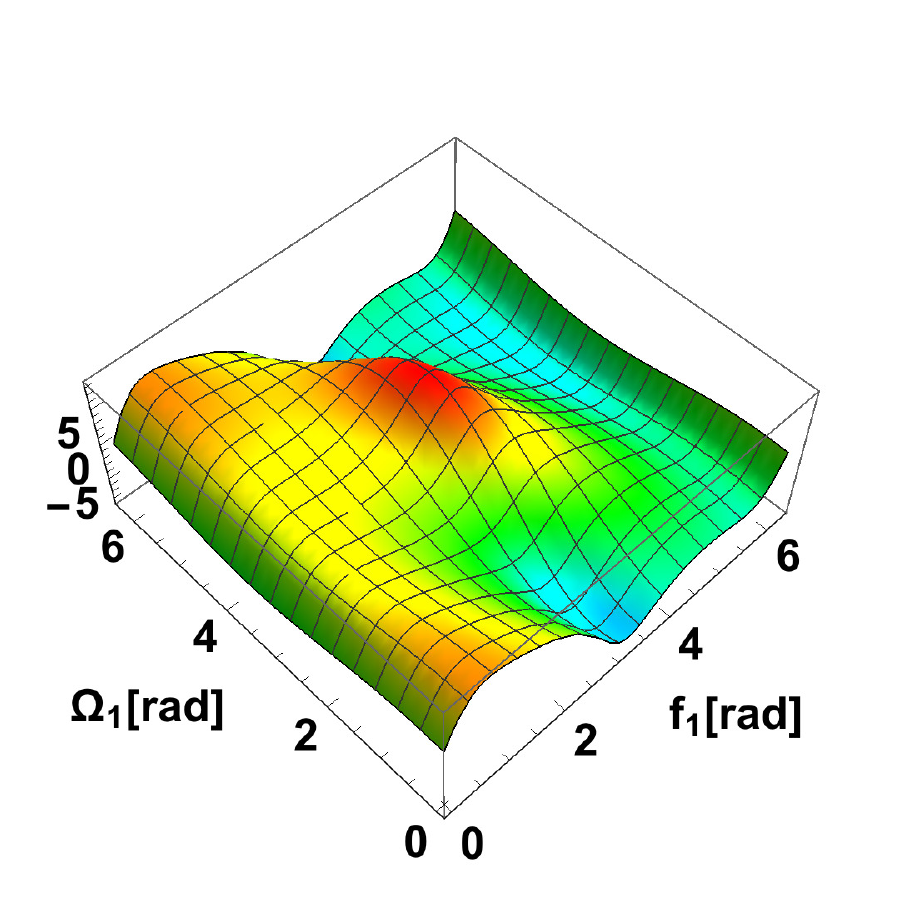} 
		\includegraphics[trim=0 80 20 115,clip,width=0.5\textwidth]{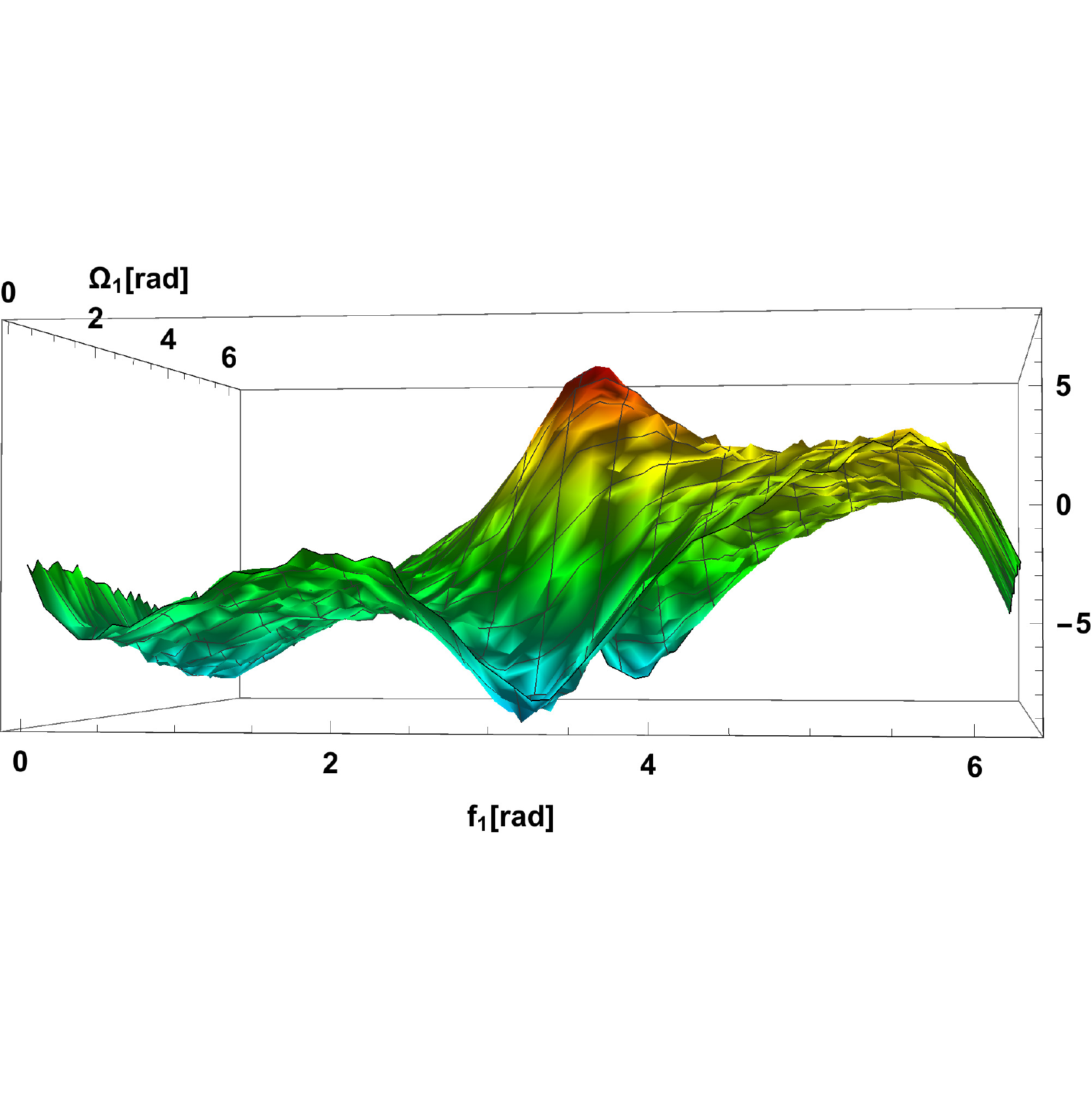} 
		\caption{ The surface mesh of the values of  the first parameter in Eq.  \ref{eq:complex6} over a grid of orbital elements of larger SMBH. The ascending node $\Omega_1$  and true anomaly $f_1$ are given in the XY coordinate plane, and values of the calculated parameter are given on the Z-axis.
			Upper left plot: Case I - parameters are listed in Table \ref{tab:A:1a}.
				Upper right plot: Case II - parameters are listed in Table \ref{tab:A:1b}.
				Bottom left: Case III - parameters are listed in Table \ref{tab:A:1c}.
			Bottom right: Case IV -  parameters are listed in Table \ref{tab:A:1d}.
		}
		\label{fig:ilustris2}
		\hfill
	\end{figure*}

	Let us take a brief look at  Eq. \ref{eq:complex6} and describe some general effects of orbital elements of clouds in both BLR in a CB-SMBH system on  characteristics of the composite differential phase of. For simplicity’s sake, we omit fluxes $F_{k},k=1,2$ in the following derivation:
	\begin{equation}
	\boldsymbol{u}\cdot({\boldsymbol{\sigma_{1}}- \boldsymbol{\sigma_{2}}})\sim\boldsymbol{u}\cdot(\boldsymbol{r}^{c}_{ 1}-\boldsymbol{r}_{\bullet 1} 
	-{\boldsymbol{r}^{c}_{2}}+{\boldsymbol{r}^{c}_{\bullet 2}})
	\label{eq:posled}
	\end{equation}
	
	\noindent where  ${\sigma_{k}}, k=1,2$ are vectors of relative positions of clouds around $k$-th SMBH in the barycentric frame.  The  vector ${\sigma_{k}}$  equals to the difference between SMBH barycentric position ${\boldsymbol{r}_{\bullet k}}$ and cloud barycentric  positions ${\boldsymbol{r}^{c}_{ k}} $ for $k=1,2$.   Similar   to the case of a  single SMBH, we  calculate the  surface mesh of approximate values of the composite differential phase (see Fig. \ref{fig:ilustris2}) using
	Eq. \ref{eq:posled}. 
	We plot it  on a grid defined by orbital ascending node and the true anomaly of the more massive SMBH. 
		To test the differential phase  sensitivity to parameters, we varied the parameters' values from within their ranges as reported in Table \ref{tab:A:1}.

	\begin{table*}[h!]
			\caption{ Parameter sets used in testing the differential phase sensitivity for CB-SMBH based on Eq. \ref{eq:posled}. For clarity, subscripts of parameters that relate to SMBHs components  are set to 1 and 2; and for clouds they are set to  $ci, i=1,2$ indicating  the clouds of respective BLRs of  SMBHs. }
		\centering
		
		\subtable[Case 1]{

			\small
			\centering
		
			\begin{tabular}{|c|c|c|c|c|c|c|c|c|c|}
				
					\hline
				\multicolumn{10}{|c|}{CB-SMBH parameters} \\
				
			\hline
				$a_{1}$& $e_{1}$& $\omega_{1}$& $ i_{1}$ & $a_{2}$& $e_{2}$& $\Omega_{2}$ & $\omega_{2}$& $i_{2}$ &\\
			$\mathrm{[AU]}$&&[$\mathrm{rad}$]& [$\mathrm{rad}$]&[AU]&&[$\mathrm{rad}$]&[$\mathrm{rad}$]&[$\mathrm{rad}$] &\\
				\hline
				
				5& 0.69& 6&0.148&10& 0.596&4.62& 5& 0.587 &\\
				\hline
			
				\multicolumn{10}{|c|}{Clouds parameters} \\
				\hline
				
				$a_{c1}$&$e_{c1}$&$ \Omega_{c1}$&$ \omega_{c1}$&$ i_{c1}$&
			$a_{c2}$& $e_{2}$&$ \Omega_{c2}$ & $\omega_{c2}$& $i_{c2}$\\
		
			$\mathrm{[AU]}$& & [$\mathrm{rad}$]&[$ \mathrm{rad}$]&[$ \mathrm{rad}$]&[AU]& & [$ \mathrm{rad}$]&  [$\mathrm{rad}$]& [$ \mathrm{rad}$]\\				
				
				\hline
				
			1& 0.77&4.71 & 4.6&0.1&1 & 0.428& 6& 4&0.7\\
			\hline				
			\end{tabular}
		
			\label{tab:A:1a}
				
		}
	\\
		\subtable[Case 2]{
		\small
		\centering
		
		\begin{tabular}{|c|c|c|c|c|c|c|c|c|c|}
			\hline
			 \multicolumn{10}{|c|}{CB-SMBH parameters} \\
			\hline
		$a_{1}$&$e_{1}$&$\omega_{1}$& $ i_{1}$& $ a_{2}$& $e_{2}$&$ \Omega_{2}$&  $\omega_{2}$& $i_{2}$&\\
			$\mathrm{[AU]}$& & [$\mathrm{rad}$]&[$ \mathrm{rad}$]&[AU]& & [$ \mathrm{rad}$]&  [$\mathrm{rad}$]& [$ \mathrm{rad}$]&\\	
			\hline
	5& 0.61&2.28&0.446&10&0.795& 2.638& 3& 0.395&\\
			\hline
	   \multicolumn{10}{|c|}{Clouds parameters} \\
		     
		        \hline
			$a_{c1}$&$e_{c1}$&$ \Omega_{c1}$&$ \omega_{c1}$&$ i_{c1}$&
			$a_{c2}$& $e_{2}$&$ \Omega_{c2}$ & $\omega_{c2}$& $i_{c2}$\\

			[AU]& & [$\mathrm{rad}$]&[$ \mathrm{rad}$]&[$ \mathrm{rad}$]&[AU]& & [$ \mathrm{rad}$]&  [$\mathrm{rad}$]& [$ \mathrm{rad}$]\\				
			
			\hline
			
			 1 &0.77& 3.82&2.66& 0.43& 1,&0.65&3.58& 4& -0.78\\
	
			\hline				
		\end{tabular}
		
		\label{tab:A:1b}
		}
	\\
			\subtable[Case 3]{

		\small
		\centering

		\begin{tabular}{|c|c|c|c|c|c|c|c|c|c|}
			
				\hline
			\multicolumn{10}{|c|}{CB-SMBH parameters} \\
		
			\hline
			$a_{1}$&$e_{1}$&$\omega_{1}$& $ i_{1}$& $ a_{2}$& $e_{2}$&$ \Omega_{2}$&  $\omega_{2}$& $i_{2}$&\\	
			
			[AU]&&[$\mathrm{rad}$]& [$\mathrm{rad}$] &[AU]&&[$\mathrm{rad}$]&[$\mathrm{rad}$]&[$\mathrm{rad}$] &\\
			\hline
		5&0.66&0.01& 0.0001&10& 0.66&1.62& 3& 0.0001&\\
			\hline

			\multicolumn{10}{|c|}{Clouds parameters} \\
			\hline
			
			$a_{c1}$&$e_{c1}$&$ \Omega_{c1}$&$ \omega_{c1}$&$ i_{c1}$&
			$a_{c2}$& $e_{c2}$&$ \Omega_{c2}$ & $\omega_{c2}$& $i_{c2}$\\	
			
			[AU]& &[ $ \mathrm{rad}$]&[$ \mathrm{rad}$]&[$ \mathrm{rad}$]&[AU]& &[ $ \mathrm{rad}$]&  [$\mathrm{rad}$]& [$ \mathrm{rad}$]\\				
			
			\hline
			
				1&0.33&1.81&2.66&2.33&1&0.249&1.96& 4&2.75\\

			\hline				
		\end{tabular}
		
		\label{tab:A:1c}
		
		}
	\\
	
				\subtable[Case 4]{

		\tiny
		\centering
		
		\begin{tabular}{|c|c|c|c|c|c|c|c|c|c|}
				\hline
			\multicolumn{10}{|c|}{CB-SMBH parameters} \\
			
			\hline
			$a_{1}$&$e_{1}$&$\omega_{1}$& $ i_{1}$& $ a_{2}$& $e_{2}$&$ \Omega_{2}$&  $\omega_{2}$& $i_{2}$&\\	
			
		[	AU]&&[$\mathrm{rad}$]&[ $\mathrm{rad}$] &[AU]&&[$\mathrm{rad}$]&[$\mathrm{rad}$]&[$\mathrm{rad}$] &\\
			\hline
			
			5& 0.786&5.68& 0.0001& 10& 0.836&5.32& 3& 0.0001 &\\
			\hline
				\multicolumn{10}{|c|}{Clouds parameters} \\
		
			\hline
			$a_{c1}$&$e_{c1}$&$ \Omega_{c1}$&$ \omega_{c1}$&$ i_{c1}$&
			$a_{c2}$& $e_{c2}$&$ \Omega_{c2}$ & $\omega_{c2}$& $i_{c2}$\\	
			
			[AU]& &[ $ \mathrm{rad}$]&[$ \mathrm{rad}$]&[$ \mathrm{rad}$]&[AU]& &[ $ \mathrm{rad}$]&  [$\mathrm{rad}$]& [$ \mathrm{rad}$]\\				
			
			\hline
			
		1& $rnd \mathcal{R}(1)$& 0.1& 2.66&0.433&1& $rnd \mathcal{R}(1)$& 3.58& 4& -0.785\\

			\hline				
		\end{tabular}
		
		\label{tab:A:1d}
	}
	
		\label{tab:A:1}
	\end{table*}

	The calculated surface  consits of two ridges and two valleys. The appearance of a complex system of ridges depends on eccentricities  and   ascending nodes of both SMBHs. If  eccentricities of both SMBHs are  smaller than 0.5, the second ridge  would disappear. Moreover, orbital ascending nodes  and angles of pericenter of both SMBHs  affect the position and slope  of ridges and valleys. The inclinations of SMBH orbits control the slopes of ridges, and the larger inclination values decrease slopes. The effects  of clouds  and SMBHs orbital elements  are similar, but former  effects are  smaller.
	 The system of ridges and valleys  cause  complicated differential phase shapes  similar to the  surface side intersections (upper and bottom left plots of Fig. \ref{fig:ilustris2}). 

\section{Atlas of interferometric observables for single SMBH}\label{appendix:atlas}

\renewcommand{\thesubfigure}{B\alph{subfigure}}

\begin{figure*}[ht!]
	\begin{center}

			\subfigure[$\Omega_{c} \in \mathcal{U}(10^{\circ}, 90^{\circ}), \omega_{c}=110^{\circ},\newline \hspace*{1.5em} i_{c} \in \mathcal{U} (-5^{\circ},5^{\circ}), e_{c}=0.5,\newline \hspace*{1.5em} i_{0}=45^{\circ}$]
	{%
		\label{fig:singleel11}
		\includegraphics[trim = 3.0mm 0mm 2.0mm 0mm, clip, width=0.23\textwidth]{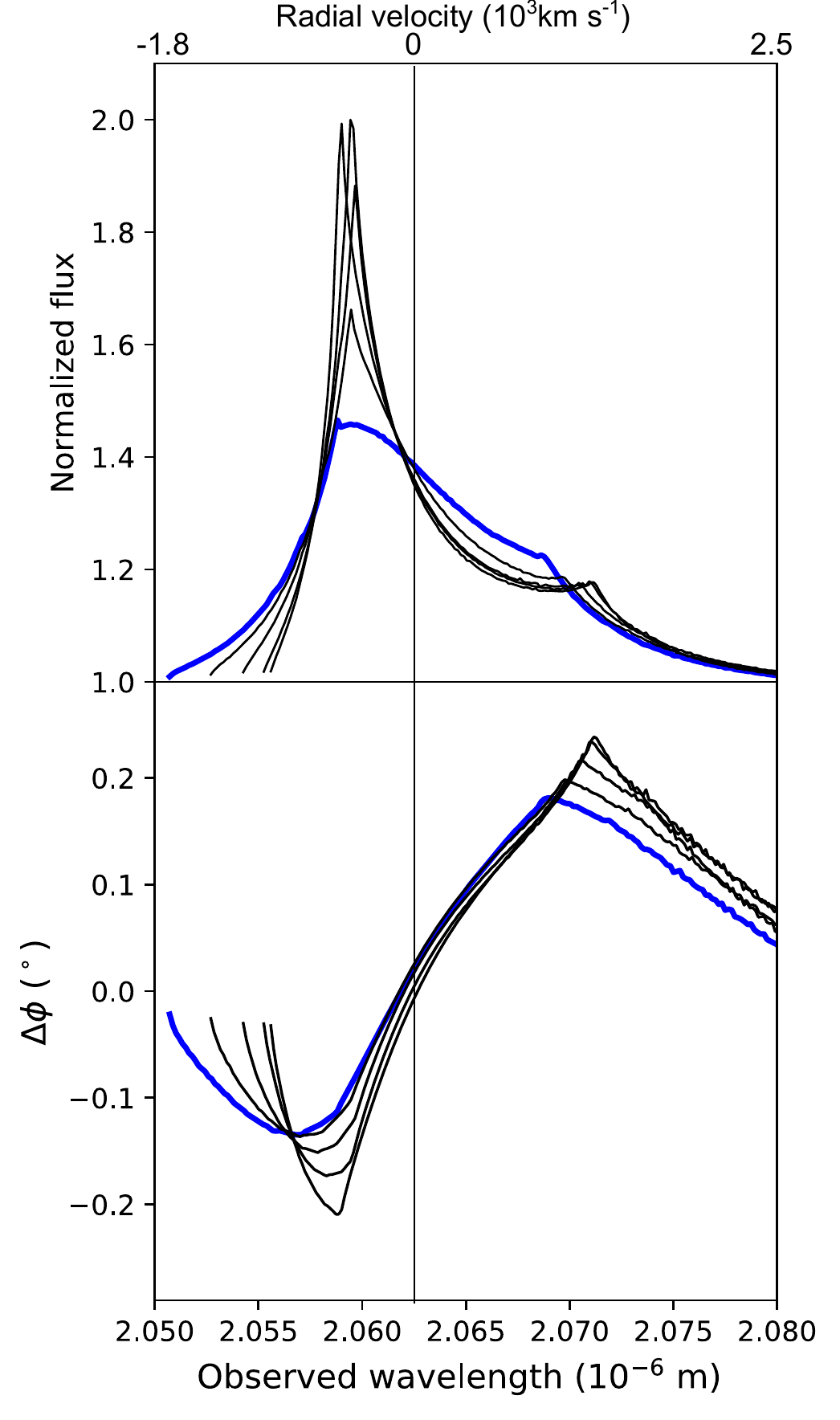}
	}%
	\hspace{-0.2em}
		\subfigure[$i_{c}=\mathcal{U}(-5^{\circ},5^{\circ})$,$\Omega_{c}=100^{\circ}$, \newline \hspace*{1.5em}  $\omega_{c}=10^{\circ}$, $i_{0}=45^{\circ}$, $e_{c}=\newline \hspace*{1.5em}  \mathcal{U}(0.1, 0.5)$]
		{%
			\label{fig:singleel14}
			\includegraphics[width=0.23\textwidth]{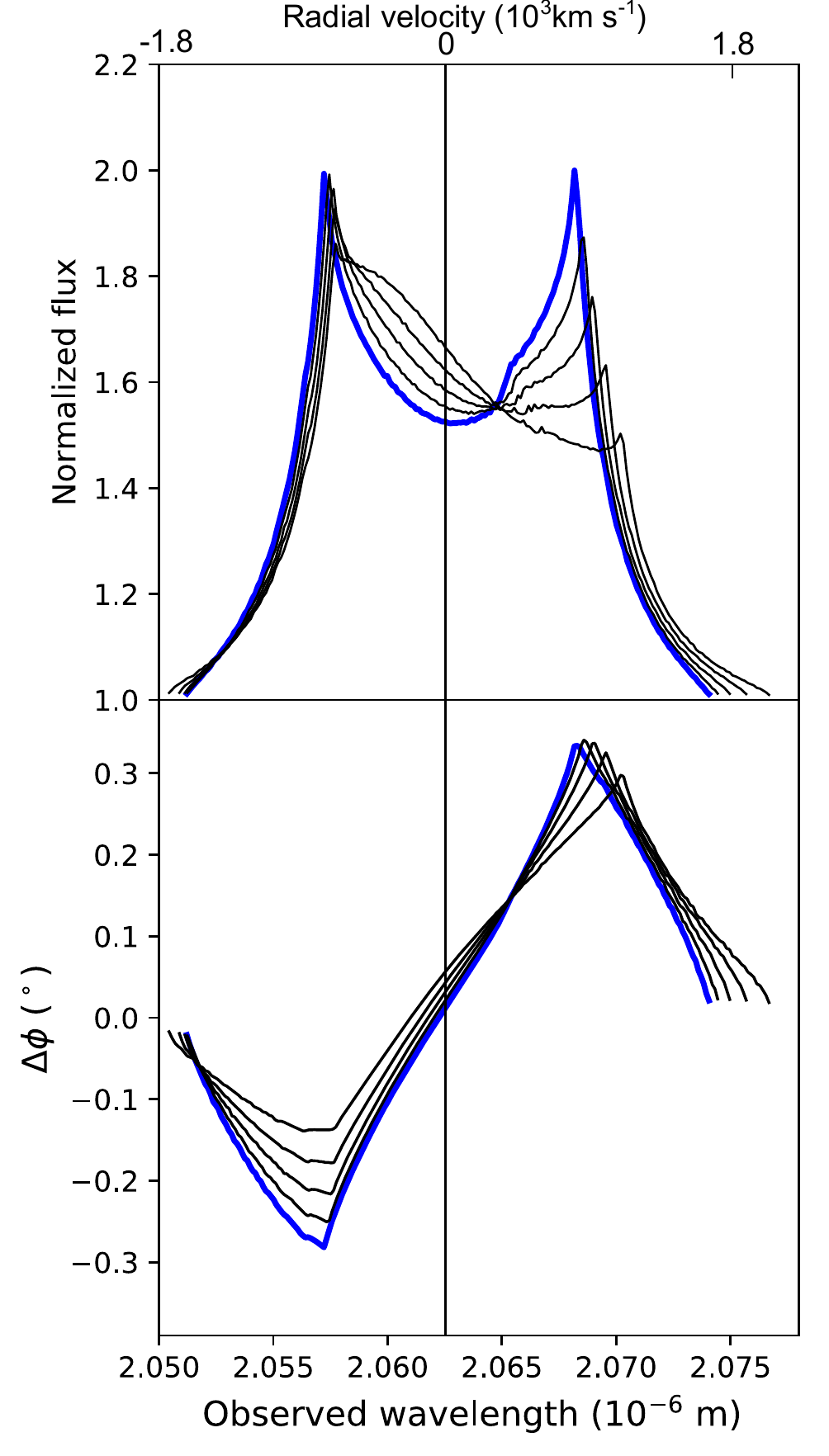}
		}%
		\hspace{-0.2em}
		\subfigure[$i_{c}=\mathcal{U}(-5^{\circ}, 5^{\circ})$, $\Omega_{c}=100^{\circ}$,\newline \hspace*{1.5em}  $\omega_{c}=110^{\circ}$, $i_{0}=45^{\circ}$, $e_{c}= \newline \hspace*{1.5em} \mathcal{U}(0.1,0.5)$
		]{%
			\label{fig:singleel15}
			\includegraphics[trim = 1.0mm 1mm 0.0mm 0mm, clip,width=0.23\textwidth]{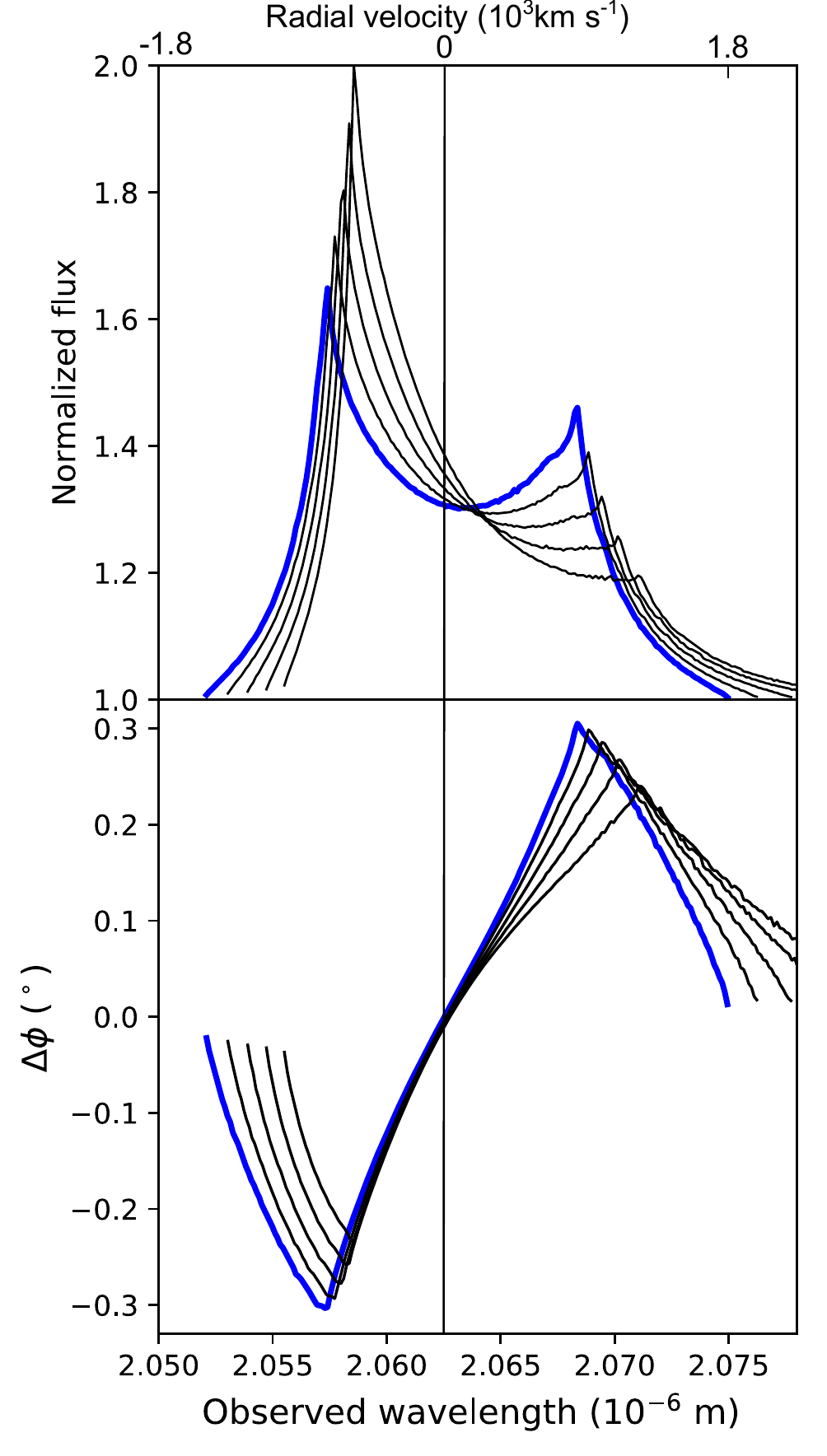}
		}
		\hspace{-0.2em}
		\subfigure[$i_{c}=\mathcal{U}(-5^{\circ}, 5^{\circ})$, $\Omega_{c}=100^{\circ}$,  \newline \hspace*{1.5em}          $e_{c}=0.5$,                                  
	  $\omega_{c}=270^{\circ}$, $i_{0}=	\newline \hspace*{1.5em} \mathcal{U}(10^{\circ},45^{\circ})$]{%
			\label{fig:singleel16}
			\includegraphics[trim = 3.0mm 0mm 2.0mm 0mm, clip,width=0.23\textwidth]{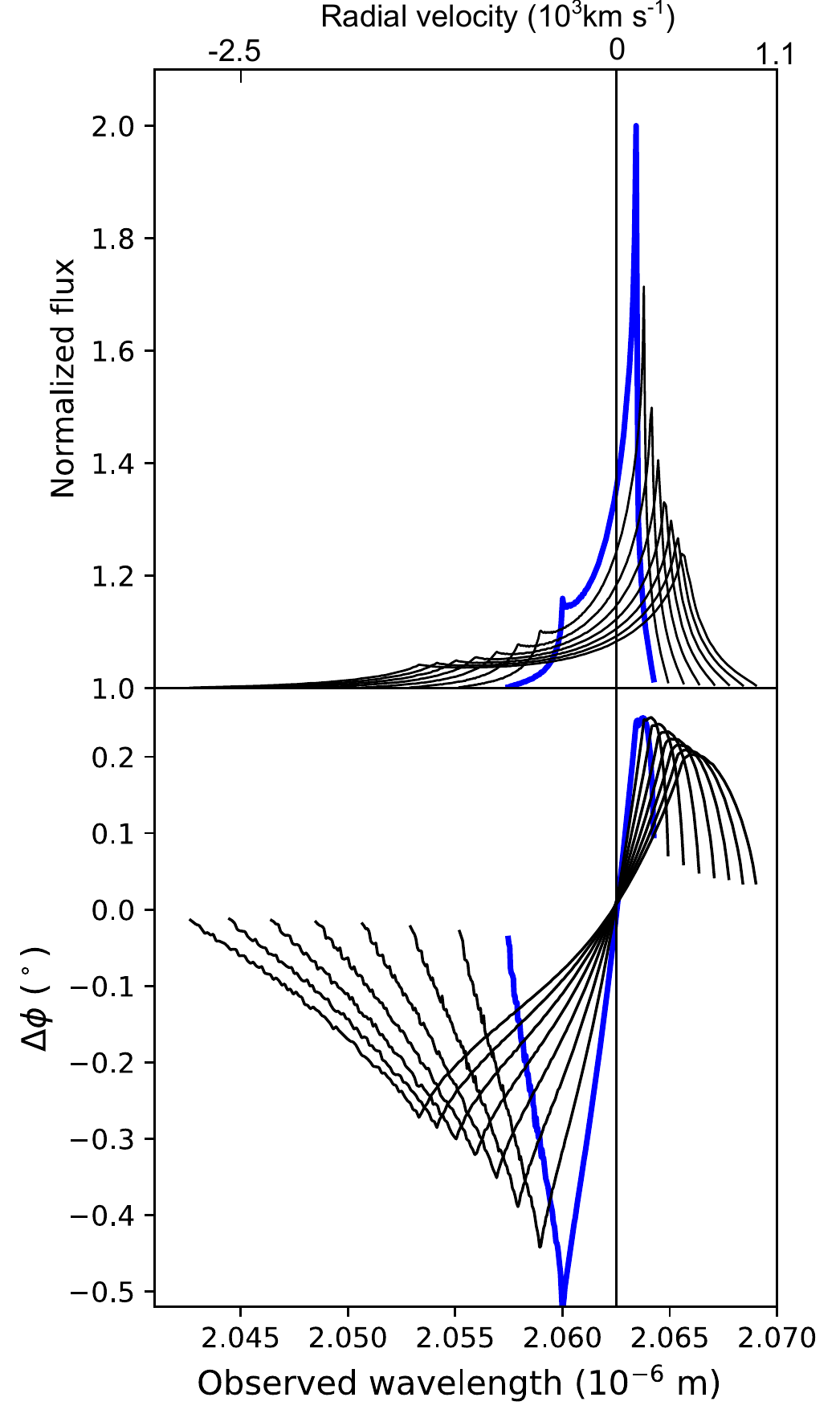}
		}%
		\hspace{-0.2em}
	
	\end{center}
	\caption{%
		Evolution of the Pa$\alpha$ emission line (upper subplots) and corresponding differential  phase ($\Delta \phi$, lower subplots) as a function of the wavelength  and radial velocity for different values of clouds orbital parameters in the model of single SMBH.  Model parameters are given in sub-captions. 
	}%
	
	\label{fig:singleel1}
\end{figure*}

  We present three atlases of observables  assuming non-randomized motion.
In the first series, shown in Fig. \ref{fig:singleel1}, the clouds orbital inclinations have uniform distribution $i_{c}=\mathcal{U}(-5^{\circ},5^{\circ})$. The  evolutions of observables depicted  in Fig. \ref{fig:singleel11} are  less dramatic then in  Figs. \ref{fig:singleel12} and \ref{fig:singleel13}. 
 Variation of $i_0$ shapes the evolution of both observables with largest effects seen when $\omega_{c}=270^{\circ}$ (see Figs. \ref{fig:singleel14}-\ref{fig:singleel16}). The left peak of the  spectral line is more prominent  when  $\Omega_{c},\,\omega_{c}\leq 180^{\circ}$ and   $e_{c}>0.3$,  but the right peak dominates when $\omega_{c}=270^{\circ}$.
 These two  orbital  parameters   have a more significant influence on the amplitude of differential phase when eccentricity  is smaller  (Figs. \ref{fig:singleel14}- \ref{fig:singleel15}).
  Notable net effects of both orbital shape angles  are found  when they have simultaneously  larger values  (compare Figs. \ref{fig:singleel15} and \ref{fig:singleel16} to Fig. \ref{fig:singleel14}).

\begin{figure*}[ht!]
	\begin{center}

		\subfigure[ $\omega_{c}=10^{\circ}$]	{%
			\label{fig:singleel21}
			\includegraphics[trim = 3.0mm 0mm 4.0mm 0mm, clip, width=0.3\textwidth]{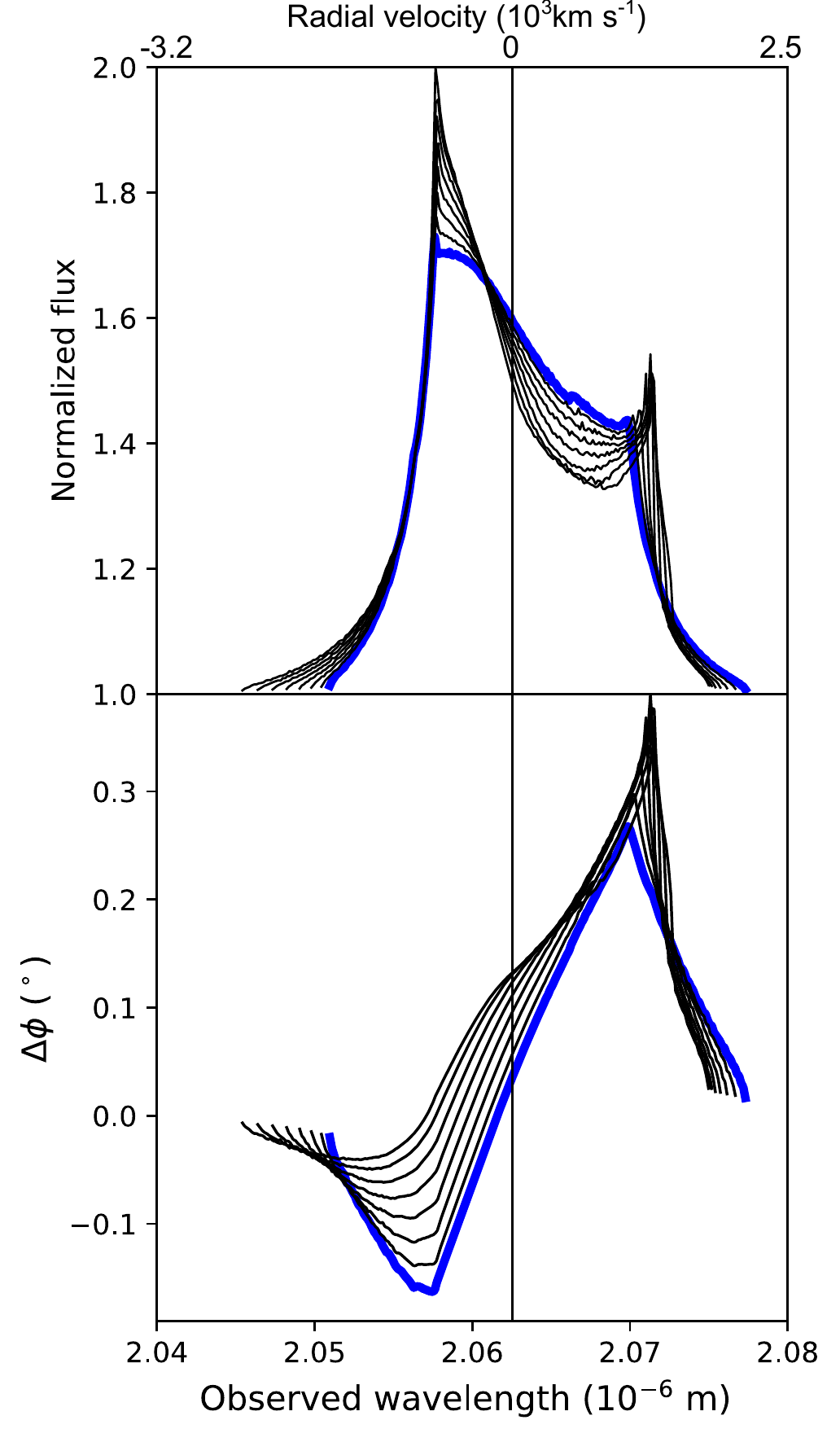}
		}%
		\hspace{-0.5em}
		\subfigure[ $\omega_{c}=110^{\circ}$]{%
			\label{fig:singleel22}
			\includegraphics[trim = 3.0mm 0mm 5mm 0mm, clip,width=0.3\textwidth]{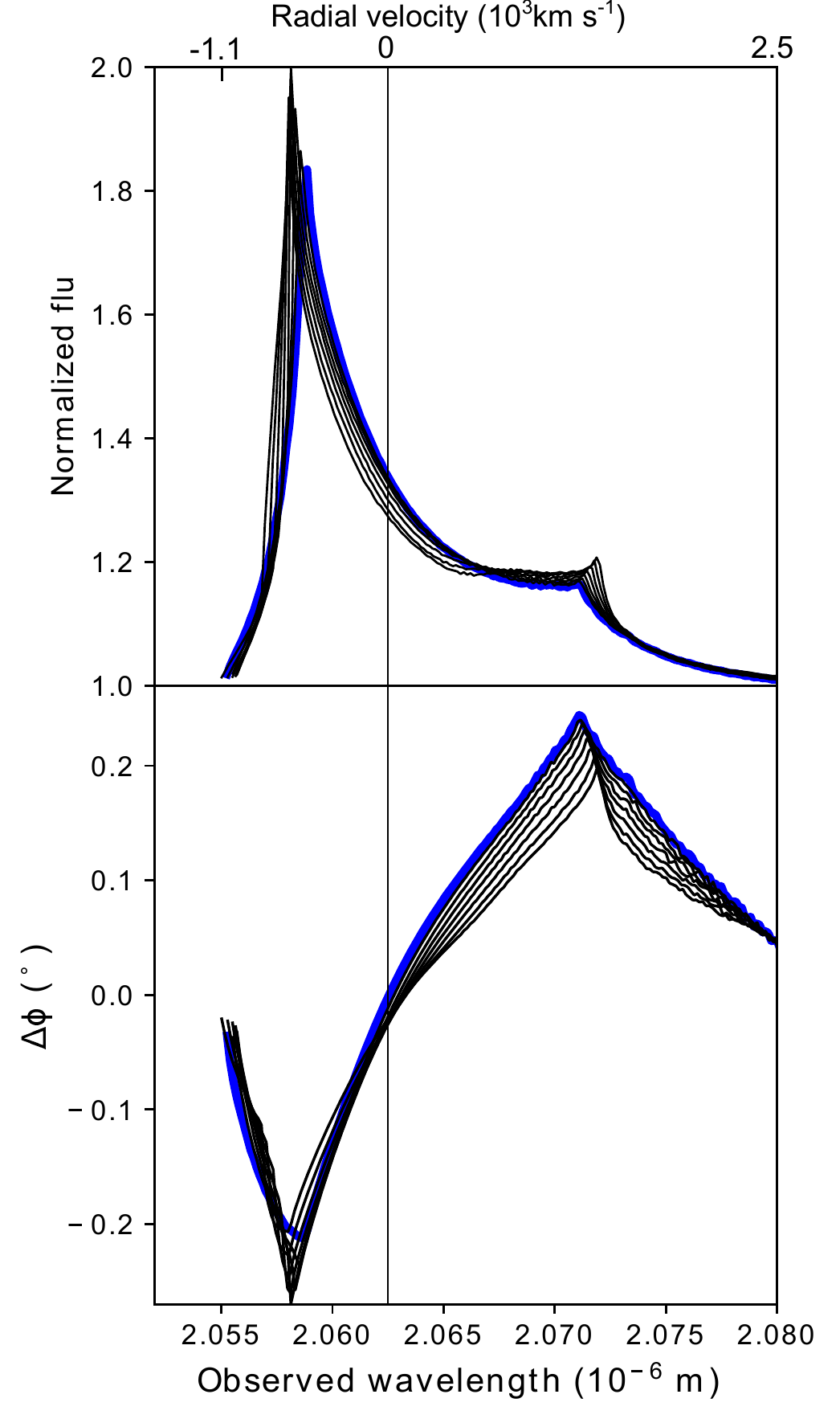}
		}
		\hspace{-0.5em}
		\subfigure[ $\Omega=210^{\circ},\omega_{c}=10^{\circ}, e_{c}=0.5, i_{c}=\newline \hspace*{1.5em}\mathcal{U}(-7.5^{\circ},7.5^{\circ}), i_{0}=\mathcal{U}(10^{\circ},45^{\circ})$]{%
			\label{fig:singleel23}
			\includegraphics[trim = 3.0mm 0mm 1mm 0mm, clip, width=0.3\textwidth]{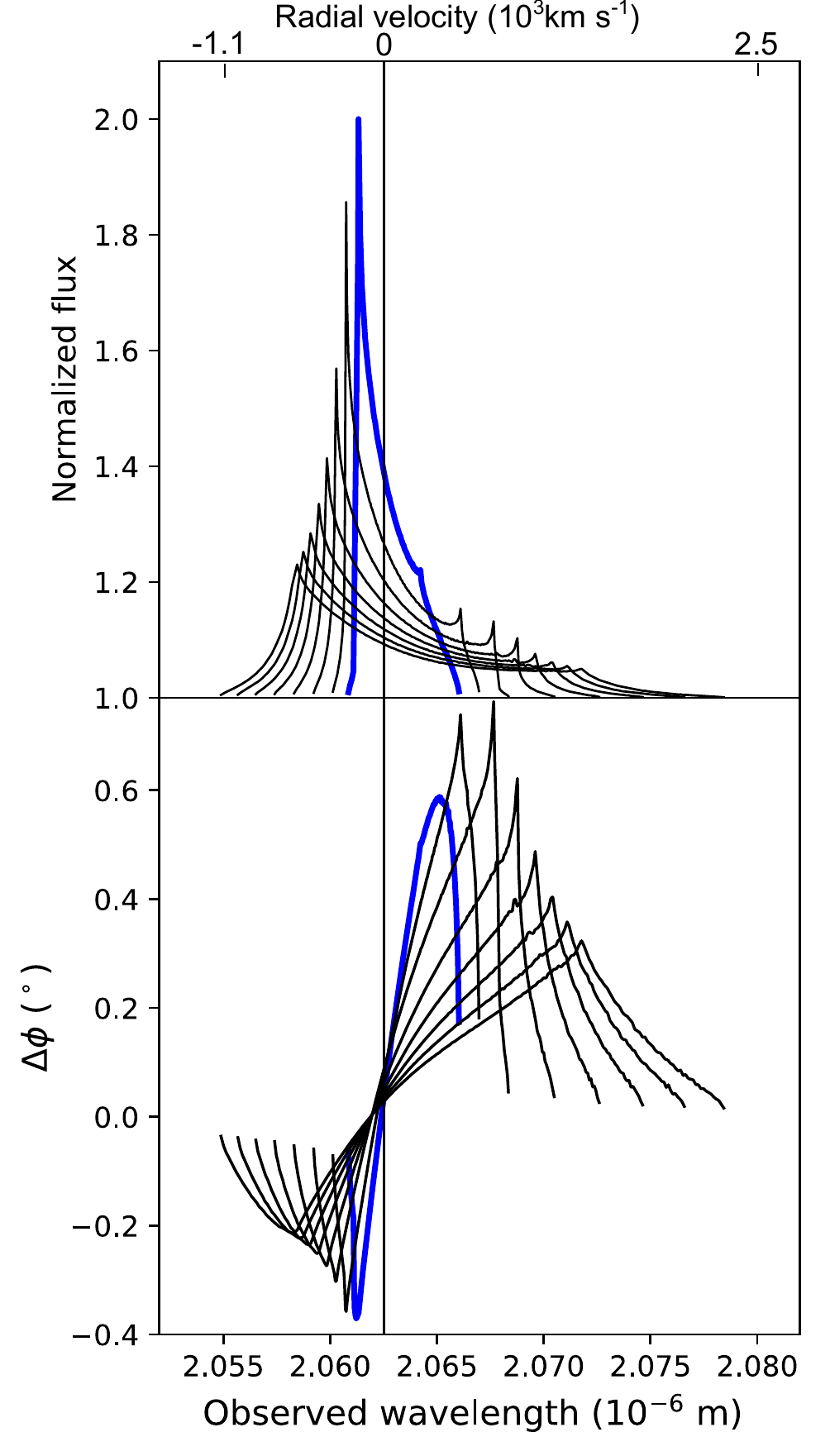}
			
		}	
		\vspace{-1.9em}
	\end{center}
	\caption{%
		Same as 	Fig. \ref {fig:singleel1} but for fixed  values of  clouds' orbital eccentricities  $e_{c}=0.5$ and  angular position of observer $i_{0}=45^{\circ}$. (a), (b) $\Omega_{c}=100^{\circ}$ and  the range of clouds' orbital inclinations for each model is from 
		 $\mathcal U(-5^{\circ}, 5^{\circ})$  to $\mathcal U(-20^{\circ}, 20^{\circ})$.
	}%
	
	\label{fig:singleel2}
\end{figure*}

\begin{figure*}[ht!]
	\begin{center}
		\begin{minipage}[s][10.5cm]{0.45\textwidth}
			\subfigure[ $\omega_{c}=10^{\circ}$]	{%
				\label{fig:singleel31}	
				\includegraphics[trim =1.0mm 3mm 2mm 3.5mm, width=6cm,height=5cm]{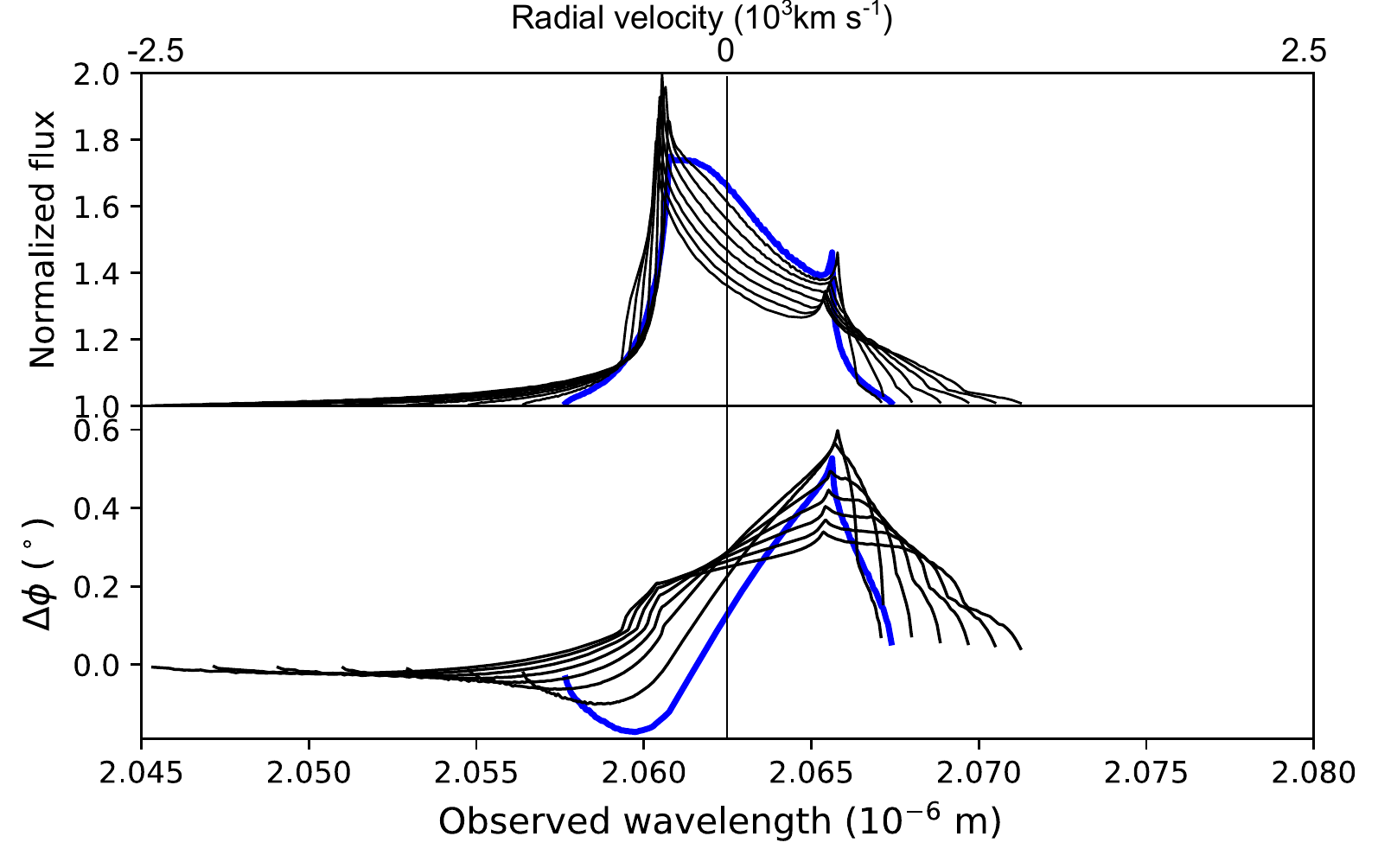}
			}\vfil
			\subfigure[ $\omega_{c}=110^{\circ}$]
			{%
				\label{fig:singleel32}
				\includegraphics[trim = 1.0mm 3mm 2mm 3.5mm, clip,width=6cm,height=5cm]{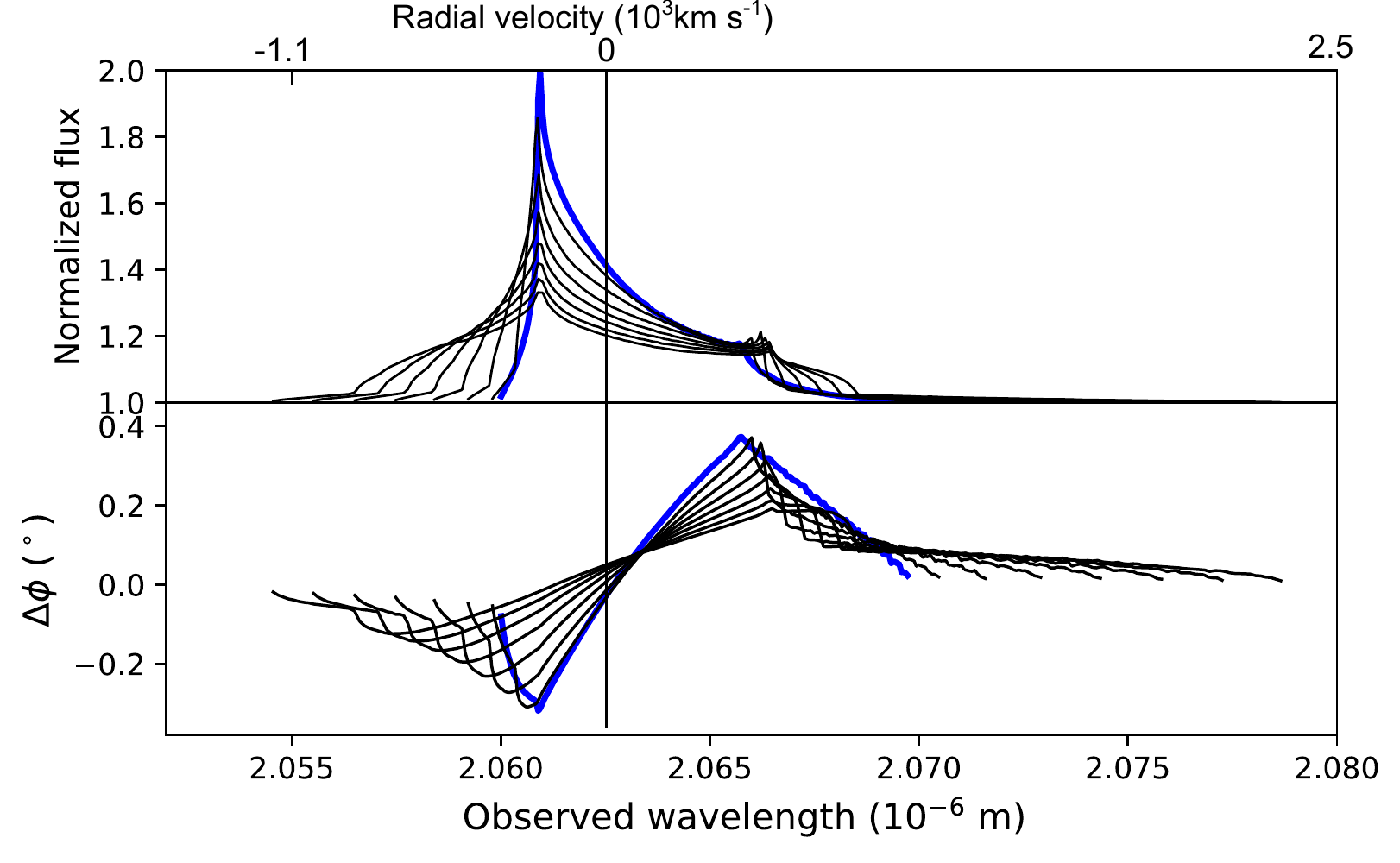}
				
			}
		\end{minipage}
		\hspace{-5em}
		\begin{minipage}[s][7.9cm]{0.45\textwidth}
			\subfigure[$i_c=\mathcal{U}(-40^{\circ}, 40^{\circ})$, $i_{0}=45^{\circ}$, 
			$\omega_{c}=\newline \hspace*{1.5em}(10^{\circ}, 180^{\circ})$ ]{%
				\label{fig:singleel33}
				\includegraphics[trim = 1.5mm 0mm 2mm 0mm, clip,width=5cm,height=9.7cm]{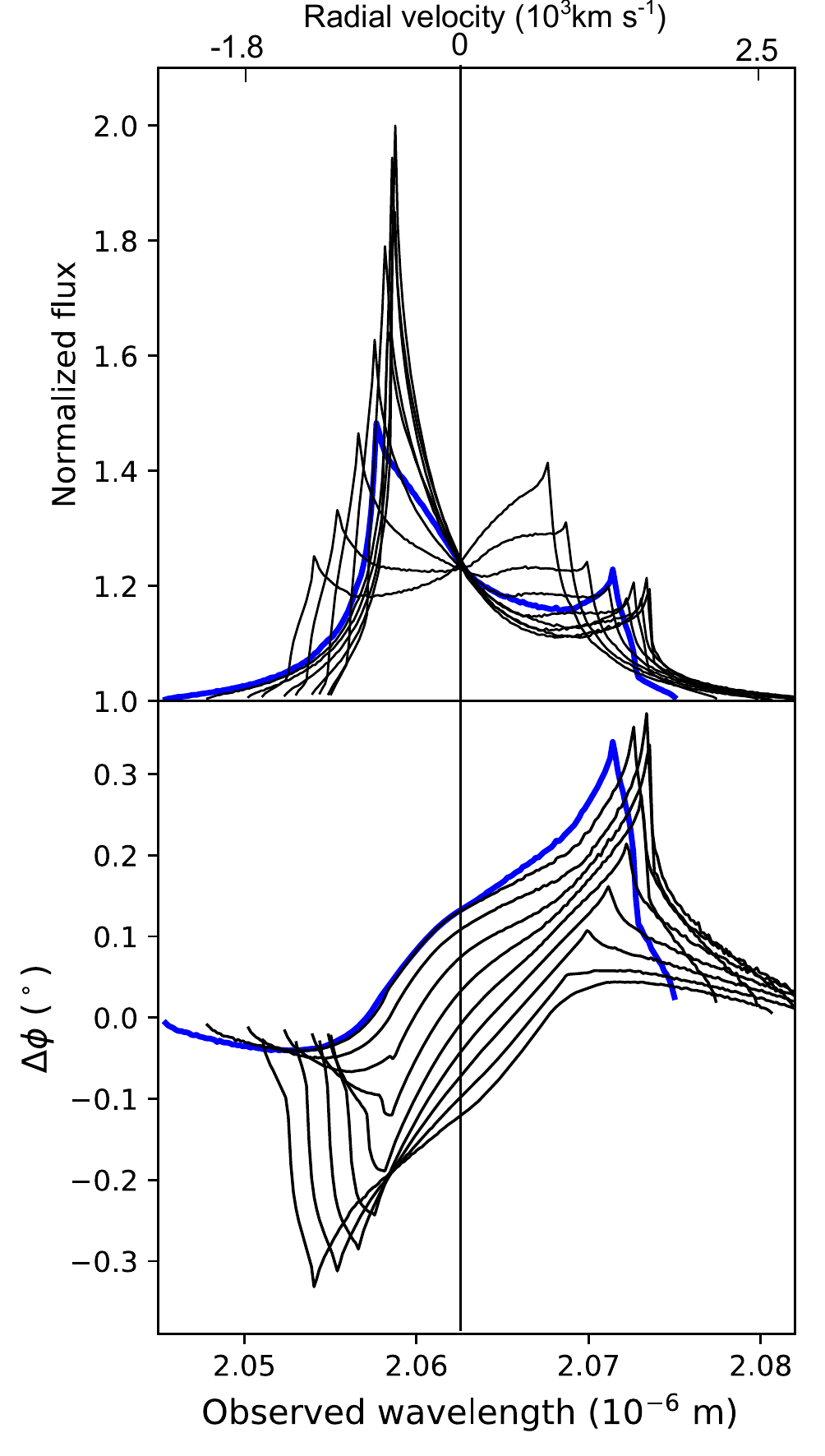}
			}
		\end{minipage}
		
	\end{center}
	\caption{%
		Same as 	Fig. \ref {fig:singleel1} but for the  range of clouds' orbital inclinations from 
		$\mathcal U(-5^{\circ}, 5^{\circ})$  to $\mathcal U(-20^{\circ}, 20^{\circ})$, fixed  values of clouds orbital eccentricities $e_{c}=0.5$, and $i_{0}=15^{\circ}$. The rest of the parameters are given in the sub-captions of the plots.
	}%
	
	\label{fig:singleel3}
\end{figure*}

\begin{figure*}[ht!]
	\begin{center}
		
		\hspace{-0.8em}
		\subfigure[$  i_{0}= \mathcal {U} (-10^{\circ},45^{\circ}),\delta i_{0}=5^{\circ}, i_{c}=\newline \hspace*{1.5em}\mathcal {U}(-20^{\circ},20^{\circ}),  \Omega_{c}=  100^{\circ},  \omega_{c}=270^{\circ},    \newline \hspace*{1.5em}  e_{c}\in\Gamma_{s}  (0.3,1)$
		]{%
			\label{fig:singledistr42}
			\includegraphics[trim = 2.5mm 0mm 1mm 0mm, clip,width=0.3\textwidth]{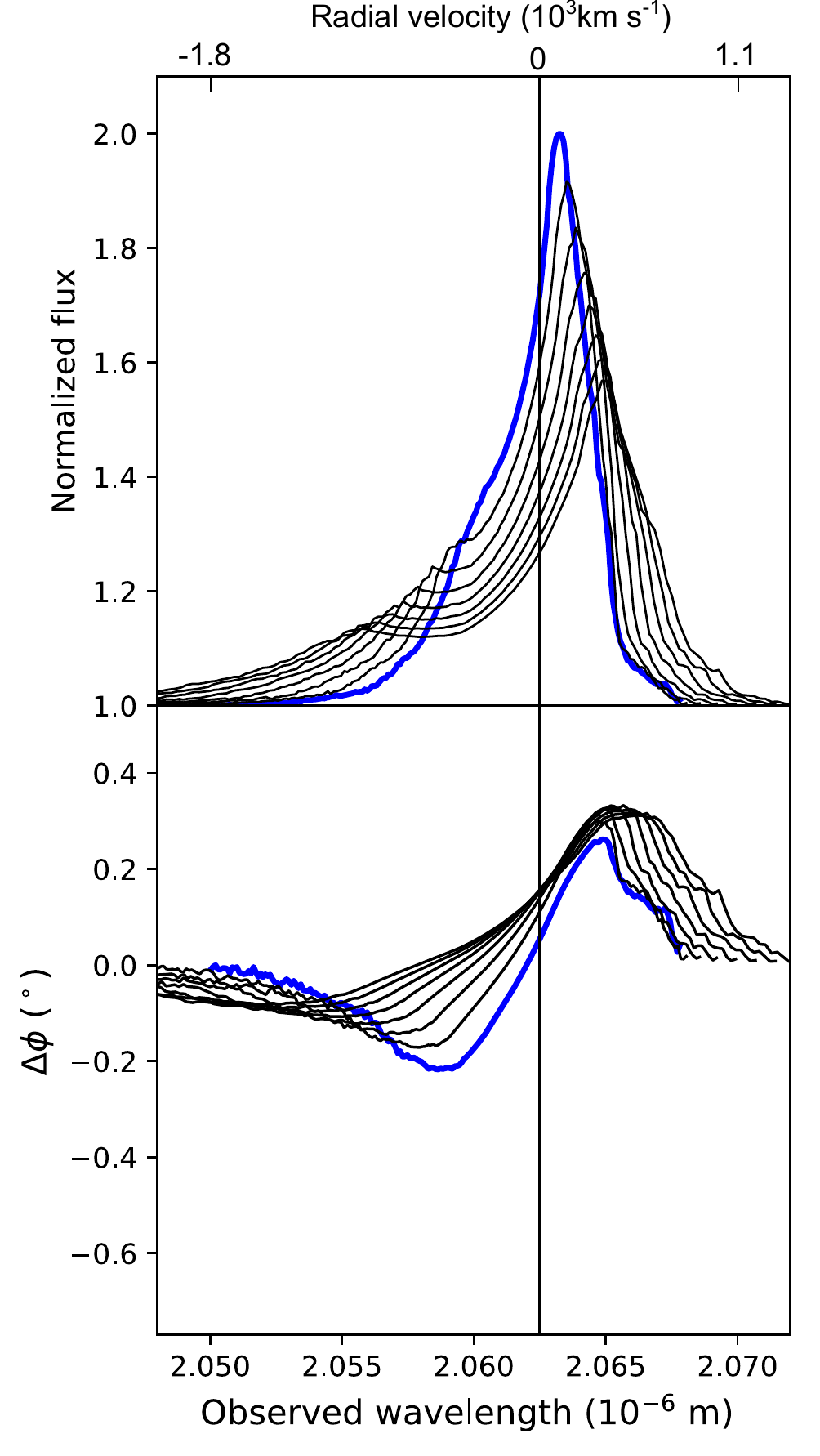}
		}
		\hspace{-0.8em}
		\subfigure[$i_{0}=\mathcal {U}(-10^{\circ},45^{\circ}),\delta i_{0}=5^{\circ},  \mathcal{C}, \Omega_{c}= \newline \hspace*{1.5em}  100^{\circ},   \omega_{c}=270^{\circ},  e_{c}\in \mathcal{R}(1)$
		]{%
			\label{fig:singledistr43}
			\includegraphics[trim = 2.5mm 0mm 1mm 0mm, clip,width=0.3\textwidth]{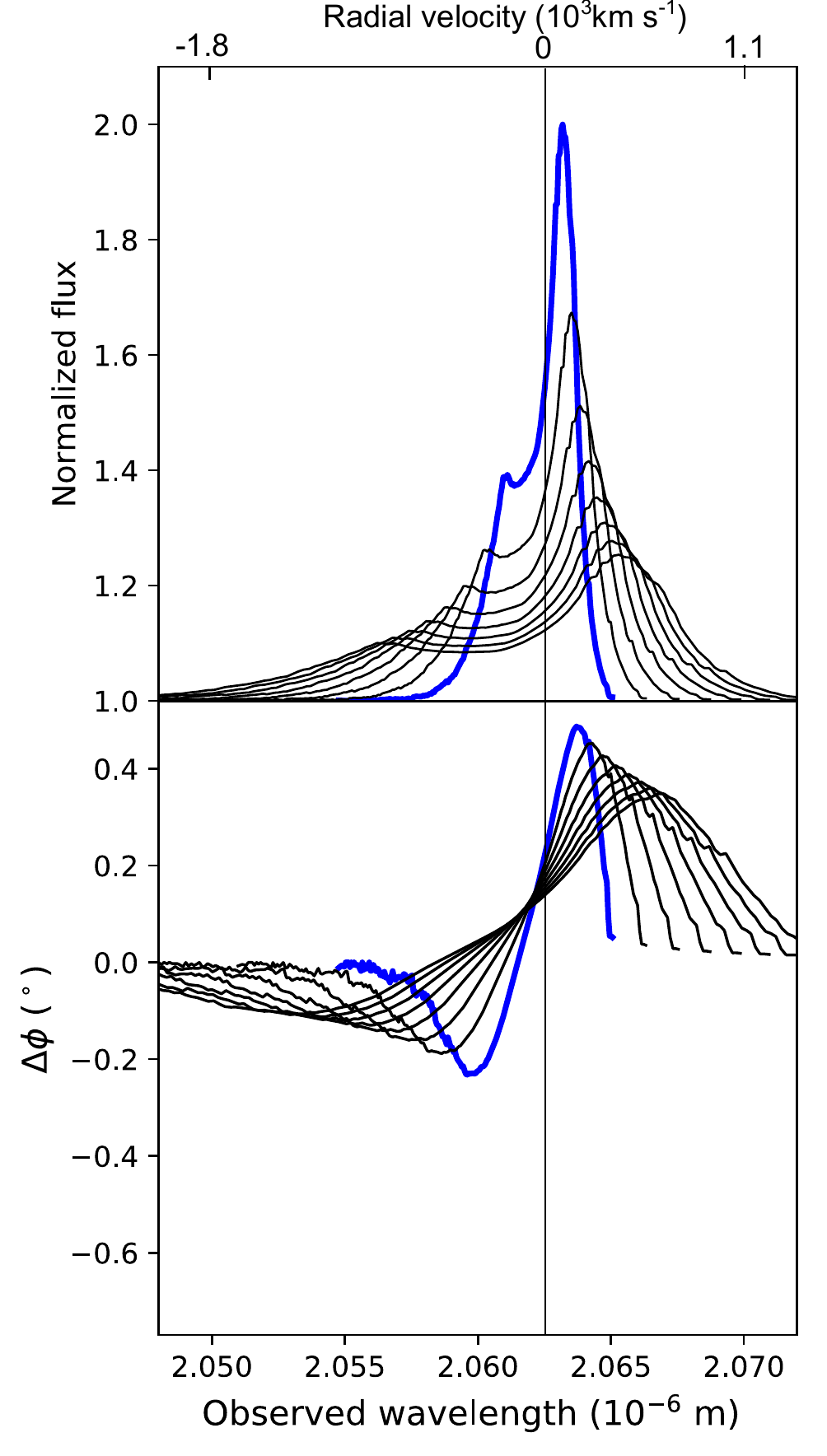}
		}%
			\hspace{-0.8em}
		\subfigure[$ i_{0}=\mathcal {U}  (-10^{\circ},45^{\circ}),\delta i_{0}=5^{\circ}, i_{c}=\newline \hspace*{1.5em} \mathcal {U} (-10^{\circ},10^{\circ}),  \Omega_{c}=100^{\circ}, \omega_{c}=10^{\circ},  \newline \hspace*{1.5em}     e_{c}\in \mathcal{R}(1)$
		]{%
			\label{fig:singledistr48}
			\includegraphics[trim = 2.5mm 0mm 1mm 0mm, clip,width=0.3\textwidth]{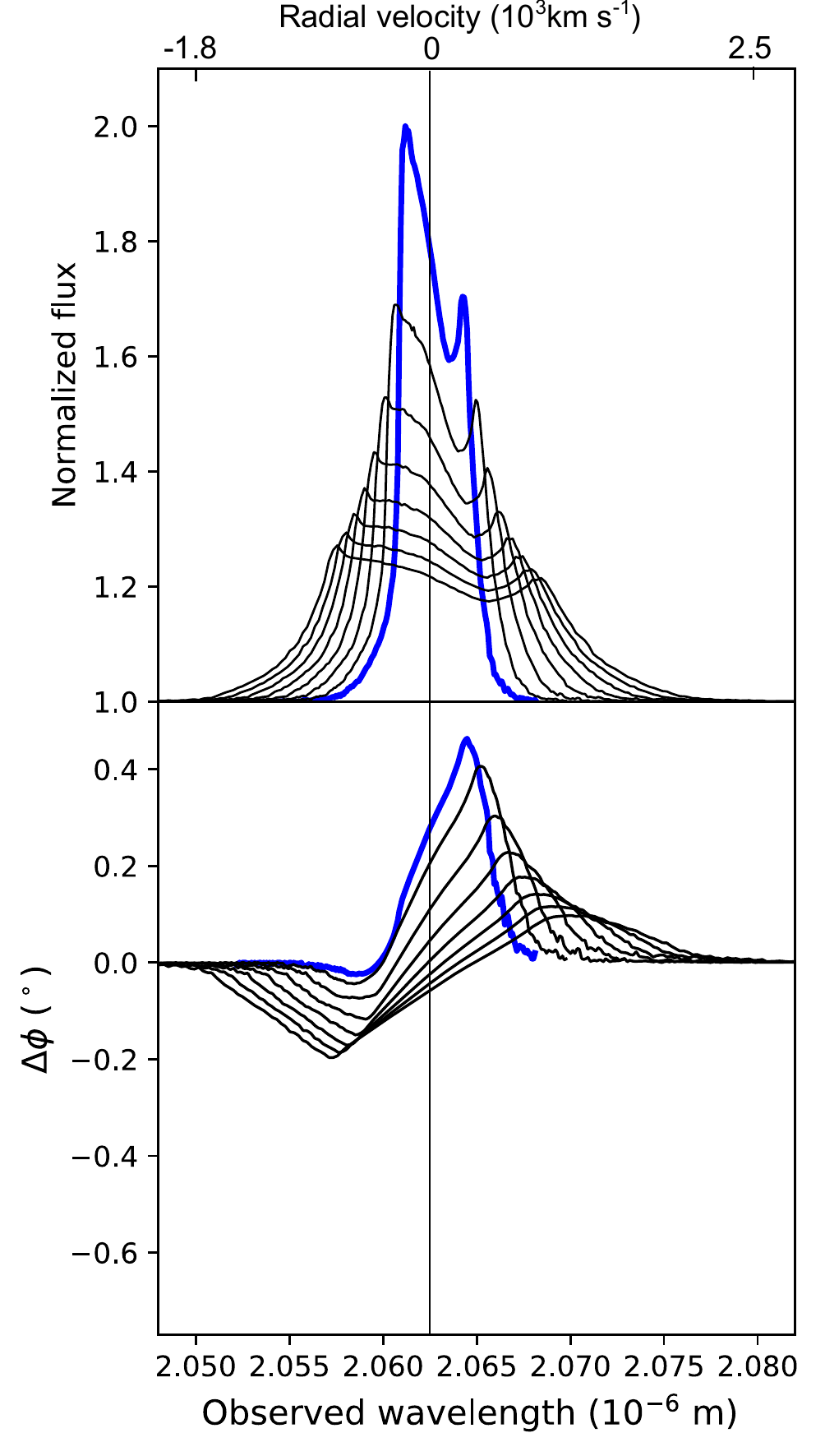}
		}%
		\\ 
		\vspace{-1.em}
		\subfigure[$i_{0}=\mathcal {U}(-10^{\circ},45^{\circ}),\delta i_{0}=5^{\circ},  \mathcal{C}, \Omega_{c}=  \newline \hspace*{1.5em} 100^{\circ},   \omega_{c}=10^{\circ},  e_{c}\in\Gamma_{s}  (0.3,1)$]
		{%
			\label{fig:singledistr45}
			\includegraphics[trim = 2.5mm 0mm 1mm 0mm, clip, width=0.3\textwidth]{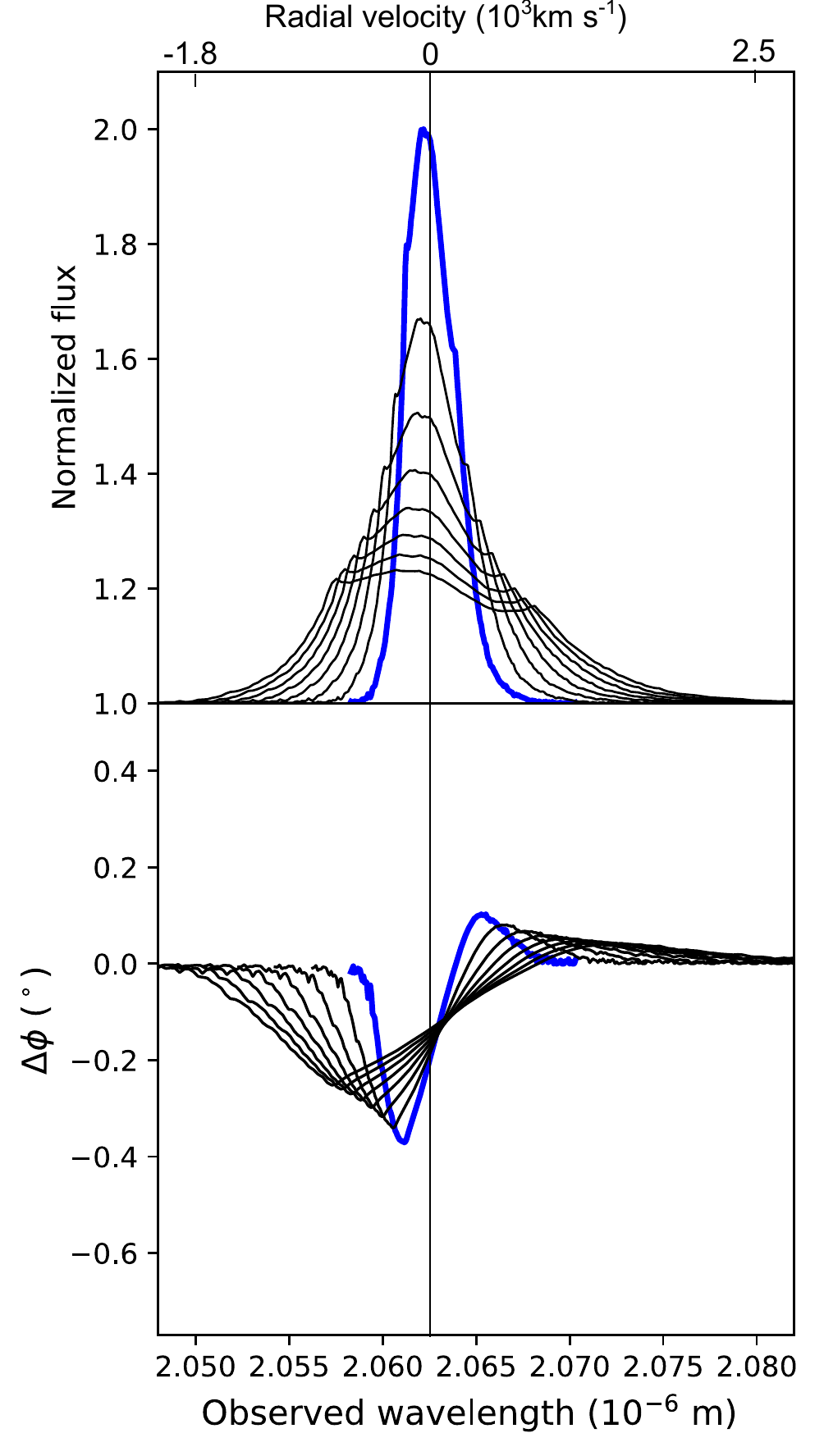}
		}%
		\hspace{-0.8em}
		\subfigure[$  i_{0}= \mathcal {U}(-10^{\circ},45^{\circ}),\delta i_{0}=5^{\circ}, i_{c}=\newline \hspace*{1.5em} (-7.5^{\circ},7.5^{\circ}), \Omega_{c}=  100^{\circ},   \omega_{c}=10^{\circ}, \newline \hspace*{1.5em}    e_{c}\in\Gamma_{s}  (0.3,1)$
		]{%
			\label{fig:singledistr46}
			\includegraphics[trim = 2.5mm 0mm 1mm 0mm, clip,width=0.3\textwidth]{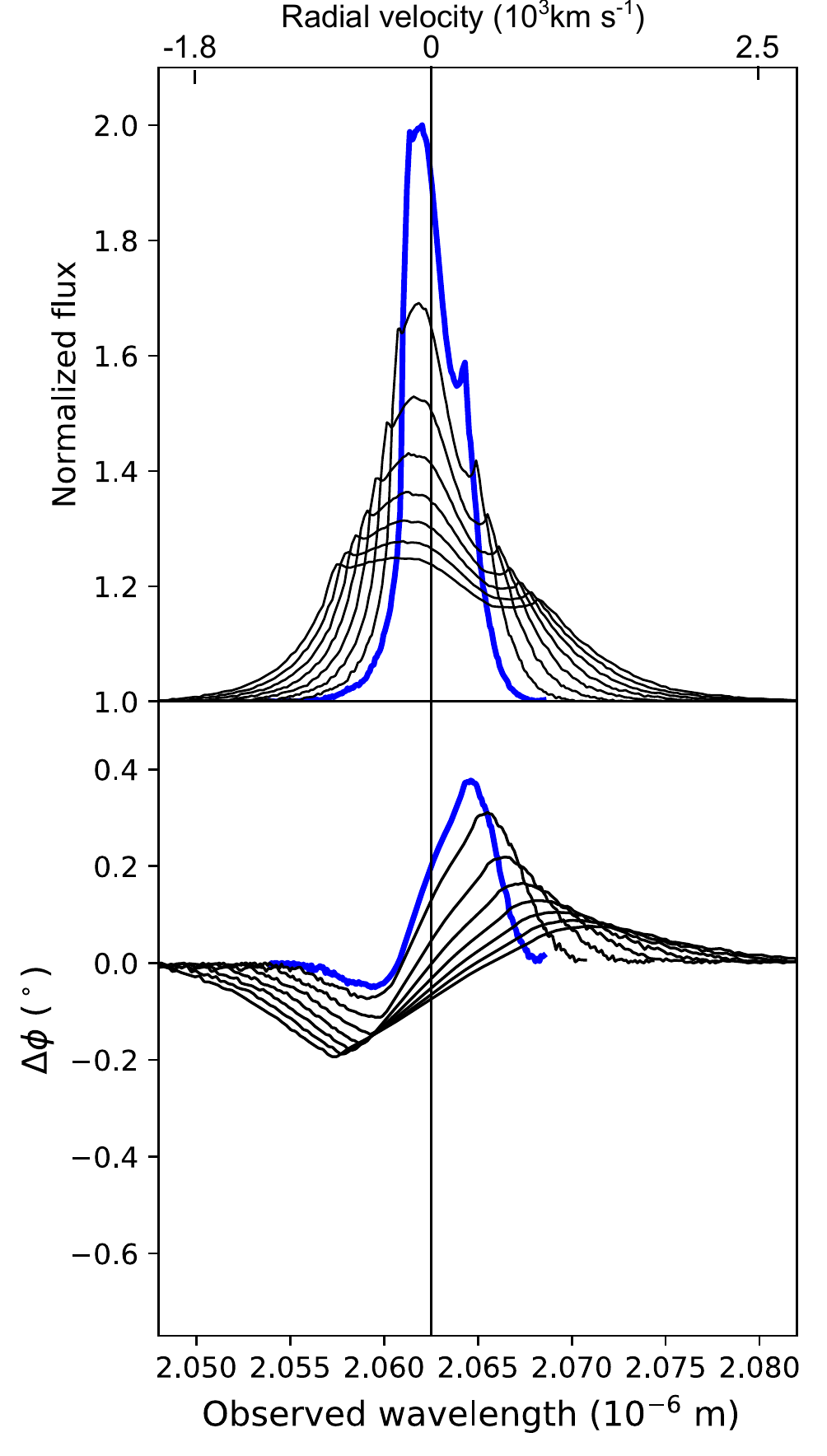}
		}
		\hspace{-0.8em}
		\subfigure[$i_{0}=\mathcal {U}(-10^{\circ},45^{\circ}),\delta i_{0}=5^{\circ},  \mathcal{C},\newline \hspace*{1.5em}  \Omega_{c}=  100^{\circ},   \omega_{c}=10^{\circ}, e_{c}\in \mathcal{R}(1)$
		]{%
			\label{fig:singledistr47}
			\includegraphics[trim = 2.5mm 0mm 1mm 0mm, clip,width=0.3\textwidth]{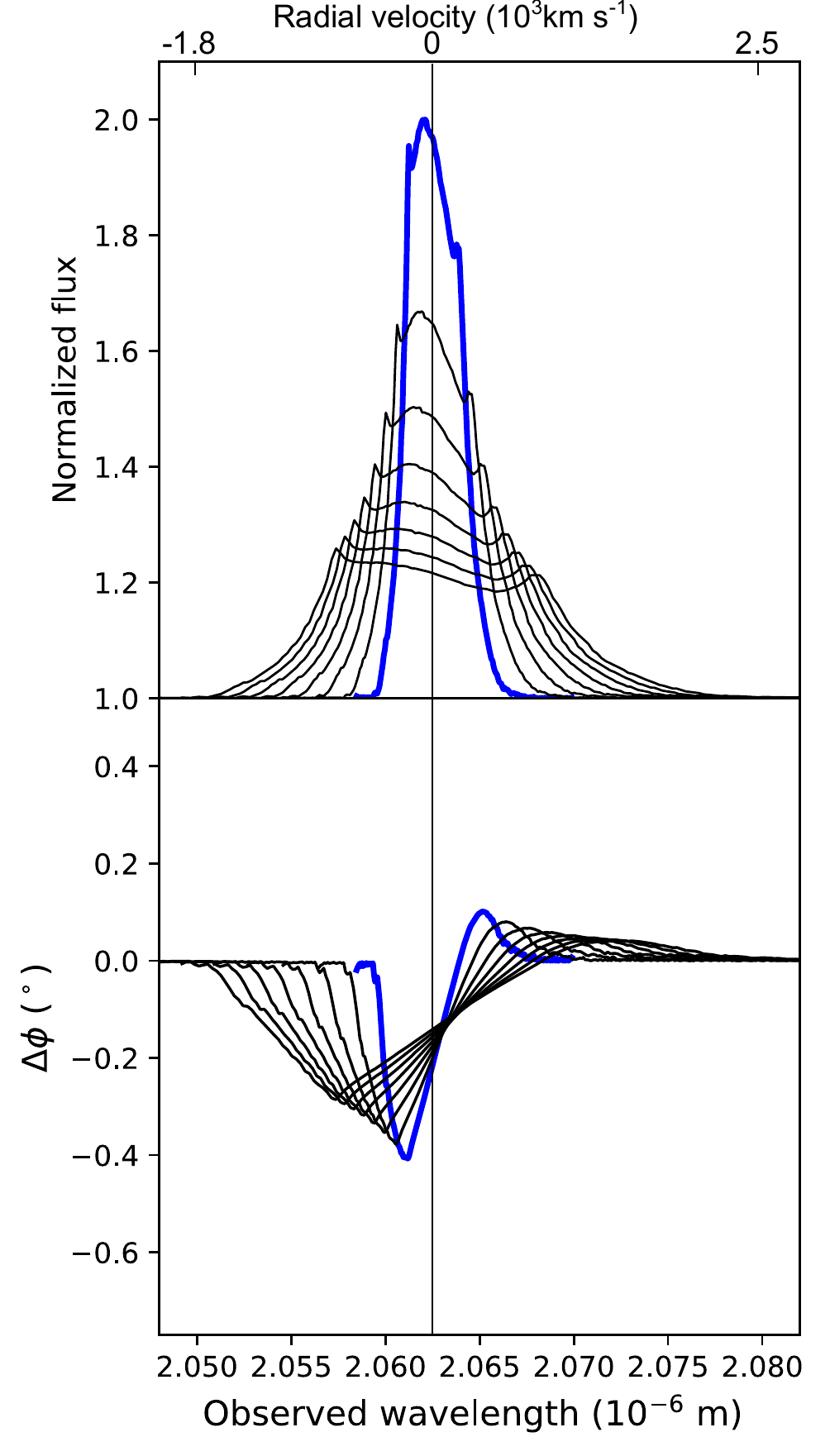}
		}%
	
		\vspace{-1.9em}
	\end{center}
	\caption{%
		Same as 	Fig. \ref {fig:singleel1} but for random samples of clouds' orbital eccentricities generated from  $\Gamma_{s}  (0.3,1)$ and $\mathcal{R}(1)$ distributions. 
		Varying parameters are listed in sub-captions,  $\mathcal{C}$ stands for coplanar clouds orbits.   
	}%
	
	\label{fig:singledistr4}
\end{figure*}

In the second series of plots (see Fig. \ref{fig:singleel2}),  simulations show the flat-topped spectra. 
 An increase in clouds' orbital inclination results in a decrease in the slope between two peaks of the differential phase (Figs. \ref{fig:singleel21}- \ref{fig:singleel22}).
The inclination of the observer $i_0$  affects the slope of the differential phase  similarly as clouds' orbital inclination (see Fig. \ref{fig:singleel23}).  Also,  an alternation of the amplitudes in the left and  the right wing of differential phase  occurs because of  variation of  the clouds' orbital pericenter $\omega$.
Smaller $\omega$ causes larger amplitudes in the right wing of differential phases, while larger $\omega$ has the opposite effect (see  Figs. \ref{fig:singleel1} and \ref{fig:singleel3}). Also, Fig. \ref{fig:singleel33} shows gradual evolution of the asymmetry of differential phases amplitudes  for variation of the angle of pericenter.  The slight variation of the slope between the differential phase peaks  occurs when $\omega\sim \pi$.
\newline Finaly,   we compare the  evolution of  the spectral lines and differential phases   for the two  non-uniform clouds' orbital  eccentricity distributions  (see Fig. \ref{fig:singledistr4}).
When $\omega\lesssim50^{\circ}$ and $i_{0}<30^{\circ}$ the spectral lines  and differential phases are narrower.
For coplanar cases,  differential phases'  deformation occurs regarding an invariant point in its right wing (compare Figs. \ref{fig:singledistr43}, \ref{fig:singledistr45}, \ref{fig:singledistr47} versus  Figs. \ref{fig:singledistr46},  \ref{fig:singledistr48}).

\section{Atlas of interferometric observables for aligned CB-SMBH}\label{appendix:aligned}
\renewcommand{\thesubfigure}{C \alph{subfigure}}

\begin{figure*}[ht!]
	\begin{center}

		\subfigure[$i_{0}=45^{\circ}$;$\mathcal{C},\Omega_{k}=100^{\circ},  \omega_{k}=110^{\circ}, \newline \hspace*{1.5em}  e_{k}=\mathcal {U}(0.05,0.1-0.4,0.45), k=1,2,\newline \hspace*{1.5em} \delta e_{k}=0.1$; $i_{c}=\mathcal{U}(-5^{\circ},5^{\circ}),\Omega_{c}=10^{\circ}$ ,\newline \hspace*{1.5em}  $\omega_{c}=180^{\circ},e_{c}=\tilde{e}_{k}$]	{%
			\label{fig:doubleel11}
			\includegraphics[trim = 3.0mm 0mm 2.0mm 0mm, clip, width=0.3\textwidth]{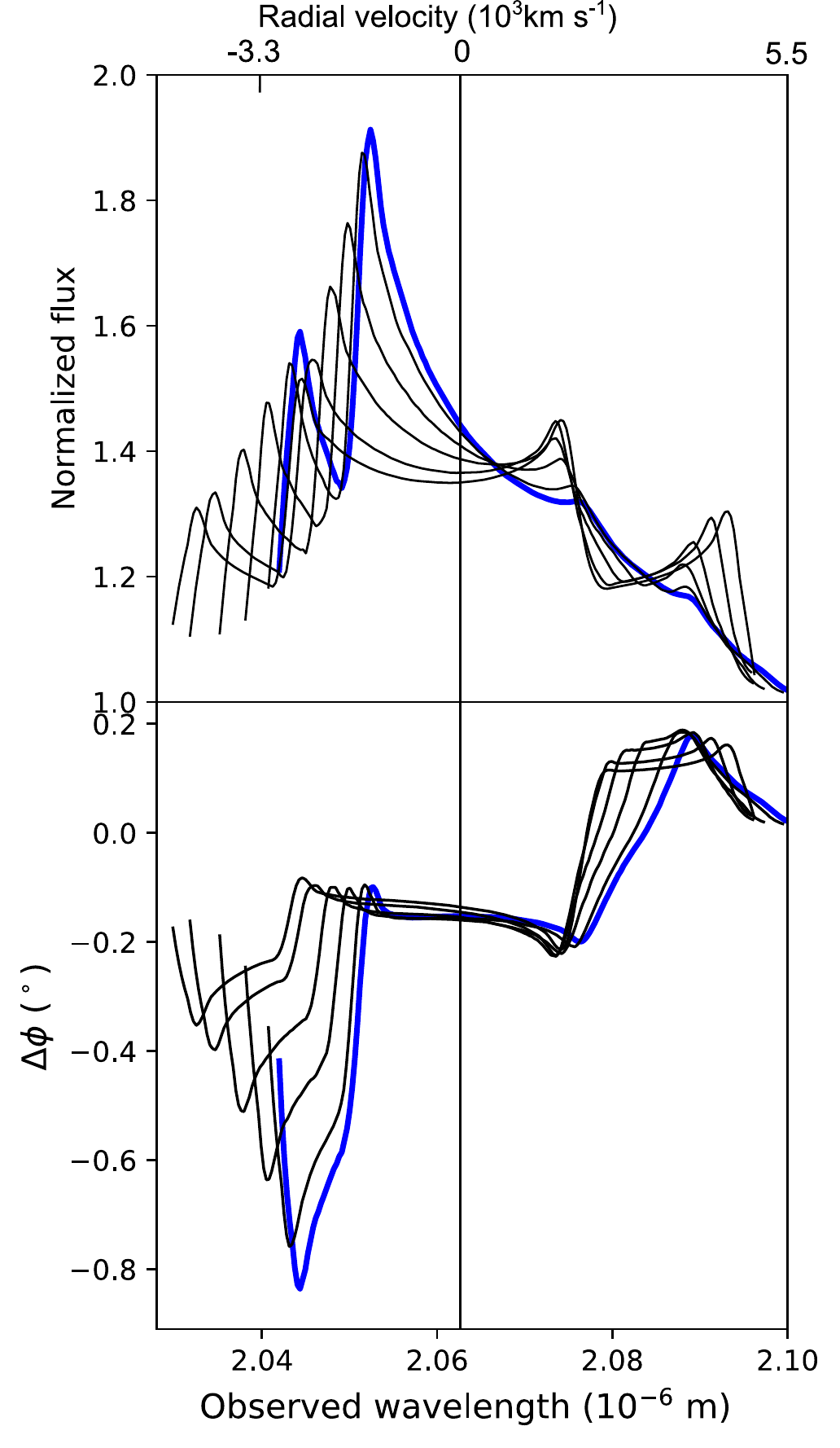}
		}%
		\hspace{-1.25em}
		\subfigure[$i_{0}=45^{\circ}$;$i_{k}=\mathcal {U}(0^{\circ},45^{\circ}),\delta i_{k}=5^{\circ},\newline \hspace*{1.5em} \Omega_{k}=  100^{\circ},  \omega_{k}=110^{\circ},  e_{k}=\mathcal {U}(0.05,0.1- \newline \hspace*{1.5em} 0.4,0.45), \delta e_{k}=0.1, k=1,2$, $i_{c}= \newline \hspace*{1.5em} \mathcal{U}(-5^{\circ},5^{\circ}),\Omega_{c}=10^{\circ}$,  $\omega_{c}=50^{\circ},  \newline \hspace*{1.5em} e_{c}=0.25$]{%
			\label{fig:doubleel12}
			\includegraphics[trim = 0.0mm 0mm 5mm 0mm, clip,width=0.3\textwidth]{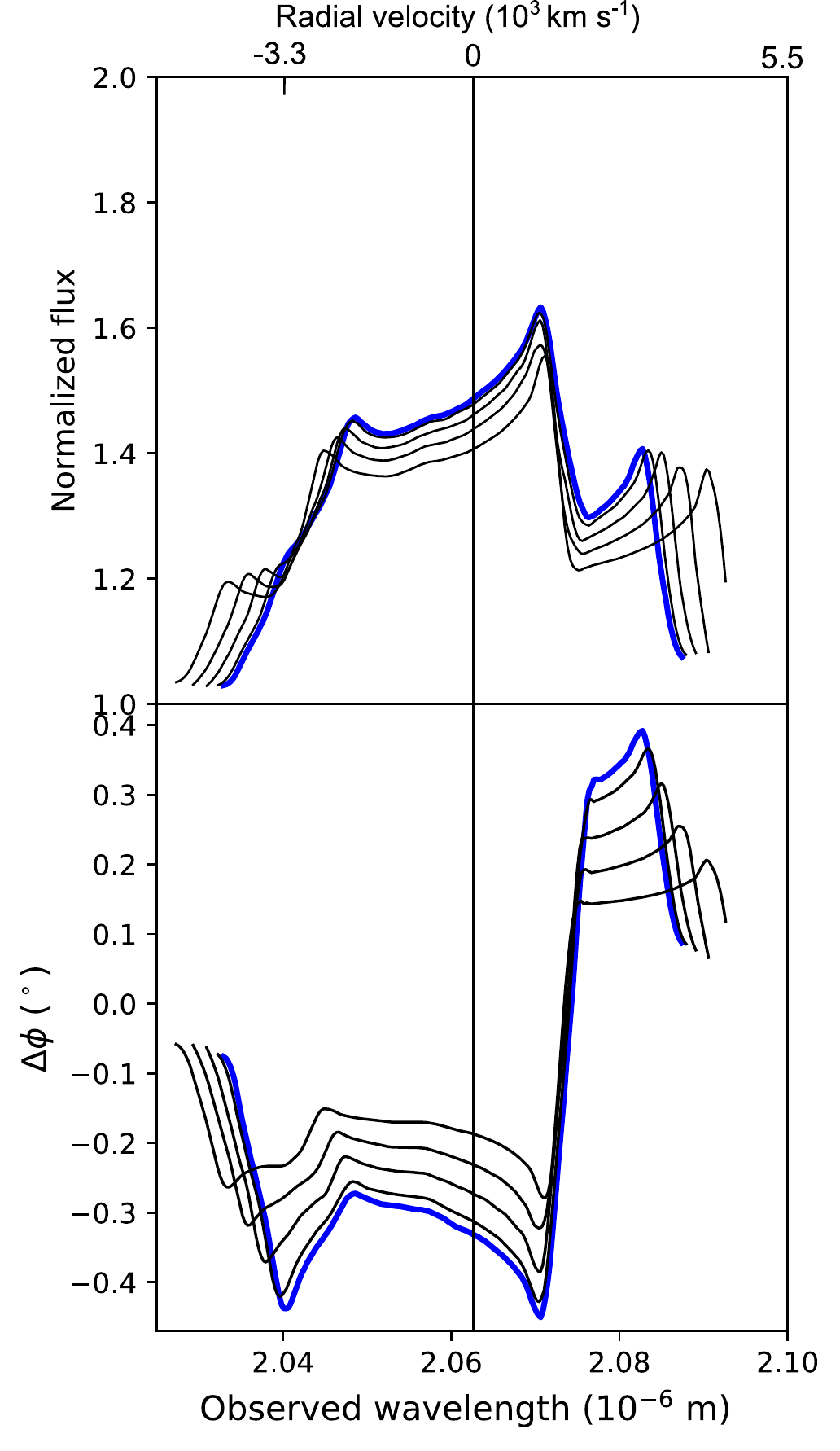}
		}
		\hspace{-1.2em}
		\subfigure[$i_{0}=45^{\circ}$;$i_{k}=\mathcal {U}(0^{\circ},45^{\circ}),\delta i_{k}=5^{\circ},	\Omega_{k}= \newline \hspace*{1.5em}100^{\circ},  \omega_{k}=110^{\circ},  e_{k}=\mathcal {U}(0.05,0.1-  0.4, \newline \hspace*{1.5em}0.45), \delta e_{k}=0.1, k=1,2$, $i_{c}= \mathcal{U}(-5^{\circ},5^{\circ}),\newline \hspace*{1.5em} \Omega_{c}=10^{\circ}$ ,  $\omega_{c}=50^{\circ},e_{c}=0.45$]{%
			\label{fig:doubleel13}
			\includegraphics[trim = 0.0mm 0mm 0mm 0mm, clip, width=0.315\textwidth]{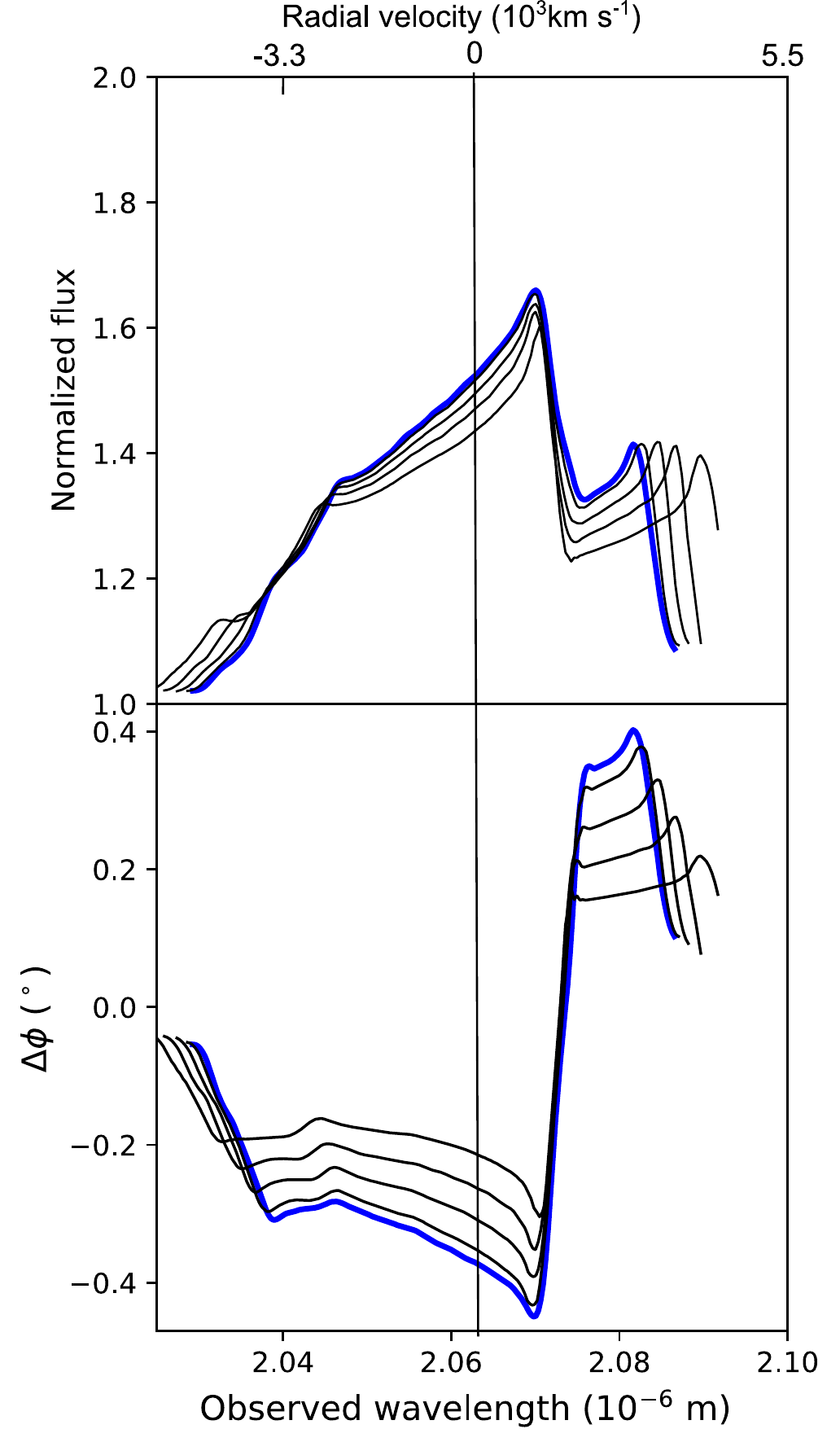}
			
		}	\\ 
	\vspace{0.0em}
		\subfigure[$i_{0}=\mathcal {U}(10^{\circ},45^{\circ}),\delta i_{0}=5^{\circ}$;$\mathcal{C}$, $\Omega_{k}=  \newline \hspace*{1.5em} 100^{\circ},  \omega_{k}=110^{\circ},  e_{k}=\mathcal {U}(0.05,0.1-  0.4, \newline \hspace*{1.5em} 0.45), \delta e_{k}=0.1, k=1,2$; $i_{c}=\newline \hspace*{1.5em}  \mathcal{U}(-5^{\circ},5^{\circ}), \Omega_{c1}=200^{\circ}$,  $\omega_{c1}=150^{\circ},\newline \hspace*{1.5em}  \Omega_{c2}=10^{\circ}$,  $\omega_{c2}=50^{\circ},e_{c}=\tilde{e}_{k}$]
		{%
			\label{fig:doubleel14}
			\includegraphics[width=0.3\textwidth]{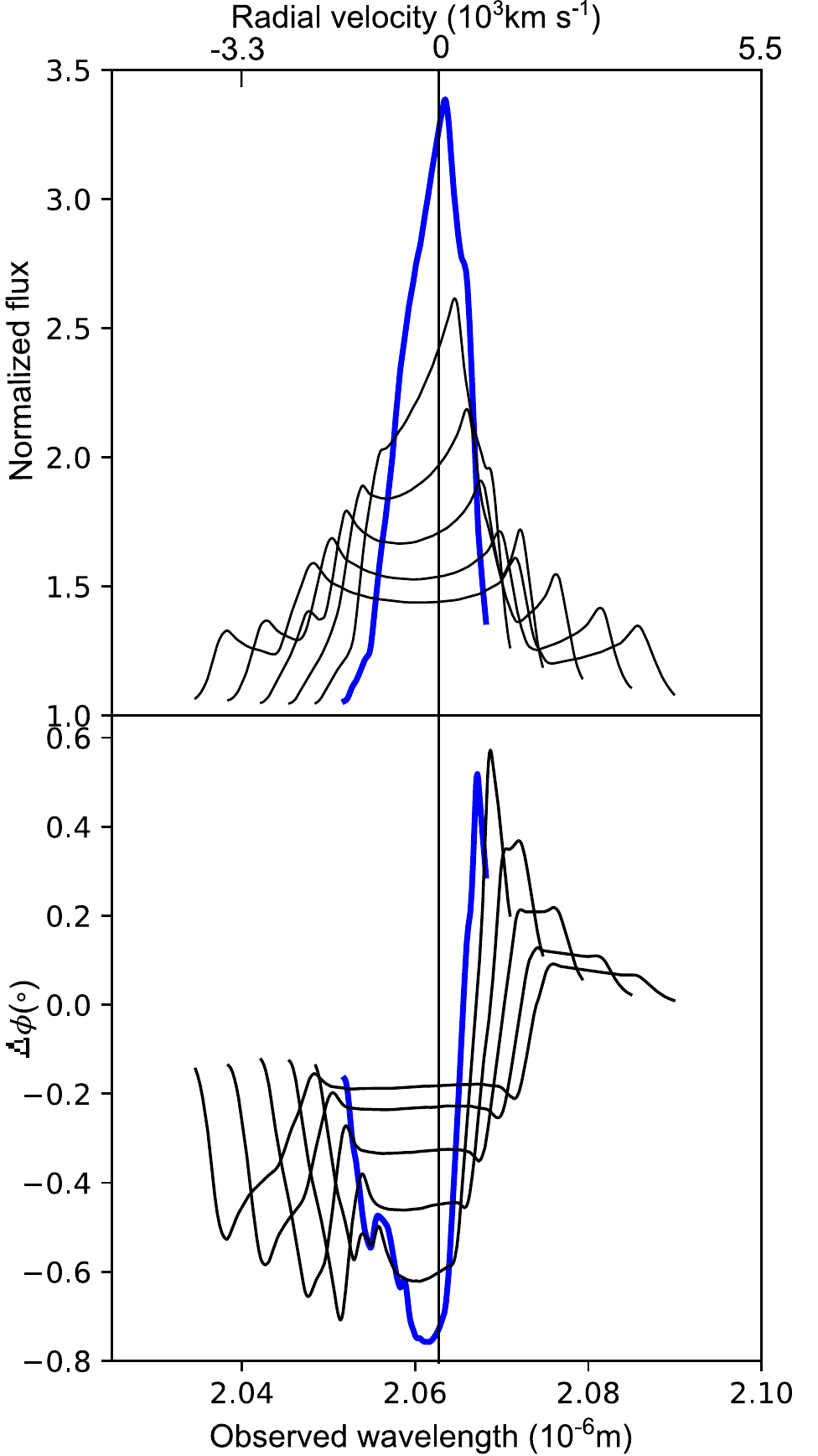}
		}%
		\hspace{-1.22em}
		\subfigure[$i_{0}=\mathcal {U}(45^{\circ},10^{\circ}),\delta i_{0}=-5^{\circ}$;$\mathcal{C}$, $\Omega_{k}=  \newline \hspace*{1.5em} 100^{\circ},  \omega_{k}=110^{\circ},  e_{k}=\mathcal {U}(0.05,0.1-0.4,  \newline \hspace*{1.5em}  0.45), \delta e_{k}=0.1, k=1,2$; $i_{c}= \newline \hspace*{1.5em} \mathcal{U}(-5^{\circ},5^{\circ}), \Omega_{c1}=200^{\circ}$ ,  $\omega_{c1}=150^{\circ}, \newline \hspace*{1.5em} \Omega_{c2}=10^{\circ}$,  $\omega_{c2}=180^{\circ},e_{c}={e}_{k}$
		]{%
			\label{fig:doubleel15}
			\includegraphics[width=0.3\textwidth]{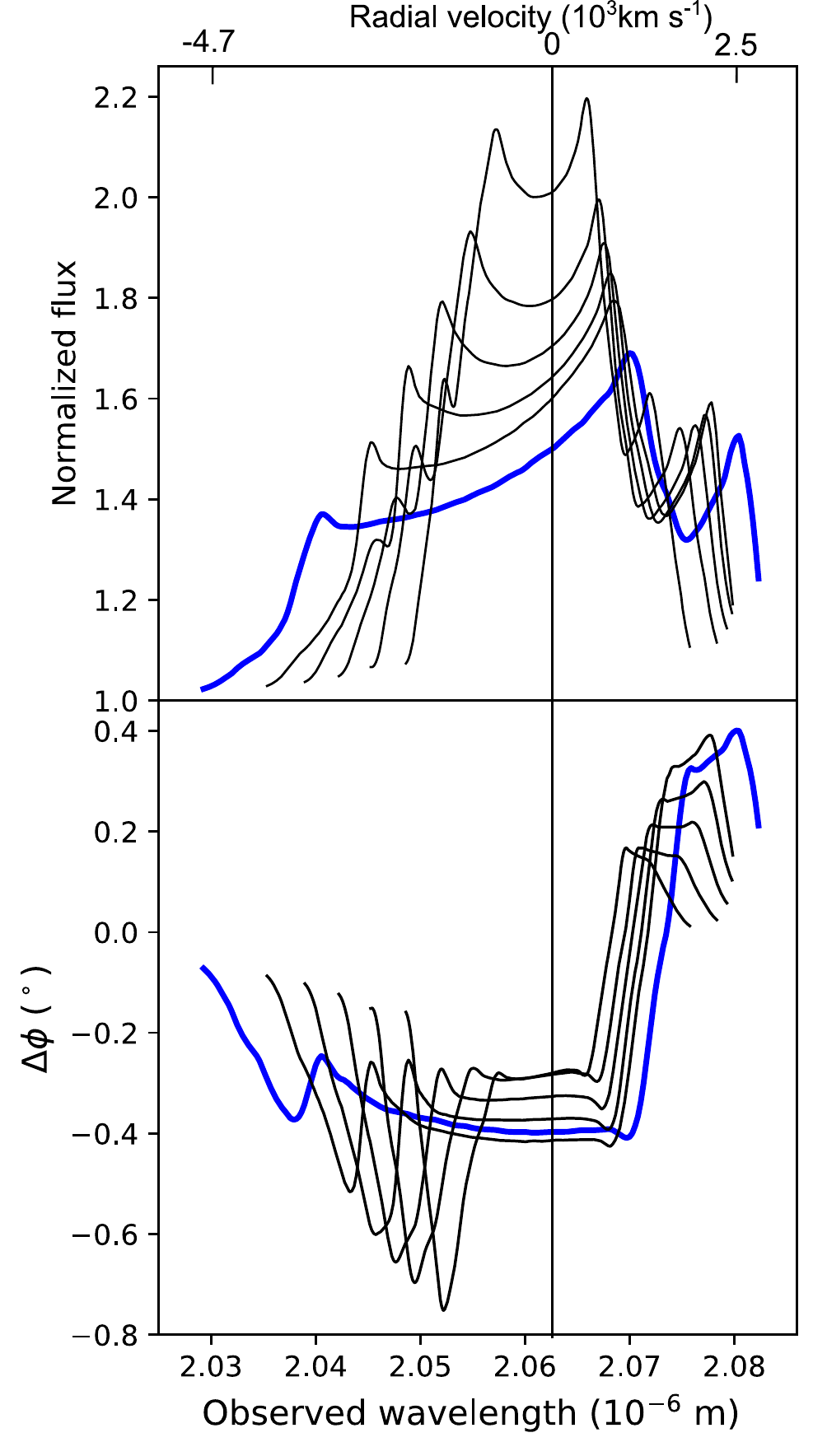}
		}
		\hspace{-1.2em}
		\subfigure[$i_{0}=45^{\circ}$;$\mathcal{C}$, $\Omega_{1}=200^{\circ}, \Omega_{2}=100, \omega_{1}=\newline \hspace*{1.5em}\omega_{2}=110^{\circ},  e_{k}=0.5, k=1,2$; $\Im= \newline \hspace*{1.5em}\mathcal {U} (5^{\circ},45^{\circ}),\delta \Im=5^{\circ}, \Omega_{c1}=200^{\circ}$,  $\omega_{c1}=\newline \hspace*{1.5em}150^{\circ}, \Omega_{c2}=10^{\circ}$, $\omega_{c2}=180^{\circ},e_{c}=0.5$]{%
			\label{fig:doubleel16}
			\includegraphics[width=0.308\textwidth]{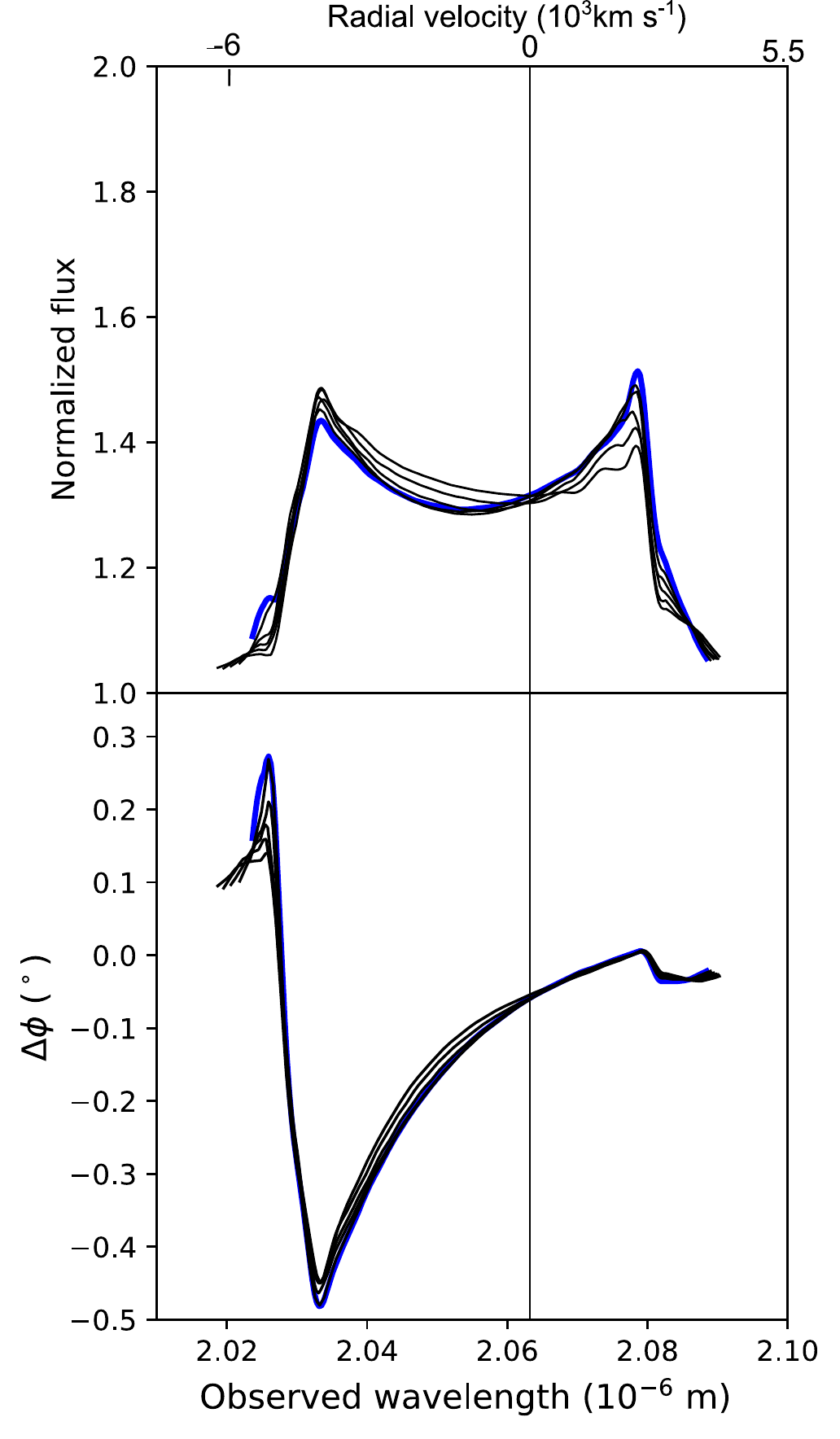}
		}%
		\vspace{-1.9em}
	\end{center}
	\caption{%
		Evolution of the Pa$\alpha$ emission line (upper subplots) and corresponding differential  phase ($\Delta \phi$, lower subplots) as a function of the wavelength radial velocity for different models of  aligned CB-SMBH system.  Common clouds' orbital elements  are designed by subscript c.  $\mathcal{C}$ stands for the coplanar  CB-SMBH system.  $\Im=\mathcal {U} (l,r)$ stands for inclination ranges from $\mathcal{U}(-l,l)$ up to $\mathcal{U}(-r,r)$.}	%
	
	\label{fig:doubleel1}
\end{figure*}

\begin{figure*}[ht!]
	\begin{center}

		\subfigure[$i_{0}=45^{\circ}$;$i_{k}=\mathcal {U}(10^{\circ},90^{\circ}), \delta i_{k}=10^{\circ}, \newline \hspace*{1.5em}\Omega_{k}=    200^{\circ},  \omega_{k}=210^{\circ},   e_{k}=0.5,\newline \hspace*{1.5em} k=1,2$; $\Im=\mathcal {U}(5^{\circ},45^{\circ}), \delta \Im=5^{\circ},  \Omega_{c}=\newline \hspace*{1.5em}200^{\circ}$,$\omega_{c1}=  290^{\circ}, \omega_{c2}=210^{\circ},  e_{c}=0.5$
		]	{%
			\label{fig:doubleel21}
			\includegraphics[trim = 3.0mm 0mm 2.0mm 0mm, clip, width=0.32\textwidth]{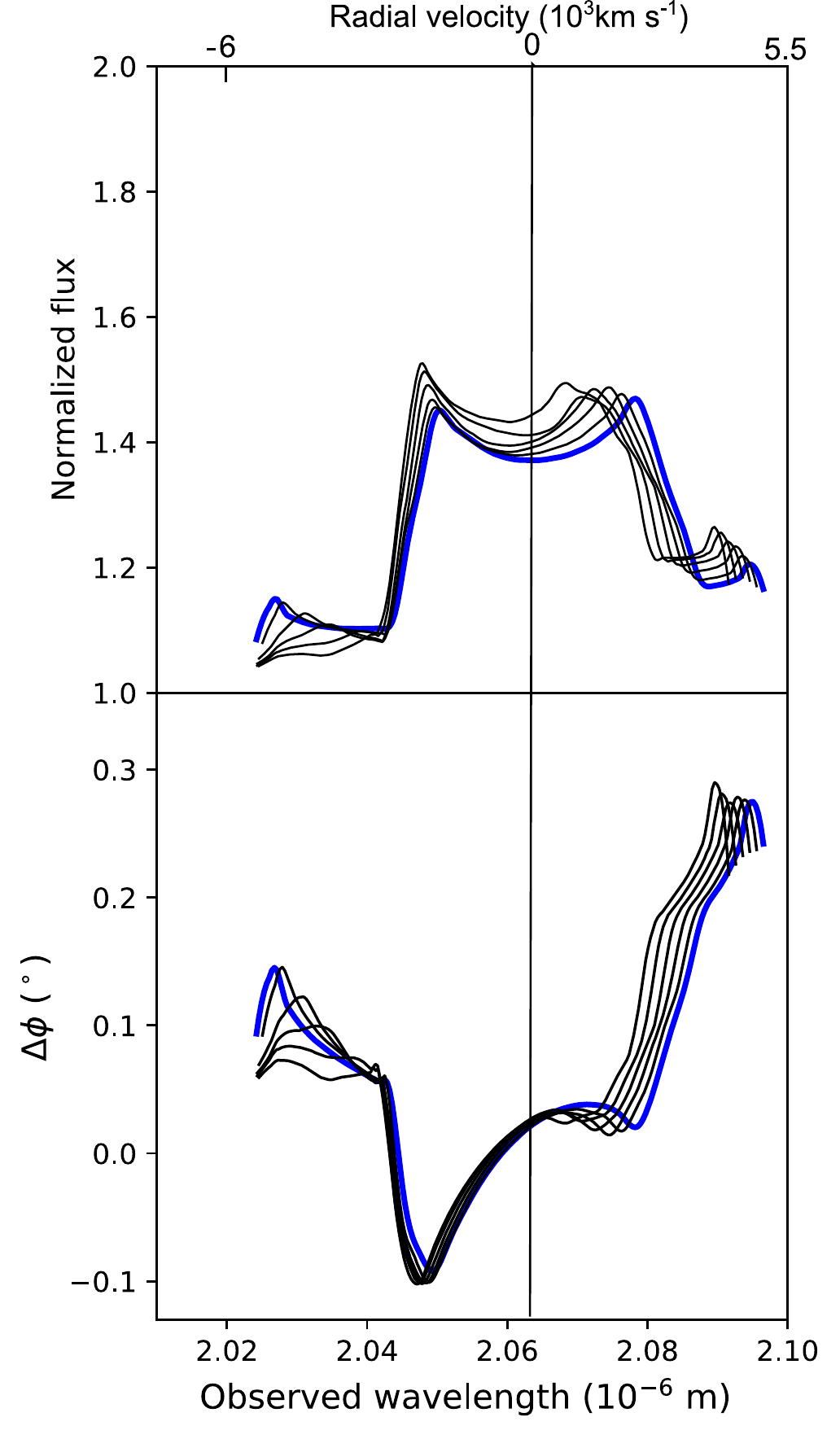}
		}%
		\hspace{-1.25em}
		\subfigure[$i_{0}=45^{\circ}$;$i_{k}=\mathcal {U}(90^{\circ},10^{\circ}), \delta i_{k}=-10^{\circ},\newline \hspace*{1.5em}   \Omega_{k}=  200^{\circ},  \omega_{k}=210^{\circ},   e_{k}=0.5, k=1,2$;\newline \hspace*{1.5em}   $\Im= \mathcal {U} (5^{\circ},45^{\circ}), \delta \Im=5^{\circ},\Omega_{c}=200^{\circ}$,\newline \hspace*{1.5em} $\omega_{c1}=  290^{\circ}, \omega_{c2}=210^{\circ},  e_{c}=0.5$
		]{%
			\label{fig:doubleel22}
			\includegraphics[trim = 0.0mm 0mm 5mm 0mm, clip,width=0.32\textwidth]{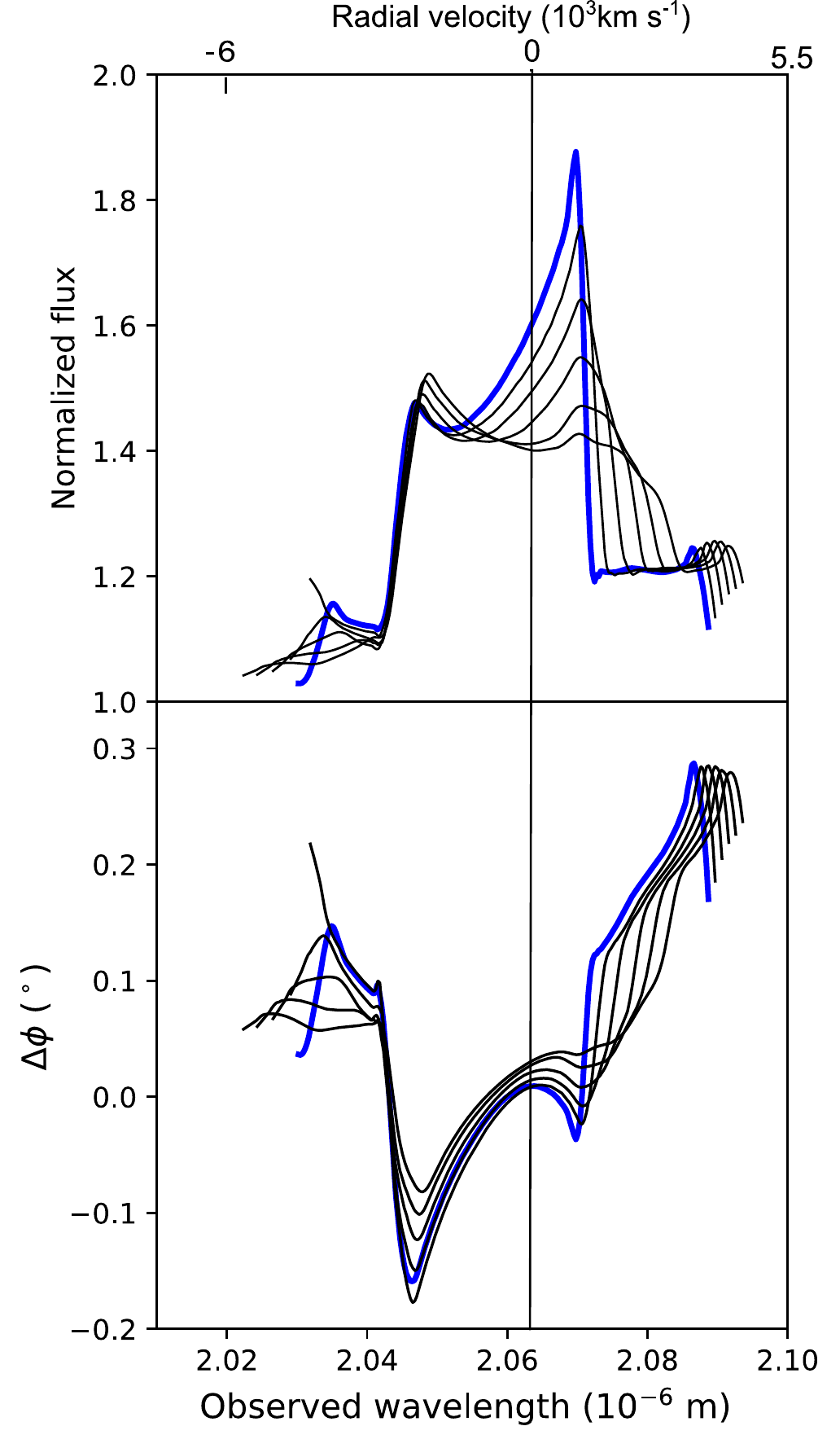}
		}\\ `
		\subfigure[$i_{0}=45^{\circ}$;$i_{k}=\mathcal {U}(10^{\circ},90^{\circ}), \delta i_{k}=-10^{\circ}, \newline \hspace*{1.5em}\Omega_{1}=    200^{\circ},   \Omega_{2}=  100^{\circ}, \omega_{k}=210^{\circ},  \newline \hspace*{1.5em} e_{k}=0.5,k=1,2$; $\Omega_{c}=200^{\circ}$,\ $\omega_{c1}= \newline \hspace*{1.5em} 150^{\circ}, \omega_{c2}=210^{\circ}, i_{c1}=rnd(5^{\circ},45^{\circ}),  \newline \hspace*{1.5em} i_{c1}=(5^{\circ},45^{\circ}), e_{c}=0.5$
		]{%
			\label{fig:doubleel23}
			\includegraphics[trim = 0.0mm 0mm 0mm 0mm, clip, width=0.315\textwidth]{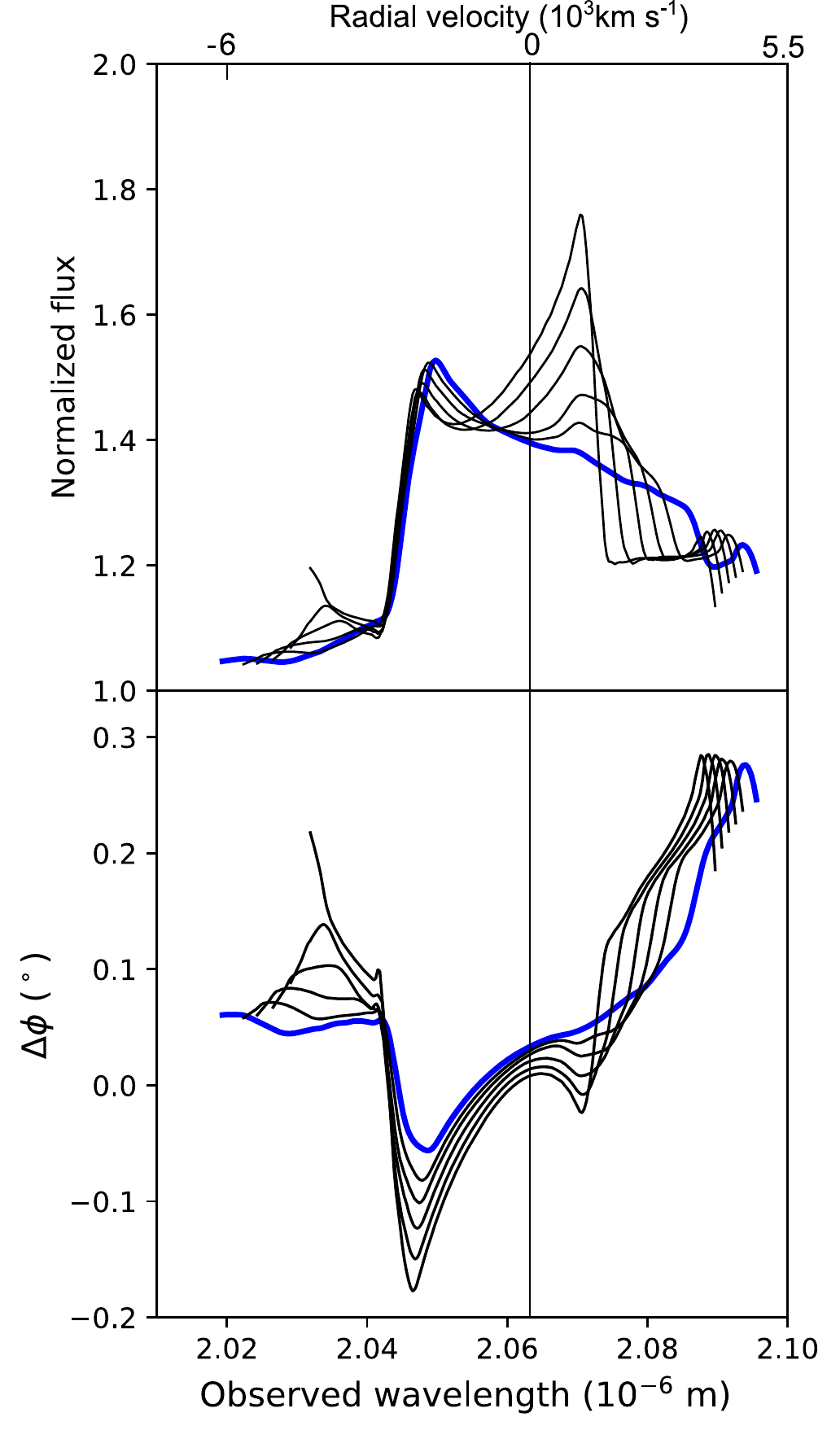}
					}
		\hspace{-1.22em}
		\subfigure[$i_{0}=45^{\circ}$;$i_{k}=\mathcal {U}(90^{\circ},10^{\circ}), \delta i_{k}=-10^{\circ}, \newline \hspace*{1.5em} \Omega_{k}=5^{\circ},   \omega_{1}=5^{\circ}, \omega_{2}=180^{\circ},   e_{k}=0.5,\newline\hspace*{1.5em} k=1,2$; $\Im=  \mathcal {U} (5^{\circ},45^{\circ}), \delta \Im=5^{\circ},\newline\hspace*{1.5em}e_{c}=0.5, \Omega_{c}=100^{\circ}$, $\omega_{c1}=  5^{\circ},\newline\hspace*{1.5em} \omega_{c2}=180^{\circ} $
		]{%
			\label{fig:doubleel25}
			\includegraphics[width=0.306\textwidth]{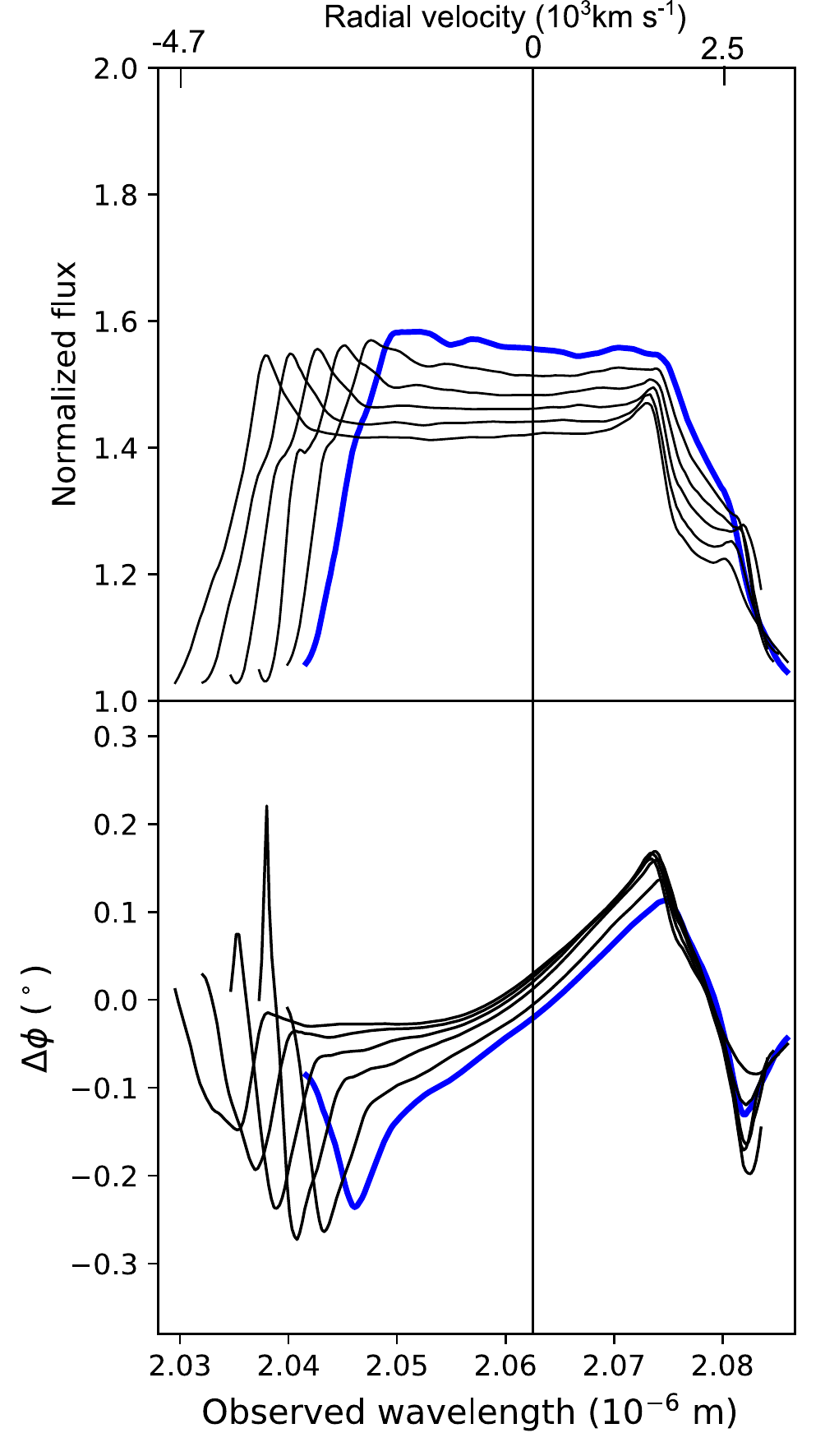}
		}
		\hspace{-1.2em}
		\subfigure[$i_{0}=45^{\circ}$;$i_{k}=\mathcal {U}(5^{\circ},45^{\circ}), \delta i_{k}=10^{\circ}, \Omega_{k}=  5^{\circ},   \omega_{1}=5^{\circ}, \omega_{2}=180^{\circ},   e_{k}=0.5, k=1,2$;   $\Im= \mathcal {U} (5^{\circ},45^{\circ}), \delta \Im=5^{\circ},\Omega_{c}=100^{\circ}$, $\omega_{c1}=  5^{\circ}, \omega_{c2}=180^{\circ},  e_{c}=0.5$
		]{%
			\label{fig:doubleel26}
			\includegraphics[width=0.306\textwidth]{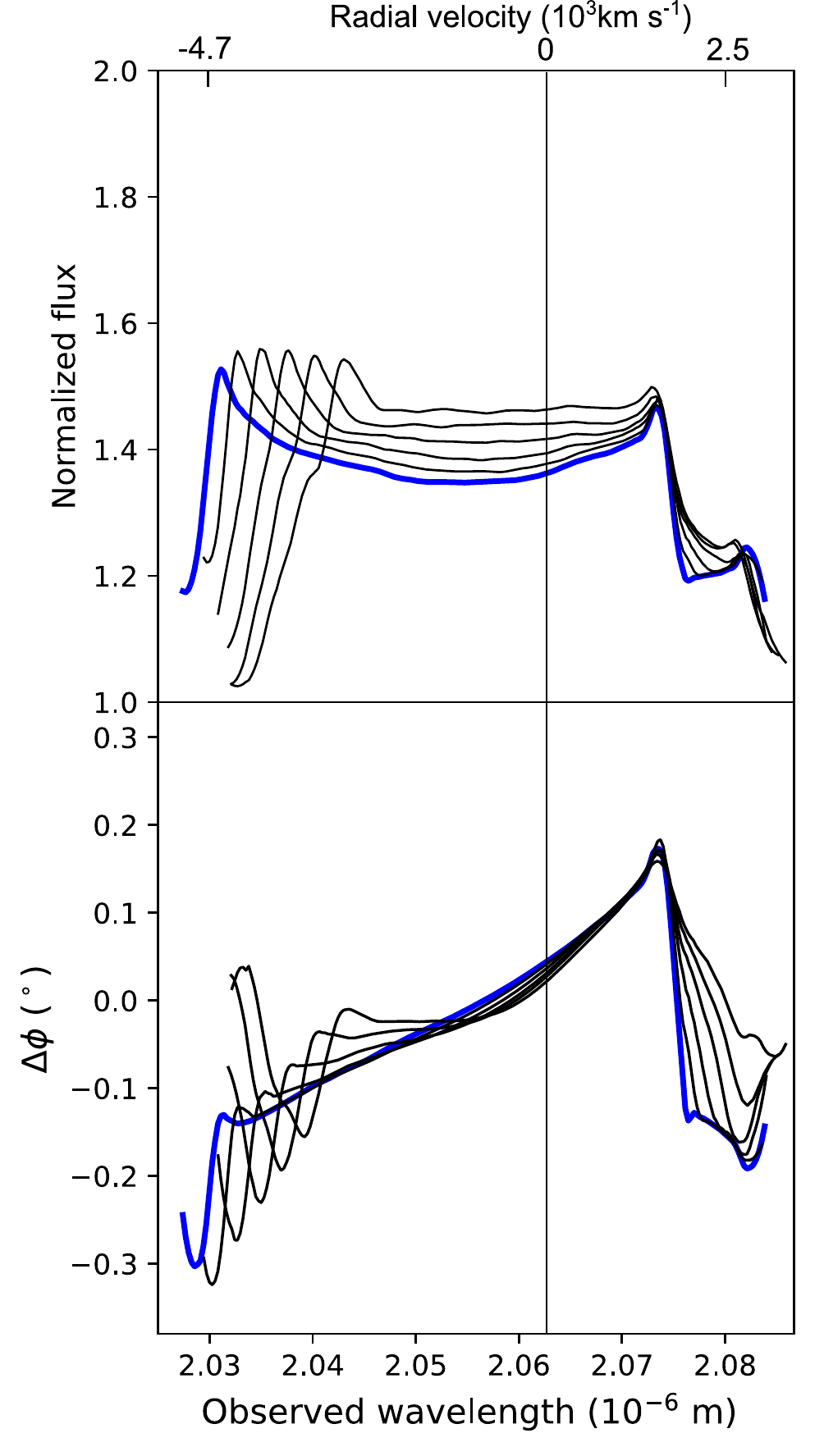}
		}%
		\vspace{-1.9em}
	\end{center}
	\caption{%
		Same as Fig. \ref{fig:doubleel1} but for different SMBHs and clouds' orbital parameters. Plot (c): $rnd$ stands for randomly chosen samples of orbital elements.  $\Im=\mathcal {U} (l,r)$ stands for inclination ranges from $\mathcal{U}(-l,l)$ up to $\mathcal{U}(-r,r)$.	
	}%
	
	\label{fig:doubleel2}
\end{figure*}

\begin{figure*}[ht!]
	\begin{center}
			\subfigure[$i_{0}=\mathcal {U}(10^{\circ},45^{\circ}), \delta  i_{0}=5^{\circ}$;$i_{k}=(10^{\circ},90^{\circ}),\newline \hspace*{1.5em}   \delta i_{k}=-10^{\circ}, \Omega_{k}=5^{\circ}, \omega_{1}=5^{\circ},  \omega_{2}=180^{\circ}, \newline \hspace*{1.5em}  e_{k}=0.5, k=1,2$; $\Im= \mathcal {U} (10^{\circ},45^{\circ}), \newline \hspace*{1.5em}\delta \Im=5^{\circ},\Omega_{c}=100^{\circ}$, $\omega_{c1}= 5^{\circ},\newline \hspace*{1.5em} \omega_{c2}=180^{\circ},  e_{c}=0.5$]
		{%
			\label{fig:doubleel34}
			\includegraphics[trim = 1.0mm 0mm 4mm 0mm, clip,width=0.36\textwidth]{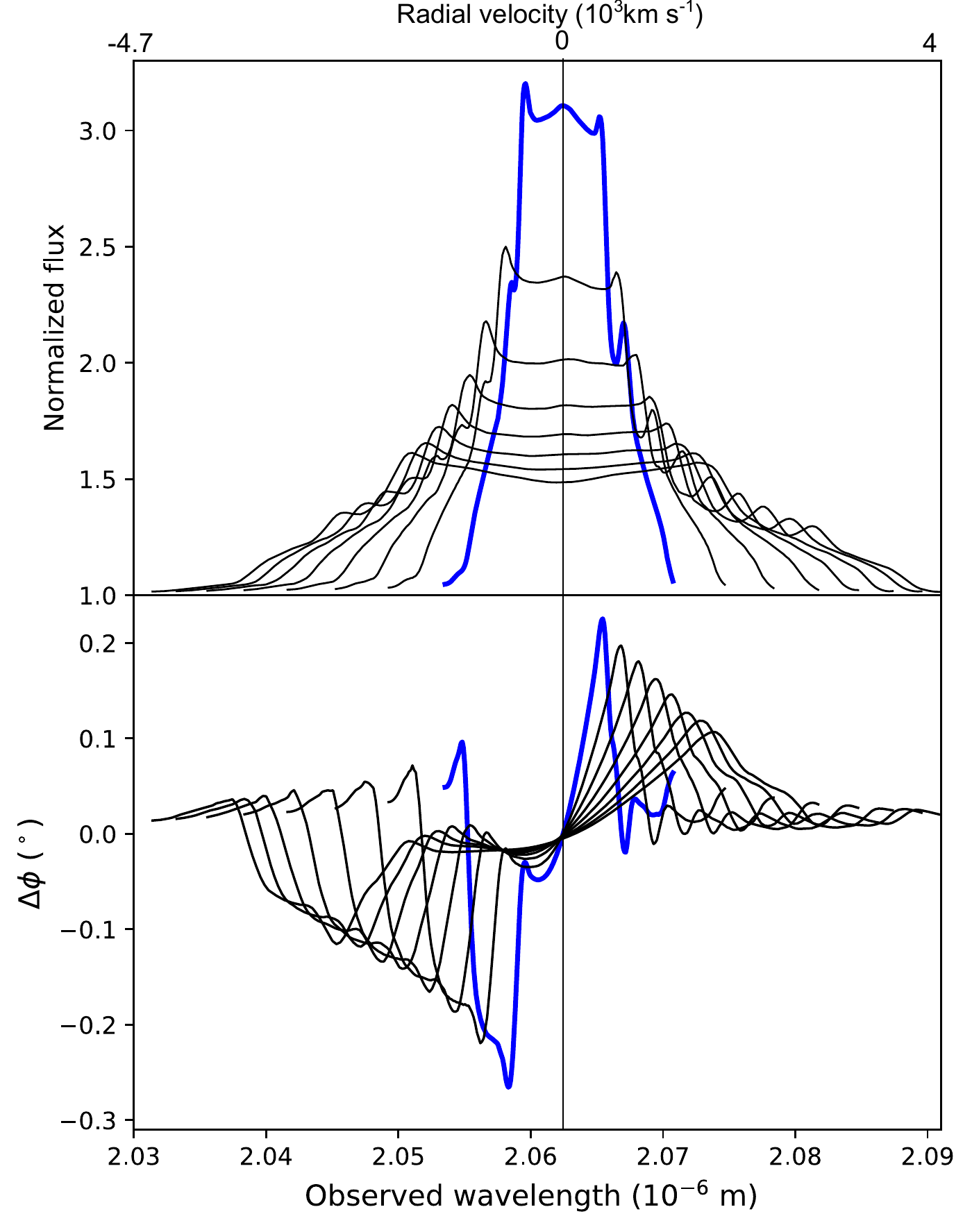}
		}%
		\hspace{-0.1em}
		\subfigure[$i_{0}=\mathcal {U}(15^{\circ},45^{\circ}), \delta i_{0}=5^{\circ}$;$i_{k}=\mathcal {U}(80^{\circ},20^{\circ}), \newline\hspace*{1.5em}\delta i_{k}=-10^{\circ}, \Omega_{k}=\mathcal {U}(230^{\circ}, 330^{\circ}), \delta \Omega_{k}=20^{\circ}$,\newline\hspace*{1.5em} $ \omega_{k}=\mathcal {U}(270^{\circ}, 170^{\circ}),\delta \omega_{k}=20^{\circ}, e_{k}=0.5, \newline \hspace*{1.5em}k=1,2$;$\Im= \mathcal {U}(5^{\circ},45^{\circ}), \delta \Im= 5^{\circ},\Omega_{c}=\mathcal {U}(80^{\circ},180^{\circ}),\newline\hspace*{1.5em} \delta \Omega_{c}=20$,\ $\omega_{c}=\mathcal {U}(120^{\circ}, 20^{\circ}),  \delta \omega_{c}=20^{\circ}, e_{c}=0.5$
		]{%
			\label{fig:doubleel35}
			\includegraphics[trim = 2.0mm 0mm 4mm 0mm, clip,width=0.39\textwidth]{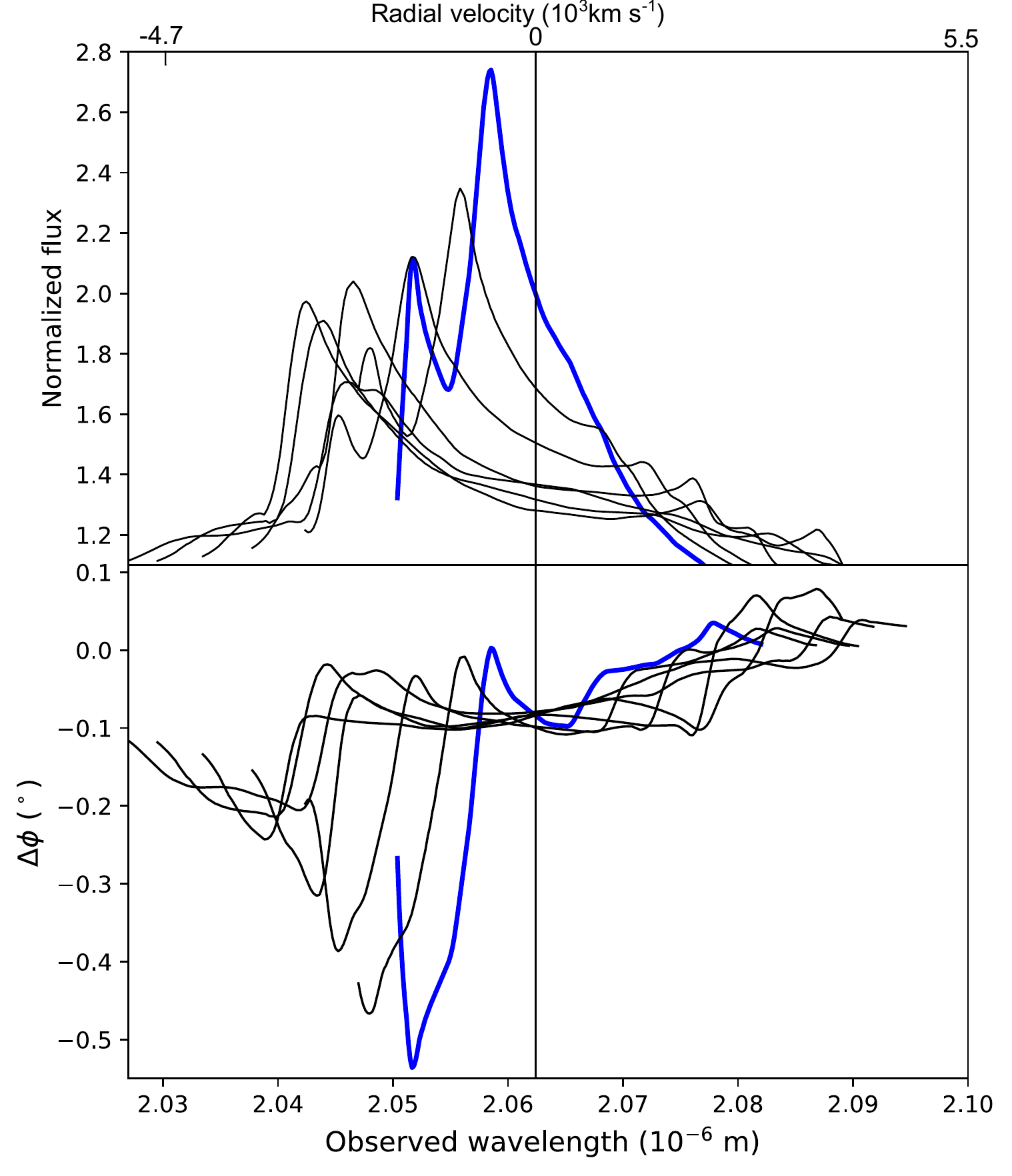}
		}\\ 
		\subfigure[$i_{0}=45^{\circ}$;$i_{k}=\mathcal {U}(10^{\circ},90^{\circ}), \delta i_{k}=10^{\circ}, \newline \hspace*{1.5em}   \Omega_{k}=  5^{\circ},  \omega_{1}=5^{\circ}, \omega_{2}=180^{\circ},  e_{k}=\newline \hspace*{1.5em} 0.2, k=1,2$; $\Im= \mathcal {U} (5^{\circ},45^{\circ}),  \delta \Im=\newline \hspace*{1.5em} 5^{\circ},\Omega_{c}=200^{\circ}$,\ $\omega_{c1}=290^{\circ}, \omega_{c2}=\newline \hspace*{1.5em}210^{\circ},  e_{c}=0.2$
		]	{%
			\label{fig:doubleel31}
			\includegraphics[trim = 3.0mm 0mm 2.0mm 0mm, clip, width=0.3\textwidth]{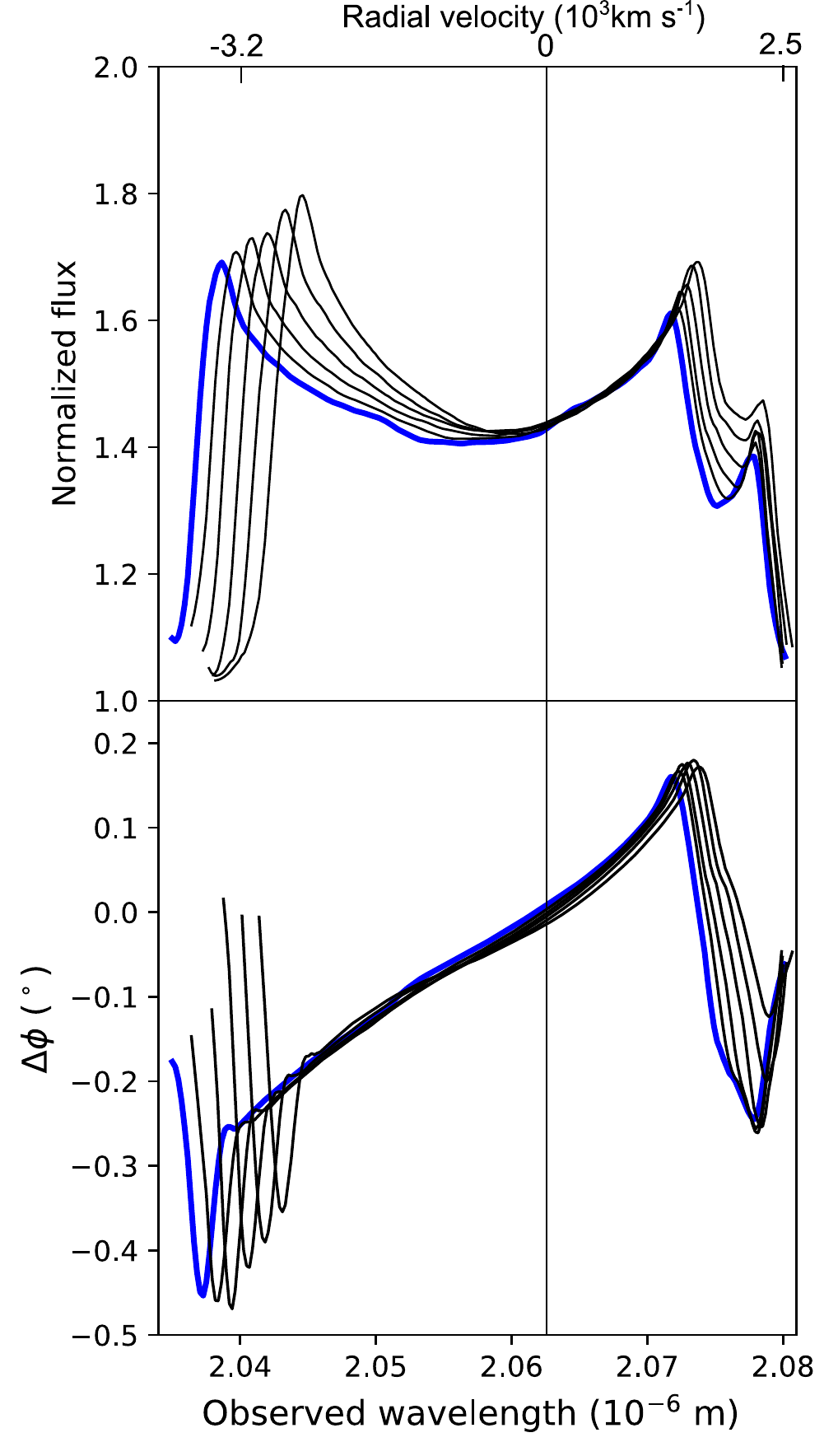}
		}%
		\hspace{-0.5em}
		\subfigure[$i_{0}=45^{\circ}$;$i_{k}=5^{\circ}, \Omega_{k}=\mathcal {U}(20^{\circ}, 180^{\circ}), \newline \hspace*{1.5em}   \delta \Omega_{k}=20^{\circ}, \omega_{k}=\mathcal {U}(180^{\circ}, 20^{\circ}).   \delta \omega_{k}=\newline\hspace*{1.5em}-20^{\circ},  e_{k}=0.5, k=1,2$; $i_{c}=5^{\circ},\Omega_{c}=\Omega_{k}$,\newline \hspace*{1.5em}\ $\omega_{c1}= \omega_{k},  e_{c}=0.5$
		]{%
			\label{fig:doubleel32}
			\includegraphics[trim = 3.0mm 0mm 1mm 0mm, clip,width=0.3\textwidth]{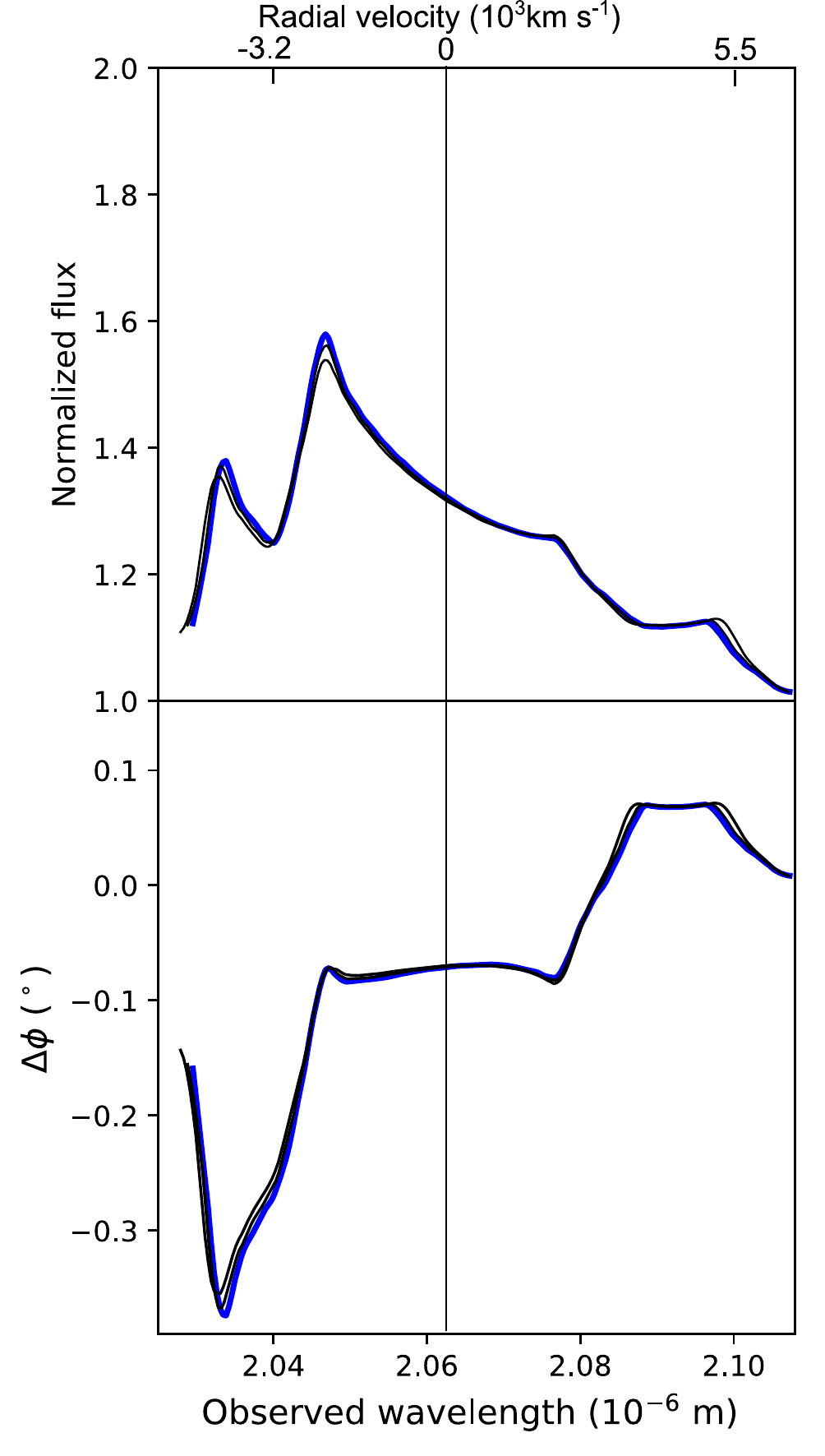}
		}
		\hspace{-0.5em}
		\subfigure[$i_{0}=\mathcal {U}(20^{\circ},45^{\circ}), \delta i_{0}=5^{\circ}$; $i_{k}=10^{\circ},\newline \hspace*{1.5em}  \Omega_{k}=\mathcal {U} (230^{\circ}, 330^{\circ}), \delta \Omega_{k}=20^{\circ},  \omega_{k}=\newline \hspace*{1.5em}\mathcal {U} (270^{\circ}, 190^{\circ}),  \delta \omega_{k}=-20^{\circ},  e_{k}=0.5, \newline \hspace*{1.5em}  k=1,2$; $i_{c}=
		5^{\circ},  \Omega_{c}=\mathcal {U}(80^{\circ}, 180^{\circ}), \newline \hspace*{1.5em}\delta \Omega_{k}=20^{\circ},$\ $\omega_{c1}= \mathcal {U}(120^{\circ}, 20^{\circ}),\newline \hspace*{1.5em}   \delta \omega_{k}=-20^{\circ},  e_{c}=0.5$  		
		]{%
			\label{fig:doubleel33}
			\includegraphics[trim = 2.0mm 0mm 0mm 0mm, clip, width=0.315\textwidth]{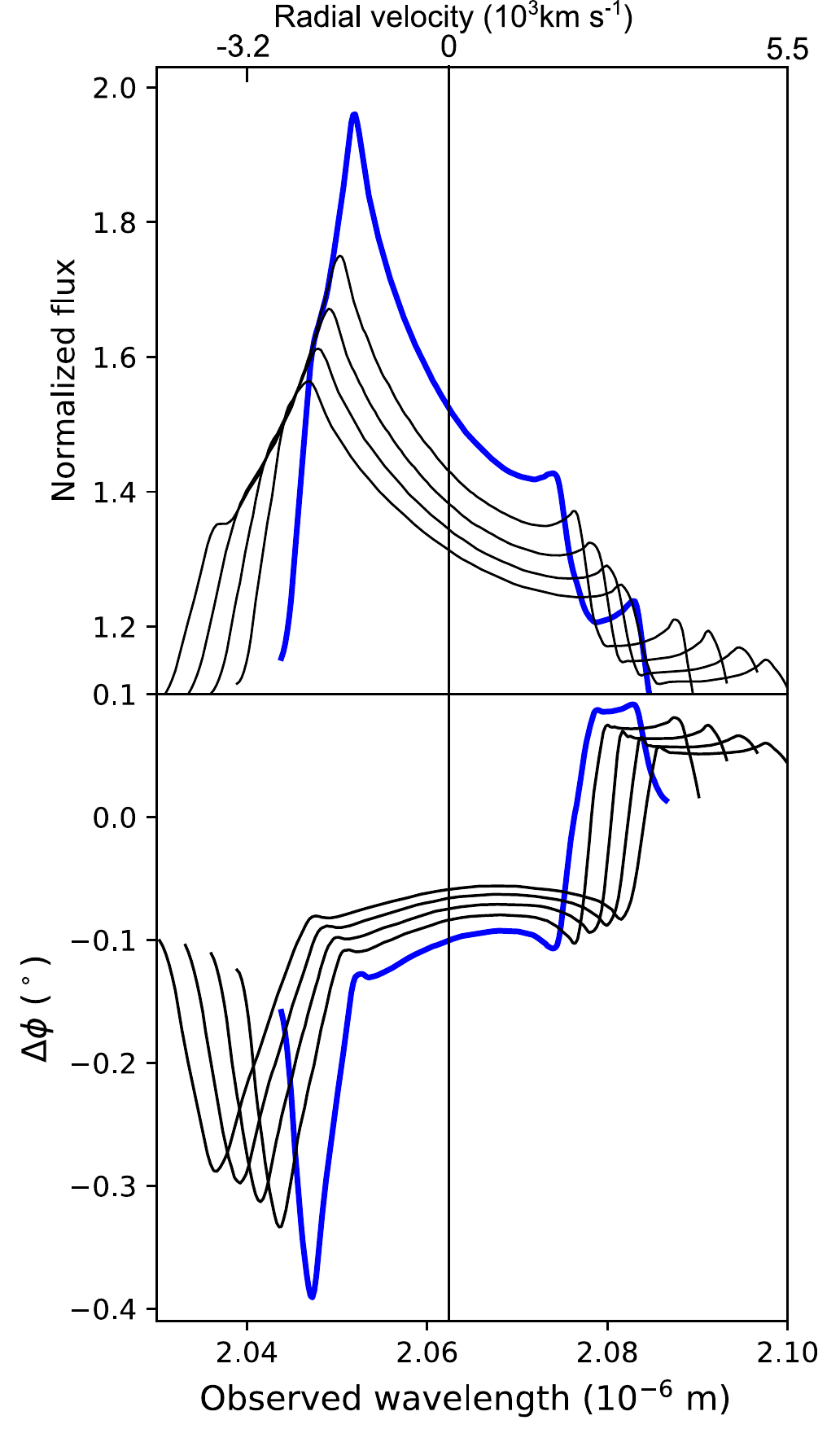}
			
		}

		\vspace{-1.9em}
	\end{center}
	\caption{%
		Same as Fig. \ref{fig:doubleel1} but for different SMBHs and clouds' orbital parameters.
	}%
	
	\label{fig:doubleel3}
\end{figure*}

The typical differential interferometry data for CB-SMBH systems are presented in Figs. \ref{fig:doubleel1}-\ref{fig:doubleel3} with partially resolved multi-peaks  spectra. 
Even when the the spectral lines  of both SMBHs are well blended as in blue-coloured models in Figs. \ref{fig:doubleel14} and \ref{fig:doubleel25},  their differential phases have two peaks. 
The  profiles and differential phases   in Figs. \ref{fig:doubleel23} and \ref{fig:doubleel32}  are less dependent on angles $\Omega$ and $\omega$ of SMBHs and clouds.  In this case, eccentricity is the same for SMBHs and clouds in the BLRs. These and other examples illustrate that information from the observables suffices to remove the ambiguities between the orbital elements (see e.g. Figs. \ref{fig:doubleel12}  and \ref{fig:doubleel13}).

For fixed observer position $i_0$ and  with increasing  SMBHs orbital eccentricity, the amplitude of differential phase decreasing,  but with distorted  left and right  wings (see Figs. \ref{fig:doubleel11} and  \ref{fig:doubleel15}). Also, large angles of pericenter of clouds' orbits in both BLRs $\omega_{c}>\pi/2$ deform core of  the differential phase below 0 value.
However, an increasing of  the angular position of the observer $i_0$ and SMBH's orbital inclination   decreases   the slopes of the differential phase (see Figs. \ref{fig:doubleel14}-\ref{fig:doubleel15}). In contrast,  Fig. \ref{fig:doubleel16}  shows obscured effects of  clouds' orbital inclination for  the fixed  observer angle and eccentricities of both SMBHs. 
For fixed eccentricities of SMBHs and clouds' orbits,  different   ranges of clouds' orbital inclinations       influence appearance of  ridges in the wings of differential phases  (compare Figs. \ref{fig:doubleel21}-\ref{fig:doubleel23}, and \ref{fig:doubleel25}). 
For  the smaller angle of pericenter  of larger SMBH and clouds in its BLR, plateaus between peaks of differential phases are less deformed (see  Figs. \ref{fig:doubleel25}- \ref{fig:doubleel26}).

If both SMBHs orbits have small   ascending nodes, but larger SMBH orbit  has a  smaller angle of pericenter, then varying SMBHs and clouds' orbital inclinations affect the amplitude of differential phase wings  (Fig. \ref{fig:doubleel31}).

\section{Atlas of interferometric observables for anti-aligned CB-SMBH}\label{appendix:nonaligned}

\renewcommand{\thesubfigure}{B \alph{subfigure}}

\begin{figure*}[ht!]
	\begin{center}

		\subfigure[$\mathcal{C}. \Omega_{k}=  100^{\circ},  \omega_{1}=250^{\circ}, \omega_{2}=70^{\circ},\newline \hspace*{1.5em}  e_{k}= 0.5, k=1,2$; $i_{c1}= \mathcal{U}(-5^{\circ},5^{\circ}), i_{c2}=\newline\hspace*{1.5em}\mathcal{U}(90^{\circ},175^{\circ}), \Omega_{c}=rnd(0.1^{\circ},359^{\circ})$,$\omega_{c1}=\newline\hspace*{1.5em}120^{\circ}, \omega_{c2}=300^{\circ}, e_{c}=0.5$
		]	{%
				\label{fig:doubleel41}
			\includegraphics[trim = 0.0mm 0mm 0mm 0mm, clip, width=0.315\textwidth]{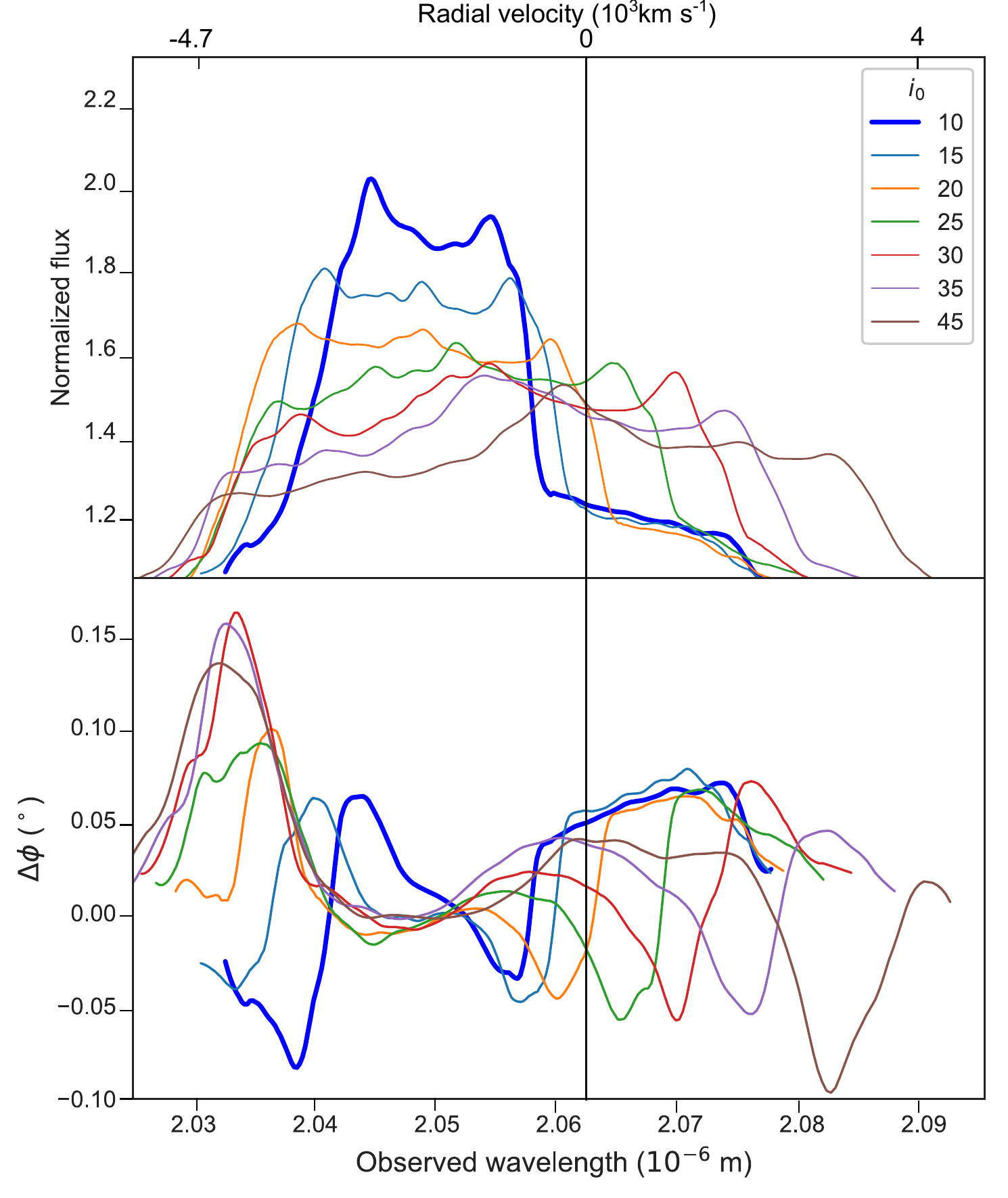}
		
		}%
		\hspace{-0.8em}
		\subfigure[$\mathcal{C}, \Omega_{k}= 100^{\circ},  \omega_{1}=250^{\circ}, \omega_{2}=70^{\circ},  e_{k}\newline \hspace*{1.5em}= 0.5, k=1,2$; $\Im=\mathcal{U}(10^{\circ},45^{\circ}),  \delta \Im=5^{\circ},\newline \hspace*{1.5em} i_{c2}=\mathcal{U}(90^{\circ}-175^{\circ}),  \Omega_{c}=\omega_{c}=\newline \hspace*{1.5em}rnd(0.1^{\circ},359^{\circ}),  e_{c}=0.5$
		]{%
			\label{fig:doubleel42}
			\includegraphics[trim = 2.0mm 0mm 0mm 0mm, clip,width=0.315\textwidth]{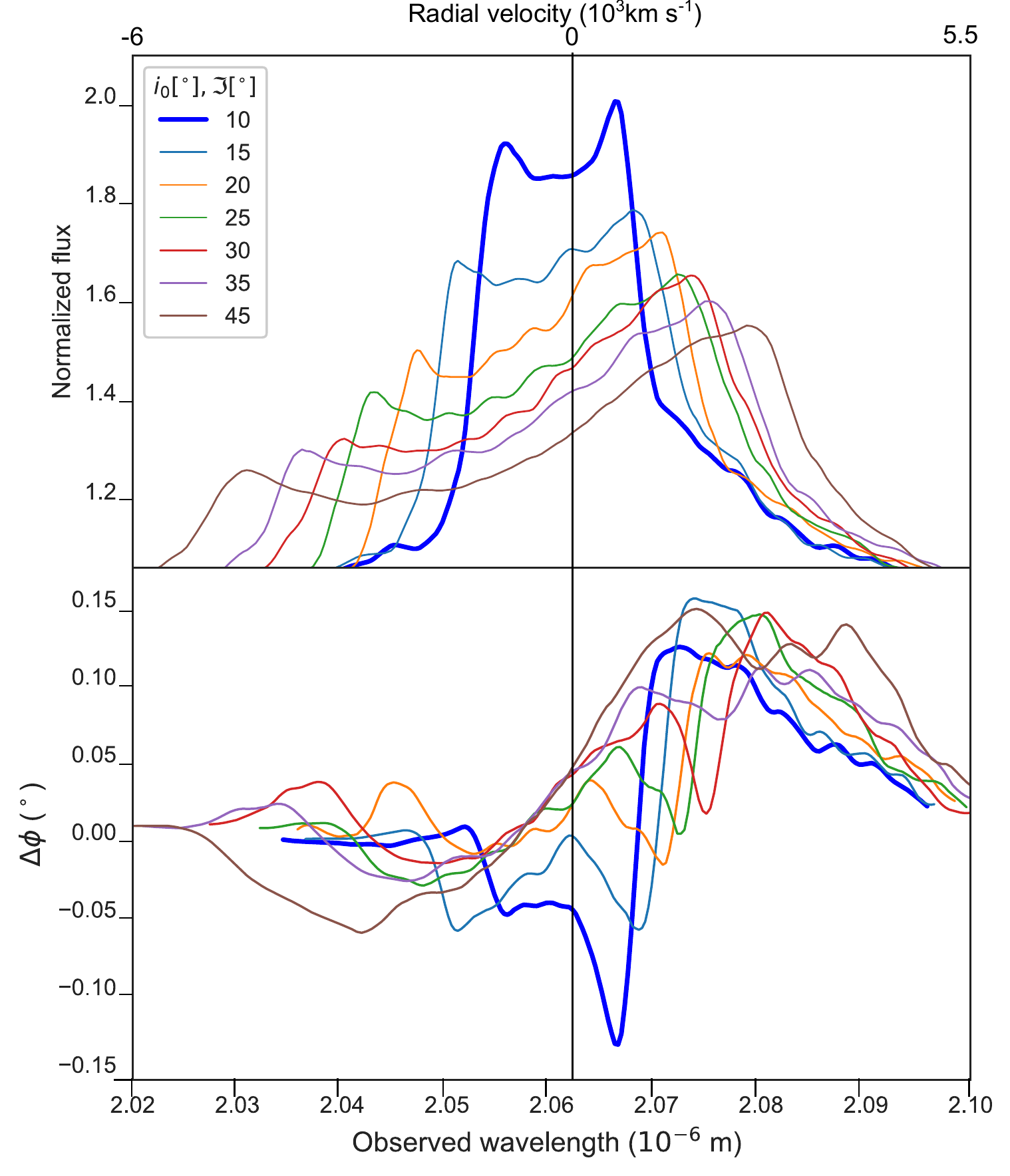}
		}
		\hspace{-1.4em}
		\subfigure[$\mathcal{C}, \Omega_{1}=  100^{\circ}, \Omega_{2}=  300^{\circ}, \omega_{1}=100^{\circ},\newline \hspace*{1.5em} \omega_{2}=180^{\circ},  e_{k}=0.5, k=1,2$; $\Im=\newline \hspace*{1.5em}(10^{\circ},45^{\circ}),   \delta \Im=5^{\circ},i_{c2}=\mathcal{U}(90^{\circ}175^{\circ}), \newline \hspace*{1.5em}\Omega_{c}= rnd(0.1^{\circ},359^{\circ}$,$\omega_{c1}=220^{\circ}, \omega_{c2}=\newline \hspace*{1.5em} 40^{\circ}, e_{c}=rnd(0.1,0.5)$		
		]{%
			\label{fig:doubleel43}
			\includegraphics[trim = 2.0mm 0mm 0mm 0mm, clip, width=0.315\textwidth]{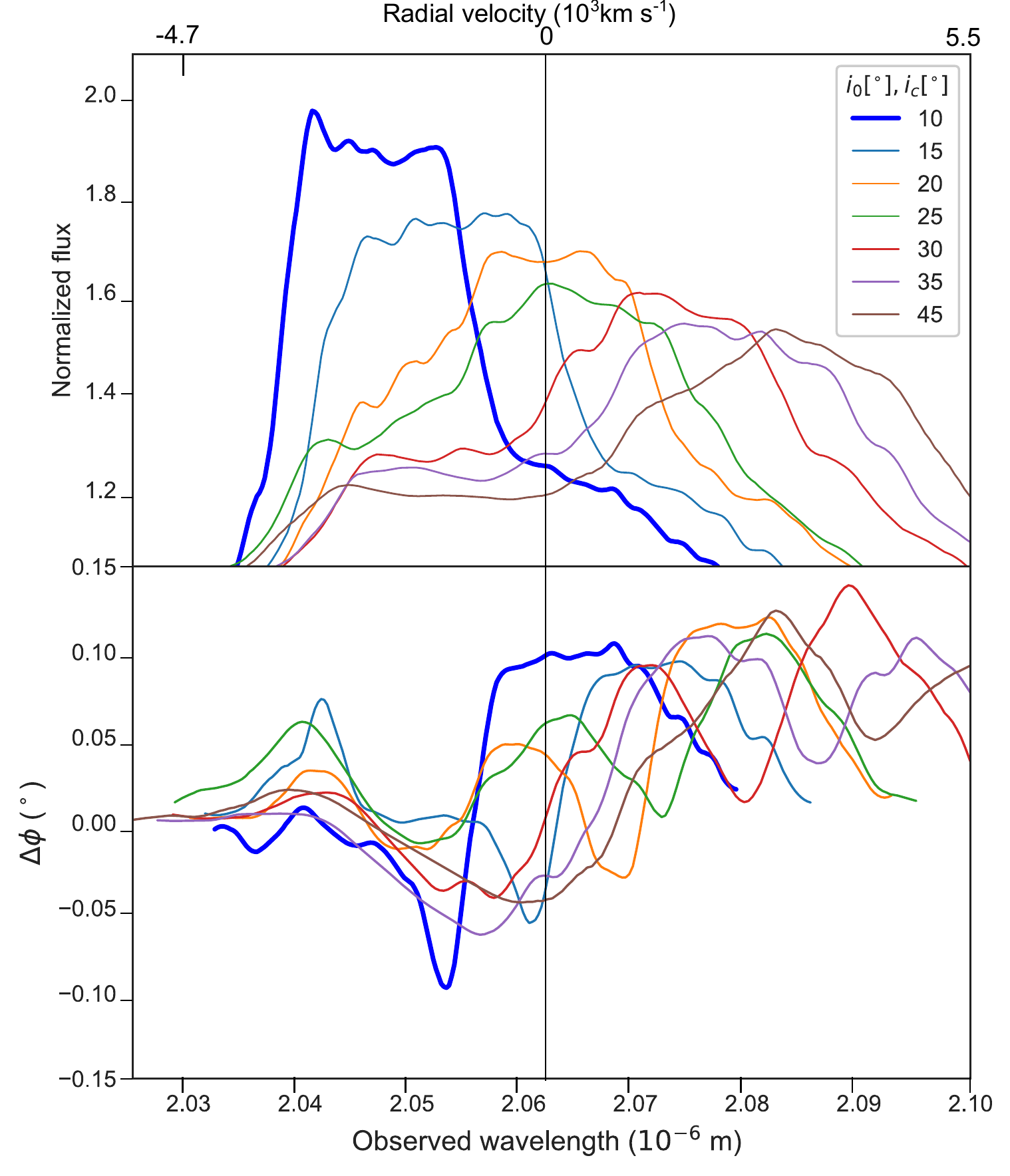}
		}
		\vspace{-1.9em}
	\end{center}
	\caption{%
		Same as Fig. \ref{fig:doubleel1} but for anti-aligned angular momenta of clouds in the BLR of less massive SMBH. 
		}%
	\label{fig:doubleel4}
\end{figure*}

 We first look at the case of coplanar   CB-SMBH,  clouds  of less massive  SMBH have anti-aligned angular momenta, and the inclinations of orbits of clouds are linearly spaced between $90^{\circ}$ and $175^{\circ}$ (see Figs. \ref{fig:doubleel41}- \ref{fig:doubleel43}).  
 It is also  clear  from the grid of models presented in Fig. \ref{fig:doubleel4} that if a more substantial number parameters vary simultaneously, the shapes of differential phases will be more complicated. The randomization of  nodes and apocenters of  clouds' orbits in both BLRs affects the forms of both observables (see  upper row of subplots in Fig. \ref{fig:doubleel4}).   A central whirl is  a prominent feature in Figs. \ref{fig:doubleel42}-\ref{fig:doubleel43}).
Before proceeding further, we make a digression to an auxiliary consideration. For $i_{0}=10^{\circ}$, spectral lines have concave wings and narrow core.  As  $i_0$ increases, the line shapes   broadens with convex sides. Still differential phases  vary drastically in their   amplitudes, widths, and forms. Wings of line shapes are arched if we randomize the inclinations of orbits of clouds  in the  BLR of more massive SMBH (see lower subfigures in Fig. \ref{fig:doubleel4}).

\begin{figure*}[ht!]
	\begin{center}
		\subfigure[$\mathcal{C},i_{0}=45^{\circ}, \Omega_{1}=  100^{\circ}, \Omega_{2}=  300^{\circ},  \newline\hspace*{1.5em} \omega_{1}= 250^{\circ},\omega_{2}=70^{\circ},  i_{c1}=  rnd(90^{\circ},175^{\circ}),\newline\hspace*{1.5em}  i_{c2}=  rnd(-45^{\circ},45^{\circ}),\Omega_{c1}=300^{\circ}, \Omega_{c2}=\newline\hspace*{1.5em}100^{\circ},  $$\omega_{c1}=170^{\circ}, \omega_{c2}=350^{\circ}, e_{c}=0.5$ 		
		]	{%
			\label{fig:doubleel51}
			\includegraphics[trim = 0.0mm 0mm 0mm 0mm, clip, width=0.315\textwidth]{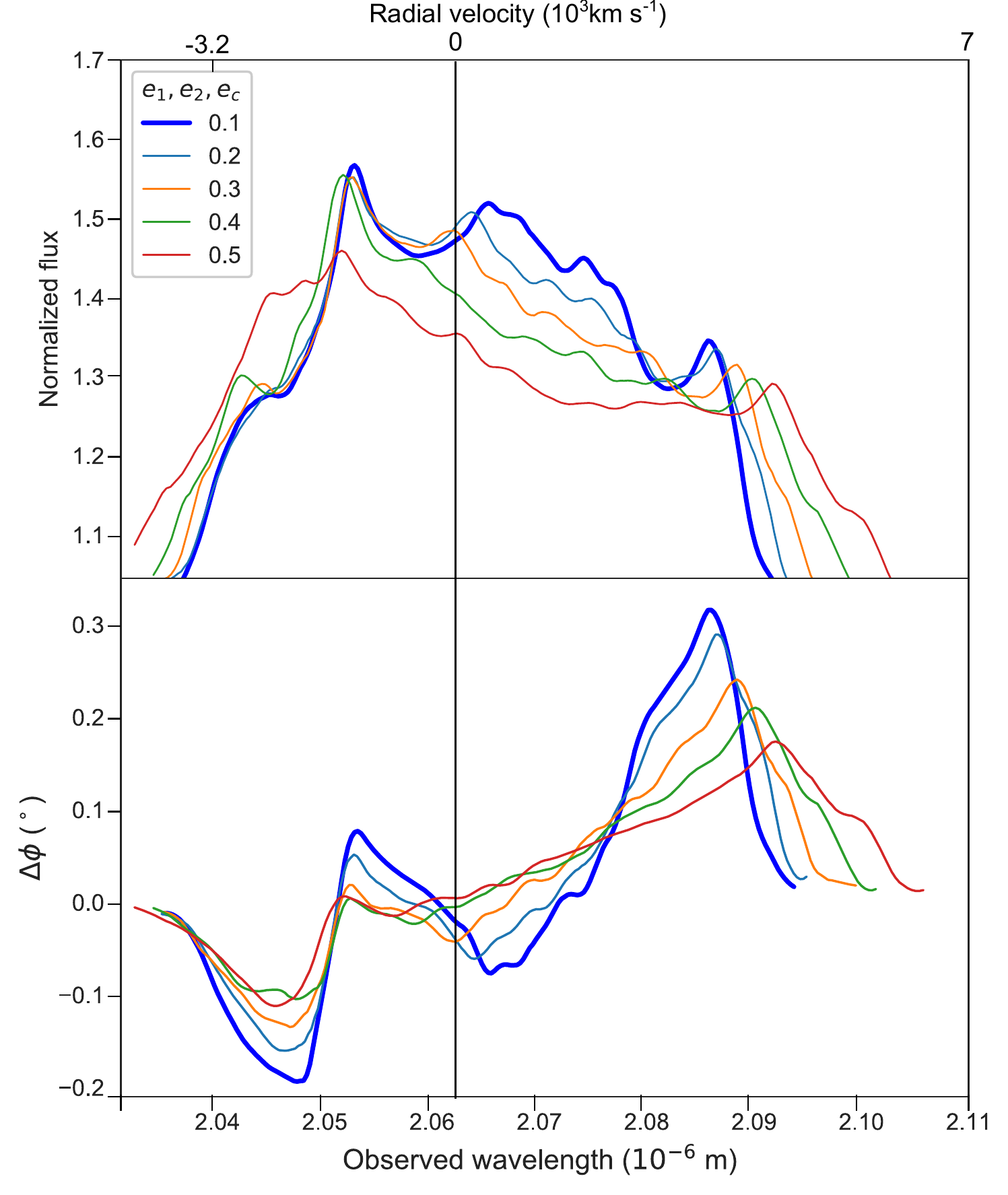}
		}%
		\hspace{-0.8em}
		\subfigure[$\mathcal{C}, i_{0}=45^{\circ}, \Omega_{1}=  100^{\circ}, \Omega_{2}=  300^{\circ},   \omega_{1}\newline\hspace*{1.5em}=250^{\circ}, \omega_{2}=70^{\circ},  e_{k=1,2}=(0.5,0.1),\newline\hspace*{1.5em} \delta e_{k=1,2}=-0.1,  i_{c1}=  rnd(90^{\circ},175^{\circ}),   i_{c2}= \newline\hspace*{1.5em} rnd(-45^{\circ},45^{\circ}),\Omega_{c1}=300^{\circ}, \Omega_{c2}= \newline\hspace*{1.5em}  100^{\circ} $,$\omega_{c1}=170^{\circ}, \omega_{c2}=350^{\circ}, \newline\hspace*{1.5em} e_{c}=(0.1,0.5), \delta e_{c}=0.1$
		]{%
			\label{fig:doubleel52}
			\includegraphics[trim = 2.0mm 0mm 0mm 0mm, clip,width=0.315\textwidth]{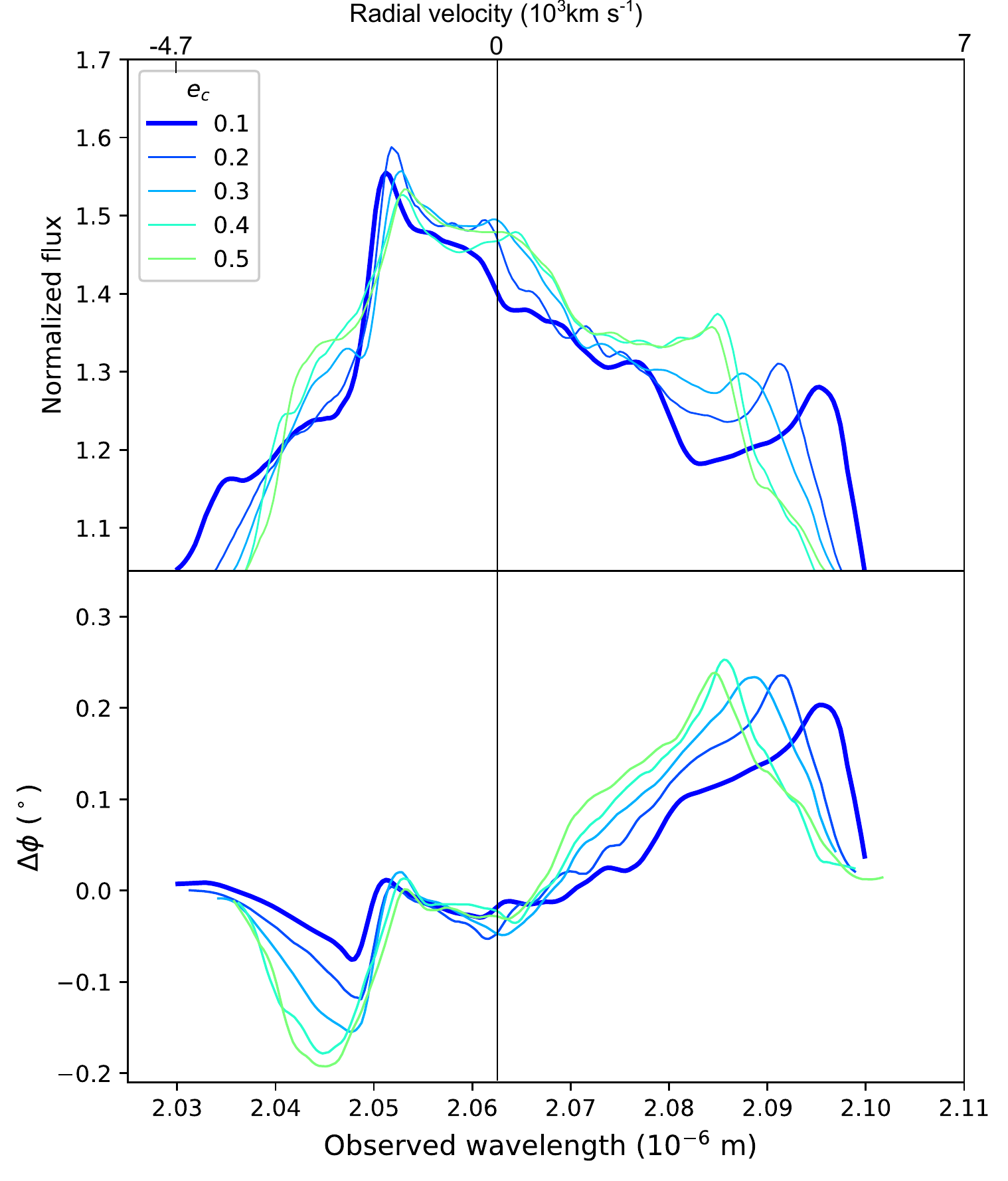}
		}
		\hspace{-1.4em}
		\subfigure[$i_{0}=45^{\circ}, i=10^{\circ},\Omega_{1}=  100^{\circ}, \Omega_{2}=  300^{\circ},  \newline\hspace*{1.5em} \omega_{1}=250^{\circ}, \omega_{2}=70^{\circ},  e_{k=1,2}=\mathcal{U}(0.5,0.1),\newline\hspace*{1.5em} \delta e_{k=1,2}=-0.1, i_{c1}=  rnd(90^{\circ},175^{\circ}),     i_{c2}= \newline\hspace*{1.5em} rnd(-45^{\circ},45^{\circ}),\Omega_{c1}=300^{\circ}, \Omega_{c2}=\newline\hspace*{1.5em}  100^{\circ} $,$\omega_{c1}=170^{\circ}, \omega_{c2}=350^{\circ}, \newline\hspace*{1.5em} e_{c}==\mathcal{U}(0.1,0.5), \delta e_{c}=0.1	$
		]{%
			\label{fig:doubleel53}
			\includegraphics[trim = 2.0mm 0mm 0mm 0mm, clip, width=0.315\textwidth]{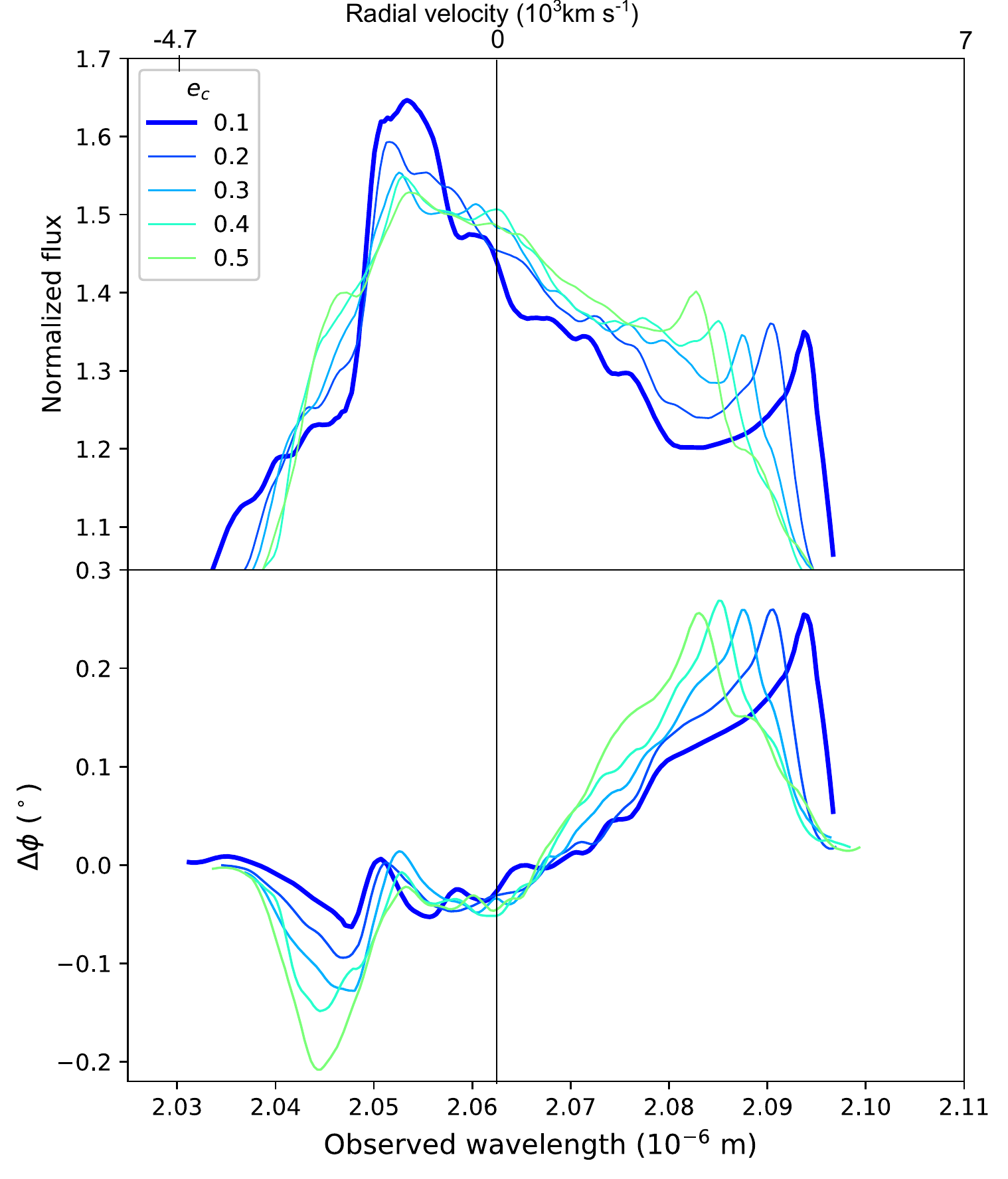}
			
		}	\\ 
		\vspace{-1.em}
		\subfigure[$ i=10^{\circ},\Omega_{1}=  100^{\circ}, \Omega_{2}= \newline\hspace*{1.5em}300^{\circ}, \omega_{1}=250^{\circ}, \omega_{2}=70^{\circ},  e_{k}=\newline\hspace*{1.5em}0.5, k=1,2, i_{c1}=  rnd(90^{\circ},\newline\hspace*{1.5em}175^{\circ}),  i_{c2}=  rnd(-45^{\circ},45^{\circ}),\newline\hspace*{1.5em}\Omega_{c1}= 300^{\circ}, \Omega_{c2}= 100^{\circ}, \omega_{c1}=,\newline\hspace*{1.5em}170^{\circ} \omega_{c2}=350^{\circ},  e_{c}=0.5$]
		{%
			\label{fig:doubleel54}
			\includegraphics[trim = 5.0mm 0mm 0mm 0mm, clip, width=0.24\textwidth]{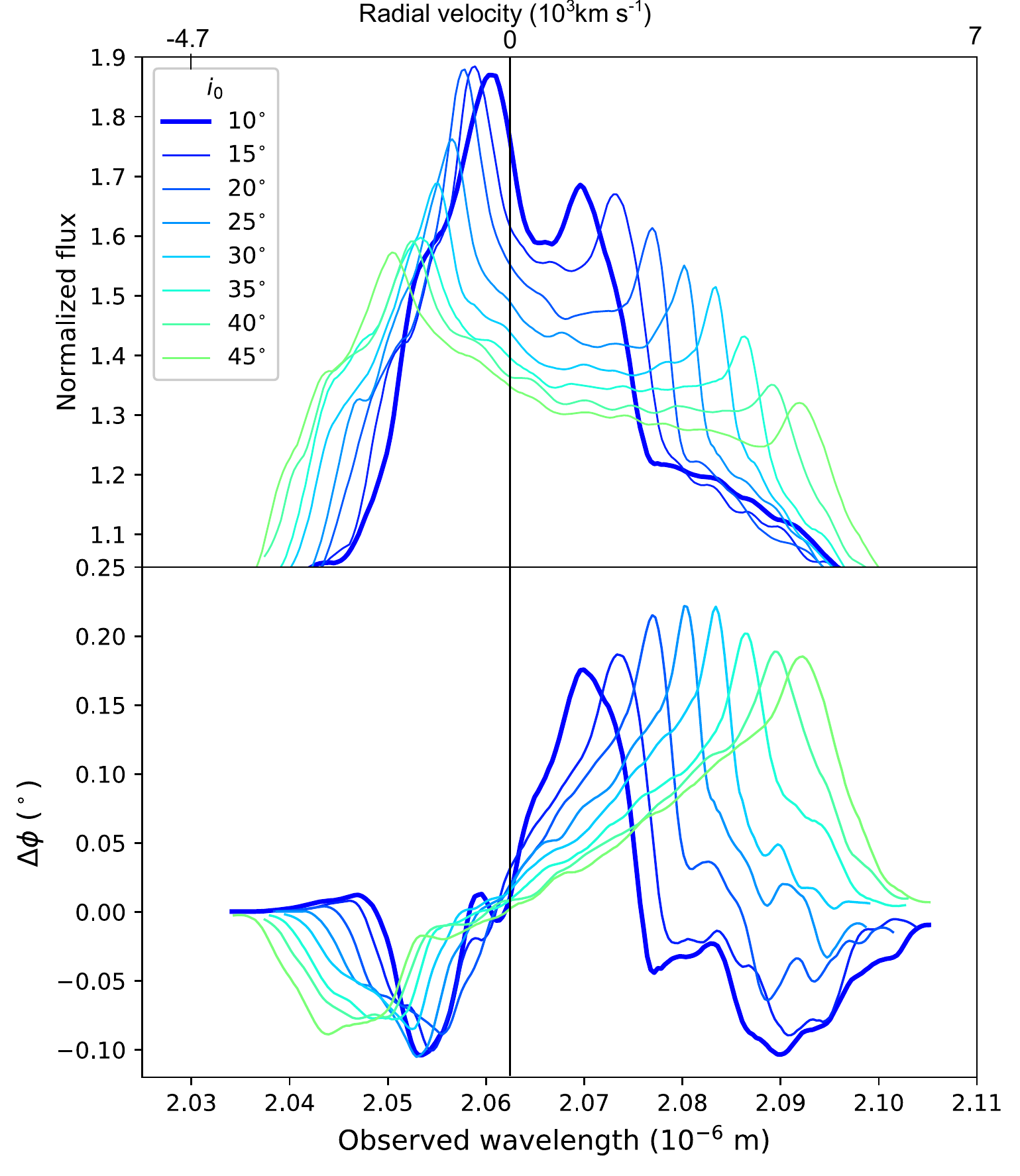}
		}%
		\hspace{-0.8em}
		\subfigure[$ \Omega_{1}=  100^{\circ}, \Omega_{2}=  300^{\circ},   \omega_{1}=\newline\hspace*{1.5em}250^{\circ}, \omega_{2}=70^{\circ},  e_{k=1,2}=0.5,\newline\hspace*{1.5em}, i_{c1}=  rnd(90^{\circ},175^{\circ}), i_{c2}= \newline\hspace*{1.5em} rnd(-45^{\circ},45^{\circ}),\Omega_{c1}=300^{\circ},\newline\hspace*{1.5em} \Omega_{c2}=100^{\circ},\omega_{c1}=170^{\circ},\newline\hspace*{1.5em} \omega_{c2}=350^{\circ},  e_{c}=0.5$
		]{%
			\label{fig:doubleel55}
			\includegraphics[trim = 5.0mm 0mm 0mm 0mm, clip,width=0.24\textwidth]{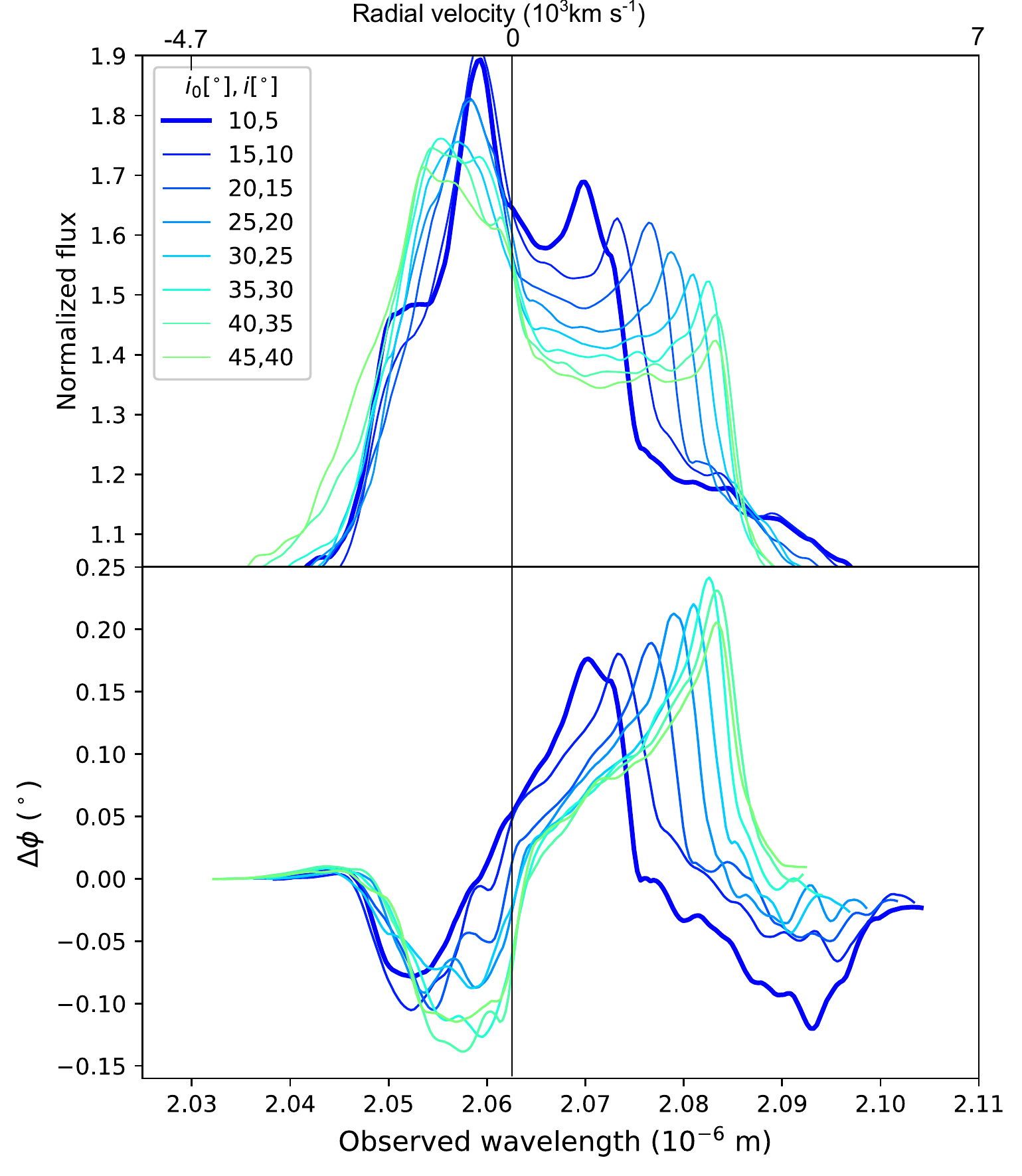}
		}
		\hspace{-0.8em}
		\subfigure[$ \Omega_{1}=  100^{\circ}, \Omega_{2}=  300^{\circ},   \omega_{1}=\newline\hspace*{1.5em}250^{\circ}, \omega_{2}=70^{\circ}, e_{k}=\newline\hspace*{1.5em}rnd(0.1,0.5), k=1,2,  i_{c1}=\newline\hspace*{1.5em}  rnd(90^{\circ},175^{\circ}), i_{c2}=  rnd(-45^{\circ},\newline\hspace*{1.5em}45^{\circ}),\Omega_{c1}=300^{\circ}, \Omega_{c2}=100^{\circ},\newline\hspace*{1.5em}\omega_{c1}=170^{\circ},\omega_{c2}=350^{\circ}, \newline\hspace*{1.5em} e_{c}=rnd(0.1,0.5)$
		]{%
			\label{fig:doubleel56}
			\includegraphics[trim = 3.0mm 0mm 0mm 0mm, clip,width=0.24\textwidth]{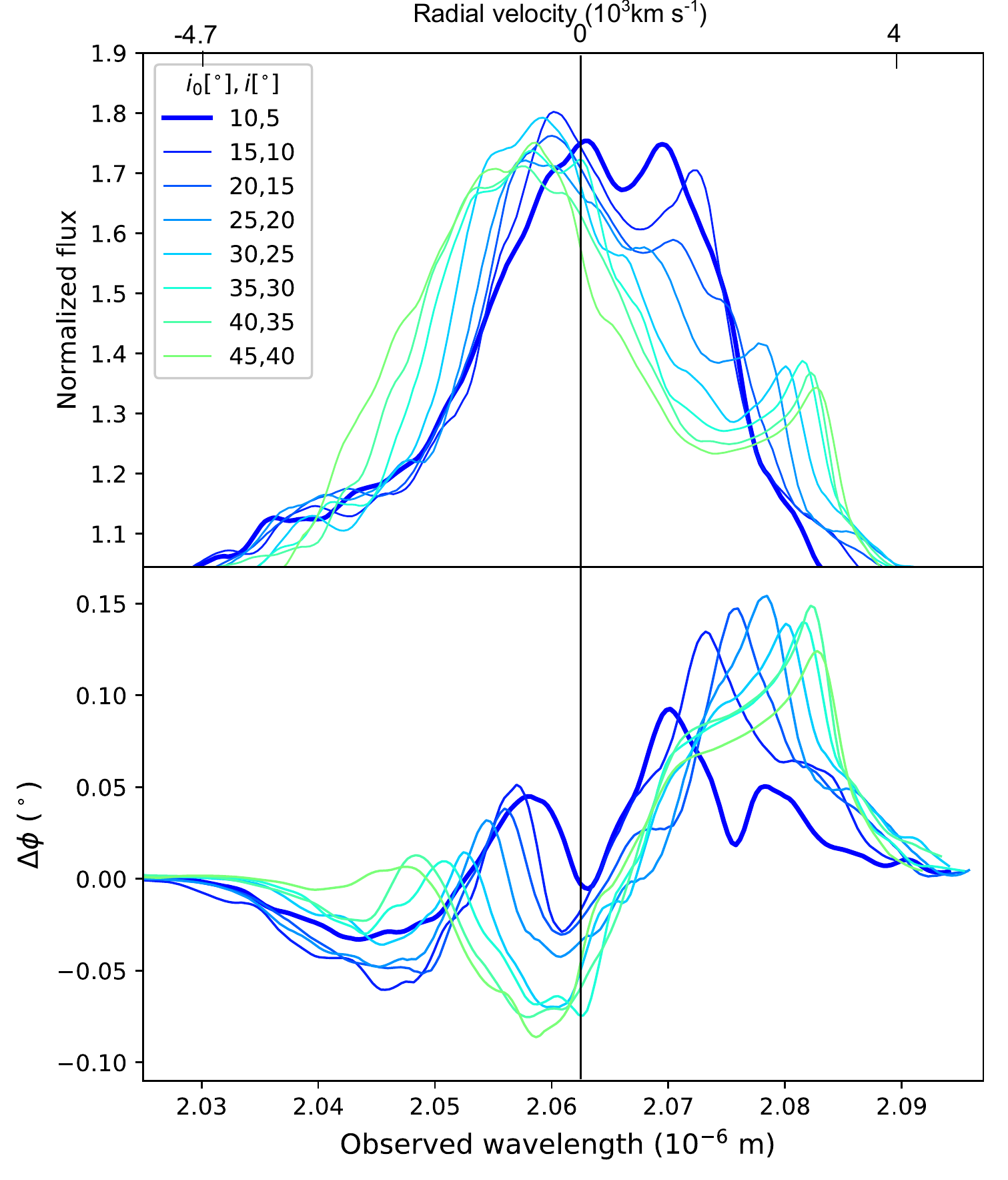}
		}%
		\hspace{-0.8em}
		\subfigure[$ \Omega_{1}=  100^{\circ}, \Omega_{2}=  300^{\circ},   \omega_{1}=\newline\hspace*{1.5em}250^{\circ}, \omega_{2}70^{\circ}$,$e_{k}=rnd(0.1\newline\hspace*{1.5em},0.5), k=1,2, i_{c1}=  rnd(90^{\circ},\newline\hspace*{1.5em}175^{\circ}), i_{c2}=  rnd(-45^{\circ},45^{\circ})$,\newline\hspace*{1.5em}$\Omega_{c1}=300^{\circ}, \Omega_{c2}=100^{\circ},\omega_{c1}=\newline\hspace*{1.5em}170^{\circ},\omega_{c2}=350^{\circ},e_{c}=e_{k=1,2}$
		]{%
			\label{fig:doubleel57}
			\includegraphics[trim = 5.0mm 0mm 0mm 0mm, clip,width=0.24\textwidth]{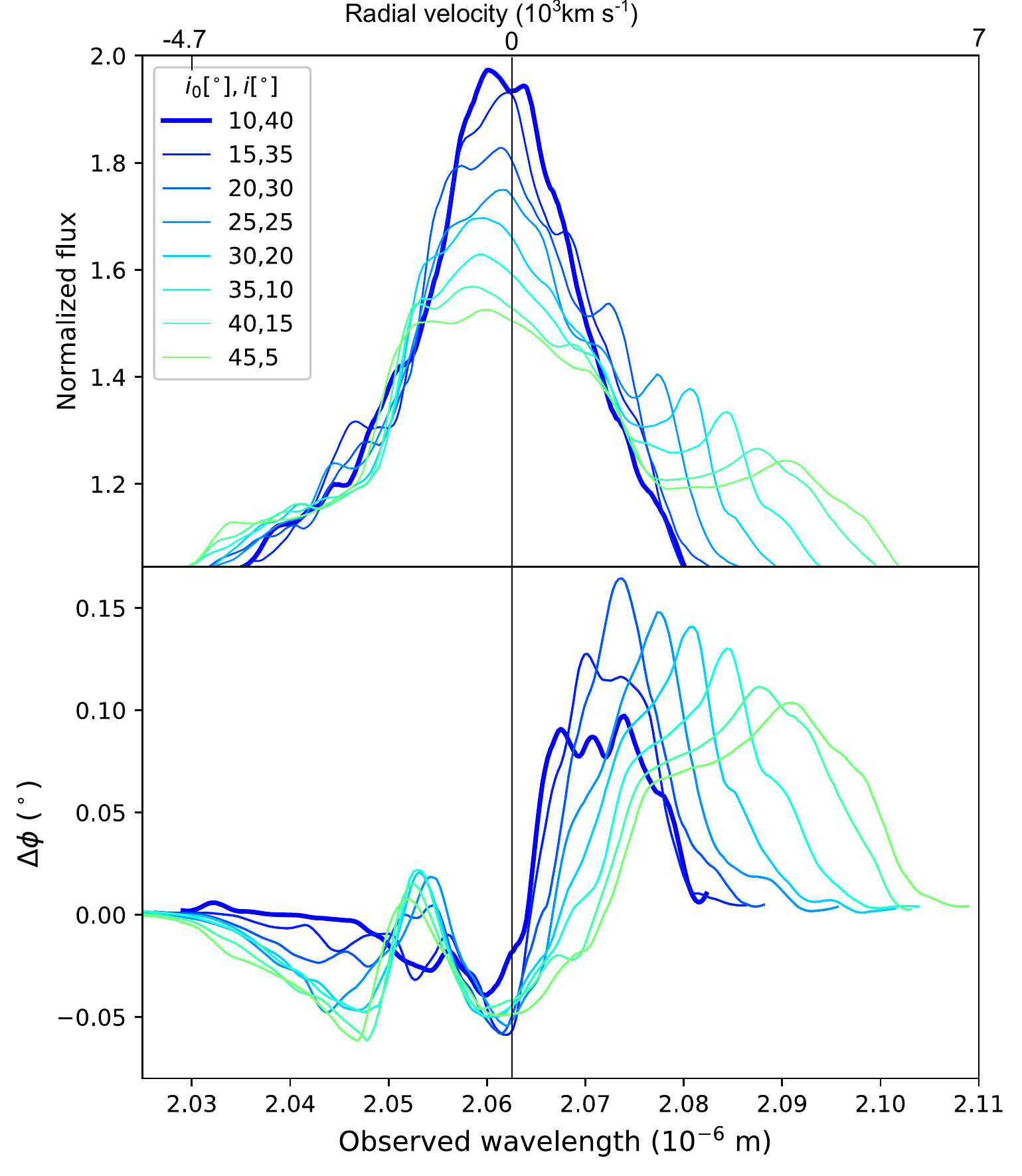}
		}%
		\vspace{-1.9em}
	\end{center}
	\caption{%
		Same as Fig. \ref{fig:doubleel4} but for anti-aligned angular momenta of clouds' orbits  in the BLR of  larger SMBH. 
	}%
	
	\label{fig:doubleel5}
\end{figure*}

For a simultaneous variation of orbital eccentricities of clouds and SMBHs, and anti-aligned angular momenta of clouds in the  BLR of larger SMBH,  the  convex shape of the line core appears (see blue line model in Fig. \ref{fig:doubleel51}). The dip between the peaks is filled  because of  a more dispersed velocity field. However, concavity  appears when  orbital eccentricities of clouds in both BLRs and SMBHs   change in  opposite directions (seen line profiles in  Figs. \ref{fig:doubleel52} and \ref{fig:doubleel53}).
Further, simultaneous  variation of  inclinations of observer and mutual inclinations of SMBHs orbits are investigated  (see bottom panels in Figs. \ref{fig:doubleel54} - \ref{fig:doubleel57}).  Spectral lines are narrower, with  a broken wing. The form of the spectral line marked with blue colour  in  Fig. \ref{fig:doubleel57} resembles some  well  known examples of 'Eiffel Tower' shapes of Pa$\alpha$ seen  in 3C 273 and Mrk 110 \citep[][see their Fig.3]{10.1086/522373}. We observe  that  top of spectral line became convex when $i_{0}$ is increasing  and mutual SMBH orbital inclination is decreasing.
Differential phases evolution  is remarkable in  their right wings and   plateau variations (see  Fig. \ref{fig:doubleel5}).  Even more drastic effects are  given in  Figs. \ref{fig:doubleel54} - \ref{fig:doubleel57}. 

\begin{figure*}[ht!]
	\begin{center}

		\subfigure[$\Omega_{k}=  300^{\circ}, e_{k}=rnd(0.1,0.5), k=1,2 ,\newline\hspace*{1.5em} \omega_{1}=250^{\circ} \omega_{2}=70^{\circ},  \Omega_{c1}=300^{\circ}, \Omega_{c2}=\newline\hspace*{1.5em}300^{\circ}, $$\omega_{c1}=170^{\circ},  \omega_{c2}=350^{\circ},\newline\hspace*{1.5em} e_{c}=rnd(0.1,0.5)$ 	
		]	{%
			\label{fig:doubleel61}
			\includegraphics[trim = 5.0mm 0mm 0mm 0mm, clip, width=0.315\textwidth]{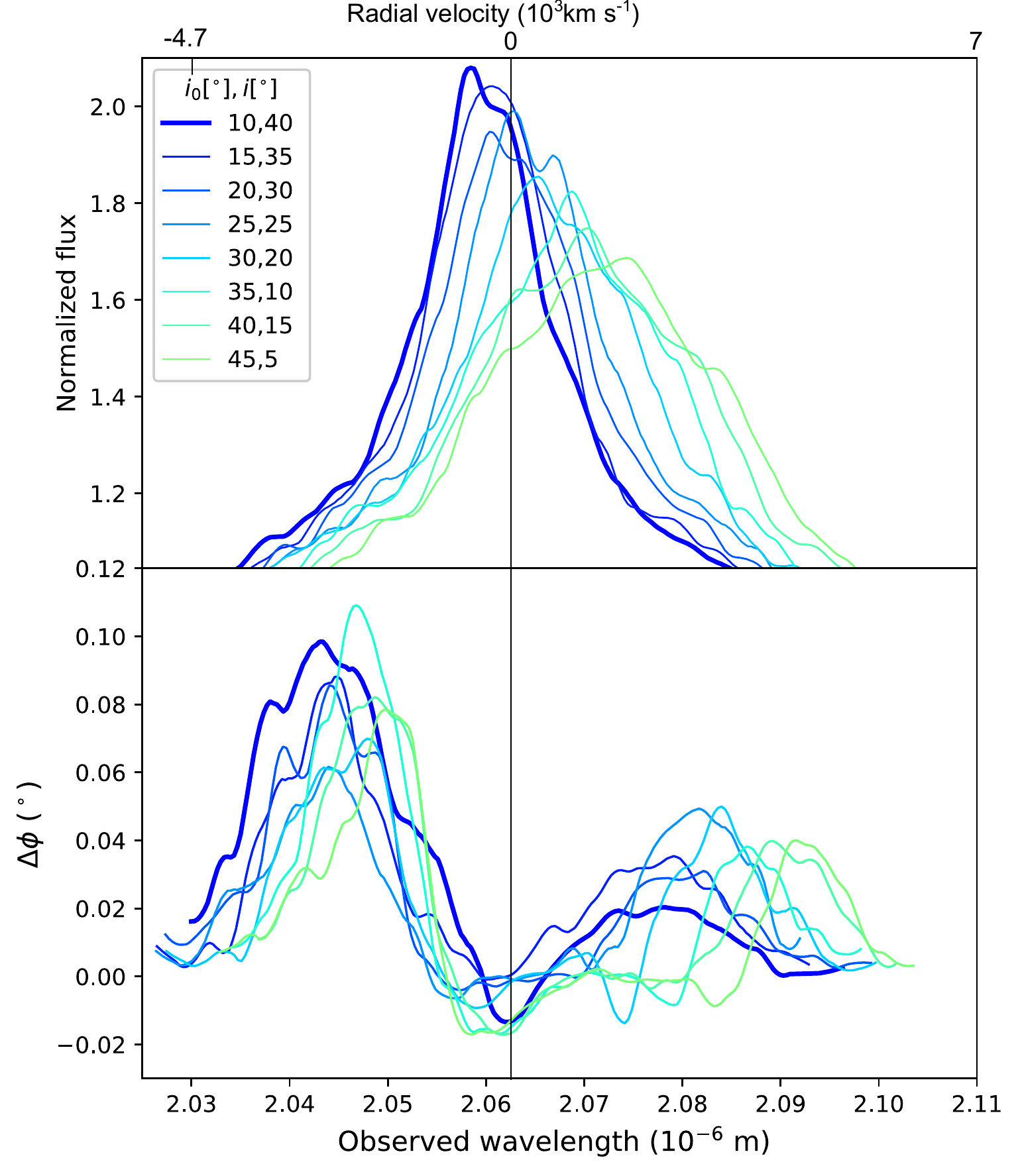}
		}%
		\hspace{-0.8em}
		\subfigure[$\Omega_{1}=  300^{\circ}, \Omega_{2}=  100^{\circ},  e_{k}=0.5, k=1,2 ,\newline\hspace*{1.5em} \omega_{1}=250^{\circ} \omega_{2}=70^{\circ},  \Omega_{c1}=100^{\circ}, \Omega_{c2}=\newline\hspace*{1.5em}300^{\circ} $,$\omega_{c1}=350^{\circ},  \omega_{c2}=170^{\circ},e_{c}=0.5$ 
		]{%
			\label{fig:doubleel62}
			\includegraphics[trim = 0.0mm 0mm 0mm 0mm, clip,width=0.315\textwidth]{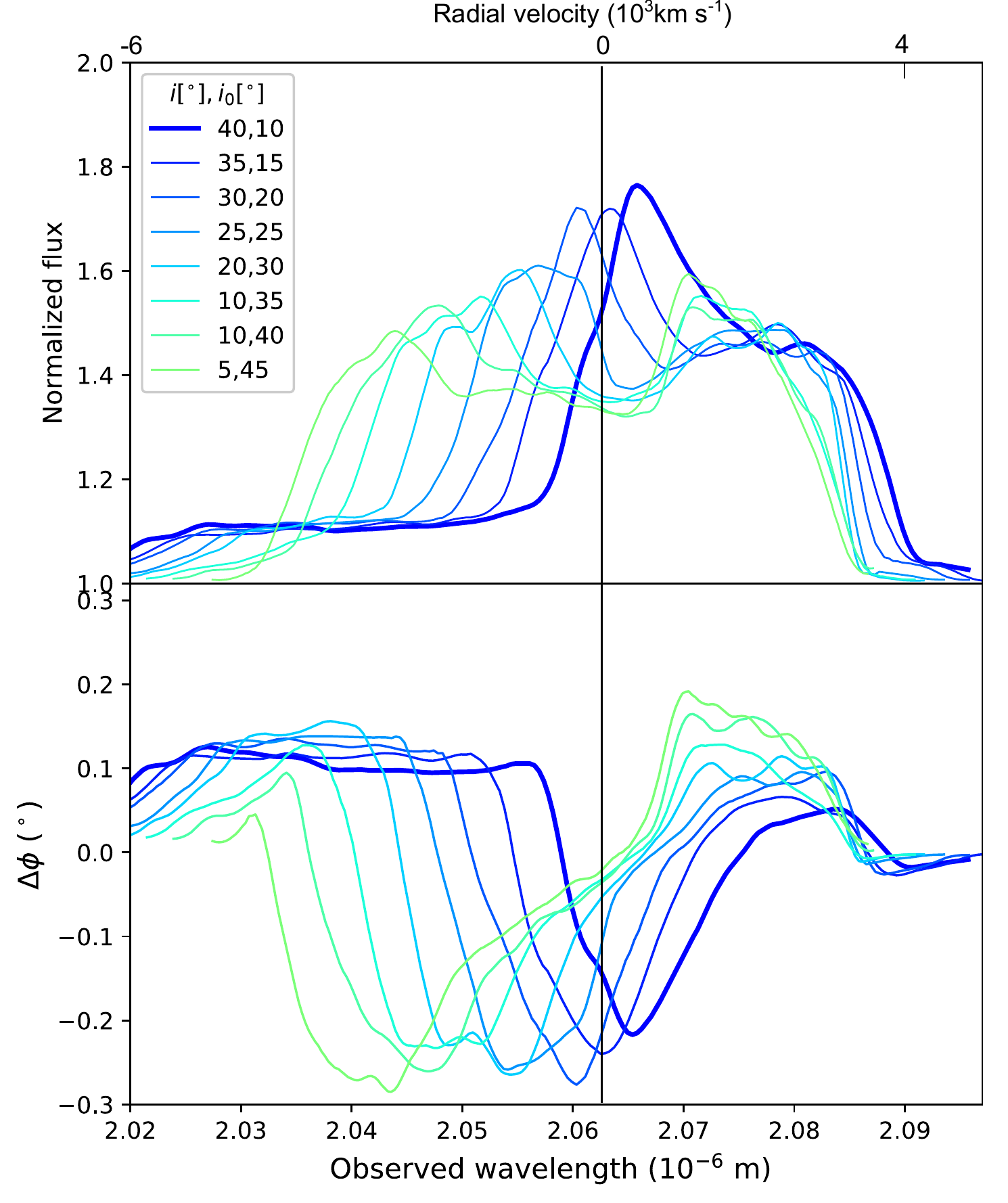}
		}
		\hspace{-0.8em}
		\subfigure[$i_{0}=45^{\circ}, \Omega_{1}=  100^{\circ}, \Omega_{2}=  300^{\circ},  e_{k}=\newline\hspace*{1.5em}(0.1,0.5), k=1,2 ,\omega_{1}=250^{\circ}, \omega_{2}=70^{\circ}, \newline\hspace*{1.5em}  \Omega_{c1}=100^{\circ}, \Omega_{c2}=300^{\circ}, $$\omega_{c1}=350^{\circ},  \newline\hspace*{1.5em}\omega_{c2}=170^{\circ}$ 
		]{%
			\label{fig:doubleel63}
			\includegraphics[trim = 2.0mm 0mm 0mm 0mm, clip, width=0.315\textwidth]{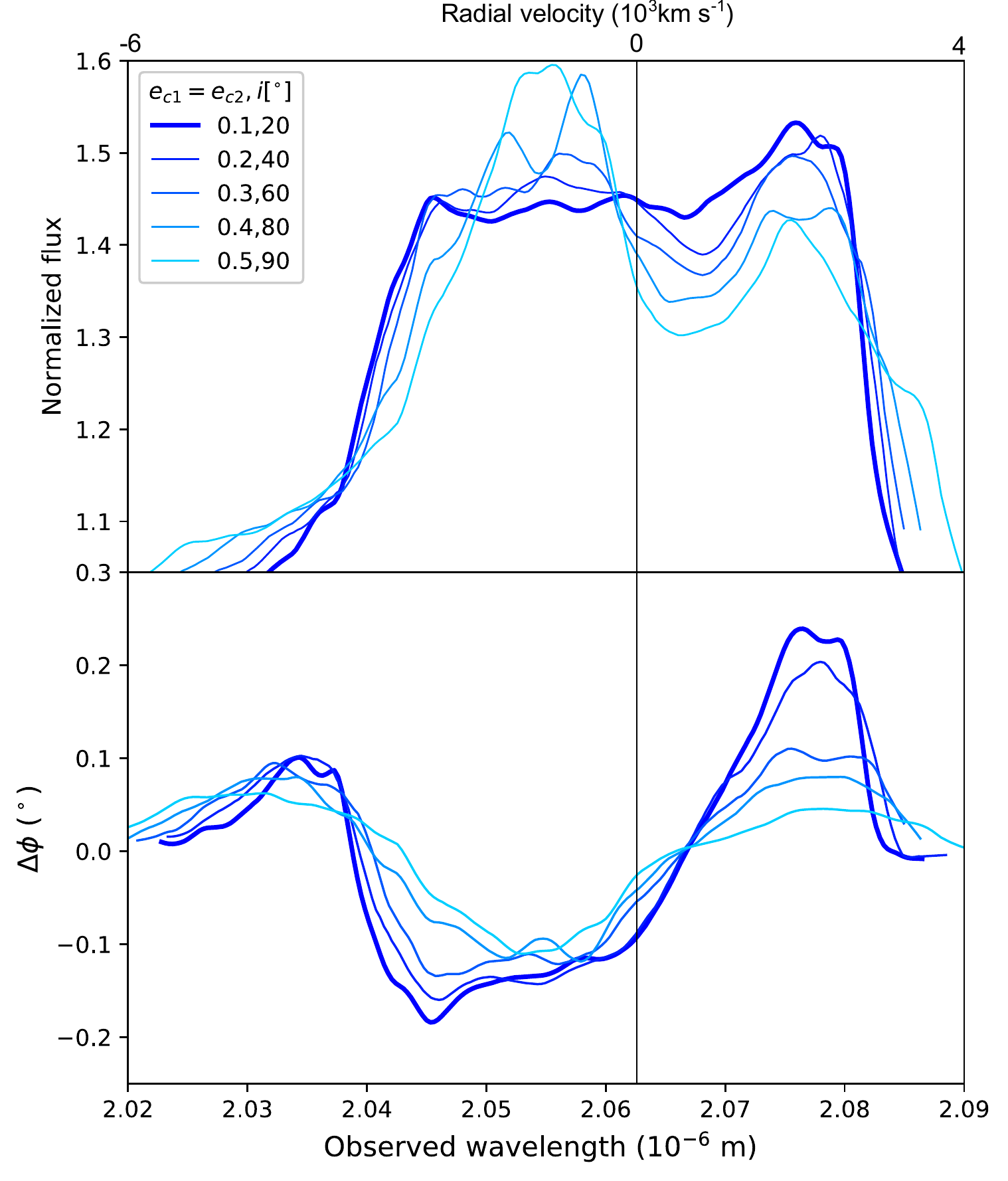}
			
		}	\\ 
		\vspace{-1.em}
		\subfigure[$ i_{0}=45^{\circ}, \Omega_{1}=  100^{\circ}, \Omega_{2}=  300^{\circ}, \omega_{1}=\newline\hspace*{1.5em} 250^{\circ}, \omega_{2}=70^{\circ},  \Omega_{c1}=100^{\circ}, \Omega_{c2}=300^{\circ}\newline\hspace*{1.5em} , \omega_{c1}=350^{\circ},\omega_{c2}=170^{\circ}$ ]
		{%
			\label{fig:doubleel64}
			\includegraphics[trim = 0.0mm 0mm 0mm 0mm, clip, width=0.315\textwidth]{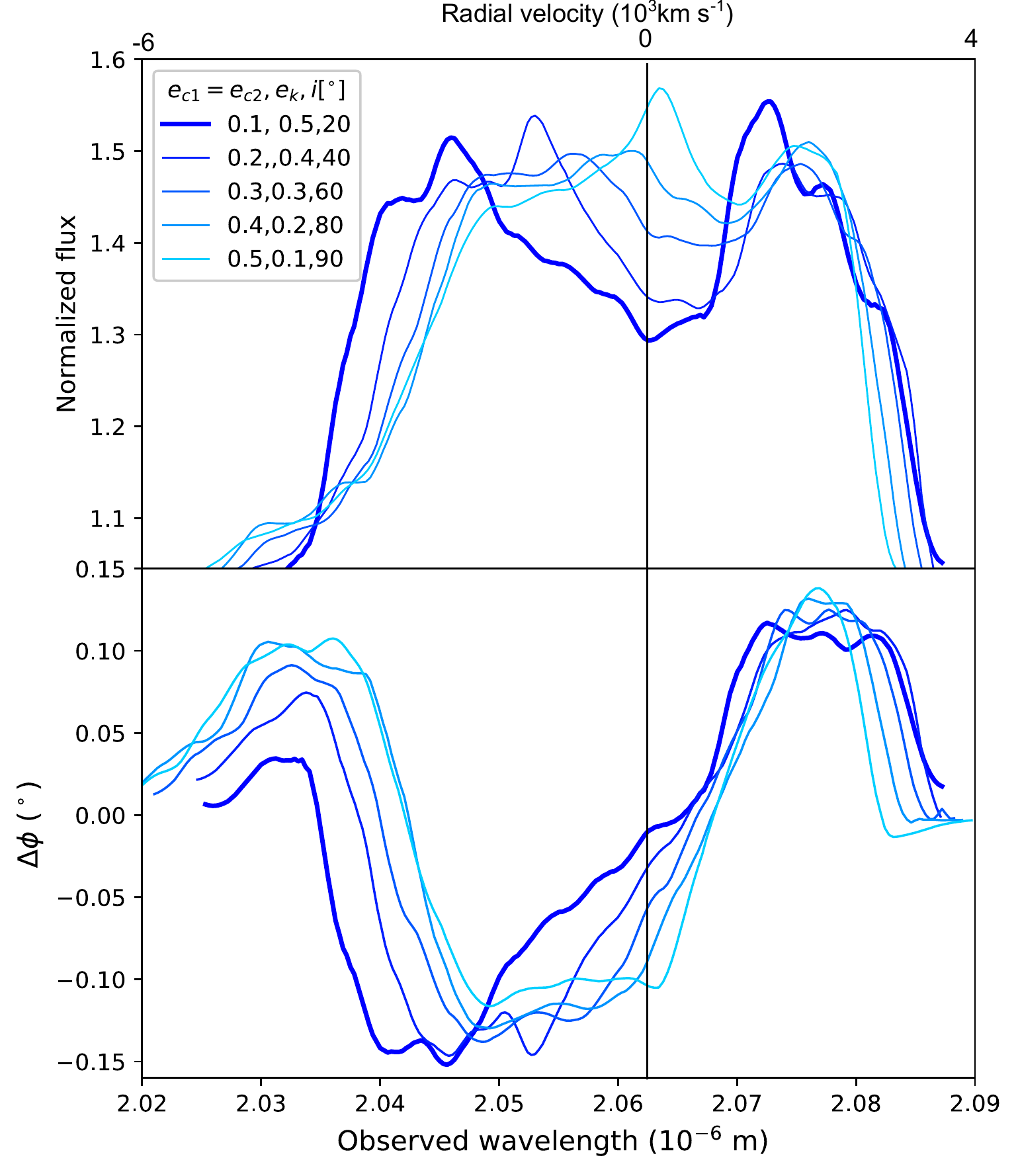}
		}%
		\hspace{-0.8em}
		\subfigure[$ \Omega_{1}=  350^{\circ}, \Omega_{2}=  100^{\circ}, \omega_{1}=230^{\circ}, \omega_{2}=\newline\hspace*{1.5em} 50^{\circ}, e_{k}=0.5, k=1,2,e_{c}=rnd(0.1,0.5),\newline\hspace*{1.5em} \Omega_{c}=\omega_{c}=rnd (10^{\circ},290^{\circ}),   $ 
		]{%
			\label{fig:doubleel65}
			\includegraphics[trim = 2.0mm 0mm 0mm 0mm, clip,width=0.315\textwidth]{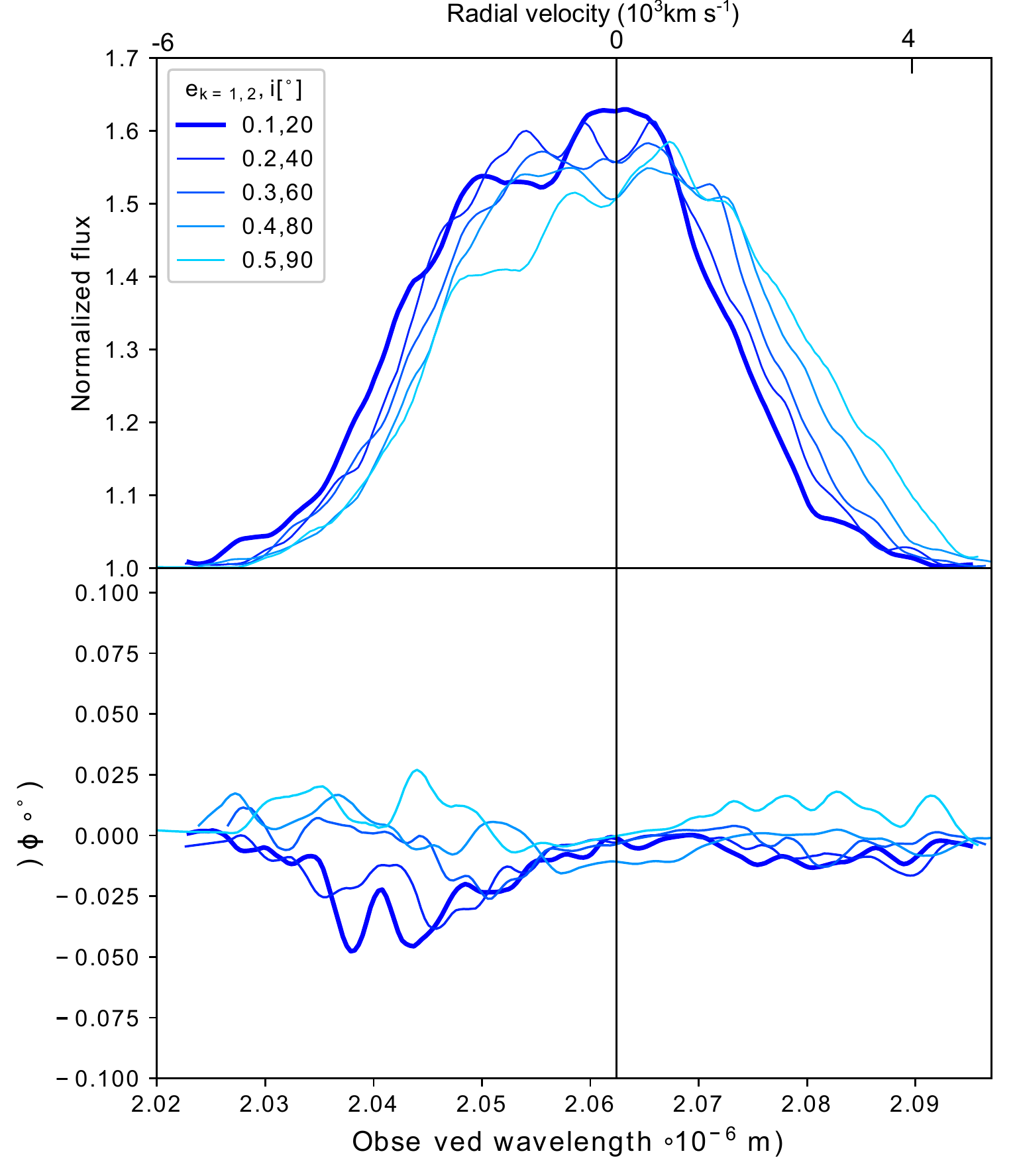}
		}
		\hspace{-0.8em}
		\subfigure[$ \Omega_{1}=  350^{\circ}, \Omega_{2}=  300^{\circ}, \omega_{1}=50^{\circ}, \omega_{2}=\newline\hspace*{1.5em}230^{\circ}, e_{k}=0.5, k=1,2,  \Omega_{c1}=100^{\circ}, \Omega_{c2}=\newline\hspace*{1.5em}rnd(10^{\circ}, 290^{\circ}), \omega_{c1}=100^{\circ}, \omega_{c2}=280^{\circ},   \newline\hspace*{1.5em} e_{c}=0.5$
		]{%
			\label{fig:doubleel66}
			\includegraphics[trim = 2.0mm 0mm 0mm 0mm, clip,width=0.315\textwidth]{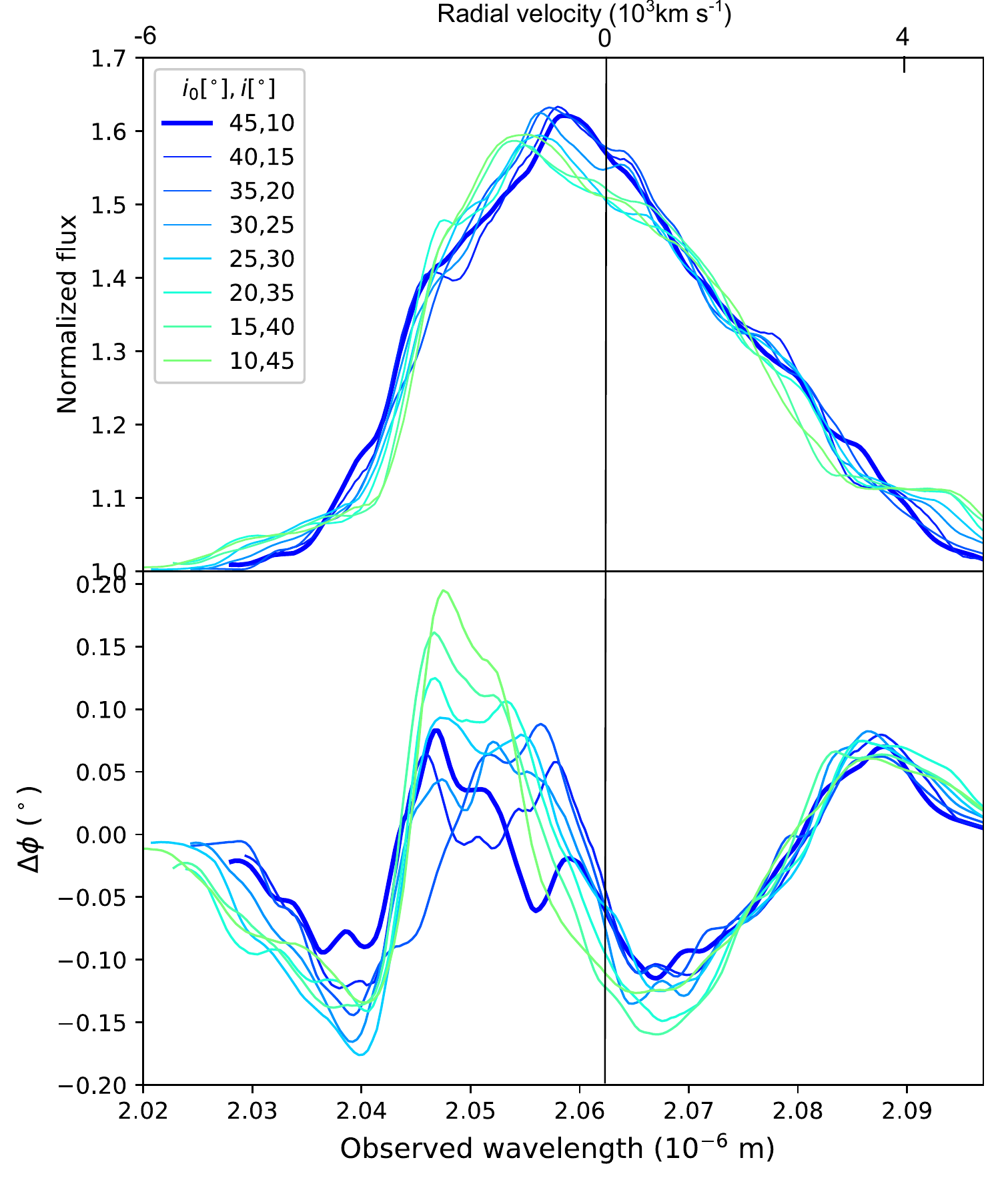}
		}%
		
		\vspace{-1.9em}
	\end{center}
	\caption{%
		Same as Fig. \ref{fig:doubleel5} but for anti-aligned angular momenta of clouds in both BLRs. Clouds' orbital inclinations are randomly distributed within the range ($90^{\circ},175^{\circ}$).
	}%
	
	\label{fig:doubleel6}
\end{figure*}

 Effects of the asynchronous orientation of the angular momenta of clouds' orbit  in both BLRs  (see Fig. \ref{fig:doubleel6}) differs from  the previous two cases.
The 'Eiffel Tower' spectral shapes  in Fig. \ref{fig:doubleel61} and  Fig. \ref{fig:doubleel57} are similar, yet  distinct in their left wings. This is reflecting a  difference in  
$\Omega$ of clouds and SMBHs orbits.

More detailed Fig. \ref{fig:doubleel9} summarizes simulations for different combinations of  $\mathcal{R}$ distributions of clouds' eccentricities and orbital parameters of SMBHs. Also, notable level of noise  is present in the differential phase. 

When  motion of clouds  in both BLRs are non-coplanar, then the differential phase is slightly deformed in the right-wing (see Fig. \ref{fig:doubleel92}).
 Also, if clouds orbital eccentricities in one of the BLRs are   high (0.5), then the differential phase shape will be smoother (see Figs. \ref{fig:doubleel94} and  \ref{fig:doubleel96}).  The same situation is for   coplanar clouds' motion (see Fig. \ref{fig:doubleel94}).
 Simulation results for $\Gamma_{s}(0.3,1)$ distribution of clouds' eccentricities are given in 	Fig. \ref{fig:doubleel10}. For   coplanar  clouds'  motion  around  a more massive component,  the differential phase  amplitude decreases (Figs. \ref{fig:doubleel103}-\ref{fig:doubleel105}). 

With the disfigured core  of the spectral lines, the net effect is that the differential phase peaks shifts away from the line center because of the smaller contribution of lower velocities of clouds. In such cases, the plateau between differential phase peaks is more prominent. The opposite  occur when there are higher contributions of projected lower velocities in spectral lines and consequently in the phase profiles.  

\begin{figure*}[ht!]
	\begin{center}

		\subfigure[$A_{0}=100\mathrm{ld},i_{0}=i=(10^{\circ}-40^{\circ},45^{\circ}), \Omega_{1}=200^{\circ}, \Omega_{2}=  150^{\circ},  e_{k}= 0.5, k=1,2,\omega_{1}=70^{\circ}, \omega_{2}=250^{\circ}, i_{c1}=i_{c2}=(10^{\circ}-40^{\circ},45^{\circ}), \Omega_{c1}=\newline\hspace*{1.5em}\Omega_{c2}=200^{\circ},  \omega_{c1}=100^{\circ},  \omega_{c2}=280^{\circ},e_{c}=0.5 $ 
		]	{%
			\label{fig:doubleel81}
			\includegraphics[trim = 3.0mm 0mm 0mm 0mm, clip, width=0.315\textwidth]{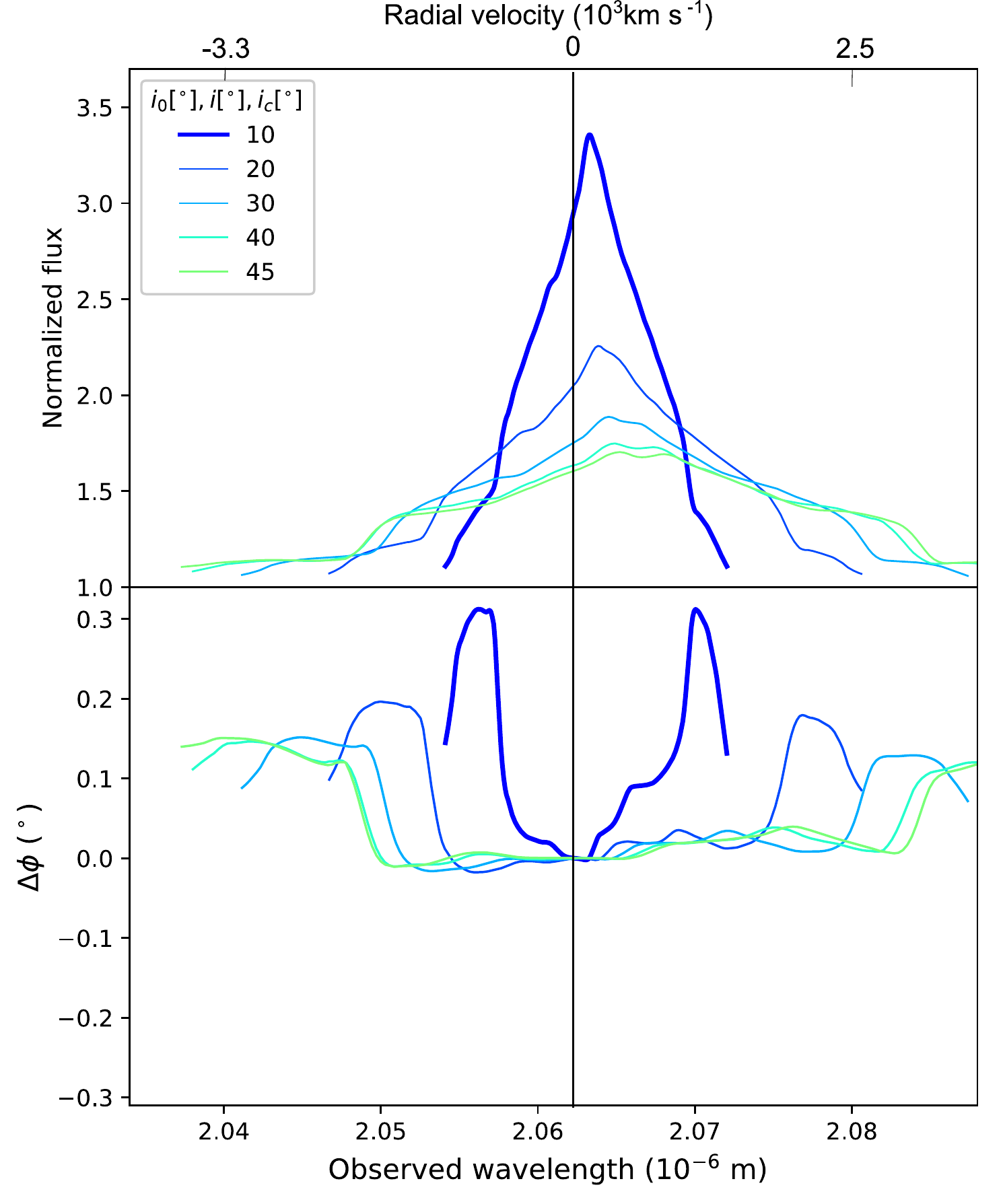}
			\includegraphics[trim = 3.0mm 0mm 0mm 0mm, clip,width=0.315\textwidth]{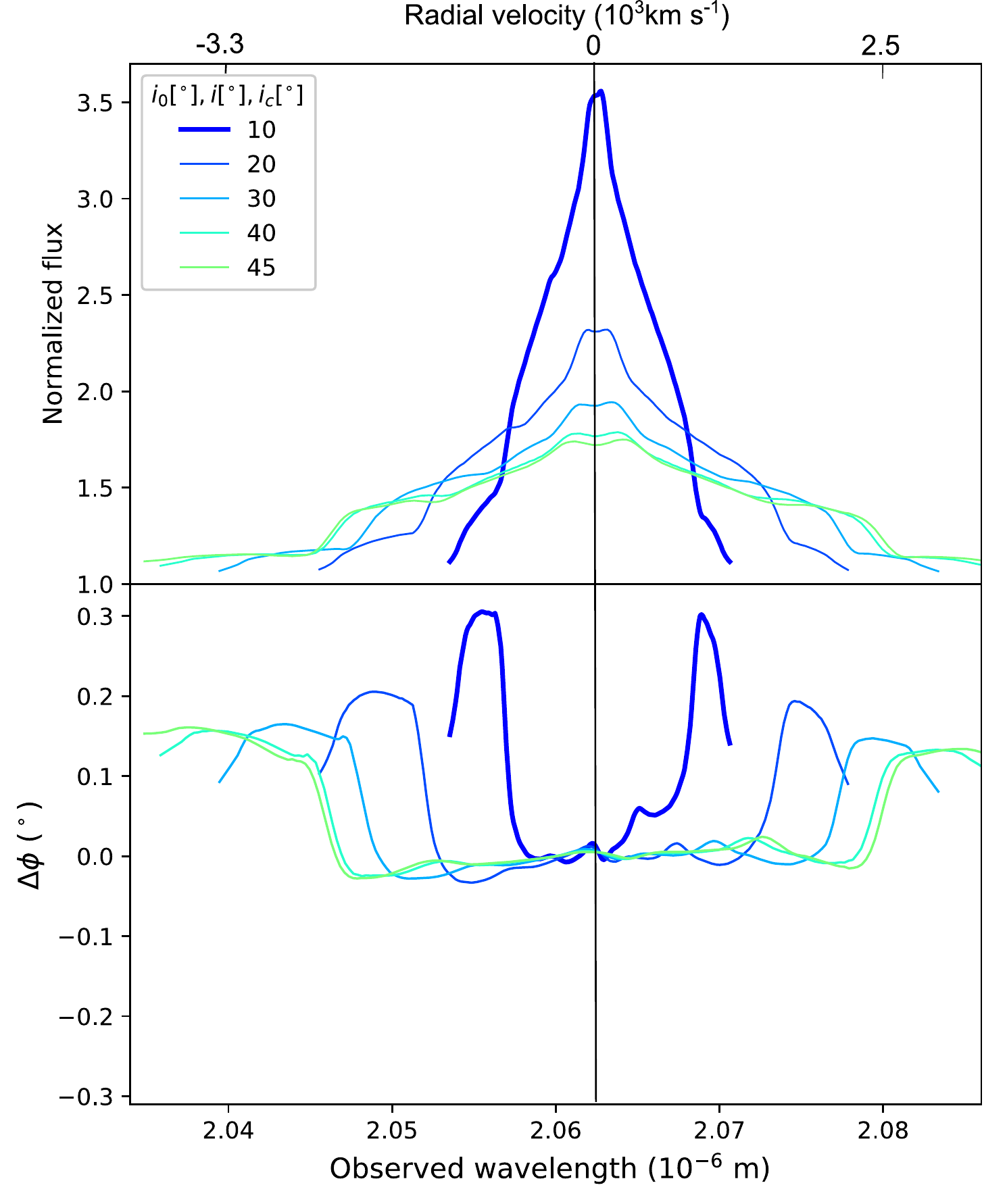}
			\includegraphics[trim = 3.0mm 0mm 0mm 0mm, clip, width=0.315\textwidth]{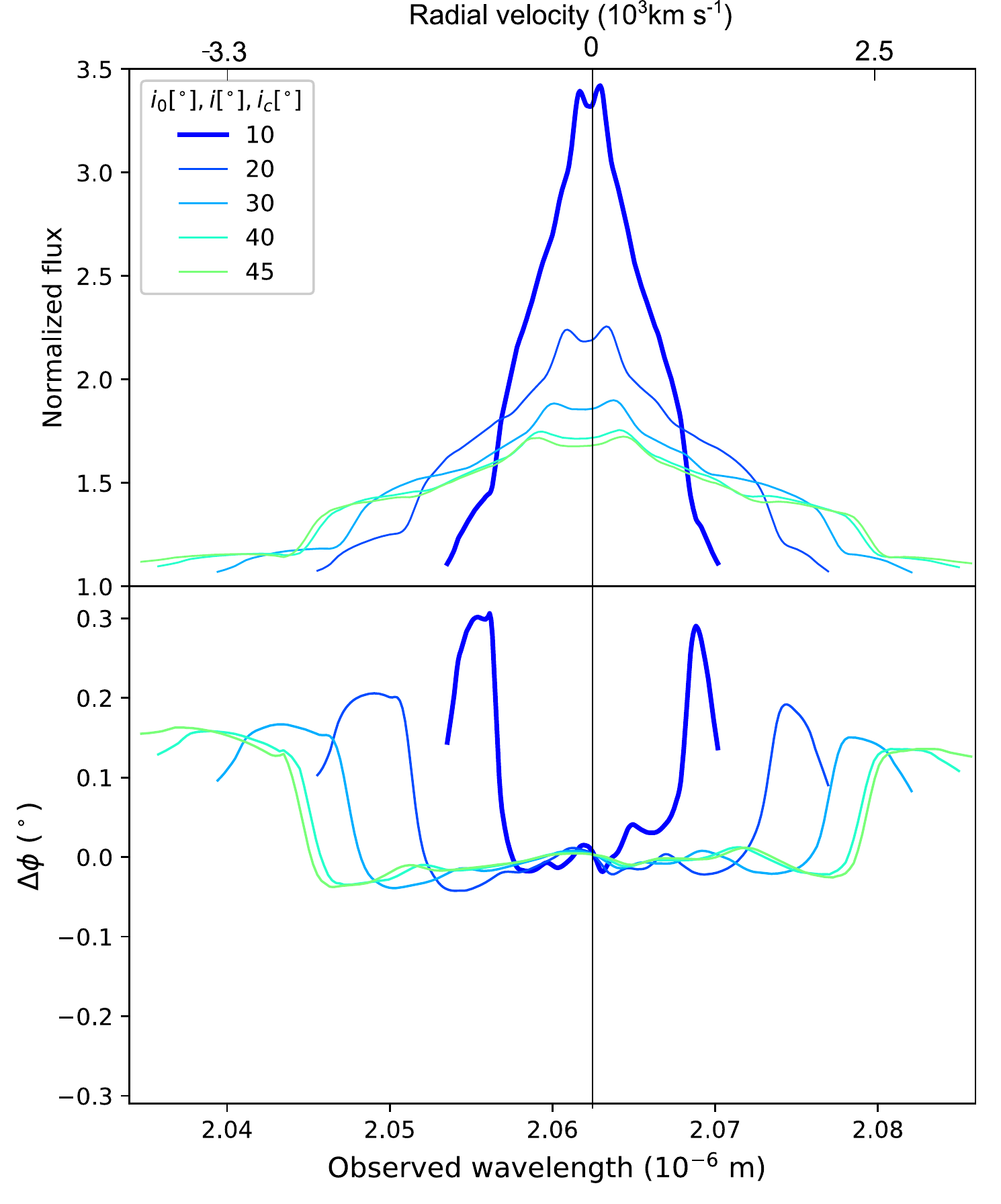}
		}%
		\\ 
		\subfigure[$A_{0}=100\mathrm{ld},i_{0}=i=(10^{\circ}-40^{\circ},45^{\circ}), \Omega_{1}=200^{\circ}, \Omega_{2}=  150^{\circ},  e_{1}= (0.1,0.5), e_{2}=(0.5,0.1),\omega_{1}=70^{\circ}, \omega_{2}=250^{\circ}, i_{c1}=i_{c2}=\newline\hspace*{1.5em}(10^{\circ}-40^{\circ},45^{\circ}), \Omega_{c1}=\Omega_{c2}=200^{\circ},  \omega_{c1}=100^{\circ},  \omega_{c2}=280^{\circ},e_{c1}=e_{2}, ,e_{c2}=e_{1} $ 
		]	{%
			\label{fig:doubleel82}
			\includegraphics[trim = 3.0mm 0mm 0mm 0mm, clip, width=0.315\textwidth]{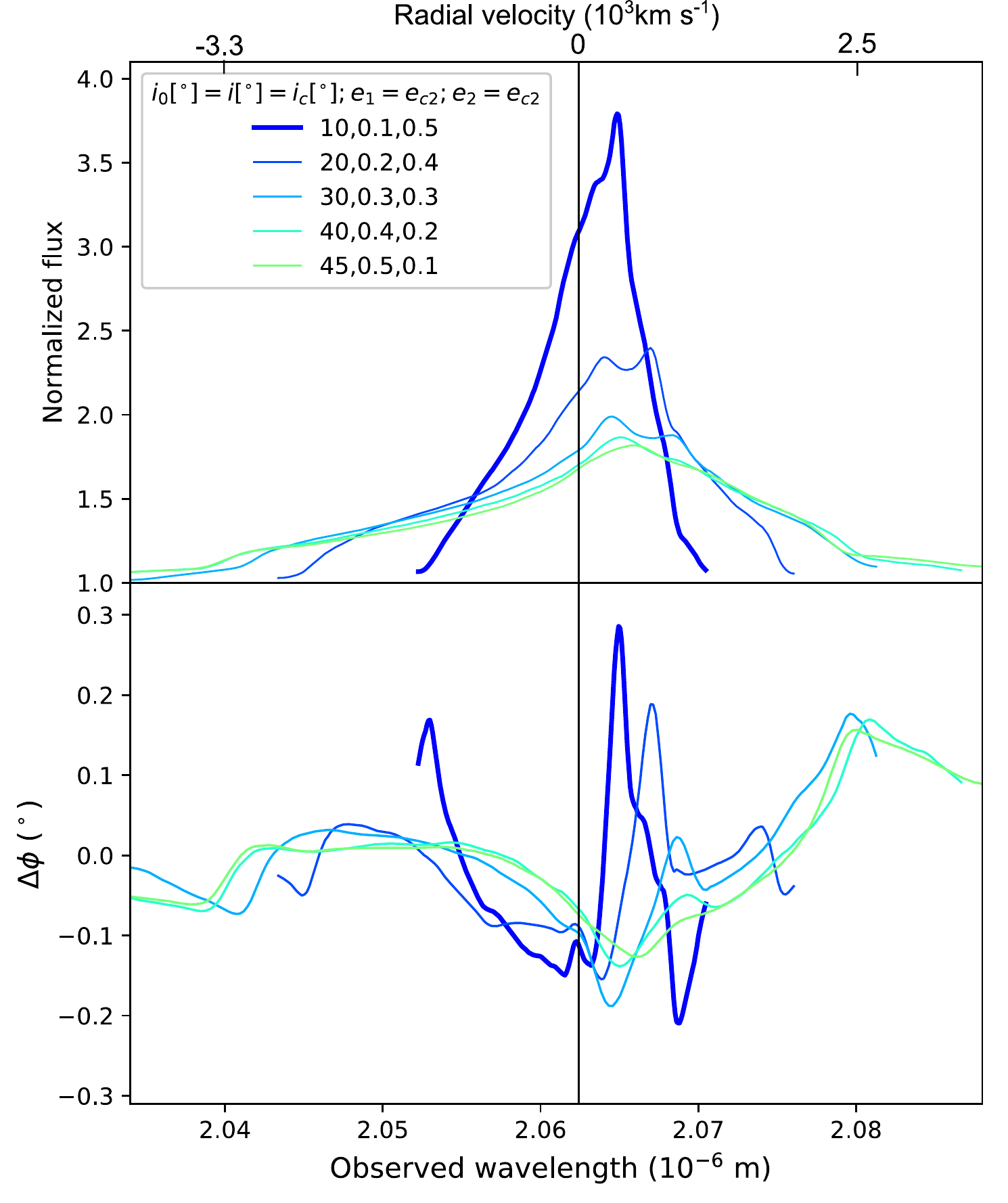}
			\includegraphics[trim = 3.0mm 0mm 0mm 0mm, clip,width=0.315\textwidth]{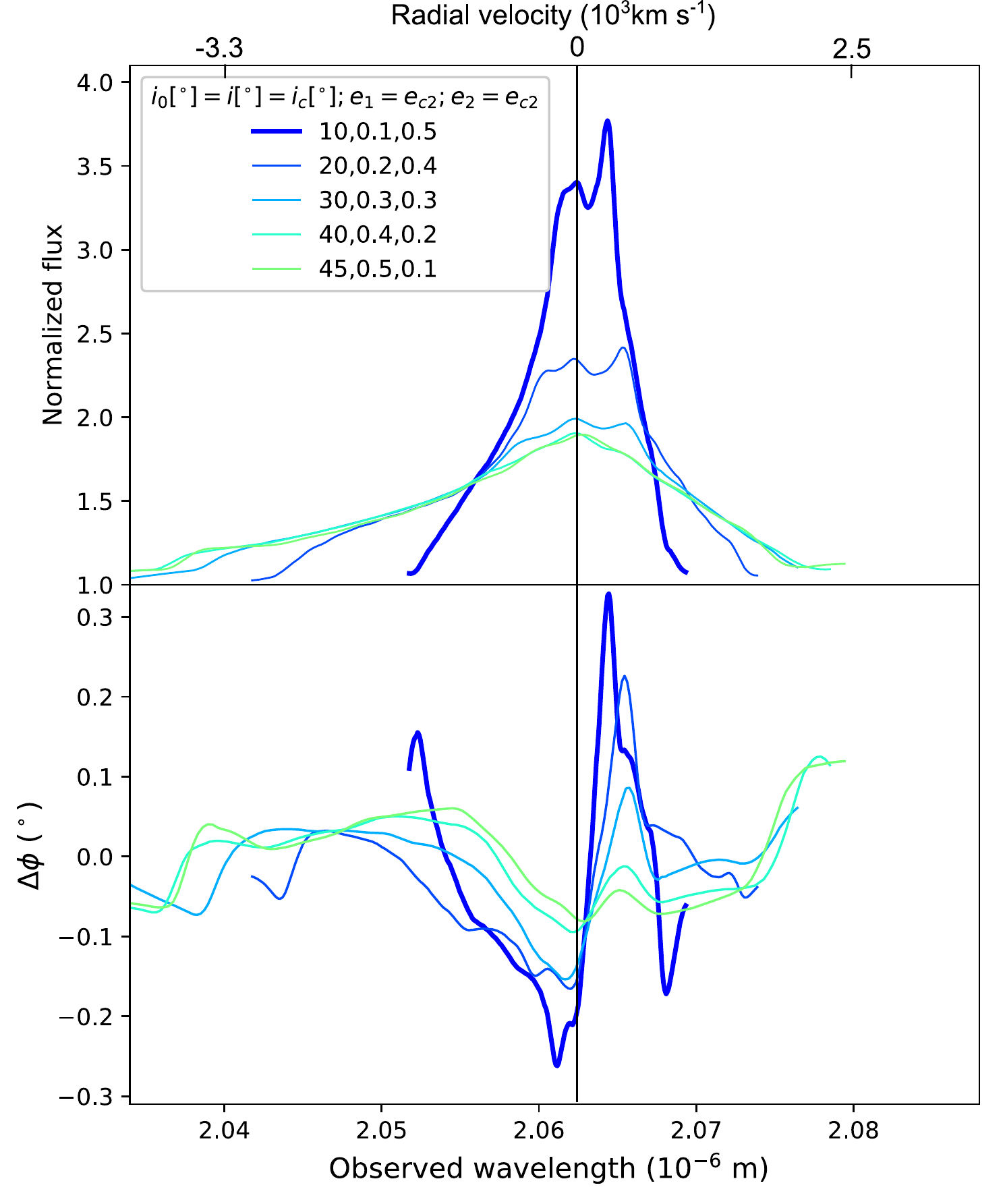}
			
			\includegraphics[trim = 3.0mm 0mm 0mm 0mm, clip, width=0.315\textwidth]{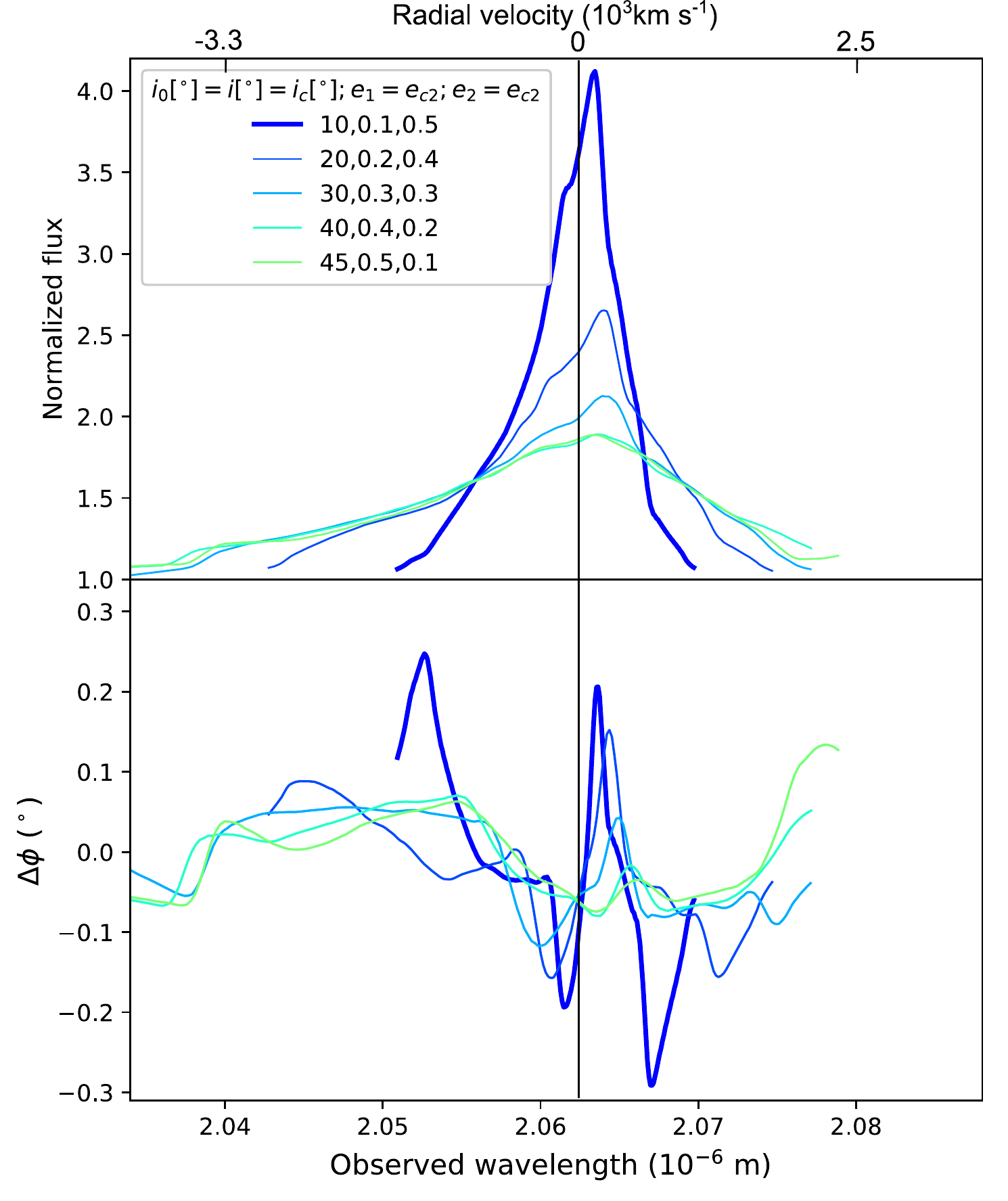}
		}%
		\\ 
		\subfigure[$A_{0}=100\mathrm{ld},i_{0}=i=(10^{\circ}-40^{\circ},45^{\circ}), \Omega_{1}=200^{\circ}, \Omega_{2}=  150^{\circ},  e_{1}= (0.1,0.5), e_{2}=(0.5,0.1),\omega_{1}=70^{\circ}, \omega_{2}=250^{\circ}, i_{c1}=i_{c2}=rnd(90^{\circ}-175^{\circ},45^{\circ}), \Omega_{c1}=\Omega_{c2}=200^{\circ},  \omega_{c1}=100^{\circ},  \omega_{c2}=280^{\circ},e_{c1}=e_{2}, ,e_{c2}=e_{1} $ 
		]	{%
			\label{fig:doubleel83}
			\includegraphics[trim = 3.0mm 0mm 0mm 0mm, clip, width=0.315\textwidth]{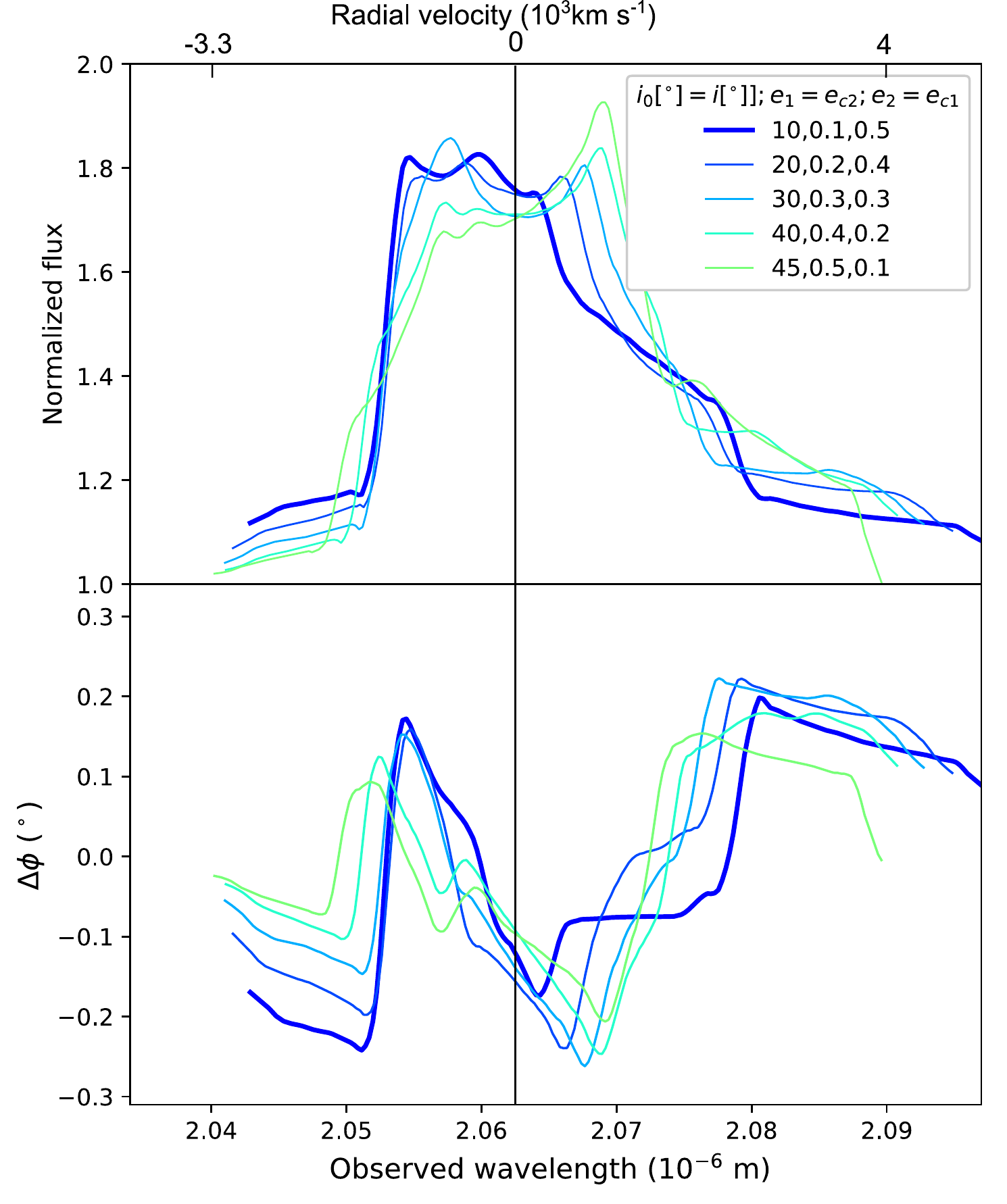}
			\includegraphics[trim = 3.0mm 0mm 0mm 0mm, clip,width=0.315\textwidth]{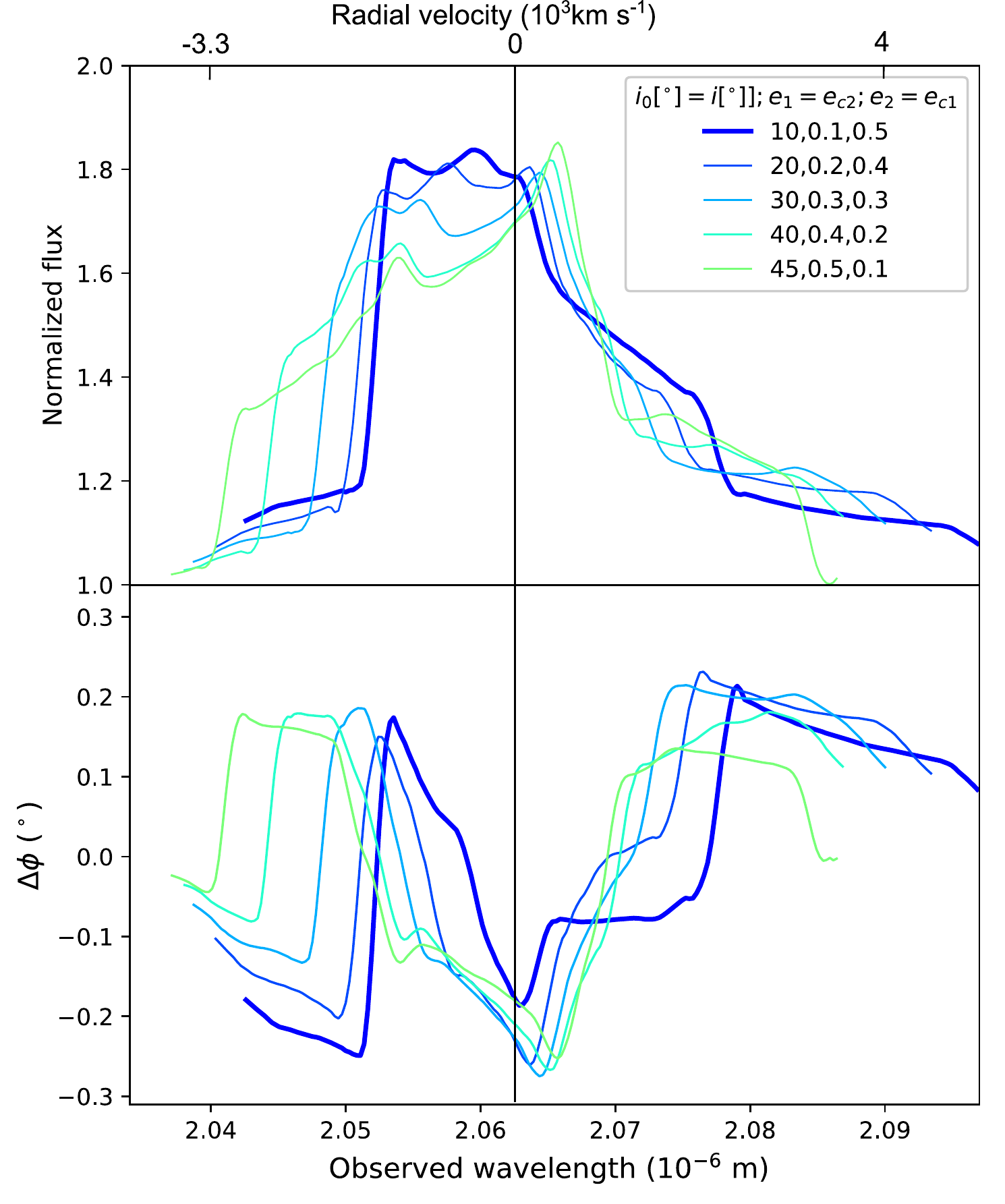}
			
			\includegraphics[trim = 3.0mm 0mm 0mm 0mm, clip, width=0.315\textwidth]{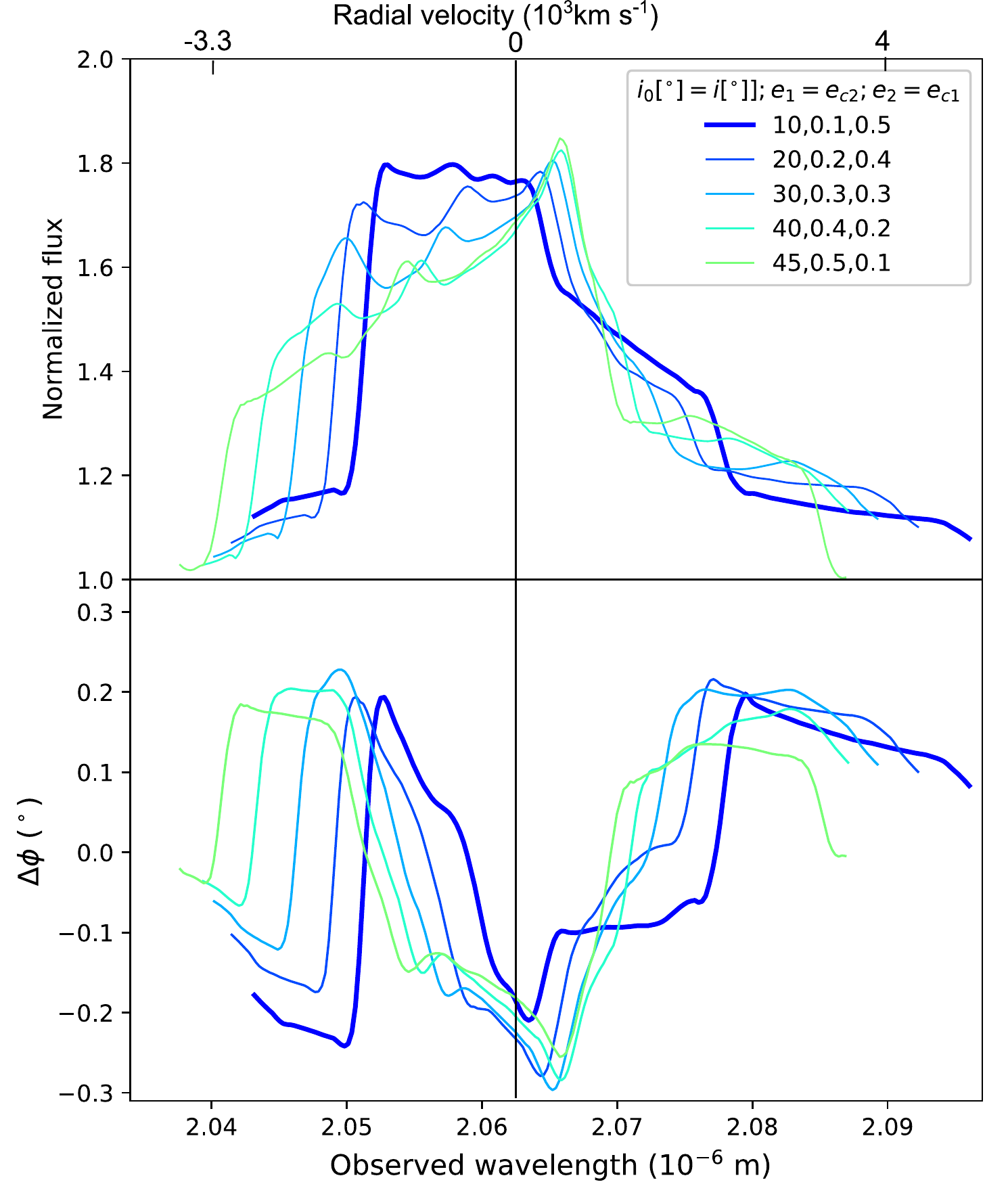}
		}%

		\vspace{-1.9em}
	\end{center}
	\caption{%
		Simulated     interferometric observables for the CB-SMBH system  during three orbital phases. From first  to third column: $\Theta(t)=(0,0.25)$ with an increment of 0.5. Varying orbital parameters are listed in legends and sub-captions. }%
	
	\label{fig:doubleel8}
\end{figure*}

\begin{figure*}[ht!]
	\begin{center}

		\subfigure[$\mathcal{C},\Omega_{k}=  100^{\circ},  \omega_{1}=110^{\circ},\omega_{2}=290^{\circ},\newline\hspace*{1.5em} i_{c}=  \mathcal{U}(-5^{\circ},5^{\circ}),  \Omega_{c1}=200^{\circ}, \Omega_{c2}=\newline\hspace*{1.5em}10^{\circ}, \omega_{c1}=150^{\circ}, \omega_{c2}=330^{\circ},\newline\hspace*{1.5em}  e_{c1}=e_{c2}=rnd \mathcal{R}(1)$
		]	{%
			\label{fig:doubleel91}
			\includegraphics[trim = 2.0mm 0mm 3.0mm 0mm, clip, width=0.27\textwidth]{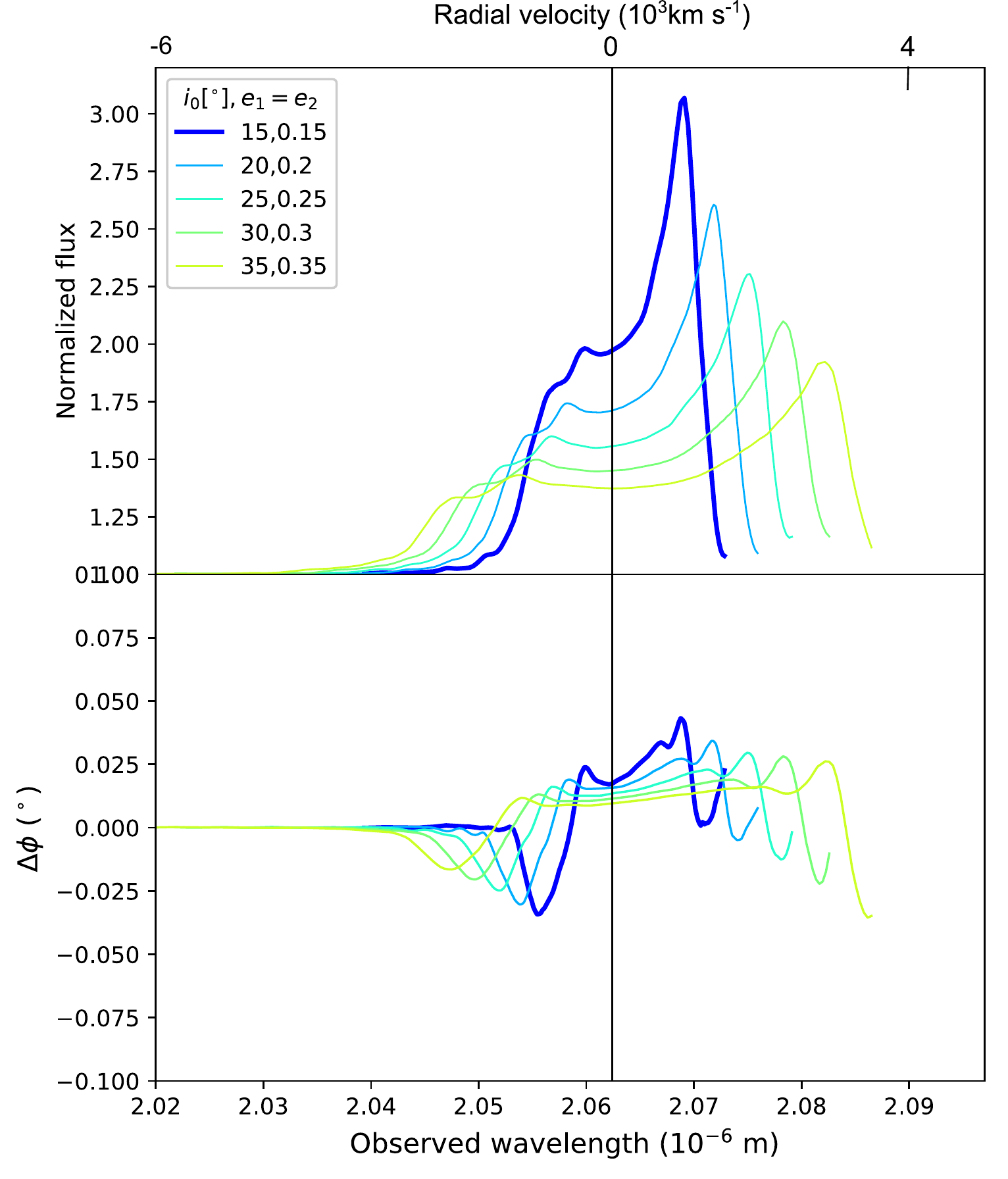}
		}%
		\hspace{-0.6em}
		\subfigure[$\mathcal{C},\Omega_{k}=  100^{\circ}, k=1,2, \omega_{1}= 110^{\circ}, \newline\hspace*{1.5em} \omega_{2}=290^{\circ}, \mathcal{C},  \Omega_{c1}=200^{\circ}, \Omega_{c2}=\newline\hspace*{1.5em}10^{\circ}, \omega_{c1}=150^{\circ}, \omega_{c2}=330^{\circ}, \newline\hspace*{1.5em} e_{c1}=e_{c2}=rnd \mathcal{R}(1)$
		]{%
			\label{fig:doubleel92}
			\includegraphics[trim = 2.0mm 0mm 3mm 0mm, clip,width=0.27\textwidth]{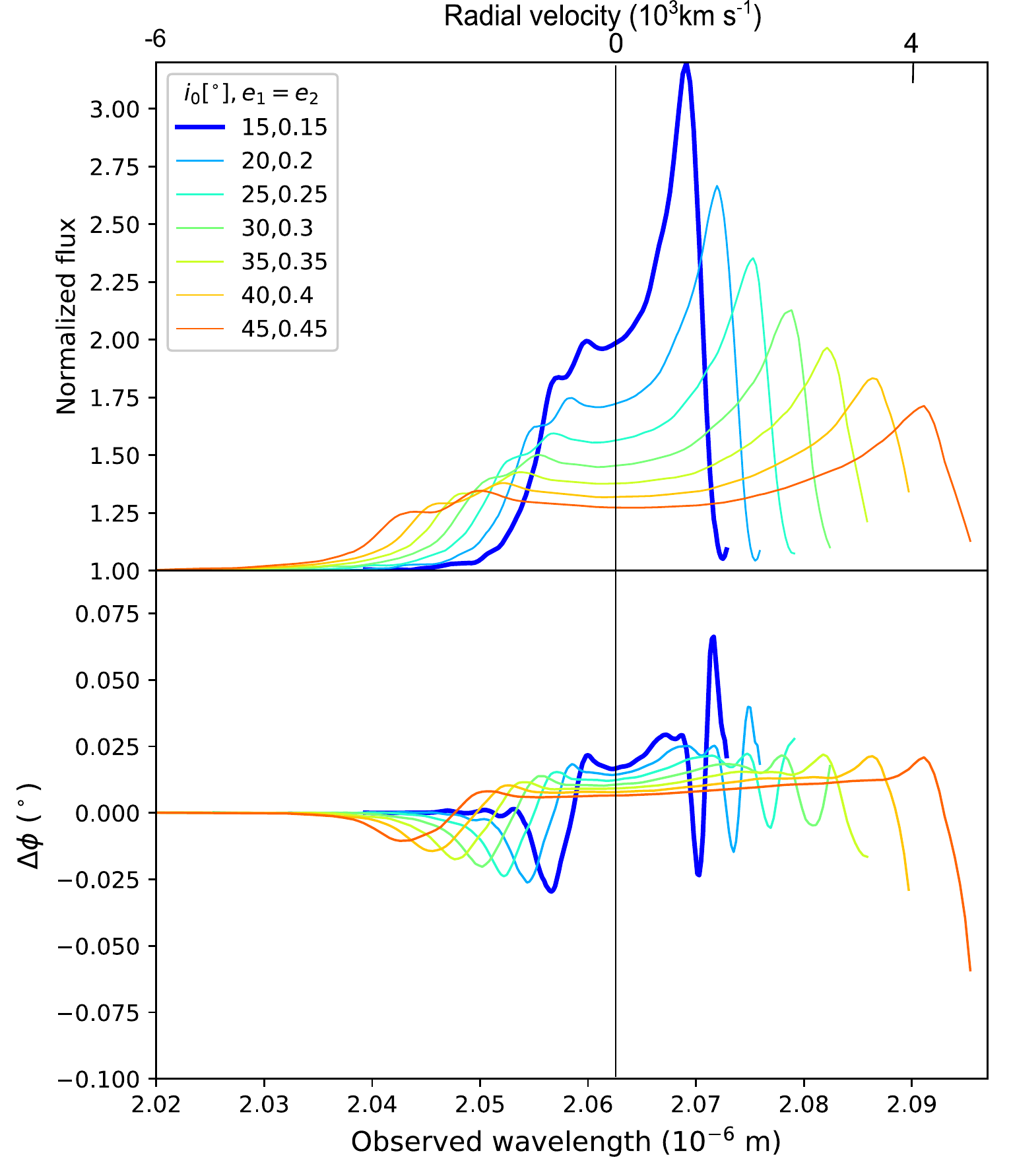}
		}		\\ 
		\subfigure[$\mathcal{C},\Omega_{k}=  100^{\circ}, k=1,2, \omega_{1}= 110^{\circ},  \newline\hspace*{1.5em} \omega_{2}=290^{\circ},i_{c1}=\mathcal{C}, i_{c2}=\newline\hspace*{1.5em}\mathcal{U}(-5^{\circ},5^{\circ}), \Omega_{c1}=200^{\circ}, \Omega_{c2}= \newline\hspace*{1.5em}10^{\circ}, \omega_{c1}=150^{\circ}, \omega_{c2}=330^{\circ},\newline\hspace*{1.5em}  e_{c1}=rnd \mathcal{R}(1), e_{c2}=0.5$
		]
		{%
			\label{fig:doubleel94}
			\includegraphics[trim = 3.0mm 2mm 3.0mm 2mm, clip, width=0.26\textwidth]{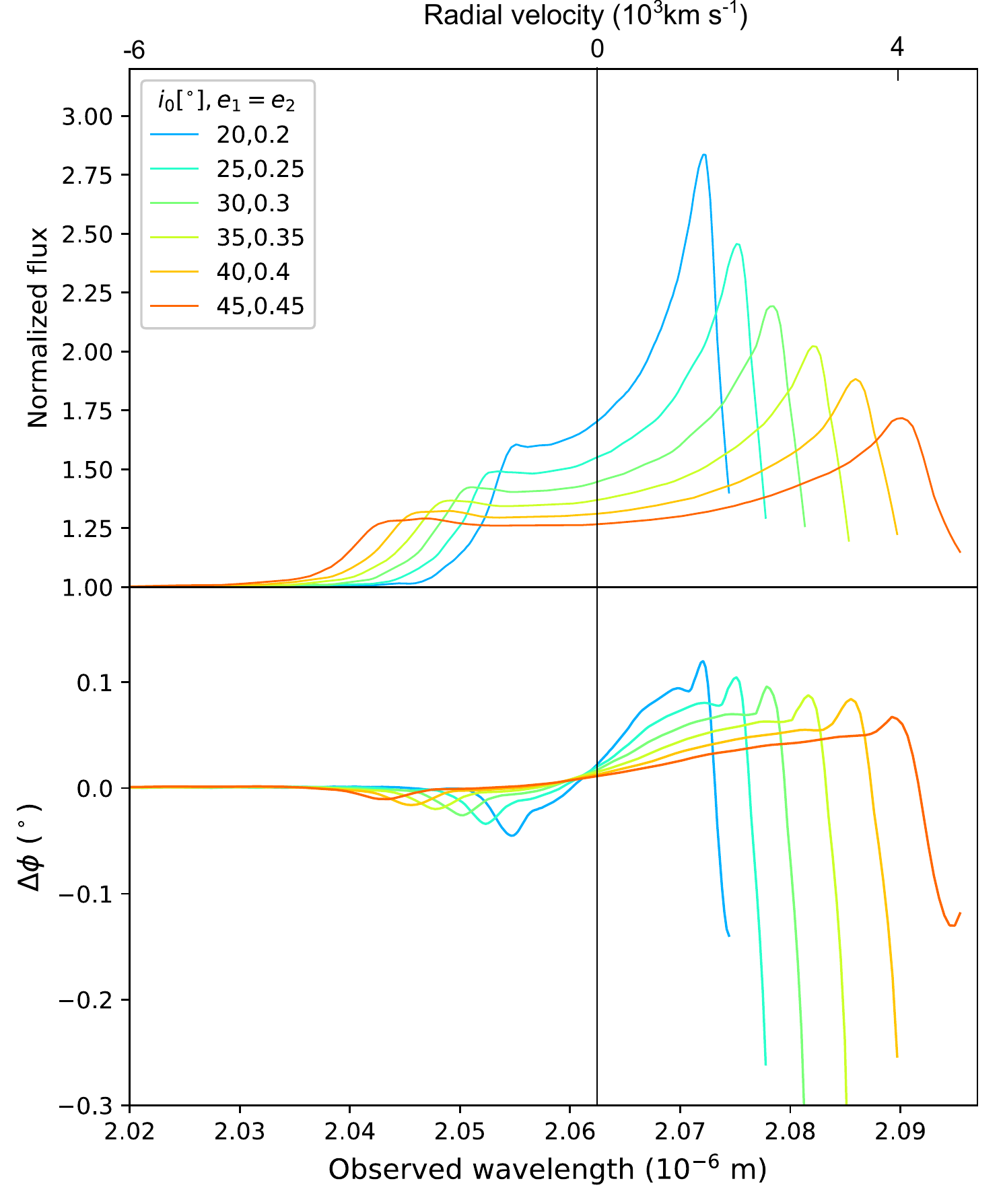}
		}%
		\hspace{-0.5em}
		\subfigure[$\mathcal{C},\Omega_{k}=  100^{\circ}, k=1,2, \omega_{1}= 110^{\circ},\newline\hspace*{1.5em}\omega_{2}=290^{\circ},i_{c1}=\mathcal{C}, i_{c2}=\newline\hspace*{1.5em}\mathcal{U}(-5^{\circ},5^{\circ}),  \Omega_{c1}=200^{\circ}, \Omega_{c2}=\newline\hspace*{1.5em}10^{\circ}, \omega_{c1}=150^{\circ}, \omega_{c2}=330^{\circ},  \newline\hspace*{1.5em}e_{c1}=e_{c2}=rnd \mathcal{R}(1)$
		]{%
			\label{fig:doubleel95}
			\includegraphics[trim = 3.mm 2mm 3.0mm 2mm, clip,width=0.267\textwidth]{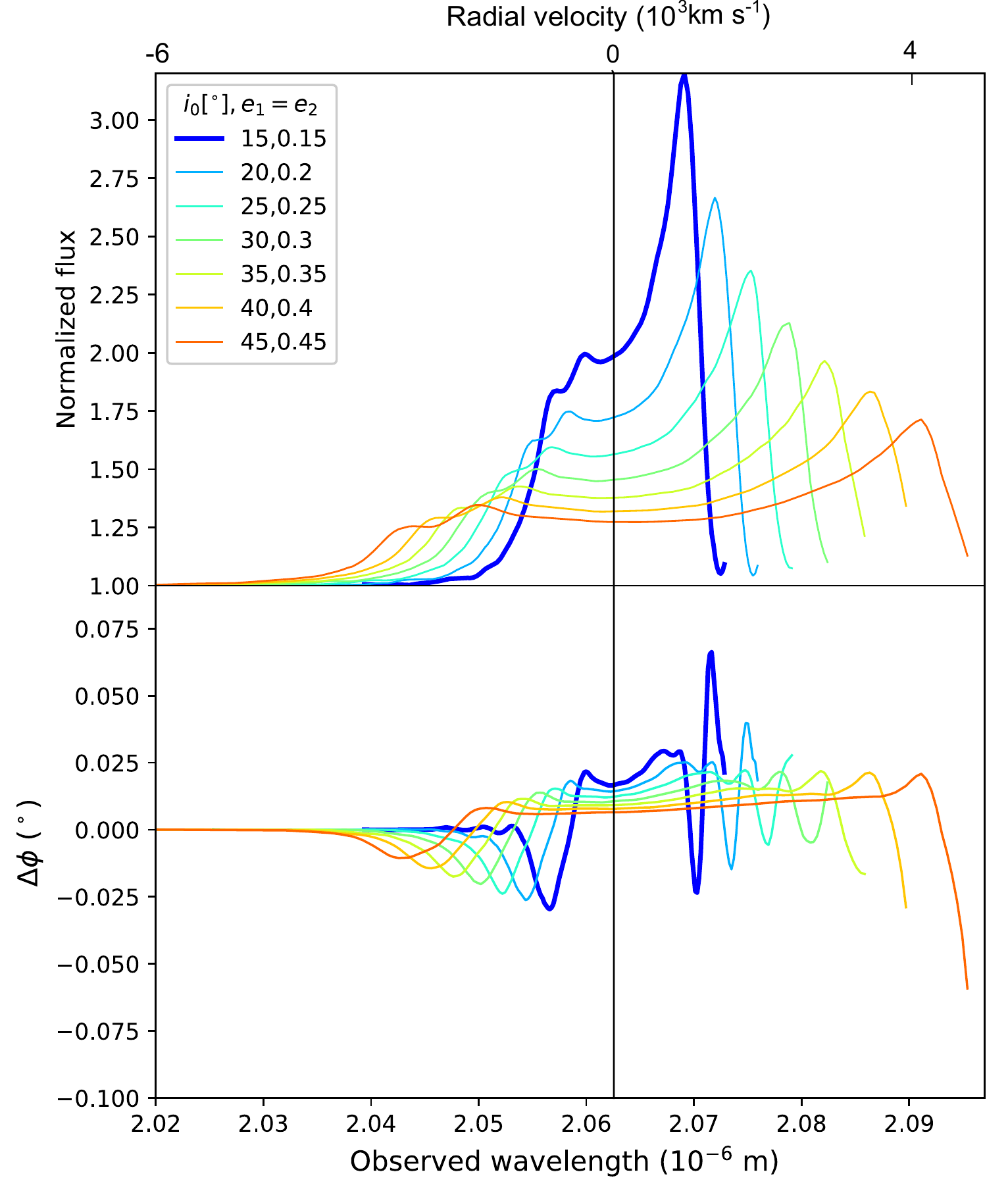}
		}
		\hspace{-0.5em}
		\subfigure[$\mathcal{C},\Omega_{k}=  100^{\circ}, k=1,2, \omega_{1}= 110^{\circ},\newline\hspace*{1.5em}\omega_{2}=290^{\circ},i_{c1}= i_{c2}=\mathcal{U}(-5^{\circ},5^{\circ}), \newline\hspace*{1.5em} \Omega_{c1}=200^{\circ}, \Omega_{c2}=10^{\circ}, \omega_{c1}=150^{\circ},\newline\hspace*{1.5em} \omega_{c2}=330^{\circ},  e_{c1}=0.5, e_{c2}=rnd \mathcal{R}(1)$
		]{%
			\label{fig:doubleel96}
			\includegraphics[trim = 1.0mm 1mm 3.0mm 2mm, clip, width=0.266\textwidth]{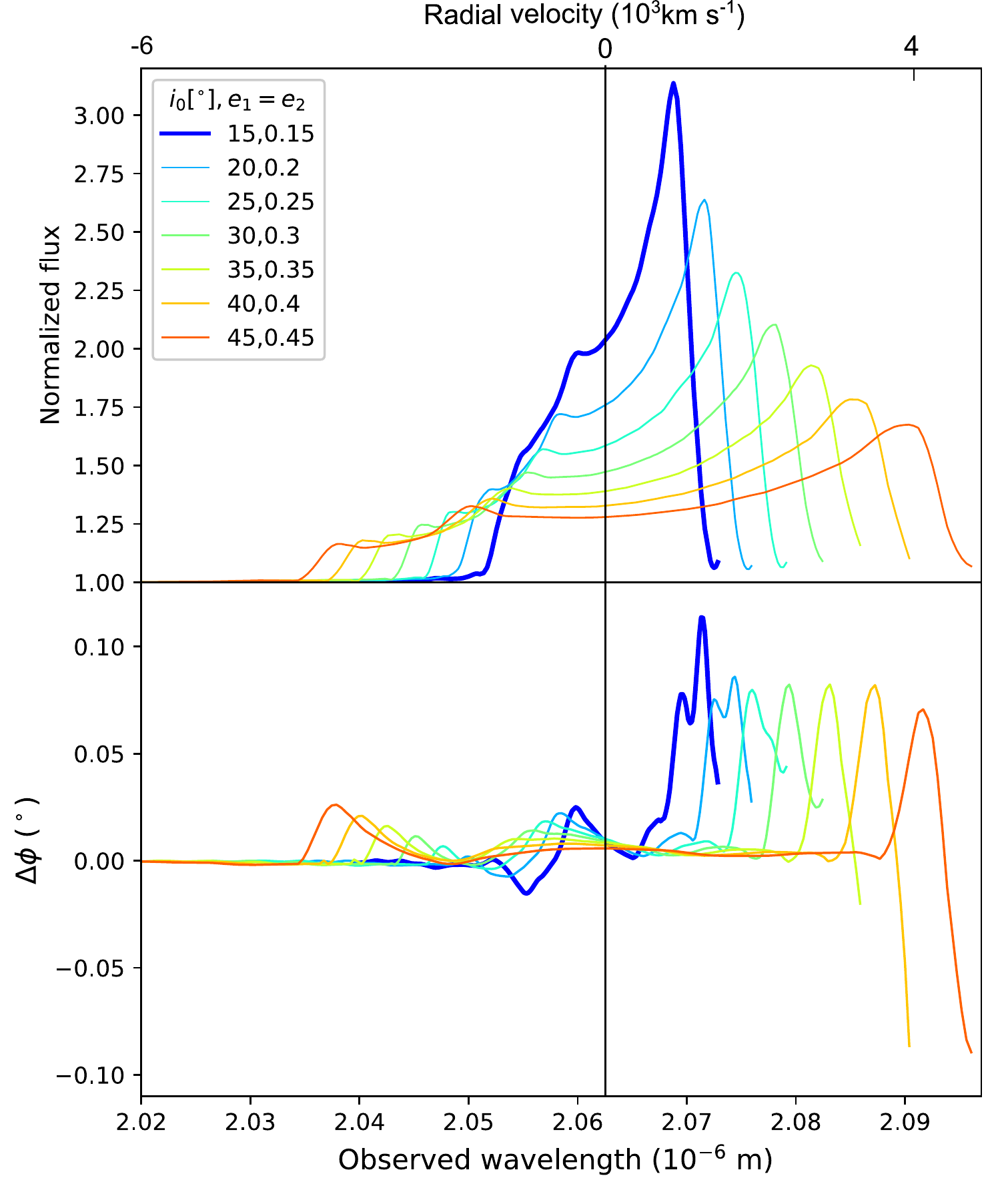}
		}%
		\vspace{-0.5em}
	\end{center}
	\caption{%
		Evolution of the Pa$\alpha$ emission line (upper subplots) and corresponding differential phase ($\Delta \phi$, lower subplots) for the CB-SMBH system as a function of the wavelength and radial velocity. The clouds' orbital eccentricities are drawn from Rayleigh distribution. Varying parameters are listed in 
		sub-captions and legends, and $\mathcal{C}$ stands for coplanar orbits. 
	}%
	
	\label{fig:doubleel9}
\end{figure*}

\begin{figure*}[ht!]
	\begin{center}

		\subfigure[$\mathcal{C},\Omega_{k}=  100^{\circ},  \omega_{1}=110^{\circ},\omega_{2}=290^{\circ},\newline\hspace*{1.5em} i_{c}=  \mathcal{U}(-5^{\circ},5^{\circ}),  \Omega_{c1}=200^{\circ}, \Omega_{c2}=\newline\hspace*{1.5em}10^{\circ}, \omega_{c1}=150^{\circ}, \omega_{c2}=330^{\circ},\newline\hspace*{1.5em}  e_{c1}=e_{c2}=rnd \Gamma_{s}  (0.3,1)$
		]	{%
			\label{fig:doubleel101}
			\includegraphics[trim = 2.0mm 0mm 3.0mm 0mm, clip, width=0.32\textwidth]{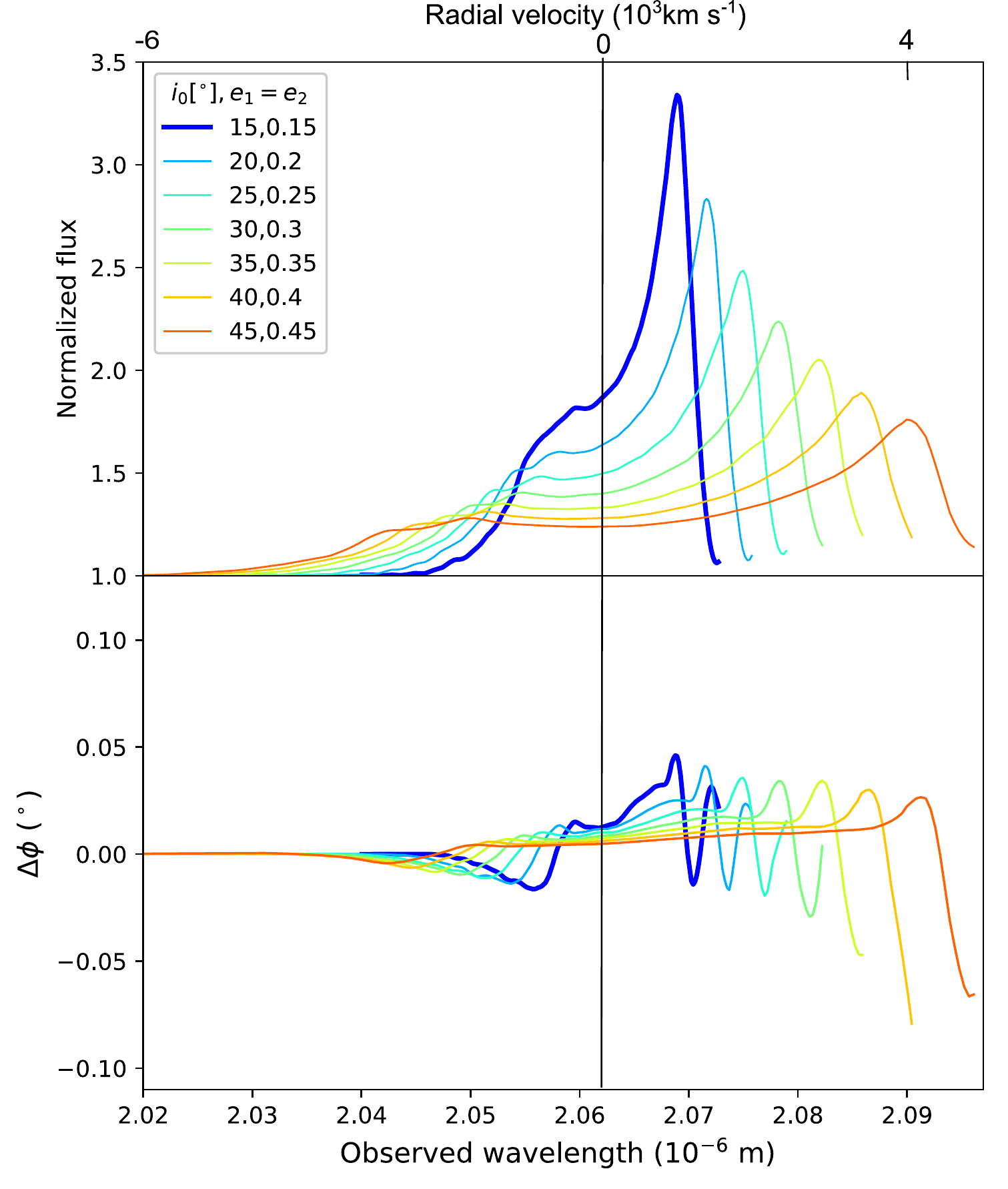}
		}%
		\hspace{-0.3em}
		\subfigure[$\mathcal{C},\Omega_{k}=  100^{\circ}, k=1,2, \omega_{1}= 110^{\circ},\newline\hspace*{1.5em}\omega_{2}=290^{\circ}, i_{c1}=\mathcal{U}(-5^{\circ},5^{\circ}), i_{c2}=\mathcal{C}, \newline\hspace*{1.5em} \Omega_{c1}=200^{\circ}, \Omega_{c2}=10^{\circ}, \omega_{c1}=150^{\circ}, \newline\hspace*{1.5em}\omega_{c2}=330^{\circ},  e_{c1}=rnd \Gamma_{s}  (0.3,1),\newline\hspace*{1.5em}e_{c2}=0.5$
		]{%
			\label{fig:doubleel103}
			\includegraphics[trim = 2.mm 0mm 3mm 0mm, clip, width=0.32\textwidth]{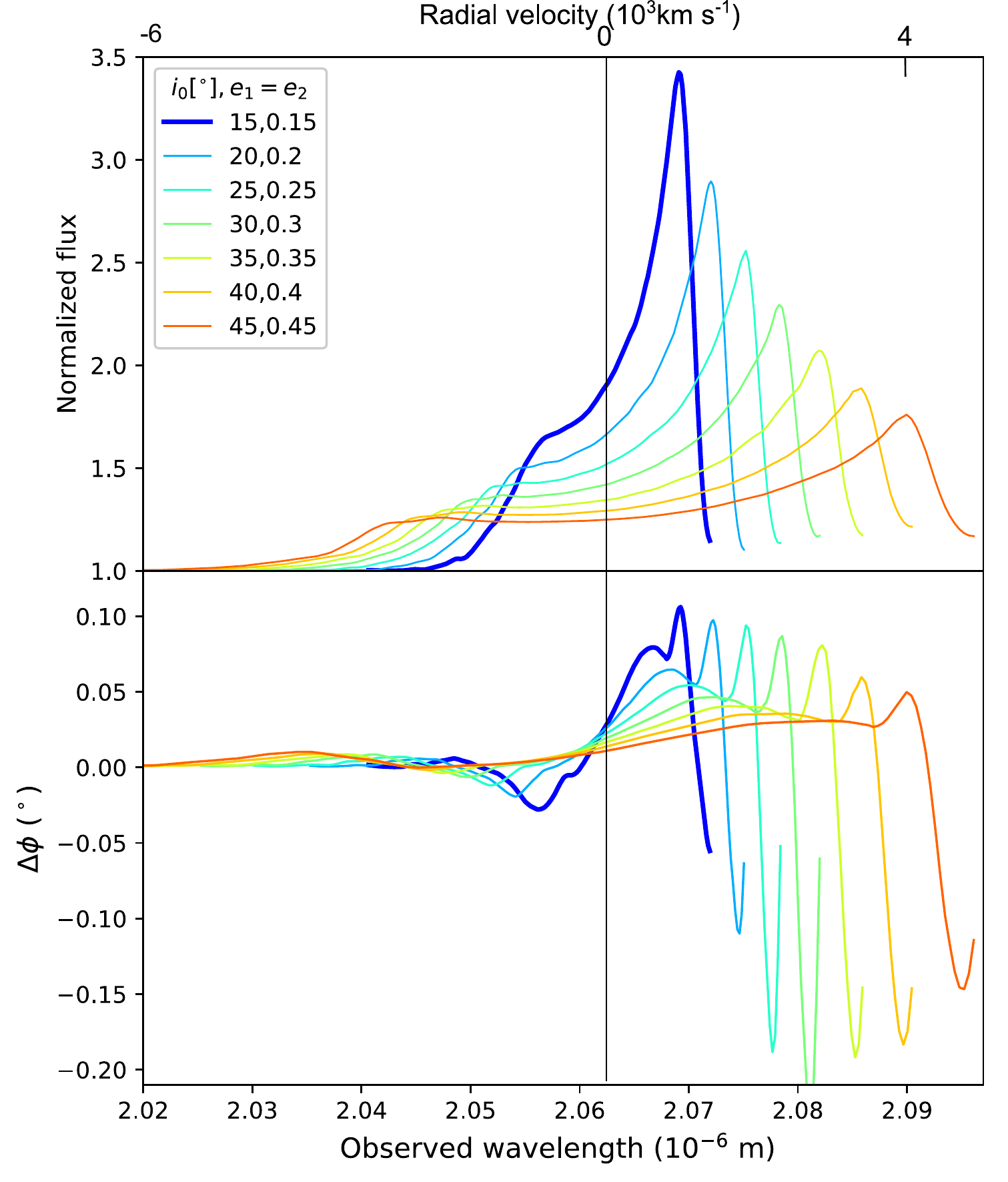}
		}	\\ 
		\subfigure[$\mathcal{C},\Omega_{k}=  100^{\circ}, k=1,2, \omega_{1}= 110^{\circ},  \newline\hspace*{1.5em} \omega_{2}=290^{\circ},i_{c1}=\mathcal{C}, i_{c2}=\mathcal{U}(-5^{\circ},5^{\circ}), \newline\hspace*{1.5em} \Omega_{c1}=200^{\circ}, \Omega_{c2}=10^{\circ}, \omega_{c1}=\newline\hspace*{1.5em}150^{\circ}, \omega_{c2}=330^{\circ},  e_{c1}=\newline\hspace*{1.5em}rnd \Gamma_{s}  (0.3,1), e_{c2}=0.5$
		]
		{%
			\label{fig:doubleel104}
			\includegraphics[trim = 3.0mm 2mm 3.0mm 0mm, clip, width=0.315\textwidth]{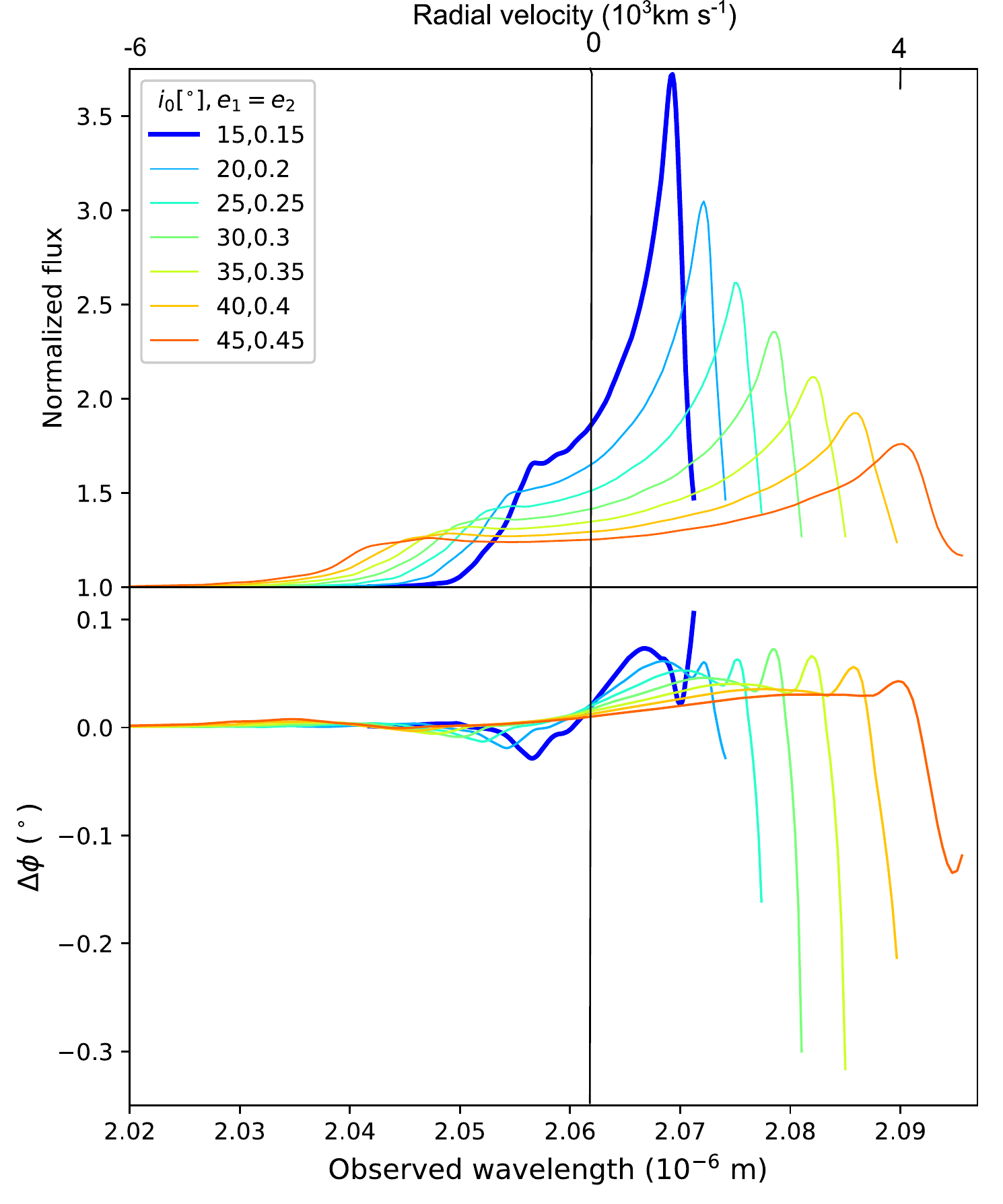}
		}%
		\hspace{-0.3em}
		\subfigure[$\mathcal{C},\Omega_{k}=  100^{\circ}, k=1,2, \omega_{1}= 110^{\circ},\newline\hspace*{1.5em}\omega_{2}=290^{\circ},  i_{c1}=\mathcal{U}(-5^{\circ},5^{\circ}),  i_{c2}=\mathcal{C}, \newline\hspace*{1.5em} \Omega_{c1}=200^{\circ}, \Omega_{c2}=10^{\circ}, \omega_{c1}=150^{\circ},\newline\hspace*{1.5em} \omega_{c2}=330^{\circ},  e_{c1}=e_{c2}=\newline\hspace*{1.5em}rnd \Gamma_{s}  (0.3,1)$
		]{%
			\label{fig:doubleel105}
			\includegraphics[trim = 3.mm 0mm 2.0mm 0mm, clip,width=0.322\textwidth]{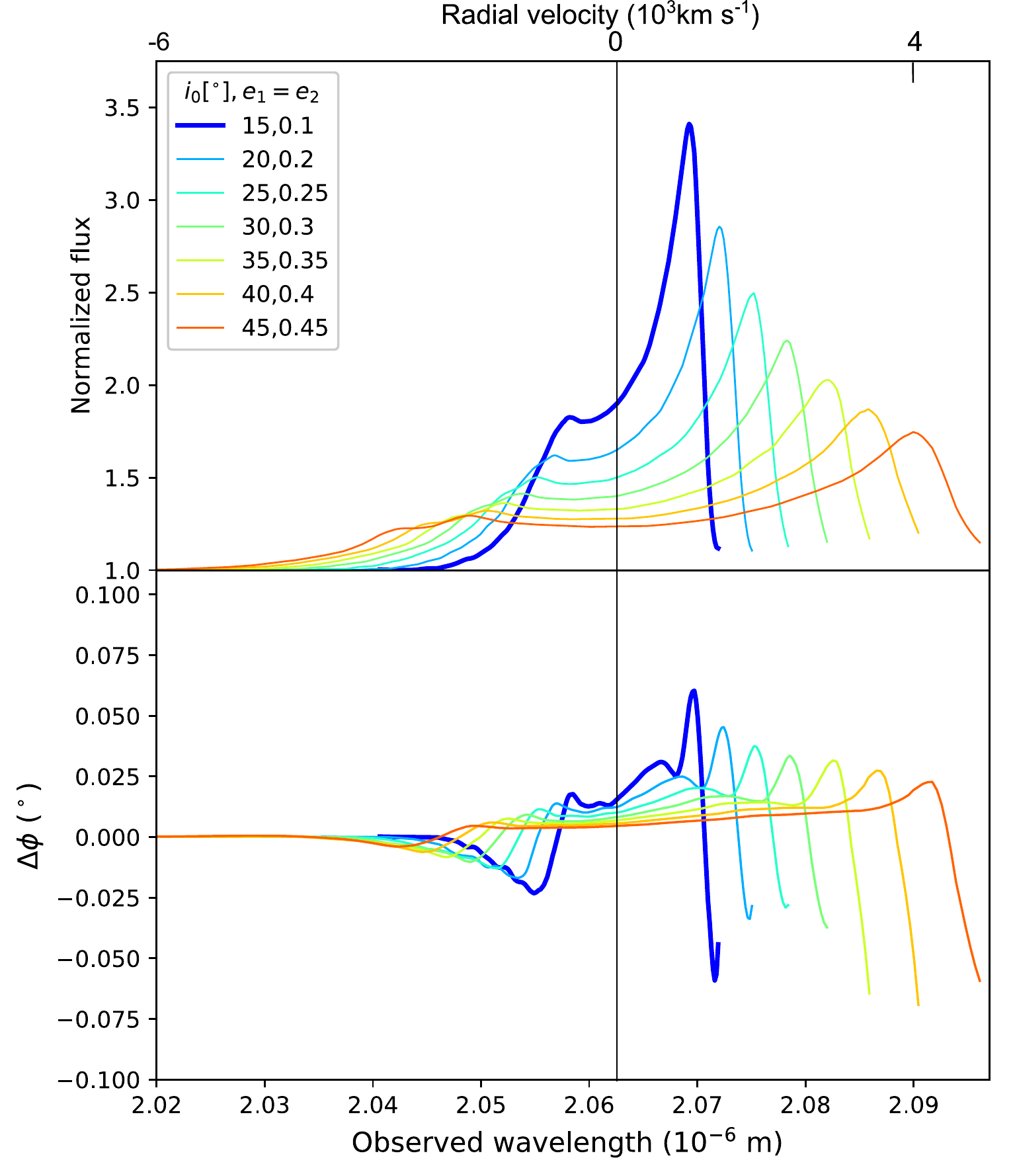}
		}
		\hspace{-0.3em}
		\subfigure[$\mathcal{C},\Omega_{k}=  100^{\circ}, k=1,2, \omega_{1}= 110^{\circ},\newline\hspace*{1.5em}\omega_{2}=290^{\circ},i_{c1}=\mathcal{U}(-5^{\circ},5^{\circ}), i_{c2}=0 \newline\hspace*{1.5em} \Omega_{c1}=200^{\circ}, \Omega_{c2}=10^{\circ}, \omega_{c1}=150^{\circ},\newline\hspace*{1.5em} \omega_{c2}=330^{\circ},  e_{c1}=0.5, \newline\hspace*{1.5em}e_{c2}=rnd \Gamma_{s}  (0.3,1)$
		]{%
			\label{fig:doubleel106}
			\includegraphics[trim = 3.0mm 2mm 3.0mm 0mm, clip, width=0.319\textwidth]{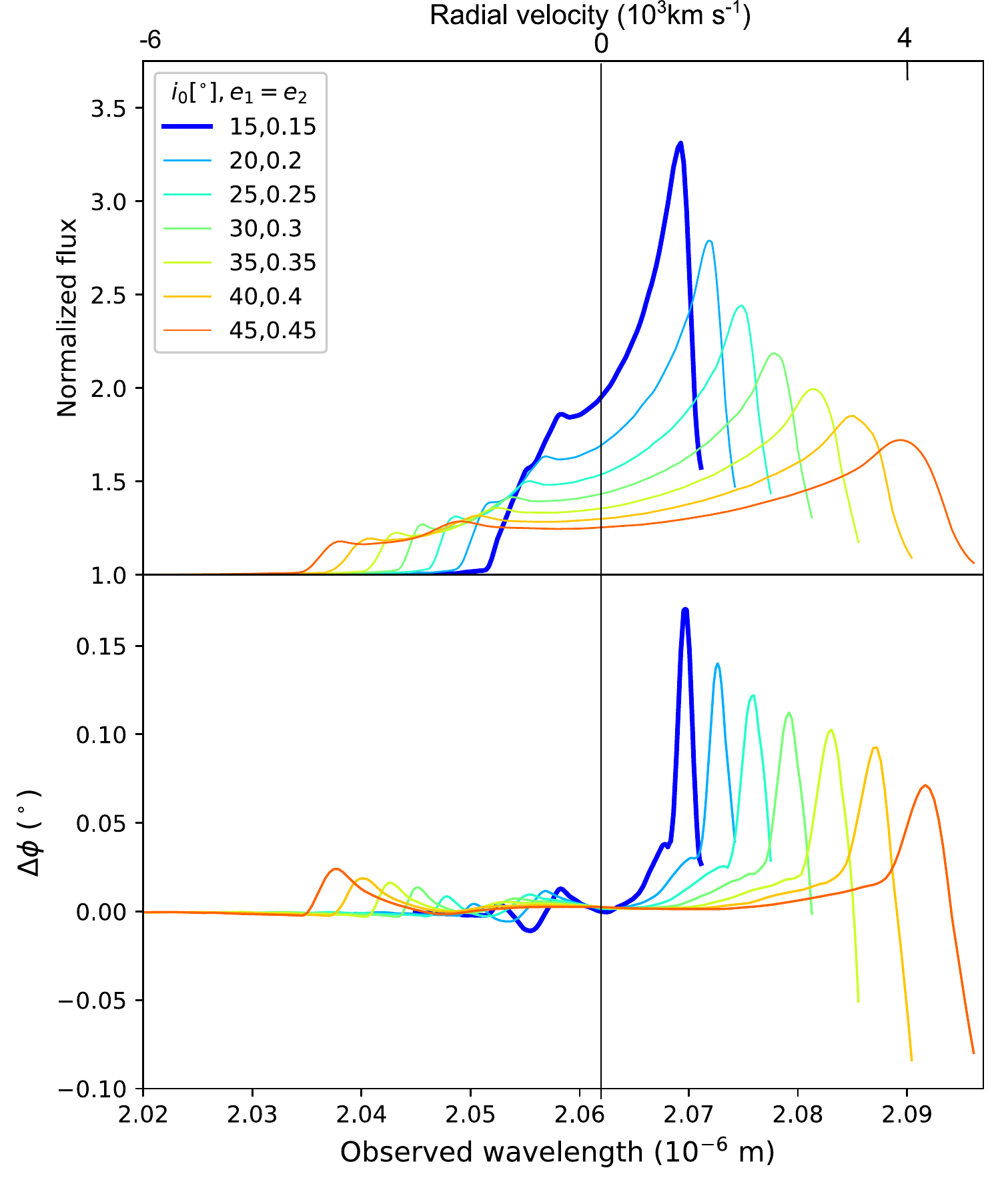}
		}%
		\vspace{-0.3em}
	\end{center}
	\caption{%
		Same as 	Fig. \ref{fig:doubleel9} but for clouds' orbital eccentricities drawn from  scaled and shifted $\Gamma_{s}(0.3,1)$ distribution.
	}%
	\label{fig:doubleel10}
\end{figure*}

\end{appendix}
\end{document}